\documentclass[epjST,usenames,dvipsnames,vecphys]{svjour}
\usepackage{amsmath,amssymb}
\usepackage{float}
\usepackage{hyperref}
\usepackage[framemethod=TikZ]{mdframed}
\usepackage{footnote}
\makesavenoteenv{tabular}
\makesavenoteenv{table}
\makesavenoteenv{figure}
\usepackage{xcolor}
\usepackage{graphicx}

\newcommand{\req}[1]{Eq.\,(\ref{eq:#1})}
\newcommand{\rss}[1]{Sec.~\ref{#1}}
\newcommand{\rf}[1]{Fig.~\ref{#1}}
\newcommand{\mydoi}[2]{\href{http://dx.doi.org/#2}{#1}}
\newcommand{\E}{\mathrm{e}}
\newcommand{\ie}{{\em i.e.}} 
\newcommand{\ms}[1]{\rule[-#1mm]{0mm}{0mm}} 
  
\begin{document}
%
\title{\phantom{.}\\\phantom{.}\\Discovery of Quark-Gluon-Plasma:\\[0.3cm]
Strangeness Diaries}
\author{Johann Rafelski\inst{1,2}\fnmsep\thanks{\email{Rafelski@Physics.Arizona.EDU}}}
\institute{CERN-TH, CH-1211 Geneva 23 
 \and Department of Physics, The University of Arizona, Tucson, AZ 85721
}
\abstract{We look from a theoretical perspective at the new phase of matter, quark-gluon plasma (QGP), the new form of nuclear matter created at high temperature and pressure. Here I retrace the path to QGP discovery and its exploration in terms of strangeness production and strange particle signatures. We will see the theoretical arguments that have been advanced to create interest in this determining signature of QGP. We explore the procedure used by several experimental groups making strangeness production an important tool in the search and discovery of this primordial state of matter present in the Universe before matter in its present form was formed. We close by looking at both the ongoing research that increases the reach of this observable to LHC energy scale $pp$ collisions, and propose an interpretation of these unexpected results.
} 
\maketitle
\vskip -15.1cm \phantom{.}\hfill \hspace*{1cm} Preprint CERN-TH-2019-138 \ \ 
\vskip -0.3cm\phantom{.}\hfill \hspace*{0.8cm} submitted to EPJ Special Topics \ \ \\
\phantom{.}\hfill \hspace*{2cm} November 4/12, 2019 \ \ \\[-0.7cm]
\begin{mdframed}[linecolor=gray,roundcorner=12pt,backgroundcolor=GreenYellow!15,linewidth=1pt,leftmargin=0cm,rightmargin=0cm,topline=true,bottomline=true,skipabove=12pt]\relax%
\label{Luciano}
\textit{It is very appropriate that you did reconstruct your version of the QGP discovery. Your quotations concerning me are correct and reproduce well my opinion, which I have not changed. CERN found good evidence for deconfinement,  and it was at all
appropriate to say that in public, independently from the status of RHIC at the time. 
}\\
\textbf{Luciano Maiani} CERN Director General 1 January 1999--31 December 2003.
\end{mdframed}
\vskip 0.5cm 

\vskip 10cm
\subsection*{Dedications} (alphabetic):
\addcontentsline{toc}{section}{Dedications}%
\begin{enumerate}
\item 
{\bf Rolf Hagedorn}, who would have been 100 years old on 20th July 2019. His influence at CERN was essential in the quest to unlock the scientific opportunities relativistic heavy ion collisions offer. I had the privilege to have been marked by Hagedorn\rq s magic wand in a decisive way. 
\item 
\textbf{Jean Letessier}, Jean\rq s 80th birthday in December of 2018 provided the initial motivation to prepare a manuscript that includes a review of the highlights of our research achievements. I worked with Jean on the topic of this review for 20 years, from a first meeting in Summer 1992, arranged by Rolf Hagedorn, to the end of 2013. After many years of fruitful collaboration, Jean is the friend and colleague with whom I have published the largest number of research papers. 
\item
{\bf Berndt M\"uller}, with whom I studied physics in Frankfurt, and with whom I published many influential works on strangeness and strong field physics. 
\item
{\bf Helga Rafelski (3 September 1949 -- 5 November 2000)}\\ she would have been 70 years old this year. 
\end{enumerate}
\vfill\eject
\markboth{Preamble}{Strangeness in QGP: Diaries}
\addcontentsline{toc}{section}{Preamble}%
This review introduces the strangeness signature of QGP in the format of a personal diary: diary means that I use some (unpublished) material buried in my history chest box. This includes some unpublished manuscripts from arXiv with some added insight into the reasons why these works are unpublished. All told, about half of this review presentation relies on earlier written personal records, the other half is freshly written with added verifiable references or records. Here is an example how this works -- in November 1993 I prepared a short write-up describing the ongoing and future work I hoped to carry out collaborating with Jean Letessier:\\

\noindent \textit{The LPTHE laboratory at the University Paris submitted this to CNRS in order to secure funding to allow me to work in Paris in the Spring 1994. This roadmap focused on the ongoing experimental program at the CERN-SPS, defining our ensuing and enduring collaboration:\/}\\[-0.7cm]
\begin{mdframed}[linecolor=gray,roundcorner=12pt,backgroundcolor=Dandelion!15,linewidth=1pt,leftmargin=0cm,rightmargin=0cm,topline=true,bottomline=true,skipabove=12pt]\relax%
%
\small
\begin{center}
\textbf{STRANGE ANTIBARYONS}
\end{center}

The high temperature phase of quantum-chromodynamics (QCD), the quark-gluon plasma (QGP) is characterized by\\ 
- color deconfinement, and\\ 
- partial restoration of chiral symmetry.\\ 
This picture is supported by numerical simulations of lattice SU(3)-gauge theory and by high temperature QCD perturbation theory. In QGP the production of strange particles is expected to be efficient due to:\\[-0.6cm] 

\begin{itemize}
\item[(a)] 
{\it Lower energy threshold:} \\
The most efficient conventional strangeness producing reactions\\[-0.3cm] 
$$p+p\to p +\Lambda+{\rm K}^+,\quad \pi+\pi\to{\rm K}+\bar{\rm K}$$
require c.m. energy of 700 MeV, in the QGP the energy for strange quark pair production\\[-0.6cm] 
$$G+G\to s+\bar s;\quad q+\bar q\to s\bar s$$
is $2m_{s}\simeq 300$ MeV. The reduction of threshold is very important in the presence of a thermalized phase space distribution, where at best temperatures $T = 160-250$ MeV can be reached today.
\item[(b)]
 {\it Increased strangeness density:} \\
In a HG the density of strange hadrons is $\simeq 0.1$ fm$^{-3}$. By contrast, in the QGP the density of strange quarks is\\[-0.3cm] 
\begin{equation}\nonumber
n_{s} = n_{\bar s} \approx 6(Tm_{s}/2\pi)^{3/2} \exp(-m_{s}/T) 
\approx 0.3 {\rm fm}^{-3} . 
\end{equation} 
\item[(c)]
 {\it Anti-strangeness can be more abundant than} $\bar u, \bar d$:\\
Since strange valence quarks are practically completely absent in the nuclear matter entering a relativistic heavy ion collision, the available phase space for strange antiquarks is not suppressed. Non-strange antiquarks, on the other hand, are suppressed due to the presence of a non-vanishing baryo-chemical potential $\mu_{\rm B}$ in a baryon-rich QGP, as it forms at AGS or SPS energies. The predicted phase space ratio in QGP:\\[-0.3cm] 
\begin{equation} \nonumber
\frac{n_{\bar s}}{n_{\bar u}+n_{\bar d}} \approx 
{1\over 2}(m_{s}/T)^2 {\rm K}_2(m_{s}/T) \exp(\mu_{B}/3T) . 
\end{equation} 
\end{itemize}
\indent Both these consequences (a), (b) of deconfinement and chiral symmetry restoration favor the production of multiply strange hadrons. The condition (c) further allows to expect that this enhancements is increasing with increasing $\bar s$ content of hadrons. This observation is certainly contrary to the normal hadronic reactions in which (e.g. in $p$ -- $p$ collisions) the yield of strange antibaryons is falling rapidly with $\bar s$ content. The corresponding experimentally measured cross sections are inputs in hadronic cascade models. Consequently, the abundance of strange antibaryons from dynamical hadronic models is expected to be opposite to the expectations from QGP, and the yields computed are considerably smaller than obtainable in QGP models.
\end{mdframed}
\vfill\eject

\addcontentsline{toc}{section}{\contentsname}
{\protect\markboth{\contentsname}{\contentsname}}
\setcounter{tocdepth}{3}
\tableofcontents
\vfill\newpage
\markboth{1. Discovery of QGP}{Strangeness in QGP: Diaries}

\section{A New Phase of Matter}
\subsection{Why we are interested in quark-gluon plasma}
\subsubsection{A new and interdisciplinary field of physics}

This review introduces the laboratory exploration of \lq quark-gluon plasma\rq\ (QGP) by means of strangeness production and strange antibaryon enhancement. QGP can be formed in relativistic heavy ion collisions. This research field has seen phenomenal growth in the past 40 years. This happened because the study of QGP is at the crossroads of several fundamental questions, connecting to several physics disciplines. 

In this spirit and in consideration of the great interest in QGP, let me first briefly describe my take on why so many theorists and even more experimentalists crowd the QGP physics field. First note that even though this research field has today a large component addressing questions arising in the context of cosmology and particle physics, it is seen as being a part of the nuclear science effort. The rationales are:
\begin{itemize}
\item
The laboratory experiments use atomic nuclei in collision to discover, study, and explore this new state of matter in the laboratory; and
\item
This new field of research evolved and absorbed the earlier exploration of the properties of nuclei in low energy heavy ion collisions. 
\end{itemize}
However, in the study of QGP we address many important and often interdisciplinary domains of physics; some are far and away from traditional nuclear physics context as is shown in \rf{WHYrhic}: 

\begin{figure}[tb]\sidecaption
\centerline{\includegraphics[width=1\textwidth]{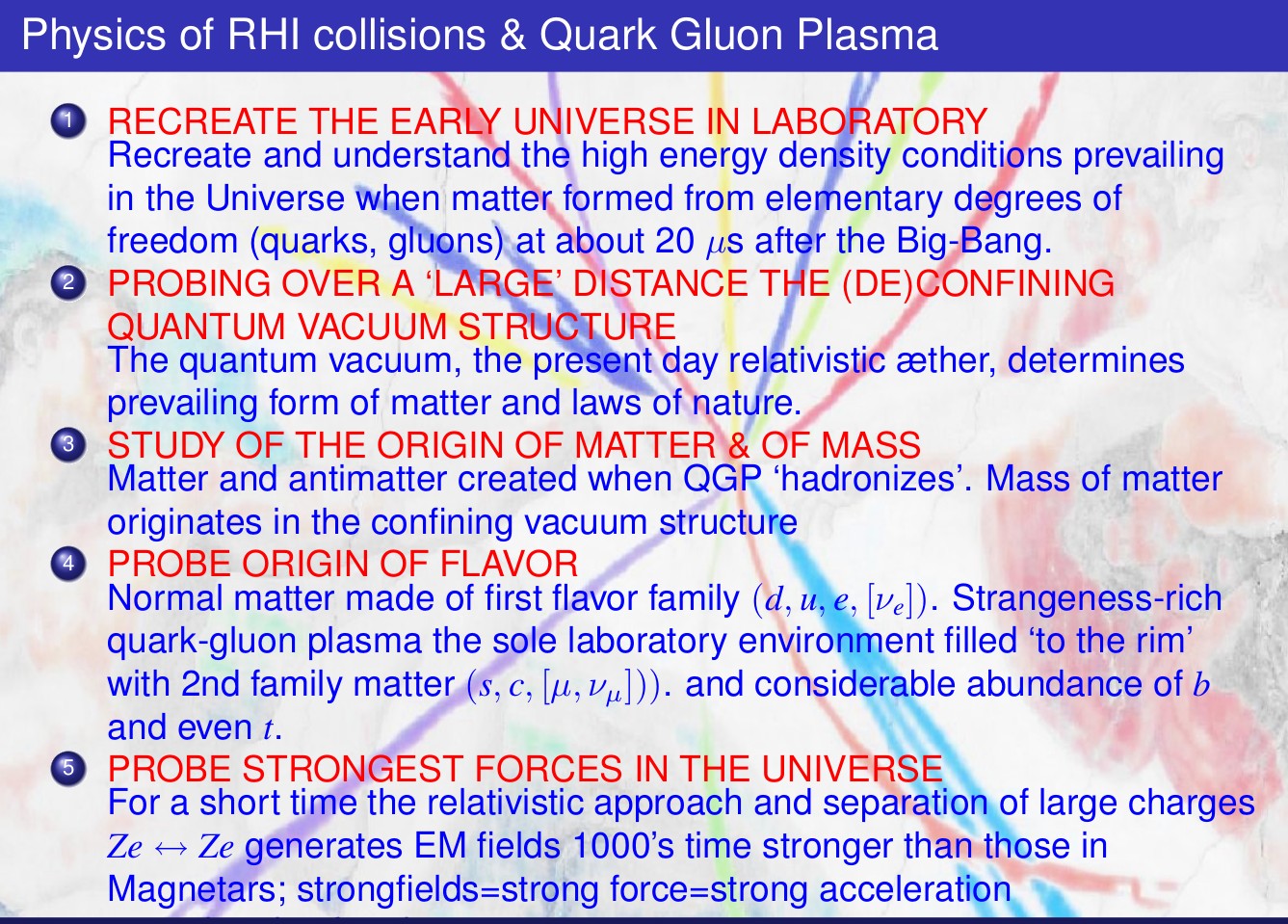}}
\caption{The five reasons to study QGP: a snapshot from a recent lecture by the author -- the background is based on the Sistine Chapel ceiling fresco \lq The Creation of Adam\rq\ by Michelangelo -- this picture ornamented the poster seen on p.138 in Ref.\cite{Rafelski:2016hnq} of the 1992 NATO Summer school \lq\lq Particle Production in Highly Excited Matter\rq\rq\ I organized with \textbf{Hans H. Gutbrod,} for proceedings see Ref.\cite{Gutbrod:1993rp}}
\label{WHYrhic}
\end{figure}

We study QGP because: 
\begin{enumerate}
\item
We desire to understand the formation of matter we are made from as it emerged from the primordial soup of quarks and gluons in the early evolution of the Universe\label{UniverseR1}.
\item
We search to improve the understanding of the mechanism of quark confinement by exploring its \lq melting\rq\ in a relatively large space-time volume into locally color deconfined quark-gluon plasma.
\item
We seek a deeper experimental and theoretical understanding of the origin of the mass of matter: how does the nuclon aquire its mass, and not another value? In other words what is the origin of the energy scales governing the vacuum structure confining quarks? 
\item 
In the laboratory QGP environment we find particles from the three flavor families: at CERN-LHC energies we find numerous $u,d,s,c$ (up, down, strange, charm) quarks, the four members of the two first quark families. There are also a few $b$ (bottom) quarks present. This may help us to explore physics phenomena that encompass all three families of particles known today allowing perhaps to study the origin of flavor.
\item
We may be able to the strong field physics phenomena: the electromagnetic (EM) forces acting in ultra relativistic collisions on the colliding heavy ions can be stronger than strong interactions. 
\end{enumerate}
 
We next discuss how these questions arose from the interest in the exploration of different possible forms of nuclear matter in heavy ion collisions. I will describe my personal perspective based on my participation in this field of physics. I was there at the birth of the ideas and I contributed to each stage of the research program development. I am convinced that the personal experience snapshots of events I present are sufficient for understanding the path into this new physics paradigm. 

\subsubsection{Quantum vacuum structure and quark confinement}\label{subsec:quarkQCD}

The Nobel prize for the quark model of hadrons was awarded in 1969. Most students of physics at that time have been influenced by this event; I was one of them\footnote{See 
\href{http://iopp.fileburst.com/ccr/archive/CERNCourier2019JulAug-digitaledition.pdf}{CERN Courier July/August 2019 } \textbf{Murray Gell-Man, Memorial Issue}.}. Less than a decade later the dynamical model of quark-quark interaction, quantum-chromodynamics~\cite{Fritzsch:1973pi} (QCD) was formulated. The first quantitative quark-bag model~\cite{Chodos:1974je,Chodos:1974pn,DeGrand:1975cf} descriptions of hadron structure ensued, followed by ever more precise quantitative models of hadrons~\cite{Thomas:1981vc,Thomas:1982kv}. We note that the quark model of hadron structure relied on quark-quark QCD based chromo-magnetic interaction.

These models postulated that quarks were constrained to reside within a small domain of space-time. This postulate, as was soon understood, required a local change of the vacuum structure to be introduced in the context of strong field physics~\cite{Rafelski:1974rh}. My immersion into the physics of local in space-time modification of vacuum structure motivated me in 1975/6 to attempt to explain quark confinement, introducing auxiliary vacuum field~\cite{Rafelski:1975ra}, followed by effort to introduce quark overcritical binding by gauge fields~\cite{Muller:1976ms}.

However, to this day a detailed model of confinement by a change in vacuum structure remains elusive. That this would be so was forecast by the bag model inventors\cite{Jaffe:1977su} R.L. Jaffe and Ken Johnson and it is elaborated in depth in the 1980 book by T.D. Lee~\cite{Lee:1980book}.\\

\noindent\textit{\textbf{R.L. Jaffe and Ken Johnson}~\cite{Jaffe:1977su} explain in one phrase in 1977:}\\[-0.7cm]
\begin{mdframed}[linecolor=gray,roundcorner=12pt,backgroundcolor=GreenYellow!15,linewidth=1pt,leftmargin=0cm,rightmargin=0cm,topline=true,bottomline=true,skipabove=12pt]\relax%
\label{JaJo}
We wish to describe here our own work on confinement which has been motivated by the belief that the starting point of conventional Lagrangian field theory is too distant from the phenomena to be useful.
\end{mdframed}
\vskip 0.5cm 

I was among a few researchers drawn into the QGP research area by our earlier consideration of the vacuum structure and strong fields; I am pretty certiain this remark applies to \textbf{Berndt M\"uller, Miklos Gyulassy,} and perhaps also to \textbf{Larry McLerran.} I described my path to QGP in a 10 year retrospective as shown below.\\

\noindent \textit{In the March 1984 inaugural lecture: \href{http://inspirehep.net/record/1750535/files/Rafelski.pdf}\lq\lq Why versus How in Theoretical Physics\rq\rq~\cite{Rafelski:1984twl} \url{http://inspirehep.net/record/1750535/files/Rafelski.pdf} at the University of Cape Town I presented the connection of the physics of strong fields with quark confinement and hadron structure:}\\[-0.7cm]
\begin{mdframed}[linecolor=gray,roundcorner=12pt,backgroundcolor=Dandelion!15,linewidth=1pt,leftmargin=0cm,rightmargin=0cm,topline=true,bottomline=true,skipabove=12pt]\relax%
\centerline{\includegraphics[width=1.0\textwidth]{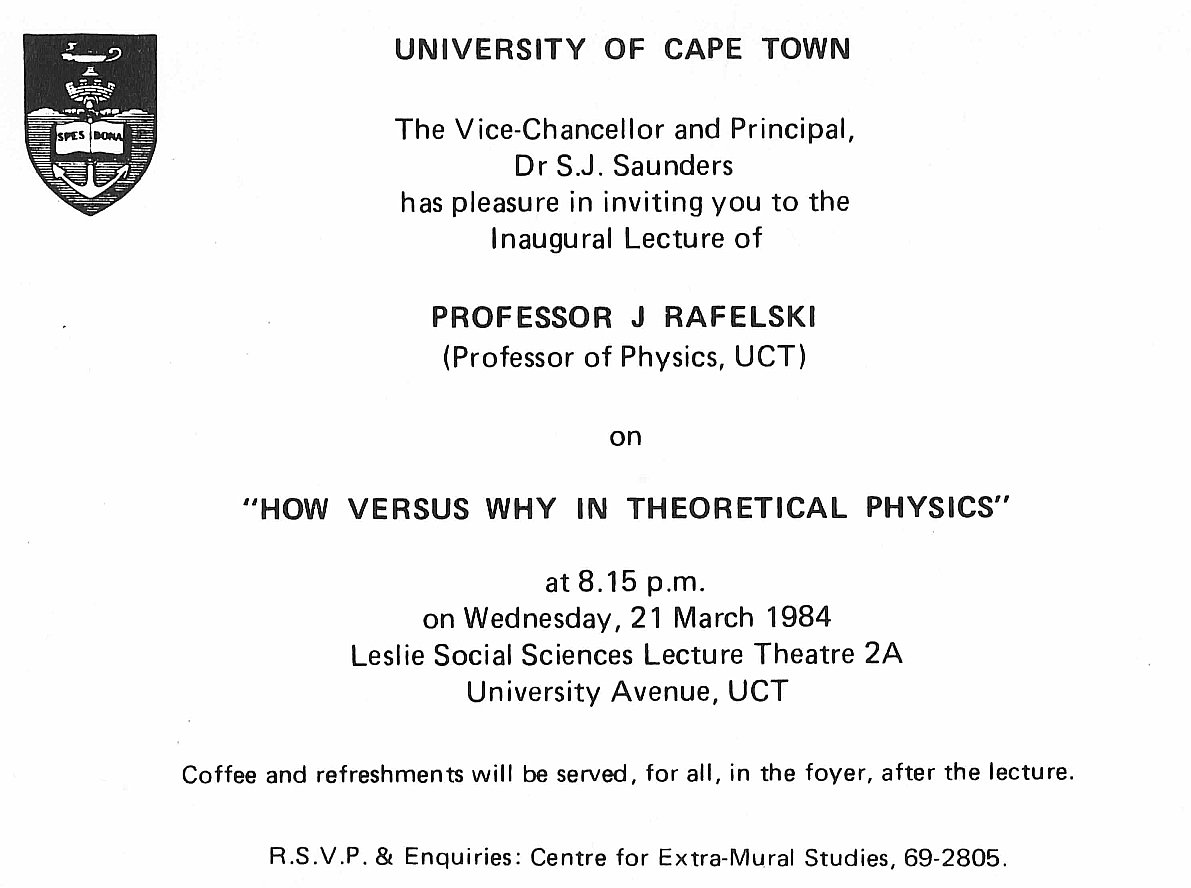}}
\noindent{\small The invitation to the Inaugural Lecture at the University of Cape Town, March 1984}\\

\noindent \ldots My own work on the vacuum\label{JRVac} begun in Frankfurt when I was a young student in 1970. I \ldots met \emph{Walter Greiner}\ldots my mentor and teacher. \ldots We studied the vacuum structure arising from properties of electrically charged particles and determined the conditions under which transitions between different vacua were expected. \ldots 

When we apply a strong electric field to it, the vacuum will spark. Virtual matter is always there and all you need to do when you apply strong electrical field to the vacuum is to create conditions to materialize what is already there. Of course you will now wonder why, if one looks around, one doesn\rq t see positrons everywhere. The point is that one has to have a very strong electric field -- and it turns out that the only way to create sufficiently strong electric fields in the laboratory right now is by \emph{bringing two heavy nuclei together.} \ldots

From the study of the vacuum of electrically charged particles arose the suggestion that vacuum structures, so established, are a general phenomenon of all charged particles, with charges now being other than electrical. In particular, if one were to pursue the substructure of atomic nuclei to the level of quarks, the well established constituents of nucleons, one encounters a new charge that these particles carry. But this charge is always neutralized -- we only detect particles neutral under strong charge and quarks are not available to be looked at individually. Why?

While one tries to understand this, the only conceptual explanation that one arrives at is the fact that actually these particles live in a different vacuum. That is, the region of space inside the nucleons in which quarks can live is, indeed, a different vacuum. But the substance around it is the kind of vacuum we live in. One usually calls this the true vacuum and the inside, the perturbative vacuum.

\ldots two different vacua can coexist, they are here simultaneously, except that the perturbative vacuum is very small. Its region is very small compared to space domains we have experienced. The radius of a proton is on the order of $10^{-13}\;$cm. Now arises the question: What do I do in order to create a large volume of this new vacuum? That is certainly the next step in order to test the consistency of the coexisting vacuum picture. I must be able to make a large box of perturbative vacuum -- a box full of different nothing. And I will come to this point below.
%
\ldots why \ldots these different nothings only come in small sizes in nature. The answer is that it takes energy to change the structure of the vacuum. The inside is a different \lq nothing,\rq\ but it takes some form of energy to get it there. This is our current understanding of what happens
\ldots you have two states -- ice and water -- they are two different structures of the same thing, exactly a parallel situation to our coexistent vacuum. We have true and perturbative vacua -- both are just different structures of the \lq nothing.\rq\ I have to supply energy to go from one to the other. 

So we must now accept that we live in an ice age! This ice age already exists for $1.5\times 10^{10}$\;years -- the lifetime of our universe. The last heat period ceased $10^{-6}$\;sec (one millionth of a second) after the birth of the universe, so we have only spent a short time in the other melted state. As the universe expanded, and temperatures dropped, the vacuum froze, leading ultimately to the present state of iced vacuum. Conceptually, this scenario is well understood, but since the birth of the universe has been a one-time event, we do not have firm experimental basis to confirm the above. While practically everybody in particle and theoretical physics believes in this picture, the belief in frozen vacuum picture is mainly supported by the fact that it is currently the only consistent explanation of all scarce experimental facts. But this hypothesis has until now not been subject to thorough experimental verification. And we recall that one negative experiment is sufficient to uproot the understanding. But there is at present no reason whatsoever for this picture of coexistent vacua not to be the correct one.

Today we can attempt to simulate the hot early universe by colliding matter -- heavy nuclei -- at high velocity. The heat generated may suffice to melt the vacuum locally and open the opportunity to study the fundamental degrees of freedom in the melted state. The concept of \lq quark-gluon plasma\rq\ is so developed. Remember, however, that subnuclear particles are investigated. So the needed particle accelerators are giant and the experimental effort quite outrageous. Temperatures and pressures thousands of times higher than in the sun would be needed. Still, the program in this research field is likely to proceed and be a fruitful one for all involved. 

\emph{Its particular importance is the undertaken test of the principle of true and perturbative vacua needed for the explanation of the elusiveness of quarks.} But we should recognize that what we learn in such experiments will not only concern the past of the universe, but also its present structure. Extreme conditions are believed to prevail in the centers of very densely collapsed stars -- neutron stars. It is possible that the interior of the star has \lq melted.\rq\ Even more exciting is the possibility that the least understood of all stellar objects, quasars, may have an energy generating core consisting of melted vacuum.
\end{mdframed}
\vskip 0.5cm

During my tenure at the University of Cape Town 1984-87, the exploration of the vacuum structure described above was accompanied by the development of the strangeness signature of the new phase of matter, the QGP, to the level of experimental usefulness. Dozens of research papers were published, and many remain well-cited to this day. The academic structures and traditions that were created to accommodate the international character of the QGP activities have endured and allowed the University of Cape Town to remain a participant in the exploration of this new phase of matter to this day.

\subsubsection{Hagedorn (temperature physics) Frontier}
The new quark paradigm that morphed into the confining vacuum paradigm we just introduced \lq happened\rq\ nearly in parallel with the proposal of thermal model of hadron production driven by Hagedorn\rq s invention of the statistical bootstrap model in 1964; for a review of Hagedorn\rq s work see Ref.\,\cite{Rafelski:2016hnq}. Hagedorn was interpreting fragmentary experimental data about particle production; these data were not in agreement with the rudimentary statistical particle production models. 

Given model difficulties that beset interpretation of multi-particle production spectra in the early 1960s it would have been easy to abandon the early thermal particle production models. This was an easy option since the majority of particle effort was devoted to other theories that have largely lost their luster today: S-Matrix bootstrap, Regge-poles, are but two examples. Hagedorn a few companions persevered. His achievements are both intuitive and imaginative: by trial and error he created a new paradigm developing thermal model and establishing the statistical physics as a new domain in the realm of strong interactions {\em before} experimental necessity arose. The concept of \lq Hagedorn temperature\rq\ is a part of current physics vocabulary. Hagedorn\rq s work is an example of a theoretical hypothesis focusing the direction of future experimental work. 

The two seminal ideas addressing the properties of strong interactions are:
\begin{itemize}
\item
Quarks and later, QCD with vacuum structurer on one side; and on another, 
\item 
Statistical bootstrap, Hagedorn\rq s temperature, thermal models of particle production. 
\end{itemize}
were proposed within a few month of each other in 1964/65. Just 15 years later these two fields merged creating the new discipline, the physics of QGP, a new phase of matter~\cite{Rafelski:2015cxa}. 

While before 1980 the deconfinement of quarks was searched in highest energy \lq elementary\rq\ collision experiments, the new QGP paradigm arises at relatively modest relativistic heavy ion collision energy. This is so since the size of the space-time domain also matters and needs to be sufficiently large. This allows the aether of modern day, the structured and confining quantum-vacuum, to be dissolved by the extreme heat generated in the large volume by colliding large atomic nuclei, which in turn melts the quark structure of strongly interacting particles called hadrons. This melting of hadrons at high temperature into the color deconfined QGP can therefore be studied in modest energy laboratory relativistic heavy ion collision experiments, and it is also found at highest energy available in $p$-$p$ interactions. 

The term quark-gluon plasma was at first a particle physics buzz-phrase. It was introduced in the exploration of relativistic proton-proton collisions in the than accessible energy range of 0.1-0.3 TeV addressing the free motion of and parton dynamics by ed Shuryak~\cite{Shuryak:1978ij}. In the context of the relativistic heavy ion (=nucleus, nuclear) collisions (RHI collisions) we instead spoke of the study of nuclear and quark matter at the two first formative meetings~\cite{Satz:1980Bil,Bock:1980GSI}. The term QGP used for parton dynamics was rapidly abandoned in particle physics, and it was adopted as more appropriate term than quark matter considering that color charged quarks were set free.

However, before the QGP term morphed to designate the new field of physics, the names of conference series were already defined. We call the primary conference series \lq Quark Matter YEAR\rq\ (QM\;YEAR) and strangeness related series \lq Strangeness in Quark Matter YEAR\rq\ (SQM\;YEAR). These terms appear often in this report. The generally accepted first QM\;1980 conference is the one held in Darmstadt~\cite{Bock:1980GSI} in October 1980, organized by \textbf{Rudolf Bock} and \textbf{Reinhard Stock.} However, some in the QGP community look at the earlier theoretical, August 1980 meeting~\cite{Satz:1980Bil} organized by \textbf{Helmut Satz} in Bielefeld, as the first in the series of QM meetings. 

At the birth of this new field of physics, the exploration of the fifth state of matter, the deconfined quark-gluon plasma, arose from the exploration of properties of nuclear matter in relativistic collisions of heavy nuclei. As noted already this implied that this novel research area is a part of nuclear science. However, I would like to argue that quark-gluon plasma as a field of research stands today on its own merit, see also \rf{WHYrhic}. It overlaps most with quark physics and hadron structure, and less with the traditional nuclear science dealing predominantly with nuclear reactions and nuclear structure. 

The Hagedorn Physics Frontier (which you will not --yet -- find mentioned on the WWW) as a research field includes, for example, the study of the Universe at the QGP epoch at the age of about 20\;$\mu$s; of quark matter in neutron stars; of the quantum vacuum structure and the deconfinement process, connecting here to the question about the origin of dark energy. Finite temperature lattice QCD is the computational method, providing insights into fully equilibrated hot QCD matter. Most of the heavy quark physics, including the discoveries of new multi-quark states relate to the \lq\lq Vacuum Structure and Quark Deconfinement\rq\rq \ novel area of research, and as noted in \rf{WHYrhic}, we accidentally touch on several other very important fundamental research topics.

\subsubsection{Superdense nuclei or QCD matter?}\label{sssec:dense}

The general interest in new types of \lq superheavy nuclei\rq\ and dense neutron star type nuclear matter was abundant in the late 60s and early 70s. There was also profound interest in some even more exotic ideas. I arrived in late Summer 1974 as a postdoc at the Argonne National Laboratory (ANL), just as \textbf{Arnold Bodmer} was promoted to be a senior scientist in \lq my\rq\ Physics Division. This coincidence is important for these diaries: I was interested to know what Arnold Bodmer did to earn this academic distinction, and this interest was my first encounter with strangness and strange quarks.

Bodmer\rq s research addressed $\Lambda(uds)$-bound in light nuclei~\cite{Ali:1967vji} (hypernuclei) and collapsed nuclei~\cite{Bodmer:1971we}. He argued that dense and more strongly bound nuclear isomer states were consistent with the then available knowledge. In the context of the quark bag model these ideas led on to strange dibaryons~\cite{Jaffe:1976yi}, strange quark matter~\cite{Farhi:1984qu,Alford:2006vz}, and morphed into strange quark drops (strangelets) formation~\cite{Greiner:1987tg,Rafelski:1987sf} in heavy ion collisions. There is an ongoing quest to discover strange quark matter in neutron stars and their fragments~\cite{Madsen:1998uh,Weber:2004kj}

Returning to the context of collapsed nuclear matter, Bodmer\rq s idea gained a lot of traction with the publication in 1974 by (T.D.) Lee and Wick~\cite{Lee:1974ma} of an effective theory model. This work occurred in parallel to the rise of the quark model and about two years after quantum-chromodynamics (QCD), a theory of strong interactions was advanced and rapidly accepted.

In their influential study of collapsed nuclear matter, Lee-Wick did not use these new, quark and gluon, degrees of freedom. However, the theoretical work on hypernuclei and collapsed nuclei by Bodmer, updated with symmetry consideration by the renowned Particle \& Fields luminaries Lee and Wick had put pressure on the nuclear community to rapidly develop experimental instruments that were suitable for the exploration of these very interesting proposals. 

The BEVALAC\label{BEVALAC} at the Lawrence Berkeley National Laboratory (LBNL) became the first operational relativistic heavy ion accelerator. It was created in 1971 by connecting two existent accelerators, the Bevatron and SuperHILAC. The at-that-time relatively high beam momentum, of about 2\;GeV/c per nucleon, opened up research into relativistic heavy ion (RHI) collisions; that is relativistic nuclear collisions: relativistic meaning that the energy available appears at the scale of multiple rest mass of colliding nuclei. This frontier experimental program got a strong boost from the work of Lee-Wick which was addressed at the so-called \lq Bear Mountain\rq-meeting~\cite{BearMountain}\label{BearMountain}. \\ 

\noindent\textit{\textbf{Gordon Baym} in his QM2001 opening remarks comments~\cite{Baym:2001in}:}\\[-0.7cm]
\begin{mdframed}[linecolor=gray,roundcorner=12pt,backgroundcolor=GreenYellow!15,linewidth=1pt,leftmargin=0cm,rightmargin=0cm,topline=true,bottomline=true,skipabove=12pt]\relax%
 \ldots at the time (Fall 1974, JR) of the Bear Mountain meeting, the idea of quark matter as the ultimate state of nuclear matter at high energy density had not taken hold.\ldots Despite suggestive hints, the experiments have not yet identified a quark-gluon plasma.
 \end{mdframed} 
%
Gordon Baym explains above why ten years after quarks were recognized as building blocks of matter, and nucleons in particular. One cannot find in the white paper that was prepared at the \lq Bear Mountain\rq-meeting~\cite{BearMountain}, or for that matter in any documents accompanying the BEVALAC scientific program from this epoch, any mention of quarks. The Bear Mountain workshop ordained the collapsed nuclei as the research objective of the early US-RHI physics effort. 

At the time of the Bear Mountain meeting, the idea of QCD as the theory of strong interactions underpinning nuclear science was whizzing around, and yet at this important meeting this topic did not get a hearing. This was so despite the fact that quark matter was already considered by several groups. In 1974 I had seen the related work of Peter Carruthers~\cite{Carruthers:74}; Carruthers was well known, soon after appointed to be the Theory Division leader at the Los Alamos Laboratory. His views were widely read in the particle physics community.

Later I learned of equally relevant quark-star work carried out in the Soviet Union~\cite{Ivanenko:1965dg} already in 1965. It is noteworthy that just at the time of the Bear Mountain meeting, the asymptotic freedom was connected with dense quark matter in neutron stars by Collins and Perry~\cite{Collins:1974ky} -- Baym misdates Collins and Perry to be the year after Bear Mountain event. The preprint existed before the meeting. I had also already lectured at Fermilab, presenting quark matter as a large scalar-bag; in that period called the \lq SLAC\rq\ quark bag, see Ref.\cite{Lee:1980book}, with quarks preferring to sit on the confinement volume surface. I did not trust in a surface of quarks as an excited or collapsed state of nuclear matter, so I never published this work.

Weinberg~\cite{Weinberg:QFT} explains the physics reasons for the remarkable rapid and universal adoption of QCD in the particle physics context. Let me add to this by clarifying why nuclear scientists did not follow: a) They could explore nuclear structure without knowing quarks existed; b) \lq People with quarks\rq\ had difficulty arguing for or even justifying the relevance of quarks in nuclear interactions. 
 
The nuclear community saw the Lee-Wick work as an alternative to working with quarks. But to me it was and is unimaginable that these authors were not aware of the newly created and widely adopted theory of strong interactions, QCD. So why would \textbf{T.D. Lee} after QCD was discovered turn to effective model of nuclear matter? The answer is seen in the quotation of Jaffe \& Johnson, see page \pageref{JaJo}. 

I believe that T.D. Lee was interested in an effective theory capable of replacing in the context of nuclear collisions the non-Abelian gauge theory, the QCD. This was also a natural step to take for him since QCD was relying on concepts developed two decades earlier by Yang and Mills~\cite{Yang:1954ek}, the same Yang who coauthored with T.D. Lee the parity violating Nobel prize winning weak interaction paper~\cite{Lee:1956qn} two years later.

The detour to the Lee-Wick effective theory may derive from another situation: I imagine that T.D. Lee was in part motivated by his relationship with C.N. Yang who was at SUNY Stony Brook, near the future Brookhaven’s RHIC machine. QCD relied on a theory invented by C.N. Yang and these two inventors of parity violation in weak interactions were not the best of friends. Maybe T.D. Lee was looking for a way to support the LBNL on distant West Coast.

Writing about \lq\lq What Fuels Progress in Science? Sometimes, a Feud\rq\rq\ \href{https://www.nytimes.com/1999/09/14/science/what-fuels-progress-in-science-sometimes-a-feud.html}{in the New York Times of Sept. 14, 1999} James Glanz uses the Lee-Yang situation as a singular counterexample to the generally science beneficial scientific feuds.\\

\noindent \textit{\textbf{James Glanz} comments on T.D. Lee and C.N. Yang as follows:}\\[-0.7cm]
\begin{mdframed}[linecolor=gray,roundcorner=12pt,backgroundcolor=GreenYellow!15,linewidth=1pt,leftmargin=0cm,rightmargin=0cm,topline=true,bottomline=true,skipabove=12pt]\relax%
\ldots\label{NYT1999} entirely destructive, with little redeeming value for scientific research-- like the bitter personal and professional split between Dr. Tsung Dao Lee and Dr. Chen Ning Yang, who won the Nobel Prize in 1957 for collaborative work on particle physics. Each now claims the lion\rq s share of credit for the work, and the two have not spoken for 35 years. \ldots
\end{mdframed} 
%
{\bf However:} 20 years later I see the situation differently; James Glanz of the NYT did not have all the facts in hand. T.D. Lee, by advising the CERN Director General (DG) Herwig Schopper in the early 80s, did help CERN to the SPS RHI collision program and by extension to the LHC program as well has created the counterbalance to the BNL forthcoming heavy ion research program. We return below to this matter, see page \pageref{SchopperRem}. 

In just the right moment I was at CERN and saw T.D. Lee in the CERN cafeteria. I do not remember the date, but the event lives strongly in my memory. I joined him as is usual at CERN with my coffee cup in hand. This was just before, as I know this today, his meetings with the DG regarding the RHI project at CERN. In this entirely random CERN cafteteria meeting I advanced the case of strange (anti)hyperons observable which benefit greatly from the SPS longitudinal boost of the unstable particle decay length. 

With T.D.s support a pivotal decision, contrarian to other advice, was taken by the CERN DG as he clarified in the introduction to Ref.\cite{Rafelski:2016hnq}, see page \pageref{SchopperRem}. So unlike James Glanz of the NYT I believe that all scientific battles without exception did advance knowledge in a decisive way -- history clarifies the question \lq How?\rq\ which in September 1999 had not yet become visible. 

Back to the early 70s: The primary outcome of the Bear Mountain meeting was to define a contextual pillar of the US-RHI community working at BEVALAC. On balance the Bear Mountain participants strengthened the future of RHI physics in USA using the Lee-Wick collapsed nuclei proposal. Being different from the QCD based understanding of strong interactions, this proposal created a well understood context for nuclear science already engaged into exotic forms of nuclear matter. 

\subsubsection{Strangeness: a natural tool to study QGP}\label{sec:ssig}

The heaviest of the three light quark flavors, strangeness, emerged as the candidate signature of QGP for the following three reasons~\cite{Koch:2017pda,Rafelski:1982ii}: 
\begin{enumerate} 
\item
When color bonds are broken, the chemically equilibrated deconfined state contains an unusually high abundance of strange quark pairs~\cite{Rafelski:1980rk,Rafelski:1980fy} leading on to strangeness (kaon, hyperon) enhancement. 
\item 
The gluon component in the QGP (rather than quark component) produces strange quark pairs rapidly, and just on the required time scale~\cite{Rafelski:1982pu} -- strangeness enhancement was now tied to the presence of gluons \lq G\rq\ in the QGP, while strangeness yield depended on size and initial conditions of the QGP fireball. 
\item The high density of strangeness at the time of QGP hadronization was a natural source of multi-strange hadrons~\cite{Rafelski:1982rq}, quantified in the coalescence picture of pre-existing quarks and antiquarks~\cite{Koch:1986ud}.
\end{enumerate}
\textit{At the inaugural lecture in June 1980 at the University Frankfurt I described the work at CERN carried out in collaboration with Rolf Hagedorn that led to the proposal of melting of nuclear matter into quark matter and introduced (strangeness as an) observable of this process:}\\[-0.7cm]
\begin{mdframed}[linecolor=gray,roundcorner=12pt,backgroundcolor=Dandelion!15,linewidth=1pt,leftmargin=0cm,rightmargin=0cm,topline=true,bottomline=true,skipabove=12pt]\relax
\centerline{\includegraphics[width=0.95\textwidth]{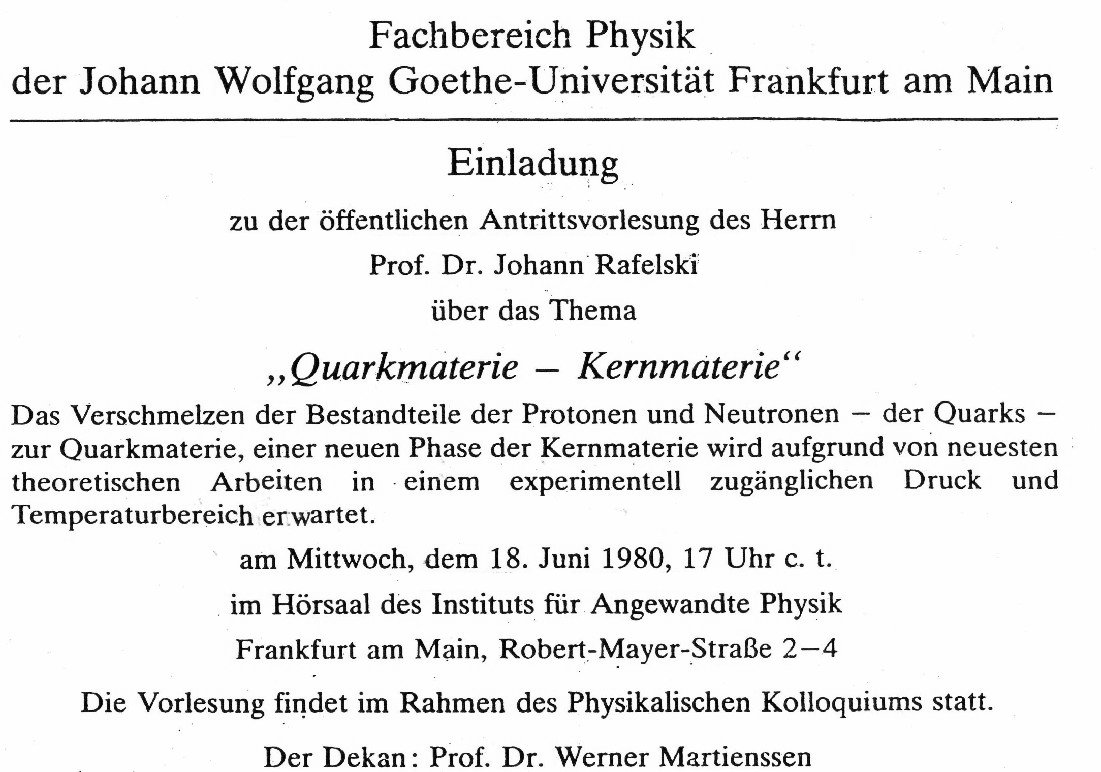}}
\noindent{\small The invitation to the Inaugural Lecture\label{FRAINAU} at University Frankfurt, 18 June 1980. The short abstract reads, translated: Recent theoretical work shows melting of the constituents of protons and neutrons -- quarks -- into quark matter, a new phase of nuclear matter. This is expected to occur in an experimentally accessible domain of pressure and temperature.}
\end{mdframed}

This event was closely followed by two conference lectures.
\begin{enumerate}
\item
I lectured on two topics in Bielefeld~\cite{Satz:1980Bil} at the end of August 1980: I) on strangeness signature of quark-gluon plasma, and II) on the strong field vacuum aspects, see page \pageref{JRVac} for this second topic discussion. Much of the work addressing the heavy ion collisions and QGP presented in the first lecture was carried out in collaboration with Rolf Hagedorn at CERN-Geneva, more details are seen in Ref.\cite{Rafelski:2016hnq}.

Rolf Hagedorn also lectured at Bielefeld, introducing our work on hot hadron matter in relativistic nuclear collisions. My talk followed. I was presenting the properties of quark-gluon plasma, the transformation between hadron and quark phases of strongly interacting matter, introducing for the first time to a international audiance strangeness as QGP observable. We submitted our papers as two parts of a joint project overlapping, but for the new strangness and QGP segments, with our joint publication~\cite{Hagedorn:1980kb} on \lq\lq Hot Hadronic Matter and Nuclear Collisions.\rq\rq\ 
\item
At the following October 1980 meeting in Darmstadt at the GSI laboratory\cite{Bock:1980GSI} we published our contributions individually. Hagedorn presented his evaluation of the status of the understanding of contemporary experimental and theoretical work on dense and hot strongly interacting matter. He was attempting an overview of heavy ion scattering experiments available at the time. This left me the task to advance stronger the phenomenology of the search for QGP and deconfinement, which compared to other contributions in Ref.\cite{Bock:1980GSI} was a true \lq progressive\rq\ effort.\\
\end{enumerate}

\noindent\textit{\textbf{Hans Specht} in his meeting summary captures the gist of my contribution as follows (see pp. 552/3 in Ref.\cite{Bock:1980GSI}):}\\[-0.7cm]
\begin{mdframed}[linecolor=gray,roundcorner=12pt,backgroundcolor=GreenYellow!15,linewidth=1pt,leftmargin=0cm,rightmargin=0cm,topline=true,bottomline=true,skipabove=12pt]\relax%
\label{Specht} 
\ldots The particular sensitivity of the production of such particles (K$^-$ and $\bar p$, JR) which do not contain any of the entrance channel quarks, to possible collective effects in this region was mentioned by J. Rafelski, who placed his emphasis on the $\overline{\Lambda^0}$. It is also here that the transition from a simple cascading picture to the full complication of the space-time development \ldots takes place.
\end{mdframed}
\vskip 0.5cm

For a long time the precision of experimental data was not at the level to allow clear recognition of the predicted $\overline{\Lambda^0}/\bar p>1$ enhancement. However, this unexpected result is clearly visible at the LHC energy scale, see \rss{sec:AliceSys}.

\noindent \textit{The argument for strangeness and strange antibaryons as a signature of QGP were already stated~\cite{Rafelski:1980rk} in 1980 and follow verbatim. These remarks cannot be expressed better today:}\\[-0.7cm]
\begin{mdframed}[linecolor=gray,roundcorner=12pt,backgroundcolor=Dandelion!15,linewidth=1pt,leftmargin=0cm,rightmargin=0cm,topline=true,bottomline=true,skipabove=12pt]\relax%
 \ldots assuming equilibrium in the quark plasma,
we find the density of the strange quarks to be (two spins and three colors)\,\footnote{I change here the notation introducing
$ s\to N_s,\ \bar s\to N_{\bar s} $ etc., adding subscript \lq B\rq\ to indicate baryo-chemical potential.}:
\begin{equation}\label{Eq1}\tag{26}
\frac{ N_s} V =\frac{ N_{\bar s}} V =6\,\int \frac{d^3p}{(2\pi)^3}e^{-\sqrt{p^2+m_s^2}/T}=3\frac{Tm_s^2}{\pi^2}K_2(m_s/T),
\end{equation}
(neglecting, for the time being, the perturbative corrections and, of course, ignoring weak decays). As the mass of the strange quarks, $m_s$, in the perturbative vacuum is believed to be of the order of 280--300 MeV\footnote{This high value of strange quark mass applies at a scale of about 0.5 GeV was obtained from hadron spectra. The tabulated value today, about 1/3 as large as I used in 1980, is at the scale of of 2 GeV.}, the assumption of equilibrium for $m_s/T\simeq 2$ may indeed be correct. In Eq.\,(\ref{Eq1}), we were able to use the Boltzmann distribution, as the density of strangeness is relatively low. Similarly, there is a certain light anti-quark density ($\bar q$ stands for either $\bar u$ or $\bar d$):
\begin{equation}\label{Eq2}\tag{27}
\frac{N_{\bar q}}{V}\simeq 6\int \frac{d^3p}{(2\pi)^3}e^{-|p|/T-\mu_q/T}=e^{-\mu_q/T}\cdot T^3 \frac{6}{\pi^2},
\end{equation}
where the quark chemical potential is, as given by Eq.(3) $\mu_q=\mu_\mathrm{B}/3$, $(\mu_\mathrm{B}$ is baryo-chemical potential). This exponent suppresses the $q\bar q$ pair production as only for energies higher than $\mu_q$ is there a large number of empty states available for the $q$. 

What we intend to show is that there are many more $\bar s$ quarks than anti-quarks of each light flavor. Indeed:
\begin{equation}\label{Eq3}\tag{28}
\frac{N_{\bar s}}{N_{\bar q}}=\frac 1 2 \left( \frac{ m_s}{T}\right)^2K_2(m_s/T)e^{\mu_\mathrm{B}/(3T)}.
\end{equation}
The function $x^2K_2(x)$ is, for example, tabulated in Ref.[15] (Abramowitz-Stegun). For $x=m_s/T$ between $1.5$ and $2$, it varies between $1.3$ and $1$. Thus, we almost always have more $\bar s$ than $\bar q$ quarks and, in many cases of interest $N_{\bar s}/N_{\bar q}\simeq 5$. As $\mu_\mathrm{B}\to 0$ there are about as many $\bar u$ and $\bar d$ quarks as there are $\bar s$ quarks. 

\label{FirstPredict}
When the quark matter dissociates into hadrons, some of numerous $\bar s$ may, instead of being bound in a $q\bar s$ Kaon, enter into a ($\bar q \bar q \bar s$) anti-baryon and in particular, a $\overline{\Lambda}$ or a $\overline{\Sigma}^{\,0}$. The probability for this process seems to be comparable to the similar one for the production of antinucleons by the (light) antiquarks present in the plasma.\ldots We would like to argue that a study of $\overline{\Lambda}$ , $\overline{\Sigma}^{\,0}$ \ldots could shed light on the early stages of the nuclear collision in which quark matter may be formed.
\end{mdframed}
\vskip 0.5cm

There are three important issues raised above, which have since seen significant elaboration:
\begin{enumerate}
\item 
The chemical equilibration of strange quarks, generalized later to address the equilibration of all QGP degrees of freedom in the deconfined phase. 
\item 
The combinant quark hadronization, generalized later to the study of hadron abundances as diagnostic tool of the hadronizing fireball.
\item 
Use of particles made entirely of newly created quarks, and in particular here the strange anti-baryon signature of QGP as introduced in 1980, and our prediction has been verified at SPS, RHIC and LHC energy range and as we will argue it is one of the key observables of the QGP phase of matter. 
\end{enumerate}

The October 1980 GSI workshop!\cite{Bock:1980GSI} was attended by \textbf{J\'ozsef Zim\'anyi,} who lectured on kinetic thery in HG: \lq\lq Approach to Equilibrium in High Energy Heavy Ion Collisions\rq\rq, which directly followed on my lecture~\cite{Rafelski:1980fy}, and was also presented in Acta Physica Hungarica~\cite{Zimanyi:1980fz}. Zim\'anyi proposed in this work the theoretical framework to check the hypothesis if chemical equilibrium can be attained in QGP. By the Summer 1981, \textbf{Tam\'as B\'{\i}r\'o,} a young graduate student working with J\'ozsef Zim\'anyi~\cite{Biro:1981zi} extended this work and obtained strangeness production rates in perturbative QCD when considering the specific processes $q\bar q\to s\bar s$. 
 
The key result of this study was that it would take much too long, about 8 times the natural lifespan of a QGP fireball, to equilibrate strangeness chemically. Unfortunately, when J\'ozsef Zim\'anyi came to present the pre-publication results in Frankfurt in late Summer 1981, set up on a short notice, I was at a meeting in Seattle. Upon my return, I believe in late October 1981, I received a copy of the Bir\'o--Zim\'anyi preprint~\cite{Biro:1981zi}. I saw an important omission: $GG\to s\bar s$ process was not considered.

During my CERN 1977--79 fellowship period I learned about QCD based charm production in $p$-$p$ reactions. I shared, for about a year, an office with \textbf{Brian Combridge,} of perturbative QCD charm production fame~\cite{Combridge:1978kx}. Brian was an extrovert who was keen to share his insights. From Brian I learned (sometimes against my will) that even if the cross sections were similar for both quark $q\bar q\to c\bar c$ and gluon $GG\to c\bar c$ fusion processes into charm, it was the gluon fusion process which dominated the production rate. 

In the Fall 1981 this CERN experience turned out to be a valuable asset; the outcome of the calculation of the strangeness production relaxation time remained open since Bir\'o--Zim\'anyi did not study glue fusion process. I described my insight to Berndt M\"uller, who was enthusiastic at the prospect of using real plasma gluons in a physical process. This was so since we had just completed a study based on virtual gluon fluctuations of the temperature dependence of the latent heat of the QCD Vacuum \cite{Muller:1980kf}. In view of this preparation the glue-based flavor producing reactions were a natural extension of Bir\'o--Zim\'anyi.

Within a few weeks of work, we confirmed in an explicit computation the hypothesis that the thermal strangeness chemical equilibrium in QGP is due to gluon fusion process~\cite{Rafelski:1982pu}. While Berndt was practicing thermal perturbative QCD, I was racing ahead preparing the manuscript seen in \rf{sProdPrep}: we see on right first corrections introduced by Berndt in red, and my second thoughts written in by pencil -- later we needed to count the words and improve the English for the PRL publication.

\begin{figure}[tb]\sidecaption
\includegraphics[width=1.0\columnwidth]{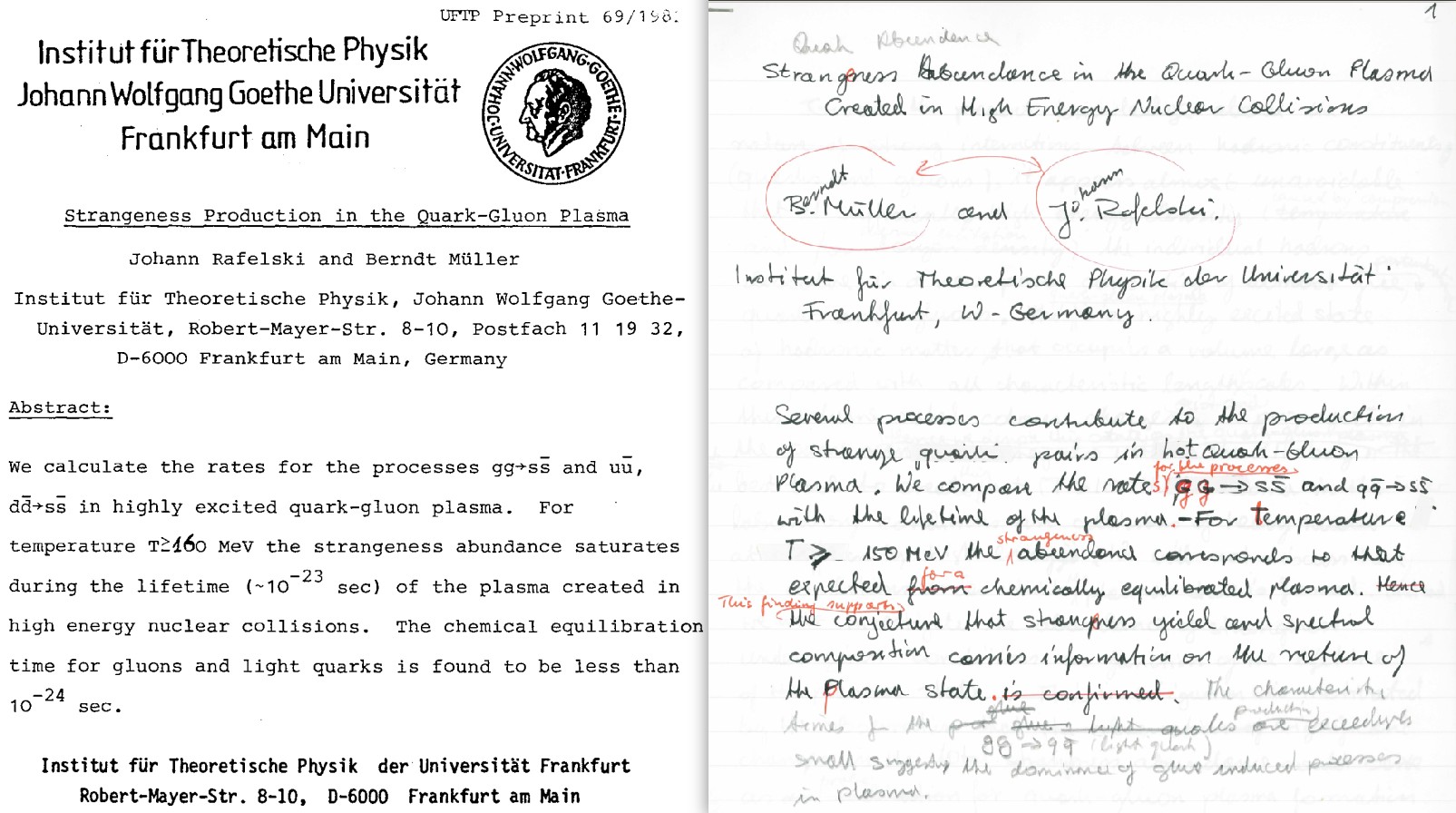}
\caption{The birth of strangness production in gluon fusion manuscript Ref.\cite{Rafelski:1982pu}; On left: the preprint page before PRL level edits; On right: the first handwritten version of November 1981, with edits: my first writing, Berndt\rq s red pen edits, my penciled-in additions: Berndt insisted I should be the first author}\label{sProdPrep}
\end{figure}

An important aspect in the evaluation of the rate of strangeness production, which Berndt and I undertook, was the choice of the value of the running strong coupling constant $\alpha_s(\Lambda)$. We knew that if one uses a 1st-order perturbative expression for a QCD process, it can only produce reasonable results if the coupling strength is chosen just at the right strength for the appropriate energy scale. 

Considering the experimental results available, we determined the value $\alpha_s$ to produce strangeness at the typical thermal collision energy $\sqrt{s_\mathrm{th}}=3$-$6T$, for $T=200$-$300$ MeV should be $\langle\alpha_s\rangle=0.6$. This turned out to be just the right choice, a value $\langle\alpha_s(0.86\;\rm{GeV})\rangle=0.60\pm0.10\pm0.07$ is appropriate -- more discussion of strangeness production allowing for the running of both $\alpha_s$ and the strange quark mass is seen in Sec.~\ref{QCDrunning}. 

However, for the following 15 years, a value $\langle\alpha_s\rangle=0.2$ was often used in literature; this smaller value is appropriate for the energy scale $\Lambda\simeq 6$ GeV. Since the reaction rates scale with $\alpha_s^2$, this seemingly small modification meant that the chemical equilibration of QGP is delayed by an order of magnitude, eliminating strangeness as signature of QGP. Thus misunderstanding of running QCD parameters explains why our results were not always trusted.

On the other hand our result, the process of chemical equilibration in QGP became an asset: the QGP chemical equilibrium yield of strangeness evolved into an indicator of the presence of mobile, free gluons required to produce strangeness. Strangeness abundance is the signature of deconfinement, since the work of B\'{\i}r\'o-Zim\'anyi showed that as long as there was no free glue, just the thermal light $u,\bar u, d, \bar d$ quarks; chemical equilibration was therefore without glue not attainable. 

The other required element of this argument is that only in deconfined QGP strangeness enhancement can be expected. Strangeness production in hadron gas was elaborated in a kinetic approach by Peter Koch a talented Franfurt graduate student. Peter computed the strangeness yields and relaxation times expected in the hadron phase~\cite{Koch:1984tz}. We also recognized that hadron processes cannot enhance strange antibaryons, detailed balance prevents excess of $\overline\Xi$ to form. 

By 1986 a detailed discussion of how a fireball of deconfined mobile (strange) quarks turns into strangeness carrying hadrons was complete~\cite{Koch:1986ud}. In this work Peter Koch, Berndt M\"uller and I proposed the nonequilibrium recombinant processes for the hadronization of the QGP fireball. These results in particular demonstratd that the high strangeness density in chemically equilibrated QGP is the source of greatly enhanced strange antibaryon yield.

With these results, we see: 
\begin{enumerate}
\item
Strangeness yield enhancement based on thermal gluon reactions has been well established theoretically as the signature for deconfinement. 
\item
The strength of this enhancement was recognized to dependent on how hot and how long the hot QGP phase would last. 
\item 
Strange antibaryons $\overline\Lambda$, $\overline\Xi$ and $\overline\Omega$ emerging in overabundance in hadronization process were understood as an unmistaken signature of QGP. No competing process was known then and now.
\end{enumerate}

These results confirmed multi-strange hadrons as the key characteristic signature of the QGP. Many detailed model predictions showed how the high density and the mobility of already produced strange and antistrange quarks in the fireball creates this signature. The expected backgrounds were explored, demonstrating that multi-strange antibaryons are by a large factor dominated by the formation mechanisms present during QGP fireball hadronization.

\subsection{Establishing (ultra)relativistic heavy ion collisions beams}\label{ref:dst}
\label{HowQGP} 
The pioneering QGP experiments were carried out at the CERN Super Proton Synchrotron (SPS) accelerator, beginning with the first beams obtained 1986/7. By the end of the last century, Pb beams with the highest energy had, to the disbelief of some of the discoverers, created the QGP phase of matter. This was announced early February 2000, see page \pageref{CERN2000}. 

The BNL laboratory (Brookhaven National Laboratory, Long Island, New York) joined in the \lq Hunting for QGP\rq, see Section~\ref{ssec:flow}. At BNL the Relativistic Heavy Ion Collider (RHIC) by means of the greater energy reach made accessible additional experimental opportunities. The SPS and RHIC results have been confirmed and elaborated at the CERN Large Hadron Collider (LHC), while the research program at SPS continued yielding additional corroborating evidence. 
 
\subsubsection{Heavy ions at CERN}\label{sec:RHI-CERN}
CERN\label{CERN} is an international, European-funded, particle physics laboratory built near Geneva across the Swiss/French border. The name derives from \lq Conseil Europ\' een pour la Recherche Nucl\' eaire\rq (European Council for Nuclear Research) established by 12 European governments in 1952. CERN\rq s RHI experimental research program was helped along by both internal interest and the German GSI laboratory research program carried out in the 70s at the Lawrence Berkley National Laboratory BEVALAC, see page \pageref{BEVALAC}. 

After GSI joined the CERN RHI effort in the early 80s, Hans H. Gutbrod and Reinhard Stock returned from their LBNL projects and became spokespersons of CERN SPS experiments, WA80 and NA35, respectively, developed at CERN. Some of the GSI paid experimental equipment was also moved from LBNL to CERN. Another CERN-SPS experiment, NA36, was prepared by LBNL researchers who chose CERN over the long wait for the to-be-built RHIC. 

At CERN the WA85 experiment under the leadership of \textbf{Emanuele Quercigh} took off focusing on strange antibaryons. All told three strangeness and antibaryon experiments were being set up to search and study quark-gluon plasma: NA35, NA36 and WA85 -- the acronyms derive in part from the location at CERN: WA is the West area on the main campus of CERN and NA is the North area at the CERN-II campus, both connected by the SPS accelerator ring. The numerical code tracks the sequential approval status of the experiment.\\

\noindent \textit{\textbf{Gra\.zyna Odyniec} writes for the proceedings of the SQM2000 meeting held in July 2000 at Berkeley \cite{Odyniec:2001}:}\\[-0.7cm]
\begin{mdframed}[linecolor=gray,roundcorner=12pt,backgroundcolor=GreenYellow!15,linewidth=1pt,leftmargin=0cm,rightmargin=0cm,topline=true,bottomline=true,skipabove=12pt]\relax%
\label{Pugh}
{\Large\bf In the memory of Howel G Pugh}\\[0.1cm]
{\bf Gra\.zyna Odyniec}\\
\ldots From the very beginning Howel (Pugh, LBNL, scientific director of BEVALAC, JR), with firmness and clarity, advocated the study of strange baryon and antibaryon production. He played a leading role in launching two of the major CERN heavy-ion experiments: NA35 and NA36, the latter being exclusively dedicated to measurements of hyperons. Strangeness enhancement predicted by theorists was discovered \ldots\\[0.2cm]
\centerline{\includegraphics[width=0.99\textwidth]{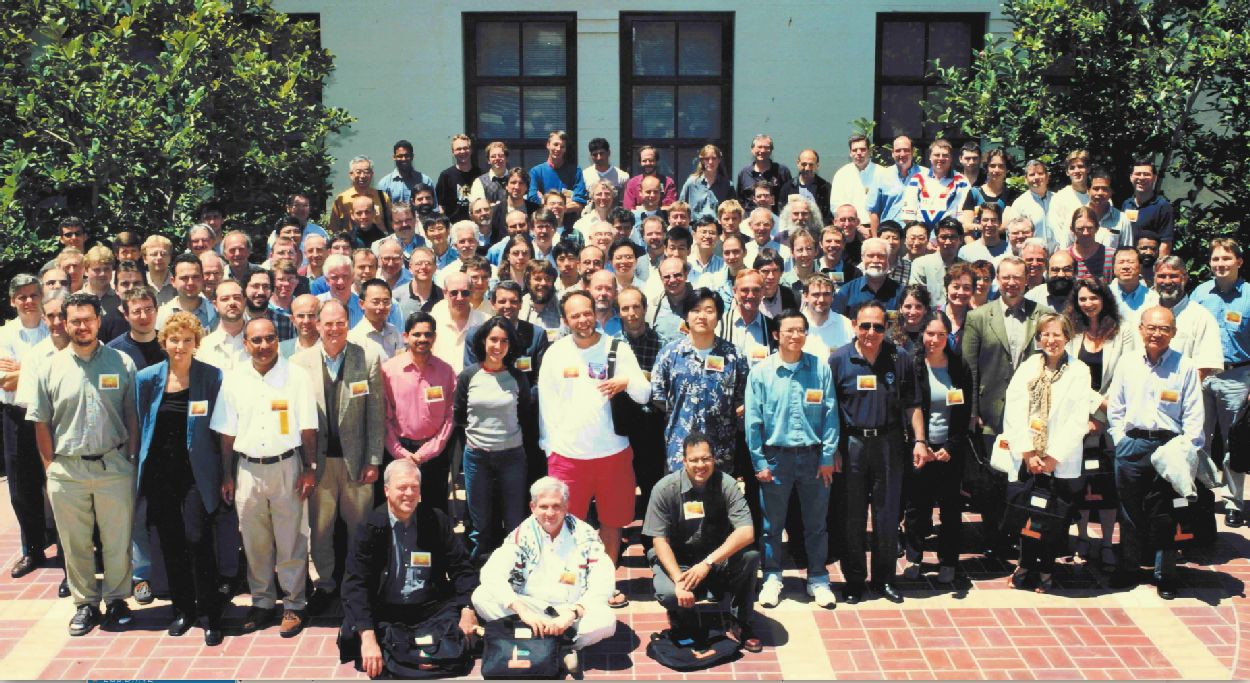}}
\noindent{\small Group picture at SQM2000\label{SQM2000Group}, the {\it 5th International Conference on Strangeness in Quark Matter}, Berkeley,July 20-25, 2000 (a B/W version is seen in the \emph{printed} proceedings volume~\cite{Odyniec:2001qb}); Gra\.zyna is second from left in front standing raw; sitting on the ground from left: Hans H. Gutbrod, Johann Rafelski and Stefan Bass}\\

\end{mdframed}
\vskip 0.5cm

My personal move to CERN preceded these developments by several years. In the mid 70s, following the Bear Mountain events, see Section~\ref{sssec:dense}, and after accidental meetings at lecture venues with both Rolf Hagedorn and Leon van Hove (soon to be DG at CERN) I understood that my interest in quark matter, vacuum structure and deconfinement would find a better home at CERN. I recived an offer in 1976 and I departed from Argonne Laboratory for Europe on leave (I was in the interim made junior staff member) in Spring 1977. Argonne was more than delighted, indeed \lq encouraging\rq\ this development, wanting to develop another field of nuclear physics. However, a few years later the ANL Physics Division turned to experimental studies in my areas of expertise: of strong fields and positron production, and later, quark matter. In these experimental efforts ANL had lost on-site theoretical expertise since I resigned from Argonne after I accepted a tenured appointment in Frankfurt, see page \pageref{FRAINAU}.

After a few months at GSI and my alma matter Frankfurt, which were useful as I met there several future RHI colleagues for whom I would later open CERN portals, I arrived in September 1977 as a Fellow at CERN. I was on the way to learn from Rolf Hagedorn, whom I met before, about thermal models of hot hadron matter, and to move him to help me study the phase transformation of strongly interacting matter into quark matter. 

Hagedorn was himself very interested in including quarks and QCD into his work on hot hadronic matter; thus our interests were well matched. He was a superb teacher; more on this can be found in the book dedicated to these events~\cite{Rafelski:2016hnq}, in which many reminiscences are collected. 

However, one input is missing in Ref.\cite{Rafelski:2016hnq} -- the contribution of Bill Willis to the creation of the RHI program at CERN. Bill passed away before I began Ref.\cite{Rafelski:2016hnq}; I found no substitute able to fill the large gap he left in the Hagedorn volume~\cite{Rafelski:2016hnq}.\\

\noindent \textit{\href{https://physics.columbia.edu/william-bill-j-willis-1932-2012}{\textbf{Bill Willis} obituary page at Columbia University} \url{https://physics.columbia.edu/william-bill-j-willis-1932-2012}reminds of his influence:}\\[-0.7cm]
\begin{mdframed}[linecolor=gray,roundcorner=12pt,backgroundcolor=GreenYellow!15,linewidth=1pt,leftmargin=0cm,rightmargin=0cm,topline=true,bottomline=true,skipabove=12pt]\relax%
Bill was a towering presence in the development of particle physics, with a career encompassing nearly the entire history of the field. \ldots he was a true renaissance figure who influenced the development of particle physics, nuclear physics and accelerator physics.\ldots

Bill made seminal contributions to nuclear physics, specifically in establishing the case for and the methods to investigate collisions of heavy nuclei at relativistic energies as a means of searching for new forms of matter. He worked\ldots to promote this new field of physics, both in early investigations at Brookhaven and CERN, and in building the case for the Relativistic Heavy Ion Collider (RHIC), which began operations at BNL in 2000.\ldots
\end{mdframed}
\vskip 0.5cm

I met Bill Willis for the first time when he stormed into Hagedorn\rq s office with a copy of our first conference manuscript in hand, a CERN preprint~\cite{Hagedorn:1978kc} created in late 1978 (Bill worked at CERN between 1973 and 1990). He was very excited and explained to us that he did not know Hagedorn was interested in colliding relativistic nuclei and that our work was helping his science case to develop this research program at the CERN-ISR (intersecting storage ring).

He was, to the best of my knowledge, the conceptual designer of collider detectors like those we use today. The first one was the AFS: it consisted of a central detector, a cylindrical drift chamber located in an axial magnetic field (hence the name Axial Field Spectrometer = AFS)\label{AFSexp}, entirely surrounded by a calorimeter. AFS became operational at the ISR and carried out the first experiments with light nuclei, $\alpha$-particles, reaching the per-nucleon center of momentum collision energy $\sqrt{S_\mathrm{NN}}=30$\;GeV. 

Bill Willis saw in our work a further justification for a CERN research program in relativistic heavy nuclear collisions at the ISR collider. He believed that a collider experiment at the ISR~\cite{Willis:1981xm} could rapidly explore the dynamic behavior of the new form of matter. However, CERN entering the LEP era had to find funding for the heavy ion research program. There was too little interest in the ISR: the ISR presented a considerable technological detector challenge, which combined with the high cost of collider operation turned out to be an insurmountable obstacle. 

A few explanations are needed, which I give from my personal perspective of this period when CERN was at the crossroads of particle and nuclear physics. The mix of a few CERN employees collaborating with a much larger number of visiting researchers from all over the world created an unusual openness to new ideas, including nuclear collisions utilizing any of the available CERN accelerators. The development of a new research program thus required a focus of interest accompanied by related new funding. This was so since at that time CERN had a specific mission to build the next big machine, the Large $e^+e^-$ Collider (LEP). Any other program needed to find external resources. 

For this reason the costly and technologically challenging ISR program championed by Bill Willis faltered. However, another CERN machine, the SPS (Super Proton Synchrotron), also an injector for the future LEP and later LHC, was offering within the realm of established technologies and existent equipment also the capability of several parallel run experiments. This situation was attracting a large population of researchers and in turn was attracting the necessary additional funding. Beginning with 5 experiments, the research program expanded to 7 large experiments, attracting several hundred external researchers. At the SPS in the fixed target mode the equivalent $\sqrt{s_\mathrm{NN}}=20$\;GeV beams of Sulfur became available in 1987/8. This energy was comparable to that ISR would have made available. 

The approval of the RHI research program was a major effort and it happened due to the decisive actions of two men, Maurice Jacob, who we will meet a few more times in these diaries, who was the CERN Theory Division leader before retirement, and Herwig Schopper, the Director General of CERN in the 80s. The situation is best understood by giving these two personalities their own voices in finding the answer to the often posed question \lq How did CERN decide to develop the heavy ion research program at the SPS while shutting down the ISR?\rq\\

\noindent \textit{Copying from CERN protocols the key elements of the Maurice Jacob presentation on 22 June 1982 to the CERN council, see Chapter 29 in~\cite{Rafelski:2016hnq}, along with the reminiscences of Herwig Schopper in the forward to this volume:}\\[-0.7cm]
\begin{mdframed}[linecolor=gray,roundcorner=12pt,backgroundcolor=GreenYellow!15,linewidth=1pt,leftmargin=0cm,rightmargin=0cm,topline=true,bottomline=true,skipabove=12pt]\relax%
{\bf Maurice Jacob} speaking for about 90 minutes to the CERN Science Policy Council, and answering questions for 20 minutes on 22 June 1982:\\

\ldots Heavy ion collisions offer the possibility to reach very high densities and very high temperatures over extended domains, many times larger than the size of a single hadron. The energy densities considered are of the order of 0.5 to 1.5 GeV/fm$^3$ and the relevant temperatures are in the 200 MeV range. The great interest of reaching such conditions originates from recent developments in Quantum Chromodynamics, QCD, which make it very plausible that, while color confinement should prevail under standard circumstances, deconfinement should occur at sufficiently high density and (or) sufficiently high temperature.\ldots 

Over an extended volume where the required density or temperature conditions would prevail, one expects that the properties of the physical vacuum would be modified. While the normal vacuum excludes the gluon field, the color-equivalent of the dielectric constant being zero (or practically zero), one would get a new vacuum state where quarks and gluons could propagate while interacting perturbatively. \ldots 

Granting the fact that a thermalized quark-gluon plasma is formed during the collision, it will very rapidly destroy itself through instabilities, expansion and cooling. One should then watch for specific signals which could be associated with its transient (but most interesting) presence. \ldots

Several signals have attracted particular 
attention.
\begin{enumerate}
\item One of them is provided by the prompt photon or lepton pairs radiated (a volume effect!) by the thermalized plasma, \ldots 
\item 
Another interesting signal may be provided by strange particles originating in relatively large number from the plasma, once it has reached chemical equilibrium. 
\item
There may also be more violent effects, with abnormal density fluctuations in the overall energy flow associated with secondaries. 
\item
Size and lifetime could be determined through pion/photon interferometry since each violent event with head on collision could produce pions in the thousands!''
\end{enumerate} 

\ldots The chairman, {\bf Prof. H. Schopper}, thanked Maurice Jacob for his presentation, and opened the discussion. 

Replying to a question from {\bf Prof. P.T. Matthews}, Maurice Jacob said that the fundamental purpose of heavy-ion collision experiments was to study matter at very high quark densities. \ldots

Replying to questions from {\bf Prof. D.H. Perkins} and the chairman, {\bf Prof. H. Schopper}, Maurice Jacob said that collisions with a projectile with a large atomic number were required because the amount of deposited energy was proportional to the number of nucleons in the incident nucleus. Estimates suggested that, in the most optimistic case of head-on uranium/uranium collisions, energy densities of the order of 2 GeV/fm$^3$ would be obtained, whereas in the case of carbon/uranium collisions, this figure would fall to 1 GeV/fm$^3$. 

Replying to a question from {\bf Prof. E. Amaldi}, Maurice Jacob said that, with regard to the question of the time necessary for the plasma to achieve equilibrium, it was expected that there was a chance that some thermalization would take place at the level of the quarks and the gluons present in the plasma, many collisions having time to take place. \ldots

Replying to a question from {\bf Prof. J. Lefran\c{c}ois}, SPS Experiments Chairman, Maurice Jacob said that at 1 GeV/fm$^3$ the temperature of the plasma would be too low for significant production of charm and beauty particles. 

In reply to a question from {\bf Prof. N. Cabibbo}, Maurice Jacob said that the great merit of the QCD calculation using the lattice over the Hagedorn model was that it made direct exploration of the system possible over and beyond the phase transition, whereas the phenomenological model had been based on a separate study of the two phases. The two approaches were, however, complementary, in many respects. What the experimenters wished to do with heavy-ion collision experiments was to ascertain whether matter existed in a different form beyond the hadron gas.

The chairman, {\bf Prof. H. Schopper}, in conclusion, said it was clear that any discussion of heavy-ion collision experiments raised as many questions as it attempted to resolve. However, before very long the Scientific Policy Committee would have to address itself to the question of heavy-ion collision experiments in a more formal way. \ldots\\

\noindent \textit{{\bf Herwig Schopper\rq s}\label{Schopper2014} reminiscences prepared in November 2014 for the forward to Ref.\cite{Rafelski:2016hnq}:}\\
\ldots in the 1970s and 80s, the study of heavy ion reactions grew out of nuclear physics and eventually became an interdisciplinary field of its own that is presently achieving new peaks. Hagedorn can rightly be considered as one of the founding fathers of this field in which the \lq Hagedorn – temperature\rq\ still plays a vital role. 

\ldots At CERN difficulties arose in the 1980s, because in order to build LEP at a constant and even reduced budget, it became necessary to stop even unique facilities like the ISR collider at CERN. Some physicists considered this an act of vandalism. 

In that general spirit of CERN physics program concentration and focus on LEP it was also proposed to stop the heavy ion work at CERN, and at the least, not to approve the new proposals for using the SPS for this kind of physics. I listened to all the arguments of colleagues in favor and against heavy ions in the SPS. I also remembered the conversations I had had with Hagedorn 15 years earlier. In the end, T. D. Lee\label{SchopperRem} gave me the decisive arguments that this new direction in physics should be part of the CERN programme. He persuaded me because his physics argument sounded convincing and the advice was given by somebody without a direct interest.

I decided that the SPS should be converted so that it could function as a heavy ion accelerator, which unavoidably implied using some resources of CERN. But the LEP construction and related financial constraints made it impossible to provide direct funds for the experiments from the CERN budget. Heavy Ion physicists would have to find the necessary resources from their home bases and to exploit existing equipment at CERN.

This decision was one of the most difficult to take since contrary to the practice at CERN, it was not supported by the competent bodies. However, the reaction of the interested physicists was marvelous and a new age of heavy ion physics started at CERN. \ldots

Since the first steps of Hagedorn and his collaborators, a long path of new insights had to be paved with hard work. The quark-gluon plasma, a new state of matter was at last identified in the year 2000. \ldots\\
\end{mdframed}

\subsubsection{(Ultra)relativistic heavy ion collisions in USA}\label{sec:RHI-US}
How did the US get the RHIC project? In Summer 1983 I was invited to lecture at a meeting at Lawrence Berkeley National Laboratory (LBNL). Ten years after Bear Mountain, quarks and gluons could be mentioned in my invited LBNL lecture~\cite{Rafelski:1983im}. This meeting had, aside of the now long forgotten \lq anomalons\rq, the objective of drumming up support for a follow-up to BEVALAC, a heavy ion collider, the VENUS.

VENUS was being designed with a size that fitted into the hillscape at Berkeley. By the landscape accident, the energy was well chosen to create and study quark-gluon plasma. I presented the status of the strangeness observable of QGP. During the lecture I was asked how many strange particles could be seen by their decay per collision. My answer, as I still like to recall, generated explosive laughter in the lecture room.\\

\noindent \textit{The answer is in the proceedings near the end of my paper~\cite{Rafelski:1983im}:}\\[-0.7cm]
\begin{mdframed}[linecolor=gray,roundcorner=12pt,backgroundcolor=Dandelion!15,linewidth=1pt,leftmargin=0cm,rightmargin=0cm,topline=true,bottomline=true,skipabove=12pt]\relax%
\ldots we can expect to have several V's (decay signature of K$_s$ and $\Lambda$) per collision, which is 100-1000 times above current observation for Ar-KCl collisions at 1.8\;GeV/Nuc kinetic energy.\\
\centerline{\resizebox{1.0\textwidth}{!}{\includegraphics{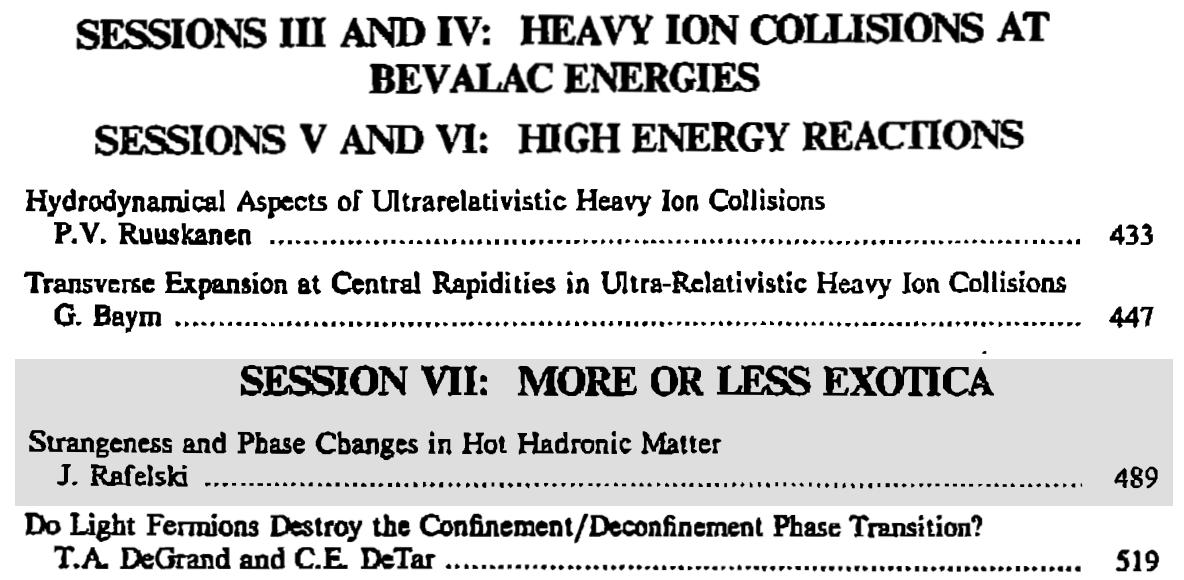}}}
\textit{The proceedings contents page show two classes of lectures: in chapter \lq High Energy Reactions\rq\ eminent theorists Vesa Ruuskanen and Gordon Baym contribute, and, there is another chapter \lq More or Less Exotica\rq.\label{exotica}}
\end{mdframed}
\vskip 0.5cm
Today we know that at the VENUS design energy the ratio K$^+/\pi^+\simeq 0.18$; hence my answer was correct. In 1983 my talk was placed (see above) in \lq More or Less Exotica\rq\ section of proceedings, rather than \lq High Energy Reactions,\rq\ where related theoretical presentations are seen above. It is good that I shared this fate with another important theoretical contributions, as we see in the contents fragment I present above: \lq\lq Do Light Fermions Destroy the Confinement/Deconfinement Phase Transition?\rq\rq\ by T. A. DeGrand and C.E. DeTar; the answer, we know, is YES. 

The LBNL project VENUS was not approved. The US heavy-ion community was served by a few times larger, and many years longer in construction RHI collider (RHIC) at the Brookhaven National Laboratory (BNL). The BNL laboratory had already built an accelerator ring for the ISABELLE project, a $p$--$p$ collider which was scrapped in view of a changing scientific landscape and technical difficulties. This civil engineering investment was handed over to Nuclear Physics and became the {\bf Re}lativistic {\bf H}eavy {\bf I}on {\bf C}ollider: RHIC. 

While the BEVALAC research program was ending, in preparation for RHIC, at a modest cost, a Heavy Ion low energy tandem accelerator at BNL was connected with a synchrotron, called AGS, and thus other investments made in the past at BNL could be reprogrammed towards a rapid creation of a heavy ion research program capable of higher energy beams compared to the Berkeley BEVALAC. However, these developments also meant that the entire heavy ion research community in the USA had to reorganize, focusing now on the East coast. 

The delay and reorganization also meant that to continue their research efforts during the years of transition from LBNL to BNL, several heavy ions groups moved on. Researchers delegated to LBNL from Germany were soon working at CERN. Moreover, my lecture of 1983 had a lasting impact on the leader of the LBNL heavy ion program, Howel Pugh, see page \pageref{Pugh}. Howel contributed to the rise of strange antibaryon CERN program at CERN in decisive way.

AGS, with its terminal heavy ion per nucleon energy of 10-15$A$\;GeV, about 15 times lower compared to CERN-SPS, was seen as a training ground for RHIC. Also, funds for the heavy-ion experimental program in US became scarce due to the ballooning construction cost of the RHIC collider. Therefore, the AGS research program included low cost searches, but not for QGP. All \lq experts\rq\ believed that the QGP formation threshold would first be breached at RHIC, justifying the large investment made. We turn in a moment to see how, without any good physics reason, this opinion came to be.

I note that AGS started delivering heavy ion beams months ahead of the CERN-SPS. To this day it is hard to tell what exactly was the AGS discovery potential of QGP, since only after 15 years the needed detector equipment became available to look for strange antibaryons. This was just before the AGS experimental program was shut, the results made available were not allowing a convincing evaluation. It is quite possible that AGS could have scooped CERN-SPS in the race for QGP had the circumstances allowed this.

For reasons that are hard to understand from today\rq s perspective, even the CERN-SPS reaction energies were by the standards of the US East Coast theorists not considered high enough. AGS was considered as totally out of the QGP league -- all RHI theorists working within a 100 miles radius of BNL\label{100miles} were advancing the view that only RHIC collider, operating at an order of magnitude higher $\sqrt{s_\mathrm{NN}}$, as compared to SPS, was capable of achieving QGP formation. 

In order to justify this view, those advancing it cited the James Bjorken scaling solution of one-dimensional hydrodynamic flow of relativistic matter~\cite{Bjorken:1982qr} proposed in Summer 1982. Bjorken, in order to illustrate the physics contents of his work, assumed initial dense matter conditions which actually were exceeded at CERN-SPS. However, at SPS \lq Bjorken longitudinal scaling\rq\ was not seen in the experimental results that emerged as early as Spring 1987: the model predicted that at sufficiently high collision energy the distribution of produced final state particles should be a very flat function of rapidity. This result confirmed in the eyes of the RHIC \lq chamber theorists\rq\ that neither AGS nor SPS would be of use in search for QGP; only RHIC had, in their perception, any chance. 

However, the absence of Bjorken rapidity scaling does not mean that there was no quark-gluon plasma formation at SPS in the S--S collisions at the equivalent $\sqrt{s_\mathrm{NN}}=20$\;GeV. Absence of scaling means that this schematic one-dimensional infinite energy hydrodynamic \lq Bjorken\rq\ solution was an irrelevant reaction model at the SPS energy range. This is so since we observe at SPS a very significant stopping of nuclear matter, a phenomenon that continues to be a topic of ongoing research, as is the question how QGP is formed at the relatively low energies available at SPS producing same observational outcome as found at much higher RHIC and CERN-LHC energy range. 

To summarize: The extra money needed for RHIC needed extra time to be \lq printed.\rq\ The additional dozen years that were tacked in the USA onto the new field of physics, the search for QGP, sent, in my opinion, the discovery of quark-gluon plasma to CERN. RHIC started a decade if not more after LBNL-VENUS would have discovered QGP. There was also plenty of time within the BNL-AGS experimental program to search for QGP, another missed opportunity. RHIC was switched on long after the experimental SPS program completed the approved experiments with maximum available energy of the maximum size lead beams, and where the experimental program included many diverse observables, including strangeness and strange antibaryons in duplicate experiments. This is why CERN alone was able to present the experimental evidence for QGP already in 1999 or even earlier, delayed to February 2000, a long time after the scheduled RHIC turn-on.

\subsection{Was quark-gluon plasma really discovered?}
\subsubsection{Strangeness is getting ready}
In the early 80s strangeness as an observable of QGP was shifting from a theoretical idea into the experimental realm, with several experiments coming on line. This research effort was supported in part by an odd couple, the University of Cape Town where I was chair of Theoretical Physics in the pivotal years, see page \pageref{JRVac}, and the CERN-TH Division which often welcomed me. In the pivotal period, 1982-1988, the Director of the Theory Division was Maurice Jacob. Maurice had assumed an important role in steering CERN into the search and discovery of QGP. At the 1988 Quark Matter Conference~\cite{Jacob:1988wt} he addressed the strangeness signature of QGP; these comments are reprinted unchanged several years later in his 1992 book \textit{Quark Structure of Matter}.\\

\noindent\textit{\textbf{Maurice Jacob} speaking at the QM1988 meeting:}\\[-0.7cm]
\begin{mdframed}[linecolor=gray,roundcorner=12pt,backgroundcolor=GreenYellow!15,linewidth=1pt,leftmargin=0cm,rightmargin=0cm,topline=true,bottomline=true,skipabove=12pt]\relax%
\underline{Is strangeness production enhanced?}\\ 
An enhanced production of strange particles has long been advocated as a signature for the formation of quark-gluon plasma$^{17}$.
A high temperature and a high chemical potential for non-strange quarks both favor the production of $s\bar s$ pairs and eventually of strange particles, as strange quarks drop out of equilibrium relatively easily. However, there are other ways to produce strange particles and a dense hadronic medium could be enough. 
\ldots
Experimental evidence for an enhanced production of strange particles will be one of the big issues at this Conference$^{18}$.
The production rate of K$^+$ and $\Lambda$ particles is typically twice as large as expected from the mere superposition of nucleon-nucleon collisions. Whether this signals the formation of a quark-gluon plasma is however still unclear (in Fall 1988, JR). In order to draw a conclusion, one would have to have a much tighter control of the strangeness production rate as a function of time. 
\ldots The clearest experimental test would be an anomalously large production of anti-hyperons which standard hadronic reactions are very unlikely to produce in large numbers. We will probably be left with too little information to draw a conclusion now (Summer 1988, JR), but this conference is likely to open up exciting perspectives on strangeness production. 
\footnotetext{\vspace*{-0.5cm}
\begin{itemize}
\item[17] J. Rafelski, Physics Reports \textbf{88} (1982) 272 
\item[\phantom{17}] J. Rafelski and B. M\"uller, Phys. Rev. Lett. \textbf{48} (1982) 1066
\item[\phantom{17}] P. Koch, J. Rafelski and W. Greiner, Phys. Lett. B \textbf{123} 91983) 151
\item[18] P. Vincent (E802), contribution to this conference
\item[\phantom{18}] E. Quercigh (WA85), contribution to this conference 
\item[\phantom{18}] M. Ga\'zdzicki (NA35), contribution to this conference 
\end{itemize}
}
\end{mdframed}
\vskip 0.5cm

After the QM1988 conference some of the participants joined the Hadronic Matter in Collision 1988 October 6-12 meeting held in Tucson, Arizona~\cite{Carruthers:1989sx}. This smaller meeting complemented the QM1988 meeting by opening up a possible future for the relativistic heavy ion collisions at the planned Superconductive Super Collider (SSC). As a second point of emphasis, it offered a comprehensive coverage of the strangeness signature of quark-gluon plasma.\\

\noindent \textit{\textbf{Berndt M\"uller\rq s} October 1988 verbal remarks on strangeness signature include~\cite{Muller:1988mj}:}\\[-0.7cm]
\begin{mdframed}[linecolor=gray,roundcorner=12pt,backgroundcolor=GreenYellow!15,linewidth=1pt,leftmargin=0cm,rightmargin=0cm,topline=true,bottomline=true,skipabove=12pt]\relax%
\textbf{Strangeness and Quark-Gluon Plasma:}\\ 
\noindent \underline{1. Abstract:} 
This rapporteur talk describes theory aspects of strangeness 
production in QGP and hot HG, with particular emphasis on signatures 
of QGP formation. 

\noindent \underline{2.1. Enhancement Mechanisms:}
According to our standard picture the QGP, i.e. the high 
temperature phase of quantum chromodynamics (QCD), is characterized by\\ 
- color deconfinement, and\\ 
- partial restoration of chiral symmetry. \ldots 
 
For both reasons, the production of strange particles
is expected to be enhanced by the QGP as compared with a 
thermalized HG, as has been proposed by Rafelski$^1$. The two main ingredients of this argument are: 
 (a){\it Lower energy threshold:} \ldots\\
(b) {\it Increased strangeness density:} \ldots\\

\noindent \underline{5. Strangeness as a QGP Signal:}
\ldots Any hadron ratio, such as K$^+/\pi^+$, $\Lambda/$N, $\Xi/$N, 
$\bar\Lambda/\Lambda$, $\bar\Xi/\bar\Lambda$ which is 
{\it much higher than in pp collisions} 
signals the intermediate presence of a QGP phase, because HG reaction 
rates (at thermal equilibrium) are too slow to allow for abundant 
secondary production.\ldots\\

\noindent \underline{6. Conclusions:}
The first experimental results$^{19,24,25,26}$, which indicate surprisingly high values in particular at central rapidity range for some of the proposed strangeness signals of QGP formation are encouraging. \ldots 

\footnotetext{\vspace*{-0.5cm}
\begin{itemize}
\item[\phantom{1}1] Rafelski, J. and Hagedorn, R., in: \textit{Thermodynamics of Quarks and Hadrons,} Satz, H. (ed.),(Amsterdam 1981), p. 253 
\item[\phantom{11}] Rafelski, J., Phys. Rep. \textbf{88}, 331 (1982)
\item[19] Steadman, S. et al. (E-802 collaboration), this volume 
\item[24] Gazdzicki, M. et al. (NA35 collaboration), this volume 
\item[25] Quercigh, E. et al. (WA85 collaboration), this volume 
\item[26] Greiner, D. et al. (NA36 collaboration), this volume 
\end{itemize}
}
\end{mdframed}

\subsubsection{Strangeness in the race for QGP}
Following on the proposal to consider strangness production in RHI collisions, the first results were reported by Anikina {\it et.al}~\cite{Anikina:1984zh}. The main merit of these DUBNA laboratory (today in Russia) effort was the development of: i) new experimental techniques; and ii) of manpower. \textbf{Marek Ga\'zdzicki,} now spokesperson of NA61 strangeness SPS experiment, started his scientific work with this DUBNA effort, which was followed by his participation in the NA35/NA35/NA49 line of CERN experiments.

The CERN NA35 experiment where Marek Gazdzicki led the analysis group in the early 90s rapidly obtained the results that I was hoping for. However, for a considerable time this large group of researchers disbelieved the implications of their results. At the QM1990 meeting in mid-May 1990 printed 11 months later, the spokesperson of NA35 Reinhard Stock conveys~\cite{Baechler:1991pp} his view about the meaning of NA35 results in the following message:\\

\noindent \textit{\textbf{Reinhard Stock} as speaker at QM1990~\cite{Baechler:1991pp}:}\\[-0.7cm]
\begin{mdframed}[linecolor=gray,roundcorner=12pt,backgroundcolor=GreenYellow!15,linewidth=1pt,leftmargin=0cm,rightmargin=0cm,topline=true,bottomline=true,skipabove=12pt]\relax%
\label{StockOAu}
In a previous NA35 experiment we reported [4] results for central $^{16}$0+Au collisions which did not exhibit spectacular (strangeness) enhancements over the corresponding $p$+Au data. \ldots we have demonstrated a two-fold increase in the relative $s + \bar s$ concentration in central S--S collisions, both as reflected in the $K/\pi$ ratio and in the hyperon multiplicities. A final explanation in terms of reaction dynamics has not been given as of yet.
\end{mdframed}
\vskip .5cm

We see that in Summer 1990 and for several years after the NA35 collaboration did not, as a group, introduce the QGP interpretation of their strangeness enhancement results, though they were aware, as unpublished NA35 documents show, of our QGP work. We note further that in the opinion of Reinhard Stock, strangeness enhancement was not present in central $^{16}$0+Au collisions, but was seen in the S--S collisions. The first inkling that the internal NA35 collaboration dynamics was evolving towards a discovery story can be seen in Ref.\,\cite{Bachler:1992js} (available at CERN preprint server in October 1992). \\

\noindent \textit{\textbf{NA35 collaboration}~\cite{Bachler:1992js}:}\\[-0.7cm]
\begin{mdframed}[linecolor=gray,roundcorner=12pt,backgroundcolor=GreenYellow!15,linewidth=1pt,leftmargin=0cm,rightmargin=0cm,topline=true,bottomline=true,skipabove=12pt]\relax%
4. Neither the FRITIOF nor the VENUS model gives a satisfactory description of the full set of the results \ldots\\
5. S--S data extrapolated to the full phase space show that the observed strangeness enhancement appears mainly as kaon-hyperon pairs which indicates that this enhancement comes from the region of nonzero baryo-chemical potential.
\end{mdframed}
\vskip 0.5cm
However, while NA35 recognizes defects of a few models presenting insights about strangness dynamics, there was no mention of their result relation to the QGP. In this NA35 comprehensive year 1992 report other theories are introduced, but QGP. The mention of QGP appears once in the first sentence of this manuscript: A motivational general comment characterizing all experimental work carried out with heavy ions. 

I am aware that in this time frame the spokesperson, Reinhard Stock, wrote an open letter to his collaboration noting that NA35 strangeness results went unnoticed. The point in the matter was that NA35 did not present a claim that was of consequence in these early years. On the topic of anti-hyperons Reinhard Stock showed preliminary results concerning $\overline{\Lambda}$ for $p_\bot>0.5$\,GeV at the QM1990 Menton meeting, and we find this picture in the NA35 publication~\cite{Bachler:1992js}. A full $4\pi$ result appeared several years later in the Summer 1994~\cite{Alber:1994tz}, and in July 1995 a direct comparison with $p$--$p$ reactions wass made made in a Ph.D. thesis for the first time~\cite{Foka:1995Thesis} (see Figure 8.24, p.271).\\

\noindent \textit{\textbf{Yiota Foka} writes (p.268) in her July 1995 thesis~\cite{Foka:1995Thesis}:} \\[-0.7cm]
\begin{mdframed}[linecolor=gray,roundcorner=12pt,backgroundcolor=GreenYellow!15,linewidth=1pt,leftmargin=0cm,rightmargin=0cm,topline=true,bottomline=true,skipabove=12pt]\relax%
The enhancement (of $\overline{\Lambda}$) at mid-rapidity is a factor 6 in S-S\ldots strange particle production that is not (due to) a simple superposition of elementary interactions. \\
\textit{Earlier on anticipating this result in the thesis resume:}\\
The question if we can conclude that QGP has been observed is the topic of hot debates and this should be considered within the context of many other observables. 
\end{mdframed}
\vskip 0.5cm

Dr. Foka worked under the direct supervision of Reinhard Stock and these comments were presumably coordinated with him. The NA35 presented the ratio $ \overline{\Lambda}/ \bar p\lesssim 1.4$ measured near mid-rapidity in Summer 1995~\cite{Alber:1996mq}, showing an enhancement by a factor 3 to 5 dependent on the collision system as compared to a measurement in more elementary reactions. This was the QGP signature/strange antibaryon signature which I proposed~\cite{Rafelski:1980rk,Rafelski:1980fy} in my first strangeness papers in 1980. $ \overline{\Lambda}/ \bar p\ > 1 $ is now well established, see page \pageref{RLam}.

While NA35 was standing at a crossroad not seeing a street sign pointing in the direction of QGP, WA85 sited at the CERN $\Omega\rq$ spectrometer under the leadership of Emanuele Quercigh took the center stage in the QGP search. The WA85 collaboration offered results on: $\Lambda$ and $\bar{\Lambda}$~\cite{Abatzis:1990cm} (available as CERN preprint 18 April 1990); on $\Xi^-$, $\overline{\Xi^-}$~\cite{Abatzis:1990gz} (available as CERN preprint 8 November 1990); and a systematic exploration of QGP characteristic behavior for both~\cite{Abatzis:1991ju} (available as CERN preprint 5 July 1991).\\

\noindent \textit{\textbf{WA85 collaboration:} A firm position in favor of QGP discovery ~\cite{Abatzis:1991ju} in 1991:}\\[-0.7cm]
\begin{mdframed}[linecolor=gray,roundcorner=12pt,backgroundcolor=GreenYellow!15,linewidth=1pt,leftmargin=0cm,rightmargin=0cm,topline=true,bottomline=true,skipabove=12pt]\relax%
The(se) results indicate that our $\overline{\Xi^-}$ production rate, relative to $\bar{\Lambda}$, is enhanced with respect to pp interactions; this result is difficult to explain in terms of non-QGP models [11] or QGP models with complete hadronization dynamics [12]. We note, however, that sudden hadronization from QGP near equilibrium could reproduce this enhancement [2].
\end{mdframed}
\vskip 0.5cm

Ref.\,[2] mentioned above is my work~\cite{Rafelski:1991rh} published in March 1991, where I invented the SHM model as an interpretative tool of experimental results. We see that at least for a year after WA85 took firmly the position that its strangeness results are QGP driven.

The WA85 (200 GeV$A$ beam S on S), and WA94 (200 GeV$A$ beam S on W) reported speedily and in definitive manner their perplexing strangeness, hyperon and in particular anti-hyperon results, giving all these results a QGP discovery interpretation as early as 1990/91. The WA85/94, focused on the more QGP characteristic multi-strange hadron ratios, for $ \overline{\Xi}/\overline{\Lambda}$ see the 1993 review~\cite{Evans:1994sg}. A full summary of all results is seen in the review of \textbf{Federico Antinori} of 1997~\cite{Antinori:1997nn} presented in Ref.\cite{Omega25y}, see \rf{SigLamCERNFig} on page \pageref{SigLamCERNFig}, with data referring to the WA85/94 reports presented at the January 1995 Quark Matter meeting~\cite{DiBari:1995cy,Kinson:1995cz}. 

We conclude: The new phase of matter reported in February 2000 (see Section~\ref{CERN2000} below) was discovered by the end of 1995, before the arrival of the Pb-beam at CERN SPS (and the new experiment names evolved into NA35/35II/NA49 and WA85/94/97). The Pb--Pb collisions provided the control showing that the sizes of S--S and S--W collision fireball were sufficient. In their writings WA85/WA94 were clear all the time in regard to the QGP interpretation of their results. 

\subsubsection{Particles from a hot fireball}\label{findingQGP}
A first theoretical analysis of the experimental particle production in RHI collisions experiment, focused on strangeness, became possible in late 1990. I presented these results at the February 1991 week-long workshop at CERN; they were published soon after~\cite{Rafelski:1991rh}. In this work, WA85/94 strange baryon and antibaryon particle production data for S--W collisions were used to determine the \lq chemical\rq\ properties of the fireball particle source, {\it i.e.} the chemical potentials $\mu_i$ and phase space occupancy $\gamma_i$. 

An important feature of these results was that despite a large observed baryon number presence in the particle source, the transverse momentum spectra for hyperons (and Kaons) predicted the shape of the pertinent anti-particles. This meant that the hadronization; that is, the particle formation process, was sufficiently fast and occurred late when particle density was low, preventing (partial) elimination by rescattering and annihilation of the low $p_\bot$ anti-particle yield. This motivated the use and further development of the sudden hadronization description of these results. The total particle ratios we study today are independent of the explosive matter flow dynamics. However, in 1990/91 results did not include all transverse momentum $p_\bot$ yields. Thus the focus at the time was on particle ratios rather than yields, evaluated for high range of $p_\bot$.

This work marked the beginning of the development of the statistical hadronization model (SHM), the present day \lq gold\rq\ standard in the study of the hadronization of QGP. In collaboration with my friend Jean Letessier, the full model including the decaying resonances, was completed, allowing many analysis results to be published in 1993/94.\\
 
\noindent {\it A few words from the abstract and conclusion of the February 1991 SHM analysis Ref.\cite{Rafelski:1991rh}:}\\[-0.7cm]
\begin{mdframed}[linecolor=gray,roundcorner=12pt,backgroundcolor=Dandelion!15,linewidth=1pt,leftmargin=0cm,rightmargin=0cm,topline=true,bottomline=true,skipabove=12pt]
\relax
Experimental results on strange anti-baryon production in nuclear S--W collisions at $200\;A$\,GeV are described in terms of a simple model of an explosively disintegrating quark-gluon plasma (QGP). \ldots We have presented here a method and provided a wealth of detailed predictions, which may be employed to study the evidence for the QGP origin of high $p_\bot$ strange baryons and anti-baryons.
\end{mdframed}

Today, we can say that with this 1990/91 analysis method and the WA85 results and claims of the period, QGP was already unmasked. More on this is also seen in a popular review I presented with the spokesman of WA85 Emanuele Quercigh shortly after the CERN announced (February 2000) QGP discovery, see Ref.\,\cite{Quercigh:2000nwx}\\

\noindent{\it Our abstract of June 2000 reads~\cite{Quercigh:2000nwx}:}\\[-0.7cm]
\begin{mdframed}[linecolor=gray,roundcorner=12pt,backgroundcolor=Dandelion!15,linewidth=1pt,leftmargin=0cm,rightmargin=0cm,topline=true,bottomline=true,skipabove=12pt]
\relax
Laboratory experiments have recreated the conditions that existed in the early universe before the quarks and gluons created in the Big Bang had formed the protons and neutrons that make up the world today
\end{mdframed}

Back to the timeline: Seeing the strangeness-WA85/WA94 CERN experiment analysis of early 90s, Marek Ga\'zdzicki from NA35 lobbied me with an inviting remark that continues to reverberate in my memory, \lq\lq \ldots would it not be nice to also apply these methods to other experiments?\rq\rq\ We began the discussion of the data available in NA35. However, I needed more data for the rudimentary SHM to be useful; NA35 was using a photographic method based on hand selected handful of events from a streamer chamber device. Analysis was a time intensive process with human biases, and for most central head-on hits on havy nuclei (central collisions), when particle track density on the photograph was high, this approach was in addition inefficient. 

On this note: The (anti)hyperon experiment NA36 at the time used pioneering time projection chamber (TPC) technology. However, it became mired in technical difficulties and did not deliver the hoped-for results before losing institutional support.

Back to the effort to analyze NA35 data: Marek\rq s and my initial objective, the confirmation data analysis study of the equivalent to WA85 S-Pb reactions, was not possible. However, we soon realized that the lighter collision system S--S experimental results were both sufficiently precise and rich in particles considered, and therefore could be analyzed. Our discussions resulted in an analysis publication of the NA35 S--S $200 A$ GeV collision results~\cite{Sollfrank:1993wn} (submitted in August 1993). Our effort was helped by a young student, \textbf{Josef Sollfrank} from Regensburg, see \rss{ss:SHARE}, introduced to us by his thesis advisor, \textbf{Ulrich Heinz}\label{UHentro}, whose own contribution to the contents of the draft manuscript was the removal of every mention of quark-gluon plasma. \\

\noindent{\it We thus read in the conclusions~\cite{Sollfrank:1993wn}:}\\[-0.7cm]
\begin{mdframed}[linecolor=gray,roundcorner=12pt,backgroundcolor=Dandelion!15,linewidth=1pt,leftmargin=0cm,rightmargin=0cm,topline=true,bottomline=true,skipabove=12pt]
\relax
This (result of analysis, JR) agrees with the notion of common chemical and thermal freeze-out following explosive disintegration of a high entropy source,\ldots
\end{mdframed}
Here read QGP=high entropy source, and see the related manuscript~\cite{Letessier:1992xd} of September 1992.

Just like the earlier WA85 analysis~\cite{Rafelski:1991rh}, this follow-up of the NA35 S--S $200 A$ GeV analysis was fully consistent with our predictions about strangeness production in QGP. At the time NA35 did not have multi-strange particles, which I always viewed to be the unique QGP signature, thus not easily subject to reinterpretation. In Fall 1992, when this work was prepared, multi-strange (anti)hyperon results were alone in the hands of the WA85/94 experiment. 

The following 25 months saw stormy and rapid development of both the experimental results and the related data analysis employing the evolving SHM model. In early 1995 strangeness enthusiasts celebrated the discovery of the QGP, a new phase of matter at a meeting in Tucson~\cite{Rafelski:1995zq}. If I had my present day experience and gravitas I would have staged a press event to announce the discovery of QGP at that event. The series of meeting initiated in Tucson continued, see Refs. \cite{Rafelski:1995zq,S96,SQM97}, and so on to this day.

Here is the reason why I should have announced in the January 1995 QGP discovery:\label{SQM95an} The anti-hyperon results from S--S and S-Pb collisions, obtained at the $\Omega$\rq-spectro\-meter by CERN experiments WA85 and WA94 are seen in \rf{SigLamCERNFig}. We note that across all reaction systems the predicted enhancement growing with the size of collision system and (anti)strangeness content was observed, in \rf{SigLamCERNFig} AFS (see page \pageref{AFSexp}) stands for Axial Field Spectrometer $p$-$p$ experiment at the ISR collider that provides for RHI collision result the baseline, supported by $p$-S and $p$-W results. 

\begin{figure}[tb]\sidecaption
\includegraphics[width=0.66\columnwidth]{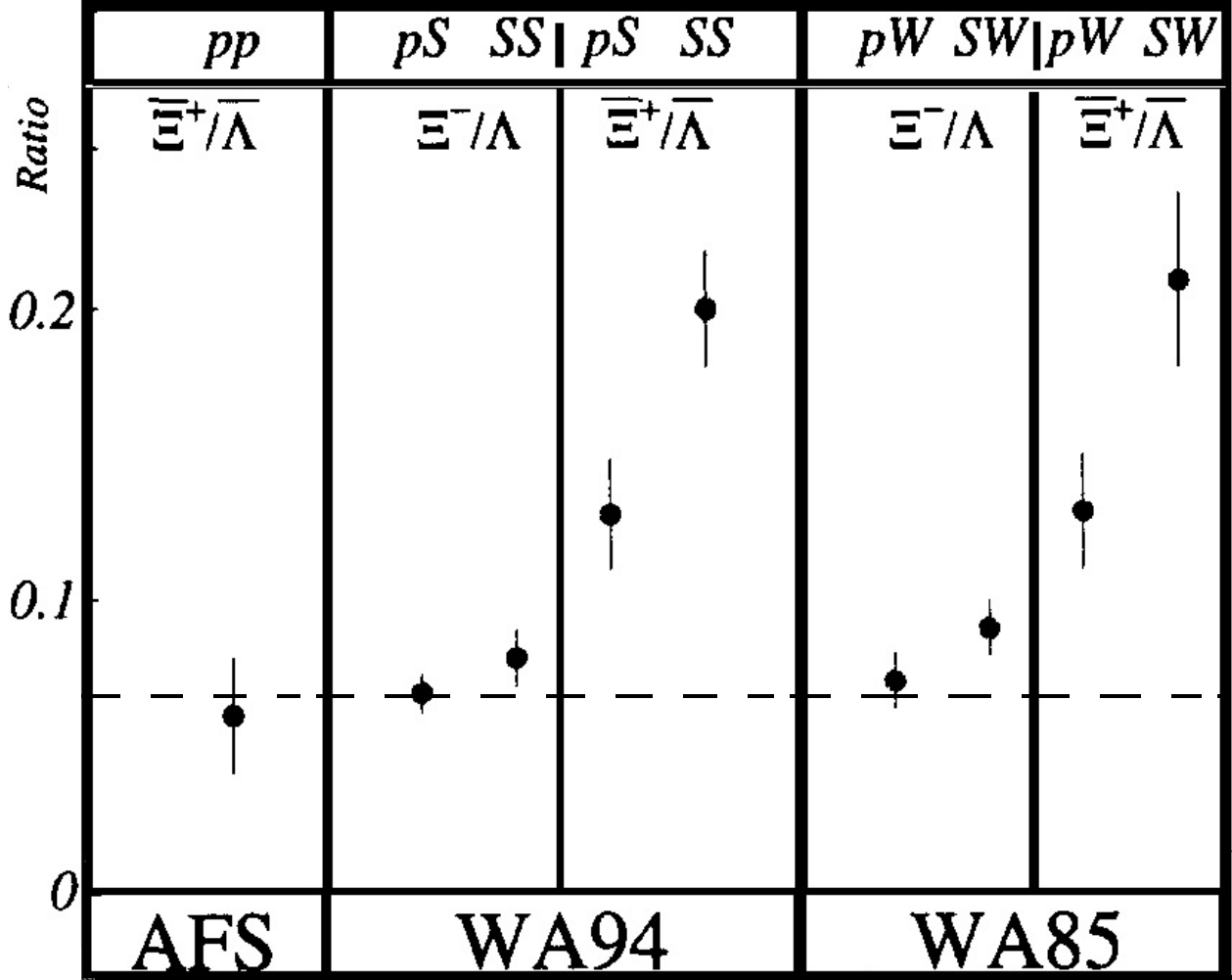}
\caption{Results obtained at the CERN-SPS $\Omega\rq$-spectrometer for $\Xi/\Lambda$-ratio in fixed target S--S and S-Pb at 200\,$A$\,GeV/$c$; results from the compilation presented in Ref.\cite{Omega25y}}\label{SigLamCERNFig}
\end{figure}

While the results in \rf{SigLamCERNFig} were compiled for the $\Omega$\rq-spectro\-meter March 19, 1997~\cite{Omega25y} 25th anniversary celebrations, they were for all to see before. To this day there is only one explanation of the large $\overline{\Xi}/\overline{\Lambda}$ ratio: {\bf quark recombinant sudden hadronization of QGP}, see Section~\ref{sec:highPT}. This model was proposed in Ref.\cite{Koch:1986ud} in 1986, and used in my following work since.

With arrival of the Pb--Pb collisions the WA85/94 experiment was redesignated WA97 (and evolved with new technologies into NA57 moving to CERN North Area when LHC beam preparations were underway). The first results from Pb--Pb collisions were reported by WA97 in December 1997 at the QM1997 conferencce~\cite{QM97}, see \rf{RSS}. We see that the enhanced $\overline{\Xi}/\overline{\Lambda}$ ratio is confirmed. The reference horizontal dashed line based on CLEO, MARKII, HSR, TPC, TASSO, UA5, and AFS experiments with collisions of $e^+$-$e^-$ and $p$-$\bar p$ is symmetric between matter and antimatter. The RHI collision data now includes NA35/NA35/NA49 results as well.

\begin{figure}[tb]\sidecaption
\includegraphics[width=0.66\columnwidth]{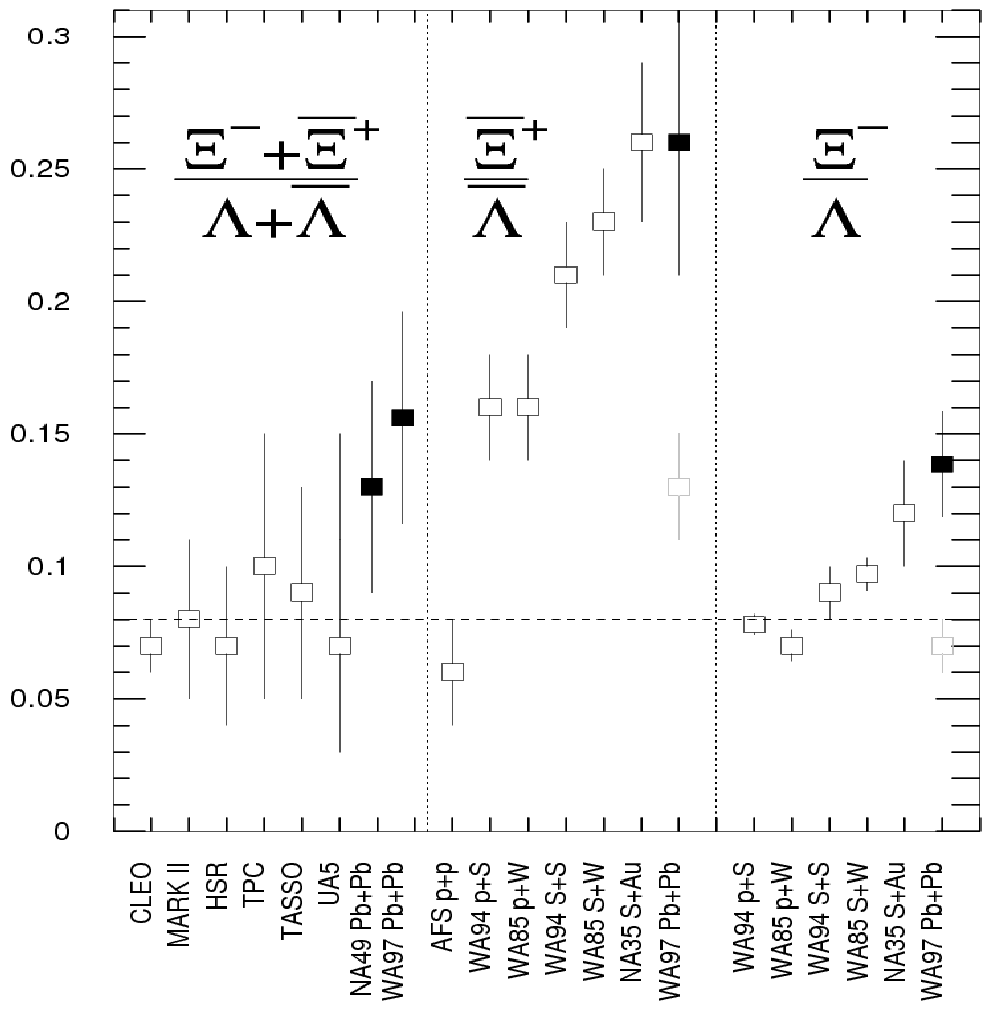}
\caption{
Sample of World results for strange (anti)baryon ratios available by the end of 1997 (as a \lq function\rq\ of experiment name). Dark squares: first 1997 Pb--Pb WA97 results announced in December 1997 Tsukuba Quark Matter conference~\cite{QM97} \label{RSS}
}
\end{figure}

The important further experimental development, in my opinion triggering the CERN February 2000 announcement (see next section), was the measurement of more than an order of magnitude $\Omega(sss)+\overline\Omega(\bar s\bar s\bar s)$ production enhancement in Pb--Pb collisions by WA97 (and confirmed by NA57 collaboration, we return to these results, see page \pageref{fig:NA57}). 

In the following years the theory evolved as well, allowing by means of SHM analysis the understanding how the QGP bulk properties depended on the size of the interaction volume. However, the process of theory testing has also set in, creating transient disarray. In the process one usually sees proposals aiming to explain within a new framework a subset of the experimental results. The reader should remember:
\begin{itemize}
\item
Enrico Fermi\rq s words: \lq the most beautiful theory is proven wrong by just one experimental result\rq.
\item 
New ideas put forward after the data is known must await controle experiments. 
\end{itemize}
This has of course been now done for QGP strangeness signature in several iterations and I intentionally do not review here \lq exotic\rq\ ideas where in the end at least one relevant data set must be excluded from analysis in order to keep this idea afloat. 

In the process of theory verification there is yet another complication that can occur, an error that another group can make in data analysis. I illustrate this problem  by looking at the discovery of full flavor chemical non-equilibrium among produced hadrons: In mid-1998 \textbf{Jean Letessier} and I recognized that the more complete S-W results available at that time would be better described allowing for all quark flavors in the hadron phase that emerges from the QGP fireball chemical nonequilirbrium 

This means that not only strange, but also light $u,d$ quark (and antiquark) yields~\cite{Letessier:1998sz} may, at the time of chemical freeze-out, need to be fitted allowing for a non-equilibrium parameter. When we speak of quarks in hadron phase we refer here to the count of valence quarks in all produced hadrons. Such a situation would be expected even if on the \lq other side\rq\ in the QGP complete chemical equilibrium prevailed. This is so since the size of the phase space differs comparing two differently structured phases. 

The reason that this idea did not enter into earlier consideration was that one expects a light quark yield to adjust more easily to chemical equilibrium. However, the equilibrium result should be an outcome of an analysis and not an input. Since the wealth of experimental data increased we attempted this test. The result was astonishing to us but there was no doubt.

We expected that the rest of the world would follow, applying our new insight which emerged generalizing the accepted strange quark yield (chemical) nonequilibrium model. I recall vividly a comment by Sollfrank who helped in the NA35 S-S data analysis, see page \pageref{UHentro}, and now was with another theory group: \lq We cannot fit the data your way, our fits become unstable when we allow full (chemical) nonequilibrium\rq. The reason slowly emerged: the numerical analysis programs used by other groups were not easily adaptable, and/or some had bugs which were innocent in one case but grossly disruptive in another. 

Sometimes one can see the error with naked eyes, see \rf{PBMletter}. I believe that in Heidelberg the assistant writing the analysis code did not know that for historical reasons the strangeness of a baryon is \emph{negative}. The leader of the analysis group checked and confirmed by return fax, concluding \lq our error does not matter.\rq\ What he was saying, and I agree, was that within the realm of this particular publication given the precision of experimental results it indeed did not matter if strangeness of hyperons is negative, as it should be, or positive as it was apparently used.

\begin{figure}[tb]
\includegraphics[width=\columnwidth]{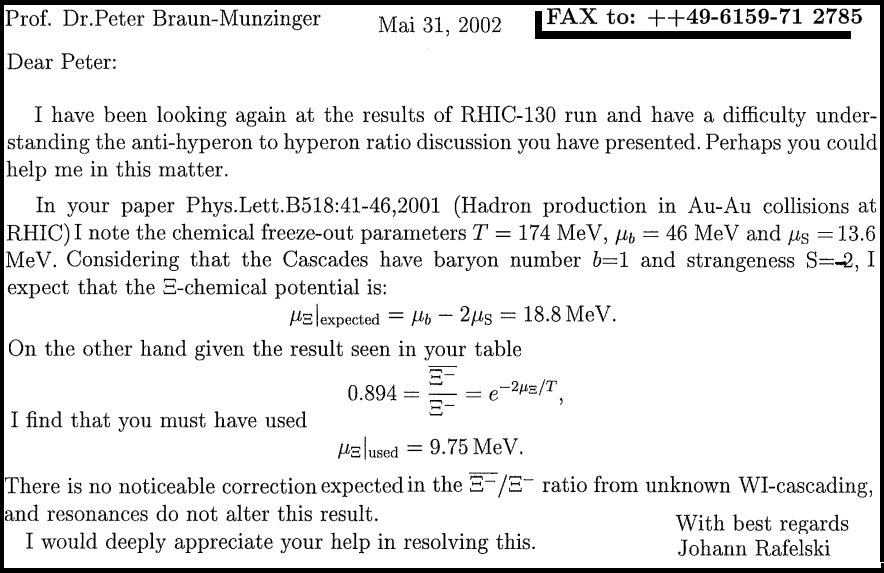}
\caption{
A FAX alert of an analyzis error, see text \label{PBMletter}
}
\end{figure}

However, the admitted error does matter in other ways. Once corrected, meaning all baryons are included with negative strangeness, the group would realize just as we did that the full nonequilibrium fit would produce a more significant description of the SPS and RHIC data available. Moreover, the value $T=174$\,MeV of the freeze-out temperature reported, was as we know today entirely wrong and it distracted for years from the reality that the fitted Temperature should be below $T=150$\,MeV. We return to this matter in \rss{ss:SHARE}; note that this fit corresponds to the second most aberrant value of $T$ which we will meet in this discussion in \rf{TEcol}. So at least for this one case we know the cause. 

Jumping ahead, we found funding to create the SHARE project, see \rss{ss:SHARE}, which provided a standarized SHM program. One could think that this would resolve the errors theory students can make when working with little supervision among experimentalists. This expectation turned out to be unrealistic. After a few years the computer programming problem returned with a wrong SHM method being used again. This time it was how SHM was incorporated into the widely used CERN ROOT platform discussed in \rss{ss:SHARE}. 

To conclude, the study of QGP in terms of strangness and strange antibaryons has suffered from \lq beyond the call of duty\rq\ scrutiny Theoretical studies to explore alternative explanations were using incomplete data sets. Analysis programs used unverified numerical approaches.

\subsubsection{The CERN February 2000 announcement}\label{CERN2000}

By the end of the last century, towards the end of the SPS Pb-Pb run, experimental results demonstrated that the enhanced production of strange antibaryon is the same irrespective of colliding systems and the collision energy as long as the size and lifespan of the fireball is tuned to assure that the fireball of matter is created in comparable conditions. It took two years after the Tsukuba conference~\cite{QM97} in December 1997 for CERN to reach consensus and to announce the evidence for QGP formation in the context of SPS relativistic collisions at a press conference~\cite{CERNPress} held on 10 February, 2000.\\

\noindent \textit{In an interview~\cite{MaianiInterview} in January 2017 with \textbf{Luciano Maiani}, Director General of CERN from 1999 to 2003 we read:}\\[-0.7cm]
\begin{mdframed}[linecolor=gray,roundcorner=12pt,backgroundcolor=GreenYellow!15,linewidth=1pt,leftmargin=0cm,rightmargin=0cm,topline=true,bottomline=true,skipabove=12pt]\relax%
\vskip 0.0cm
\hspace*{0.4cm} \textbf{Virginia Greco:} But the public announcement \textit{(you made in February 2000 as CERN DG, JR)\/} was cautious, wasn\rq t it? Was there still some doubt?

\textbf{Luciano Maiani:} I think that the announcement was quite clear. I have the text of it with me, it reads: \lq\lq The data provide evidence for color deconfinement in the early collision stage and for a collective explosion of the collision fireball in its late stages. The new state of matter exhibits many of the characteristic features of the theoretically predicted Quark-Gluon Plasma.\rq\rq\ The key word is \lq\lq evidence,\rq\rq\ not discovery, and the evidence was there, indeed.\ldots

\textbf{Virginia Greco:} The announcement came just a few months before the start of the programme of RHIC. Were there some polemics about this \lq\lq timing?\rq\rq\

\textbf{Luciano Maiani:} We were almost at the conclusion of a long and accurate experimental programme at the SPS, so making a summing up was needed. In addition, as I said, we thought there were the elements for a public announcement. And this has been proved right by later experiments.\ldots
\end{mdframed}

Reading these remarks I recall the RHIC timeline: the first physics Au-Au collisions at RHIC were recorded in June 2000, see for example Ref.\cite{Back:2000gw}. However, this happened after a long RHIC commissioning. Some readers could recall that the start-up difficulties of RHIC delayed the Quark Matter conference to mid January 2001; The organizers moved the meeting from the prior agreed to a nine-month later date in 2001. At this meeting Gordon Baym presented his view, see also page \pageref{BearMountain}, on the history how RHIC came to be.

During the long commissioning period of RHIC I heard that it was possible RHIC could never deliver Au-Au collisions. Some even suggested RHIC could be abandoned, a failed project. At that time it seemed wise for CERN to move towards QGP annoucement supporting this research program. Had the RHIC start-up been on schedule, perhaps a common BNL-CERN announcement could be made. 

Recalling this environment it is evident that CERN was coasting on its own inertia towards the QGP announcement, perhaps also aiming to strengthen the reseach field in the process. This announcement also, as Luciano Maiani explained, was set after the experiments exploring the full reach of SPS were complete. This also emerges reading the detailed timelineof the activities at CERN reported in 2008~\cite{Heinz:2008ds}.\\

Let us now look at some of the context of the CERN announcement:

\noindent \textit{Maurice Jacob before Pb--Pb CERN-SPS run was underway had set up in Summer 1996 his views in a conference report (submitted for publication on 22 July 1996)~\cite{Jacob:1991pb}:}\\[-0.7cm]
\begin{mdframed}[linecolor=gray,roundcorner=12pt,backgroundcolor=GreenYellow!15,linewidth=1pt,leftmargin=0cm,rightmargin=0cm,topline=true,bottomline=true,skipabove=12pt]\relax%
\textbf{The quest for the quark-gluon plasma}\label{Jacob1996}\\
{\bf Abstract:} Heavy-Ion Collisions at very high energy offer conditions such that QGP could be formed. \ldots New rounds of tests using the heaviest ions \ldots are proceeding. The present situation is assessed.\\

 \textbf{p4952:}\\ 
 \indent We can now look for signals\ldots J/$\Psi$ suppression has been observed \ldots\ \lq fac\-tor of two\rq\ effect \ldots one could not refrain from attempting to explain it differently \ldots evidence for \ldots very dense system \ldots evidence for something new.\ldots\\

Strangeness enhancement has long been advocated by Rafelski as evidence for (QGP)\ldots again typically a \lq factor of two\rq\ effect \ldots This is nevertheless \ldots tool to probe further, studying in particular the production of strange antibaryons. \ldots Much should be learned in that case from the increase in volume associated with lead beams.\\

\noindent {\bf Conclusions:}\\
\indent Exciting signals have been seen. The prominent ones are offered by quarkonium, \ldots strangeness production with enhancement seen \ldots information provided by interferometry\ldots There is no doubt that a new state of matter, with density of at least an order of magnitude higher than hadronic matter, is created. \ldots we have good proven tools now at hand, we can expect much from an increase in volume\ldots
\end{mdframed}


\noindent \textit{CERN document prepared for the CERN QGP February 2000 event~\cite{Heinz:2000ba} is quoted in Ref.\cite{Heinz:2008ds} as follows:}\\[-0.7cm]
\begin{mdframed}[linecolor=gray,roundcorner=12pt,backgroundcolor=GreenYellow!15,linewidth=1pt,leftmargin=0cm,rightmargin=0cm,topline=true,bottomline=true,skipabove=12pt]\relax%
\label{CERN2000} \ldots compelling evidence now exists for the formation of a new state of matter at energy densities about 20 times larger than that in the center of atomic nuclei and temperatures about 100\,000 times higher than in the center of the sun. This state exhibits characteristic properties which cannot be understood with conventional hadronic dynamics but which are consistent with expectations from the formation of a state of matter in which quarks and gluons no longer feel the constraints of color confinement.
\end{mdframed}
\vskip 0.5cm


This document was coordinated and prepared by by Maurice Jacob and also signed by a CERN theorist,  mentioned earlier in the context of our collaboration on S-S NA35 data analysis, see page \pageref{UHentro}. The second author gives his personal assesement in Summer 2000:\\

\noindent \textit{The following remarks arXiv\rq ed September 2000~\cite{Heinz:2000ba} were presented at the 7th International Conference on Nucleus-Nucleus Collisions (NN 2000) 3-7 July 2000 in Strasbourg, France:}\\[-0.7cm]
\begin{mdframed}[linecolor=gray,roundcorner=12pt,backgroundcolor=GreenYellow!15,linewidth=1pt,leftmargin=0cm,rightmargin=0cm,topline=true,bottomline=true,skipabove=12pt]\relax%
\label{Heinz2000} \ldots what is missing to claim \lq\lq discovery\rq\rq\ of the quark-gluon plasma? \\
1st: on the theoretical side, we only know that with known hadronic physics we can not describe the data, but \ldots description of strongly interacting matter and its dynamics in the neighborhood
of the phase transition is an exceedingly difficult problem \ldots \lq\lq new physics\rq\rq\ has so far not received enough careful theoretical attention \\
\ldots important experimental questions which can be answered at the SPS (and in a few cases only there) are: Assuming that we have seen quark deconfinement, where is its energy threshold? How big does the collision system have to be to establish approximate thermal equilibrium and strangeness saturation and to exhibit collective flow? \\
\ldots Some answers will be provided by data already collected at lower beam energies and with smaller nuclei and more peripheral collisions.\ldots\\
 A detailed characterization of the \lq\lq new state of matter\rq\rq\ will only be possible when the larger initial energy densities and resulting longer plasma lifetimes before hadronization provided by RHIC and LHC become available. 
\end{mdframed}
\vskip 0.5cm

These words were spoken by the only member of the CERN Theory Division (TH) conducting research in the field of relativistic heavy ion collisions  after Maurice Jacob\rq s retirement April 1998. I interpret these demands as follows:
\begin{enumerate}
\item 
More theory to build on the 20 years 1980-2000 effort: I believe that  the need for a framework to explore the quark matter flow, a topic very important to the Frankfurt school of \textbf{Walter Greiner} was here in play. The nuclear matter bounce now referred to as $v_1$ was the trademark of \textbf{Horst St\"ocker}, and there was the opportunity to study the azimuthal flow $v_2$  characterizing the transverse fireball matter explosive flow, introduced as early as 1992 in the Ph.D. thesis of \textbf{Jean-Yves Ollitrault}\cite{Ollitrault:1992bk} from Paris.
\item
More experimental verification at SPS of the CERN evidence for QGP: After the CERN announcement the strangeness SPS-QGP research program was indeed continued with experiments NA57, and NA61.
\item  The verification of CERN results at the RHIC collider: this effort was delayed by a decade due to the \lq 100 mile BNL horizon group of theorists\rq\ conviction that CERN results were unworthy of confirmation, for more details see next quote. 
\end{enumerate}

From the above we can clearly learn that this author  does not know if CERN did or not make a discovery. Moreovr, he presents in 2008 an eulogy on Maurice Jacob~\cite{Heinz:2000ba} -- for a eulogy, the contents is unusually critical of the seminal role of Maurice Jacob in the QGP discovery.  

In 1999 I was observing QGP discussions at CERN mostly from afar. However, I am sure in view of (mostly phone) interactions  see Sec.\ref{PRL2000}, that there was this  one profoundly negative voice working against CERN announcing the QGP discovery. This maybe indeed the reason why the Director General of CERN called on an already retired Maurice Jacob in early 1999 to help advance the QGP discovery announcement. 

Nearly 20 years later we find another retrospective view on the QGP discovery at CERN. It is convenient that the UniReport, a periodic of the J.W.v. Goethe University, Frankfurt, published a feature containing a transcript of a conversation  that directly relates to this matter. This article was triggered by a research visit in Frankfurt in the Summer/Fall 2019 and is signed by \href{http://www.muk.uni-frankfurt.de/36152080/anne_hardy}{Dr Anne Hardy}, a communication director at University Frankfurt, who specialized in Physics.\\ 

\noindent \textit{In Frankfurt JWGoethe University UniReport of July 11, 2019 (No. 4 issue) p.6, an article entitled\footnote{Original: Als das Universum die Gr\"o\ss e einer Melona hatte} \textit{Looking back at the Universe the Size of a Melon} we read\footnote{German: \lq\lq Man hatte bereits Mitte der 1990er Jahren Hinweise f\"ur das Quark-Gluon-Plasma in Schwerionen-Experimenten am CERN und auch am Brookhaven National Laboratory gefunden\lq\lq, erkl\"art Heinz. \lq\lq Aber wir waren damals, aufgrund der noch etwas fragmentarischen Datenlage, in unseren Schlussfolgerungen sehr vorsichtig. R\"uckblickend wissen wir, dass wir zu vorsichtig waren.\rq\rq\ Das stellte sich heraus, nachdem vor etwa 10 Jahren am RHIC Sto\ss experimente auch bei niedrigeren Energien durchgef\"uhrt wurden, um die \"alteren Experimente am CERN-SPS (Super Proton Synchrotron) zu \"uberpr\"ufen und weiter zu verbessern. \lq\lq Eigentlich hoffte man mit dieser Prozedur das Quark-Gluon Plasma kontrolliert abzuschalten, aber dieser Versuch misslang. Auch bei SPS-Energien zeige es immer noch seine (in den nun viel umfangreicheren Datens\"atzen unmissverst\"andlichen) 
Signaturen\,!\rq\rq, so Heinz.}:}
\begin{mdframed}[linecolor=gray,roundcorner=12pt,backgroundcolor=GreenYellow!15,linewidth=1pt,leftmargin=0cm,rightmargin=0cm,topline=true,bottomline=true,skipabove=12pt]\relax%
{\bf Heinz explains:}\label{Heinz2019} Already in the mid-90s there was some indication for quark-gluon plasma in heavy ion experiments at CERN and at Brookhaven National Laboratory. However, I was at that time due to fragmentary data very cautious. In hindsight I know my position was too cautious.\\ 
{\it \textbf{Anne Hardy:} This insight surfaced when about 10 years ago lower-energy collider experiments were performed at RHIC to test and further improve the older experiments at the CERN-SPS (Super Proton Synchrotron).}\\
{\bf Heinz continues:} Actually, I hoped that this procedure would switch off the quark-gluon plasma in a controlled manner, but this attempt failed. Also at SPS energies there are in the now much more extensive data records unmistakable signatures (of the QGP formation, JR).
\end{mdframed}

In the above translation I interpreted the so called royal \lq we\rq\ as \lq I, my\rq\ given that he comments on his own position in regard to the CERN QGP announcement of February 2000. We note that he is referring to \lq fragmentary results\rq\ at CERN. This wording is a mystery to me: I presented strange antibaryons results aboved; they were in my opinion not fragmentary. Nothing changed in these results in the following 20 years -- many more results were obtained corroborating these.

The phenomenon called \lq reflection\rq\ maybe at the origin of the use of the words \lq fragmentary\rq: This researchers work on QGP was at that time of CERN QGP announcement just that.  This was so since he joined this field relatively late, driven into this research area by the NATO collaborative grant with me, see page \pageref{NATOfun}. 

In the book \textit{Quark-Gluon Plasma: Theoretical Foundations -- An Annoted Reprint Collection} prepard in 2002/ by Berndt M\"uller, \textbf{Joseph Kapusta} and myself~\cite{Rafelski:2003zz} one does not encounter a work by Heinz: This book introduces the theoretical roots of QGP with a time cut-off in 1992. The rationale of the authors was to look more than 10 years back in 2002/3, since in 1992 QGP was already discovered but recognized only by a small subset of researchers, see Sec.~\ref{findingQGP}. This 2002/3 volume celebrates 10 years of QGP discovery, unofficially, of course.

To conclude: As the Summer 2019 conversation in Frankfurt shows, the one strong CERN voice  against the discovery announced by CERN in February 2000 evolves. We see  \lq mea-culpa\rq\ words, and recognition that CERN discovered QGP at the SPS energy range. This co-author of the February 2000 CERN announcement, who opposed it in the following months and years finally realized his mistake. In my interpretation of these words only when RHIC reduced the collision energy to the CERN-SPS domain, and was connecting the strangeness signatures for QGP previously seen at CERN with the nearly ideal matter (quark) flow now measured at RHIC, he became convinced that SPS had indeed discovered QGP many years earlier. 

\subsubsection{SQM2000 Meeting in Berkeley}\label{Berk2000}

The first  SQM meeting  after the CERN announcement took place in late July at Berkeley, see picture on page \pageref{SQM2000Group}.  My written summary~\cite{Rafelski:2001rj}   was received on November 2, 2000 -- due to personal events, recognized on the front page an arXive version of this work does not exist. In this short summary, I pointed out the  need to work in support of the CERN QGP announcement, which was primarily carried by the strangeness results. 

I then introduced the important results obtained now to a great precision by the WA97 collaboration~\cite{Antinori:2000sb} very shortly before the CERN announcement shown in \rf{fig:WA97Spectra}, also presented at the meeting~\cite{Antinori:2001qk}. We note that the transverse slopes for the four collision centralities considered, and the four particles $\Lambda(uds)$, $\overline\Lambda(\bar u\bar d\bar s)$, $\Xi^-(dss)$, and $\overline\Xi^+(\bar d\bar s\bar s)$ are the same. This means that the production mechanism of these particles is the same,  independent to a large degree of the quark content or  matter-antimatter nature of these particles. 

\begin{figure}\sidecaption
\centerline{
\includegraphics[width=0.75\textwidth]{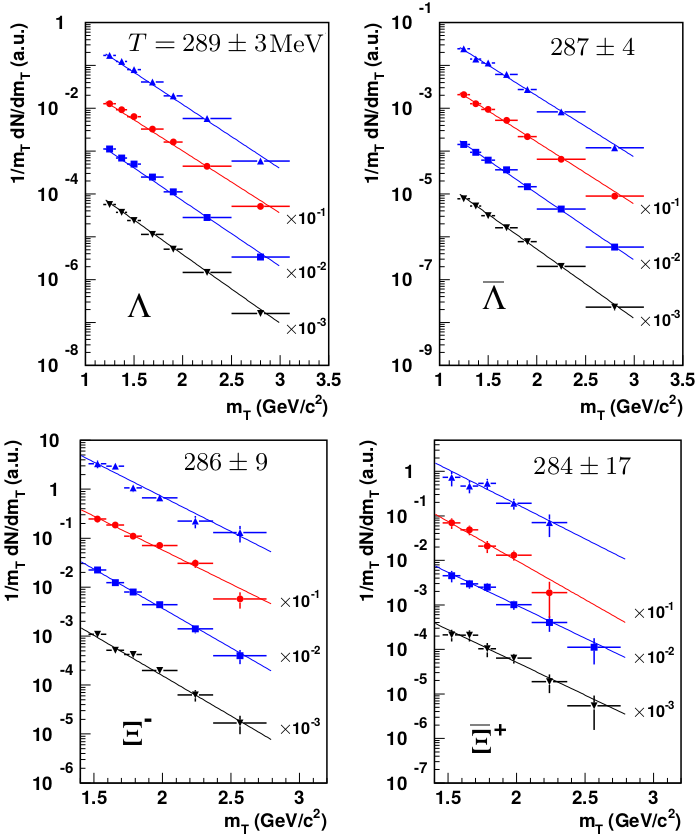}
}
\caption{The final Pb--Pb results for hyperon spectra by WA97 adapted from~\cite{Antinori:2000sb} 
}
\label{fig:WA97Spectra}
\end{figure}

This result was  direct evidence for an exploding fireball source. The  light and strange quarks and antiquarks participated in the same fashion in the formation of these particles. No rescattering after formation was visible, as the spectra of antimatter particles were just like those of matter and these particles were born in relatively baryon rich environment. These remarks were confirmeed in a more detailed study~\cite{Torrieri:2000xi} arXived a few months later. \textbf{Giorgio Torrieri}, who had just started his Ph.D. program in Tucson, demonstrated the single freeze-out of hadrons: thermal and chemical decoupling was found at the same temperature near to $T=140$ MeV, with the speed of expansion near $v_f=0.55c$, this study included all other particles presented by the WA97 collaboration, omitted  in \rf{fig:WA97Spectra}.

A   thorough least-square deviation fit analysis  was carried out by Wojtek Broniowski and Wojtek Florkowski~\cite{Broniowski:2001uk}. They also invented the appropriate name \lq single freeze-out model\rq.

\noindent{\it In conclusion we read~\cite{Rafelski:2001rj}:}\\[-0.7cm]
\begin{mdframed}[linecolor=gray,roundcorner=12pt,backgroundcolor=Dandelion!15,linewidth=1pt,leftmargin=0cm,rightmargin=0cm,topline=true,bottomline=true,skipabove=12pt]
\relax
I believe that the diligent work of CERN experimental collaborations such as WA97
and NA49 regarding hadronic and, in particular, strange particle production has clearly demonstrated the formation of a new state of matter. Considering that the results obtained are in agreement with the predictions made some 20 years ago, and the current analysis of experimental results, there is good reason to believe that this new phase is most probably the quark–gluon plasma. However, \textbf{some of the qualitative theoretical arguments} put forward in the (CERN) consensus report  are deeply flawed and thus obstruct the development of an understanding of the experimental results.  
\end{mdframed}
\vskip 0.5cm

One must see here the word  \lq flawed\rq: some of the presented  arguments were  conflicting  with the model of a fireball of QGP hadronizing out of chemical equilibrium, a reaction picture as clearly favored then and now by these original and beautiful WA97 experimental results. These flowed qualitative theoretical arguments are further discussed in Sec.\ref{PRL2000}.

\subsubsection{BNL announces ideal flow}\label{ssec:flow}
At the April 2005 meeting of the American Physical Society, held in Tampa, Florida, a press conference took place on Monday, April 18, 9:00 local time. At this event I made a few amateur pictures, a sample is shown below, probably the only photographic record of the event.\\

\noindent \textit{The \textbf{BNL public announcement} of this event was available as of April 4, 2005:}\\[-0.7cm]
\begin{mdframed}[linecolor=gray,roundcorner=12pt,backgroundcolor=GreenYellow!15,linewidth=1pt,leftmargin=0cm,rightmargin=0cm,topline=true,bottomline=true,skipabove=12pt]\relax%
\textbf{Evidence for a new type of nuclear matter:}\label{RHIC2005}
At the Relativistic Heavy Ion Collider (RHIC) at Brookhaven National Lab (BNL), two beams of gold atoms are smashed together, the goal being to recreate the conditions thought to have prevailed in the universe only a few microseconds after the big bang, so that novel forms of nuclear matter can be studied. At this press conference, RHIC scientists will sum up all they have learned from several years of observing the world’s most energetic collisions of atomic nuclei. The four experimental groups operating at RHIC will present a consolidated, surprising, exciting new interpretation of their data. Speakers will include: Dennis Kovar, Associate Director, Office of Nuclear Physics, U.S. Department of Energy's Office of Science; Sam Aronson, Associate Laboratory Director for High Energy and Nuclear Physics, Brookhaven National Laboratory. Also on hand to discuss RHIC results and implications will be: Praveen Chaudhari, Director, Brookhaven National Laboratory; representatives of the four experimental collaborations at the Relativistic Heavy Ion Collider; and several theoretical physicists.\\

\centerline{\includegraphics[width=1.0\textwidth]{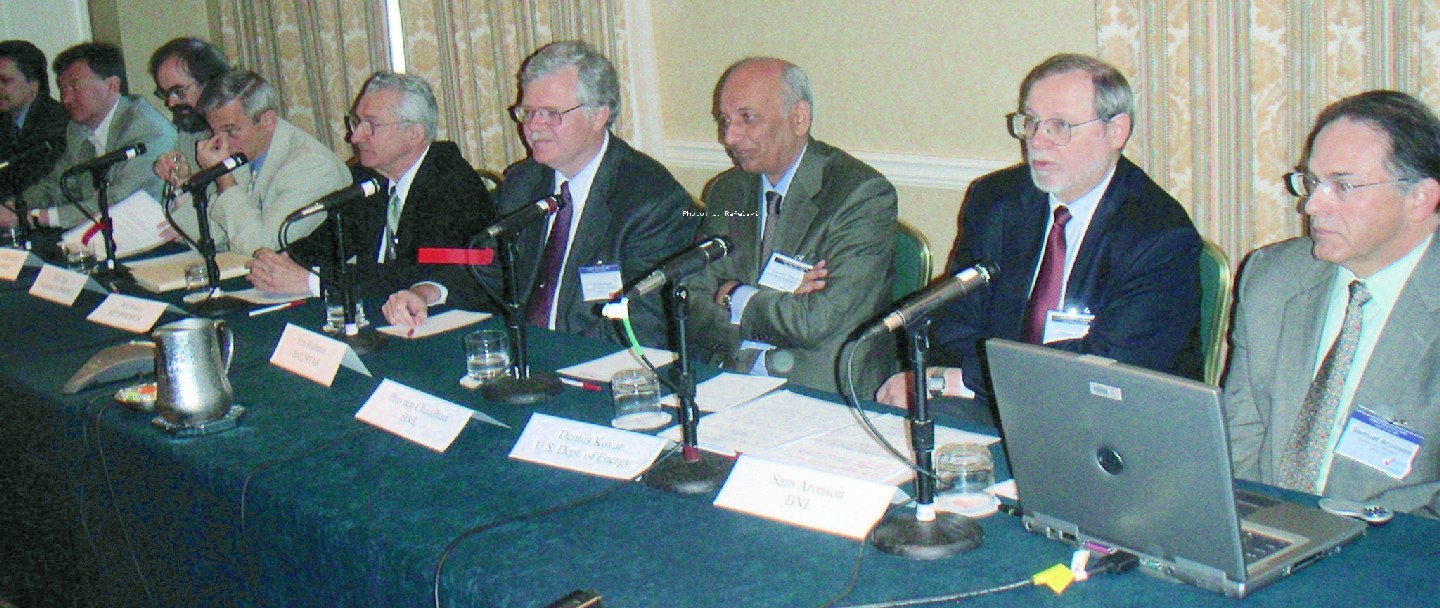}}
\vskip-35pt\phantom{.}\hfil {\color{yellow}{\small Photo: J. Rafelski\hspace*{1cm}}}\\
\centerline{\includegraphics[width=1.0\textwidth]{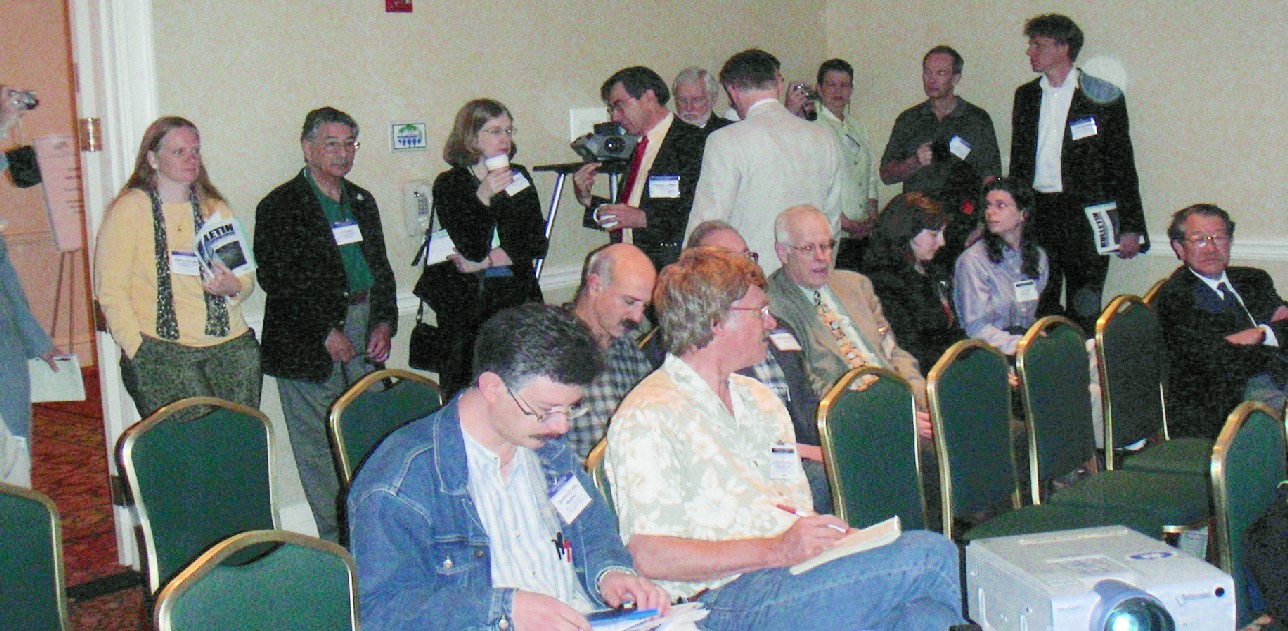}}
\vskip-24pt\phantom{.}\hfil{\color{yellow}{\small Photo: J. Rafelski}}\\
\end{mdframed}
\vskip 0.5cm

I cannot recall at this announcement any mention of the five year earlier CERN presentation of the QGP discovery. Nor do we see in the above pictures the theorist critical of CERN who since moved to USA: this I cannot understand easily as much of the BNL discovery is today claimed by Heinz to be his doing. Akin to the CERN press release of February 2000, the BNL press announcement speaks of \lq a new type of nuclear matter\rq\ (compare to CERN\rq s \lq new state of matter\rq). The participants at this event received a \textit{Hunting for Quark-Gluon Plasma} report introducing the four BNL experiments operating at the time: BRAHMS, PHOBOS, PHENIX, and STAR, which reported on BNL results obtained since Summer 2000. These four experimental reports were later published in an issue of Nuclear Physics A~\cite{Arsene:2004faB,Adcox:2004mhB,Back:2004jeB,Adams:2005dqB}. We note that the report title does not say \textit{Discovery of Quark-Gluon Plasma}. So what was discovered? There was a retrospective review of this situation at BNL:\\ 

\noindent \textit{At the RHIC Users' Meeting June 9-12, 2015 a 10 year anniversary session \lq\lq The Perfect Liquid at RHIC: 10 Years of Discovery\rq\rq\ was held, \href{https://www.bnl.gov/newsroom/news.php?a=25756}{the press release of June 26, 2015 says}:}\\[-0.7cm]
\begin{mdframed}[linecolor=gray,roundcorner=12pt,backgroundcolor=GreenYellow!15,linewidth=1pt,leftmargin=0cm,rightmargin=0cm,topline=true,bottomline=true,skipabove=12pt]\relax%
\lq\lq RHIC lets us look back at matter as it existed throughout our universe at the dawn of time, before QGP cooled and formed matter as we know it,\rq\rq\ said Berndt Mueller, Brookhaven\rq s Associate Laboratory Director for Nuclear and Particle Physics. \lq\lq The discovery of the perfect liquid was a turning point in physics, and now, 10 years later, RHIC has revealed a wealth of information about this remarkable substance, which we now know to be a QGP, and is more capable than ever of measuring its most subtle and fundamental properties.
\end{mdframed}

The press notice of 2015 says that BNL in 2005 was reporting a single specific property of a new form of nuclear matter, which could have been Lee-Wick matter, see Sec.~\ref{sssec:dense}. However, the BNL actors decided in 2015, it was QGP. The obvious questions here are:\\
\begin{itemize}
\item
Why and how is the ideal flow of matter evidence of QGP?
\item
Why is ideal flow worth scientific attention 15 years after the new phase of matter was announced for the first time at CERN?
\end{itemize}

\noindent \textit{\textbf{Berndt M\"uller}, Brookhaven\rq s Associate Laboratory Director for Nuclear and Particle Physics, responded to me as follows:}\\[-0.7cm]
\begin{mdframed}[linecolor=gray,roundcorner=12pt,backgroundcolor=GreenYellow!15,linewidth=1pt,leftmargin=0cm,rightmargin=0cm,topline=true,bottomline=true,skipabove=12pt]\relax%
Nuclear matter at \lq room temperature\rq\ is known to behave like a superfluid. When heated the nuclear fluid evaporates and turns into a dilute gas of nucleons and, upon further heating, a gas of baryons and mesons (hadrons). But then something new happens; at critical temperature $T_\mathrm{H}$ hadrons melt and the gas turns back into a liquid. Not just any kind of liquid. At RHIC we have shown that this is the most perfect liquid ever observed in any laboratory experiment at any scale. The new phase of matter consisting of dissolved hadrons exhibits less resistance to flow than any other substance known. The experiments at RHIC have a decade ago shown that the Universe at its beginning was uniformly filled with a new type of material, a super-liquid, which once Universe cooled below $T_\mathrm{H}$ evaporated into a gas of hadrons. 

Detailed measurements over the past decade have shown that this liquid is a quark-gluon plasma; \textit{i.e.\/} matter in which quarks, antiquarks and gluons flow independently. There remain very important questions we need to address: What makes the interacting quark-gluon plasma such a nearly perfect liquid? How exactly does the transition to confined quarks work? Are there conditions under which the transition becomes discontinuous first-order phase transition? Today we are ready to address these questions. We are eagerly awaiting new results from the upgraded STAR and PHENIX experiments at RHIC.
\end{mdframed}
%
We note: In the first part of the response Berndt invokes \textit{ex officio} and without introducing the work at CERN the outcome: \lq consisting of dissolved hadrons\rq. Another \textit{ex officio} statement in this paragraph is connecting RHIC to early Universe. This postulates the principle that we can in a local domain of space-time recreate the molten vacuum structure that filled \textbf{ALL} the Universe once upon a time, granting the big-bang scenario, another CERN claim. All told, he says, that ideal flow of matter is evidence for QGP because this form of primordial matter is already known to exist. Reading the first paragraph in this way the 2nd is a meaningful answer to my 2nd question.

Back to the timeline: The BNL quark matter ideal flow announced in April 2005 was neither made nor presented in the context of QGP actual discovery, and would not be related to the QGP form BNL was \lq hunting\rq\ in the following years. In his above note Berndt did not tell when BNL reached its institutional decision to accept CERN announcement as a valid scientific discovery announcement. However, he shares my view that the nearly ideal flow by quarks including in particular strange quarks supports and confirms the CERN-SPS strangeness enhancement signature. This is so since the SPS signature relies on the independent presence of quark and gluon degrees of freedom in the dense matter fireball. 
 
We see that while no other (than QGP) global and convincing explanation of the SPS strangeness and strange antibaryon QGP signature results was ever presented, the direct observation of free motion of quarks helps to reduce the psychological resistance that even today is hindering the acceptance of the QGP discovery. At this point let me step back in history a few years to better understand the reasons why QGP discovery even today is argued about. 

\subsubsection{Can QGP be experimentally recognized?}
\label{canQGP}
\label{subsubsec:hoho}
It is hard if not impossible to find someone directly involved in the quark-gluon plasma research program doubting the experimental results and their theoretical interpretation in terms of the properties of a new phase of matter comprising highly mobile deconfined quarks, the phase we call quark-gluon plasma. Nevertheless, in the last 5 years I found one person within the group of 1000 researchers making up the ALICE collaboration denying QGP was already discovered. This is the proverbial exception to the rule. 

However, books addressing particle and/or nuclear physics written years after the QGP discovery do not, as yet, introduce this field. I see QGP addressed just like one writes about unicorns, animals that exist in sagas but not in the real world. I believe that this situation relates to some doubts about QGP from years long gone:
\begin{itemize}
\item if the QGP phase of matter can in principle be observed, see for example in Ref.\cite{Muller:1991jk} Section 7;
\item which continue with the question: when, how, and by whom the discovery was completed -- I believe I did asnwer this in the above pages.
\end{itemize}

Consider a few words taken from an eminent introductory text to particle and nuclear physics, which I extract from the year 2008 English edition, repeated in that format in the more recent German 9th edition ~\cite{Povh:1995mua} (granted that this was the time window of more widespread QGP recognition).\\

\noindent \textit{It is hard to tell who among severall authors (B.~Povh, K.~Rith, C.~Scholz and F.~Zetsche) signs the following seen in the English book edition pp.\;321 and 328}:\\[-0.7cm]
\begin{mdframed}[linecolor=gray,roundcorner=12pt,backgroundcolor=GreenYellow!15,linewidth=1pt,leftmargin=0cm,rightmargin=0cm,topline=true,bottomline=true,skipabove=12pt]\relax%
\ldots this state (QGP), where the hadrons are dissolved, cannot be observed through the study of emitted hadrons \ldots\\
Such a quark-gluon plasma has, however, not yet been indubitably generated and a study of the transition to the hadronic phase is thus only possible in a rather limited fashion. 
\end{mdframed}
\vskip 0.5cm
All told, I have not seen a student level general nuclear and/or particle physics textbook that gives justice to the discovery of QGP. One could argue that the only way QGP can be discovered is that we sit out a few more decades; during this period QGP researchers of today will reach textbook writing age. 

That decades are needed for this is illustrated by a closely related anecdote. In the Summer 2019 I visited the Wigner Institute in Budapest where QGP physics is a major research direction. I was welcomed to the heavy ion research retreat at Lake Balaton. I gave a lecture loosely related to QGP addressing another topic, the compact ultra dense objects~\cite{Labun:2011wn} (CUDOs). 

All present noted that I treated QGP as discovered and, well established, and a potential source of specific quark-matter CUDO objects in the Universe. A participant at the retreat (a just tenured staff researcher from Budapest) approached me later \lq \ldots you really think QGP was discovered?\rq\ I probed and learned that my lecture triggered conversations about QGP discovery. Another outside visitor to the group, a senior meeting participant claimed that QGP was not discovered.\label{HoHo} Given his position and experience he should have known better.

While some individual \lq spectators\rq\ continue to discuss the discovery of QGP, within the interdisciplinary \lq participant\rq\ community of QGP physicists the research objectives have shifted from the QGP discovery to the exploration mode of the new deconfined phase of matter, and the study of the quantum vacuum properties of strong interactions. 

To conclude: There are many senior members of the nuclear physics community who were not directly involved in the QGP discovery, have fragmentary knowledge of experimental results, but have loud voices and distract from the objective status of the field, sometimes justifying their position by recalling early \lq stone age\rq\ period views, such as the already noted Ref.\cite{Muller:1991jk} Section 7. We may have to wait for these voices to fade into retirement homes.

\subsection{Non-strangeness signatures of QGP: J/$\Psi$ psions/charmonium}
Here just a few words about the (unrelated to our presentation) topic of more weakly coupled to QGP signatures considered over the past 50 years. These in particular include~\cite{Shuryak:1978ij}:
\begin{enumerate}
\item Dileptons and photons: In consideration of the strong background originating in secondary hadronic particle decays after QGP hadronized makes observation and interpretation difficult. 
\item The J/$\Psi(c\bar c)$ (psions or better charmonium) abundance. 
\end{enumerate}
Unlike strangeness, and strange antibaryons formed in hadro\-nization, the J/$\Psi$ yields are determined by evolution in the dense matter formed, be it QGP or other forms of strongly interacting matter. Thus we need to trust models to use this signature to distinguish between the influence on the yield by confined and deconfined matter. Another obstacle is that any breakup of J/$\Psi$ preformed in initial interactions is accompanied in kinetic models due to detailed balance by formation processes. This in turn requires precise modeling of J/$\Psi$ freeze-out process in the primordial form of matter.

Initially there was a strong case made for J/$\Psi$ suppression by QGP compared to confined matter by \textbf{Tetsuo Matsui} and Helmut Satz~\cite{Matsui:1986dk}. However, it was noted that at SPS energies the J/$\Psi$ yields could also be described in a statistical model~\cite{Gazdzicki:1999rk,BraunMunzinger:2000px}. Moreover, in a kinetic model we have shown~\cite{Thews:2000rj} that at higher RHIC and LHC energies an enhancement of J/$\Psi$ abundance is possible since the population development of charmonium J/$\Psi$ abundance in quark-gluon plasma is very complex and depends in decisive way on QGP properties and component interaction with these small and tightly bound $\bar c c $ states.

Although Helmut Satz and co-workers over past decades have refined their models considerably, there is in my opinion no convincing QGP evidence that $\bar c c $ states demonstrate. In particular, an enhancement over initial abundance can also occur in some phase space domains~\cite{Thews:2000rj}. Just about any conclusion can be reached about the suppression or enhancement of charmonium J/$\Psi$ propagating through the quark-gluon plasma fireball. The reader can best appreciate this by reading the referee evaluation of our work~\cite{Thews:2000rj} where we were first to note that in principle the kinetic models are more likely to result in charmonium J/$\Psi$ enhancement than the suppression that was the rage of the day.\\

\noindent \textit{The first rejection of our manuscript Ref.\,~\cite{Thews:2000rj} submitted on 29 August 2000(PRL reference LV7733) by Physical Review Letters arrived on 23 October 2000. We received the following two reviews:}\\[-0.7cm]
\begin{mdframed}[linecolor=gray,roundcorner=12pt,backgroundcolor=GreenYellow!15,linewidth=1pt,leftmargin=0cm,rightmargin=0cm,topline=true,bottomline=true,skipabove=12pt]\relax%
\textbf{ Referee A:}\\
\indent Strong \lq\lq anomalous\rq\rq\ J/$\Psi$ suppression is regarded as one of the most important signals of quark-gluon plasma formation in relativistic heavy ion collisions. \ldots

\ldots On the basis of the information presented in this manuscript, it is conceivable that these effects compensate completely the \lq\lq order of magnitude\rq\rq\ enhancement claimed by the authors. The argument made by the authors is - at least in its present form - not as stringent as I would require for a publication in the Physical Review Letters.\\
 
\noindent \textbf{Referee B:}\\
\indent I do not recommend the publication of this manuscript in Physical Review Letters.

The main idea and formalism have already been published in a recent paper by the same authors [Phys. Rev. C 62, 024905 (2000)] in the case of $B_c$ meson production. This is just an application to another process. \ldots 

\ldots
The increase of the relative abundance of species with higher masses (in this case J/psi) with increasing collision energy is what one would expect not knowing anything about the quark-gluon plasma. So, before the J/psi enhancement can be considered as a signature for the quark-gluon plasma formation, it must be said how it compares with this trivial phase-space expectation.
\end{mdframed}
\vskip 0.5cm

We see that the referees represent two different schools of thought: Referee A is a believer in Matsui-Satz~\cite{Matsui:1986dk} simple suppression argument, while Referee B indicates the need to review the statistical phase space arguments~\cite{Gazdzicki:1999rk}. Both referees were competent and took considerable effort to reject our manuscript that perturbed the prevailing status quo by advancing a kinetic theory model. There was incipient competition from statistical equilibrium considerations~\cite{BraunMunzinger:2000px} that due our prolonged referee battle appeared in press a half year earlier than our work.\\

\noindent \textit{This was a very difficult personal context for this author, hence it took about 6 weeks to compose a comprehensive response and modify the manuscript. Our resubmission letter to Physical Review Letters, was dated Dec. 5, 2000:}\\[-0.7cm]
\begin{mdframed}[linecolor=gray,roundcorner=12pt,backgroundcolor=Dandelion!15,linewidth=1pt,leftmargin=0cm,rightmargin=0cm,topline=true,bottomline=true,skipabove=12pt]\relax%
\underline{General comment:} At this point in time and without experimental feedback we do not attempt in this short publication a detailed phenomenology of J/$\Psi$ production at RHIC. In our rewording of the text we have emphasized that our purpose is to show that the effects of formation of J/$\Psi$ in a deconfined region to be quite large at RHIC. We support this conclusion from kinetic calculations in a simple model, supplemented by input parameters motivated by perturbative QCD and a generic picture the deconfined region space-time parameters. 

We have rewritten many of the sections and changed our presentation emphasis in order to clarify the basis of our scenario. In the revised manuscript, the major uncertainties in the formation rates are now incorporated directly into the text and shown in the results of Figures 2 and 3. These involve the initial charm quark momentum distributions, and to a lesser degree the effects of color screening vs gluon dissociation in determining the appropriate kinematic parameters.

{Key specific changes in resubmitted manuscript:} \ldots 
\end{mdframed} 

\noindent \textit{The rejection letter from Physical Review Letters arrived before the end of the year on Friday, 29 Dec 2000 10:19:45 (EST), quoting the referees as follows:}\\[-0.7cm]
\begin{mdframed}[linecolor=gray,roundcorner=12pt,backgroundcolor=GreenYellow!15,linewidth=1pt,leftmargin=0cm,rightmargin=0cm,topline=true,bottomline=true,skipabove=12pt]\relax%
\textbf{Second report of referee A:}\\
\indent My main criticism remains unchanged: On the basis of the information presented in this manuscript, it is conceivable that effects not treated sufficiently explicit in the discussion compensate completely the \lq\lq order of magnitude\rq\rq\ enhancement of J/$\Psi$ production claimed by the authors. The argument made by the authors is not as stringent and clear as I would require for a publication in the Physical Review Letters. \ldots Referee B is correct in pointing out that the main idea and formalism of this manuscript was already explored by the same authors in PRC 62, 024905 (2000) for the case of B$_c$ meson production.\\

\noindent\textbf{Second report of referee B:}\\
\indent Instead of mentioning my further objections and doubts I would like to stress that the main idea explored in this manuscript have already been published by the same authors in their recent paper [Phys. Rev. C 62, 024905 (2000)] about the B$_c$ meson production. This fact itself suggests that the right place for publishing this paper is the Brief Report section of Physical Review C. 
\end{mdframed}
\vskip 0.5cm

To conclude: It was not easy to shift the attention from suppression to enhancement of charmonium J/$\Psi$ in quark-gluon plasma. Our work~\cite{Thews:2000rj} should have been seen as a watershed event. Instead, it was shredded by two expert referees burying us in unessential details (lawyers call this \lq burying in paper\rq) while a somewhat competing but much less elaborate study~\cite{BraunMunzinger:2000px} went into press rapidly. It is interesting to note that our paper~\cite{Thews:2000rj} published in the Physical Review C earns more annual citations than, arguably, most PRL published articles.

\subsection{After the CERN quark-gluon plasma discovery}
The heavy ion community at CERN continued work to clarify QGP formation thresholds and bulk matter properties. A similar program followed soon at RHIC. We note:
\begin{itemize}
\item[\phantom{i}i.] A research program at CERN-SPS fixed target has run in parallel to the LHC operation. 
\item[ii.] The two colliders, LHC at CERN and RHIC at BNL continue as important parts of their research program, the study of relativistic nuclear collisions.
\end{itemize}
Among all CERN experiments contributing to the CERN announcement, two devoted to strangeness continued for many years: the successor to WA97 called NA57, and the successor to NA49 called NA61/SHINE. This situation reflects on the preeminent role of strangeness as characteristic signature of quark-gluon plasma.
 
\subsubsection{Enhancement of multi-strange baryons at CERN-SPS}
At the CERN-SPS the experiment NA57 continued the program of research of the WA97 experiment shown in \rf{RSS}, see page \pageref{RSS}. The final report~\cite{Antinori:2006ij} confirms the results that have been used in the CERN announcement and demonstrates an enhancement effect by more than an order of magnitude for the $\Omega$ and $\bar \Omega$, see \rf{fig:NA57}. The enhancement is shown dependent on the number of inelastically damaged nucleons called \lq wounded\rq. To this date these NA57 results are the largest \lq medium\rq\ effect observed.

\begin{figure}\sidecaption
\centerline{
\includegraphics[width=0.85\textwidth]{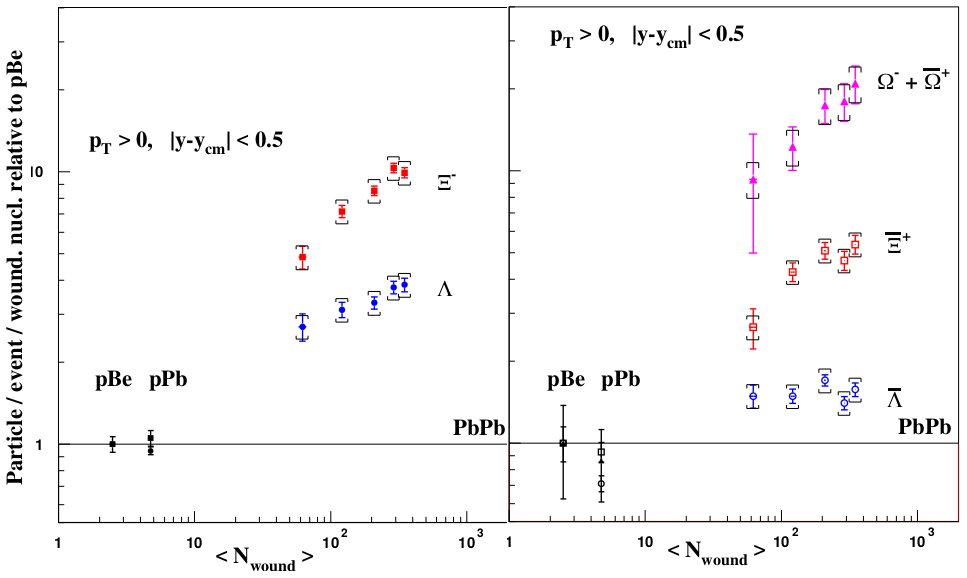}
}
\caption{The final Pb--Pb results on hyperon enhancement by NA57, Energy dependence of strangeness adapted from~\cite{Antinori:2006ij} 
}
\label{fig:NA57}
\end{figure}

The reader should note that the research program at SPS after the CERN QGP discovery was focused on strange antibaryons and strangness exclusively.

\subsubsection{NATO support for strangeness and SHARE}\label{ss:SHARE}
\label{diabolicum}
One would perhaps not see the defense organization NATO\label{NATOfun} as a source of funding for the QGP research and strangeness, and yet it happened; Hans Gutbrod and I were able to secure funding for the Summer 1992 Il Ciocco School, and a Summer 1994 Hagedorn celebration workshop. In addition, in the early 90s I obtained funding for the Tucson-Regensburg collaboration where much of the work was done with than graduate student Josef Sollfrannk, see Sec.\ref{findingQGP}, and than, for the current topic, the Tucson-Krakow SHARE collaboration early this century, where my NATO partner was \textbf{Wojtek Florkowski}. NATO itself was hardly explicitly present but there were guidelines for how we could use the funding to bring togather participants from member countries at meetings. The proceedings were part of a uniform publication series. Everyone at the meeting wore a standarized conference badge showing the flags of the NATO member countries.

The need for a standardized formulation of the hadronization of the QGP fireball was discussed at the end of Sec.~\ref{findingQGP}, see \rf{PBMletter}. This was clearly an imminent objective after the CERN QGP announcment. At the time I lectured at several Summer and Winter schools in Poland. I could see that several researchers in Krakow, in particular Wojtek Florkowski and \textbf{Wojtek Broniowski}, shared in our research interest creating their own statistical hadronization program. This laid the roots of the SHARE-1 collaboration where the acronym derives from {\bf S}tatistical {\bf HA}dronization with {\bf RE}sonances.

The Krakow and Tucson groups joined forces and in a period of two years we created and documented a web available SHM model SHARE~\cite{Torrieri:2004zz}. Our program was thoroughly vetted against the existent Tucson and Krakow programs. In the end the three programs were agreeing to the last significant digit in all benchmark tests. Aside of creating a debugged tool, another objective of our effort was to enhance the capabilities of the SHM approach. 

There were three generations of SHARE: SHARE-2 developed in the following two years in collaboration with the Montreal group~\cite{Torrieri:2006xi} incorporated the option to fit particle fluctuation results, aside of considerable update of the input of all particle data tables. SHARE-3 (SHARE with CHARM) introduced into the program hadron prodduction by charmed particles~\cite{Petran:2013dva}. While the program is fully functional there is rapidly increasing data field for charmed hadrons thus the situation is in need of active managemeent.

SHARE was designed to offer study options far beyond the previous norm:
\begin{enumerate}
\item We maximized the parameter set to be able to try new model ideas:
the full set of parameters that can be fitted to the observed hadron abundance of any directly produced elementary hadron created by a hot fireball in relativistic heavy-ion collision is:\\[0.3cm]
\begin{tabular}{p{2ex}p{13ex}p{7ex}p{44ex}}
&Symbol & SHARE Param. & Param. Description\\
\hline
1)& $V$ or $dV/dy$ & \texttt{norm} & source volume (normalization) in fm$^3$;\\
2)& $T$ & \texttt{temp} & chemical freeze-out $T$ (in MeV); \\
3)& $\lambda_q=e^{(\mu_q/T)}$ & \texttt{lamq}& light quark fugacity factor;\\
4)& $\lambda_s=e^{(\mu_s/T)}$ & \texttt{lams}& strangeness fugacity factor;\\
5)& $\gamma_q$ & \texttt{gamq }& light quark phase space occupancy;\\
6)& $\gamma_s$ & \texttt{gams} & strangeness phase space occupancy.\\
7)& $\lambda_3$ &\texttt{lmi3} & proton-neutron (isospin) $I_3$ fugacity factor\\
8)& $\gamma_3$ & \texttt{gam3} & $I_3$ phase space occupancy\\[0.3cm]
\end{tabular}
\item We allowed bulk properties of QGP source to be usable as an input that could be fitted -- one such natural bulk constraint is the count of strange and antistrange quarks. SHARE allows this constraint $\langle s-\bar s\rangle =0\pm \epsilon$ to be \lq fitted\rq\ as much as one fits a prescribed yield of produced hadrons. 
\end{enumerate}
The role of parameters 3)-- 6) is illustrated in \rf{gamlam} according to insights seen in the 1986 report~\cite{Koch:2017pda}. Parameters 7), 8) allow for $u$, $d$ light quark asymmetry. This is necessary in fits in which for example also charge $Q$ of the fireball is explored.

\begin{figure}[bt]\sidecaption
\includegraphics[width=1.0\textwidth]{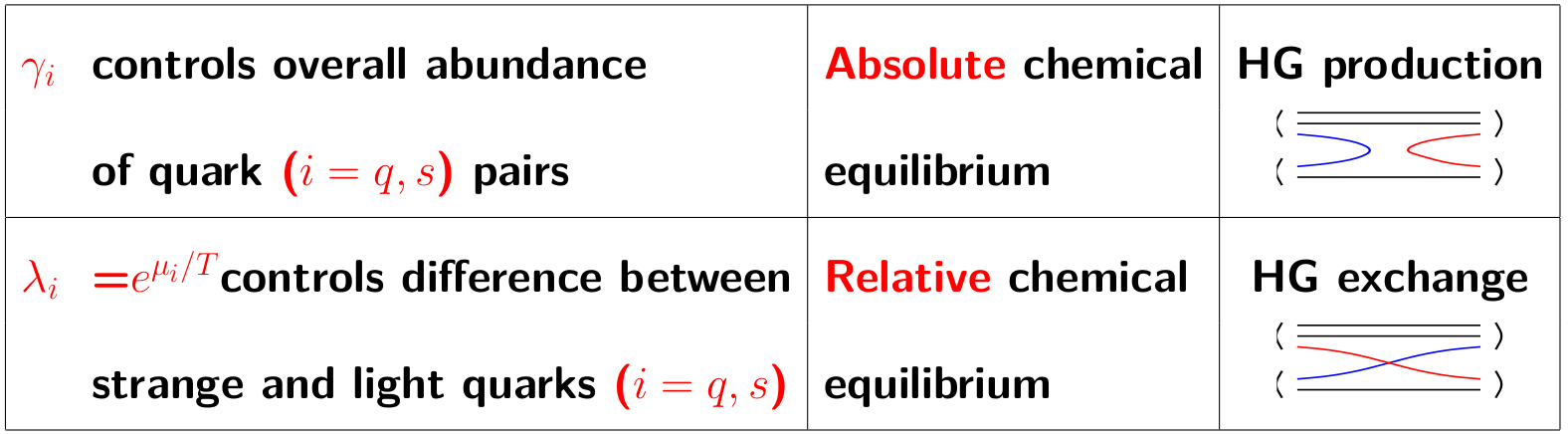}
\caption{The meaning of chemical parameters 3)-- 6), and by extension also 7), 8)}
\label{gamlam}
\end{figure}

As noted we can \lq fit\rq\ the constraint $\langle s-\bar s\rangle =0\pm \epsilon$ -- and in SHARE any bulk output property, such as energy density $\varepsilon$, entropy density $\sigma$, and pressure $P$ can also be fitted. This allows to compare the fireball source created in different collision systems.

The study of bulk properties of QGP furthermore provided an unexpected asset: when the number of parameters increases (up to 8 in SHARE, see above), finding a fit minimum in a rich data field is haphazard, as one can easily get caught in a false minimum. It turns out that by leaving parameter range unconstrained but contstraining loosely the bulk property of the hadronizing fireball is sufficent to both accelerate and make reliable the particle production data fit minimum. 

All these novel features incorporate into SHARE meant that user manuals~\cite{Torrieri:2004zz,Torrieri:2006xi,Petran:2013dva} became bulky and the program difficult to handle without in-house training. This spelled trouble; our code was \lq open source\rq\ and much of its easy to use content was thus quickly adopted in simplified student programs. This meant that the old era of SHM errors was back with a vengeance. The hearsay is that the most troublesome error is seen in the implementation on the CERN-ROOT analysis platform a simplistic SHARE version called Thermus.

The error in ROOT-Thermus version 1 and version 2 is, as I am told, that once particle resonances were read out, this meant these were stable particles. In order to understand what this means, consider, as an example, K$^*$ abundance needed in some fits that use this yield. In ROOT-Thermus that meant K$^*$ is a \lq stable\rq\ particle and thus kaons from the decay of K$^*\to \pi+$K were not included in the final kaon yields, which were of course also fitted. 

Even if different arguments float, in my opinion, it is impossible to simultanously fit both K and K$^*$ and accordingly Thermus fit results before 2019  with a hadron resonance probably need an erratum. To repeat: publications you read where Thermus is used (2004-2018) and where for example $\phi$, K$^*$ or/and any other hadronically decaying particle is fitted must be reconsidered.

To cut the story short: Soon after SHARE was created I saw again aberrantly high hadronization temperatures that were presented as the best fit. However, today there is an easy physics test of this situation: We know that free-streaming particles we analyze must emerge after freeze-out; that means, below the QGP disintegration condition. QGP fireball breaks into hadrons when temperature cools below but near to the QGP existence boundary. SHM analysis provides therefore a value of the chemical freeze-out temperature below the lattice-QCD phase cross-over boundary.

\begin{figure}[bt]
\centerline{%
\includegraphics[width=0.85\textwidth]{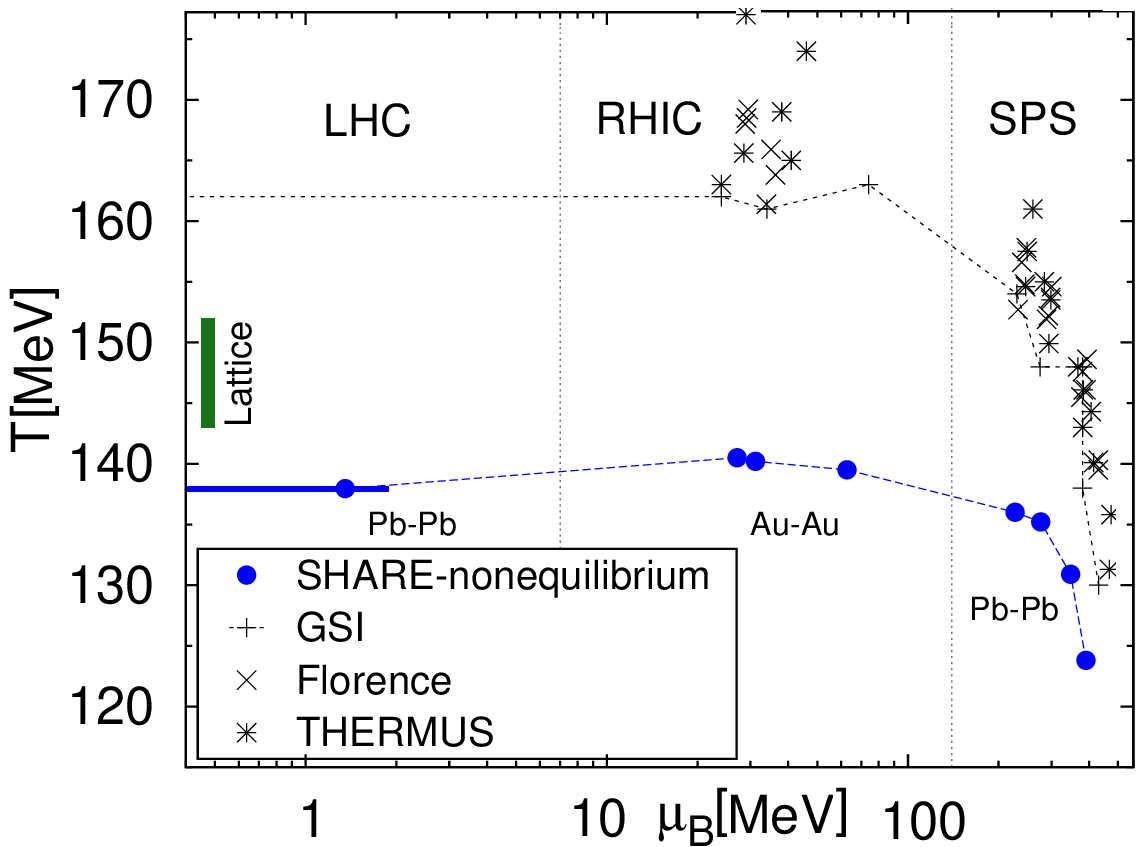}
}
\caption{The $T,\mu_\mathrm{B}$ scatter diagram showing lattice value of critical temperature $T_c$ (bar on left), as compared to SHM results of different groups for analysis performed for different collision energies as indicated. For references to these results see related discussion in Figure~9 in Ref.\cite{Rafelski:2015cxa}. Our SHARE chemical non-equilibrium results are seen as full blue circles, dashed blue line guides the eye; the GSI chemical (semi-)equilibrium results are crosses with dashed black line guiding the eye. Other results are also shown, including those obtained using the wrong decay chain of the Thermus program}
\label{TEcol}
\end{figure}

This present day situation is shown in the hadronization temperature -- chemical potential plane scatter plot in \rf{TEcol}, which is an update of Figure~9 in Ref.\cite{Rafelski:2015cxa} (see there for all pertinent references to data and lattice QCD). All model values that we see in \rf{TEcol} well above the lattice value $T_c$ on the left margin in \rf{TEcol} are either old pre-SHARE results of other groups, like the $T=174$\,MeV fit, see \rf{PBMletter}, or are using the newly recreated (but wrong) SHARE simplifications such as the ROOT-Thermus program, or in some cases, both. As can be seen in \rf{TEcol}, only the full chemical non-equilibrium results obtained using SHARE are convincingly below the phase transformation boundary between QGP and hadron phase obtained in lattice-QCD. 

As already noted some groups fit the data with a very small parameter set using the four first parameters in above list. The tacit assumption of absolute chemical equilibrium after hadronization is made in this approach. However, if such equilibrium among particles produced by QGP exists, one should find it as an output of analysis. In my personal opinion the real reason to use simplisitc chemcal equilibrium is the lack of capability to use a more complex approach such as offered by SHARE. On the other hand, by publishing SHARE program and data files we have opened a  door to eager students who realized the opportunity of creating their own reincarnations of SHM using our work and impressing their supervisors in the process. I have no doubt that this was done after looking over the shoulders of some of these students. 

\subsubsection{How does SHM work?}\label{sec:SHMwork}
To obtain a grand canoncical description of the produced particles we study the quantum Fermi and Bose phase space distributions, which maximize the entropy at a fixed particle number. In the local rest frame of the volume element, the particle spectra take the form 
\begin{equation}\label{eq:fermiDist}
 {{d^6N_\mathrm{F/B}}\over {d^3pd^3x}} = {g\over (2\pi)^3}n_\mathrm{F/B}, \qquad 
n_\mathrm{F/B} (t)= \displaystyle\frac 1 {\gamma^{-1}(t) \E^{(E \mp \mu )/T(t)}\pm 1}, 
\end{equation}
where $g$ is the statistical degeneracy and $E =\sqrt{p^2+m ^2}$ the particle energy in  a local comoving frame. The integral of the distribution \req{fermiDist} provides the particle yield. In the Boltzmann limit suitable for heavy particles we do not need to make a distinction between Fermions and Bosons and we obtain 
\begin{equation}\label{eq:fermiY}
N = \frac{g V T^3}{2\pi^2} \gamma \E^{\pm\mu/T} \left(\frac{m}{T}\right) ^2 K_2\left(\frac{m}{T}\right) \ \to \ gV (mT/2\pi)^{3/2} \gamma \E^{-(m\mp\mu)/T}+\ldots\;.
\end{equation}
In a program like SHARE the full quantum distribution is included; here we proceed to use the Boltzmann limit as it offers simplicity in presentation.

It is common to express chemical potentials related to conserved quantum numbers of the system, such as the baryon number $B$, the strangeness $s$, and the third component of isospin $I_3$ in terms of corresponding quark fugacities 
\begin{equation}\label{eq:mu}
\mu_B = 3T \log \lambda_q\;, \quad
\mu_S = T \log \lambda_q/\lambda_s\;, \quad
\mu_{I_3} = T \log \lambda_3\;. 
\end{equation} 
Notice the inverse, compared to intuitive definition introduction of $\mu_S$, which has a historical origin and is a source of frequent mistakes. 

As is a common practice we took advantage of the approximate isospin symmetry to treat the two lightest quarks ($q = u,d$) using light quark and isospin phase space occupancy and fugacity factors which are obtained via a transformation of parameters:
\begin{equation}\label{eq:q3toud}
\lambda_q = \sqrt{\lambda_u\lambda_d}\;,\quad
\gamma_q = \sqrt{\gamma_u\gamma_d}\;,\qquad 
\lambda_3=\sqrt{\frac{\lambda_u}{\lambda_d}}\;,\quad
\gamma_3=\sqrt{\frac{\gamma_u}{\gamma_d}}\;,
\end{equation}
with straightforward backwards transformation
\begin{equation}\label{eq:udtoq3}
\lambda_u = \lambda_q\lambda_3\;,\quad \gamma_u = \gamma_q\gamma_3\;,\qquad 
\lambda_d = \frac{\lambda_q}{\lambda_3}\;,\quad \gamma_d = \frac{\gamma_q}{\gamma_3}\;.
\end{equation}
Even if the electrical charge $Q=Ze$ has not appeared explicitly, it can be defined in full using the available chemical potentials considering that quarks carry a specific charge. 

The fugacity of hadron states is defining according to \req{fermiY} the yields of different hadrons is obtained from the individual constituent quark fugacities. In the most general case, for a hadron consisting of $N_u^i, N_d^i ,N_s^i$ and $N_c^i$ up, down, strange and charm 
quarks respectively and $N_{\bar{u}}^i,N_{\bar{d}}^i,N_{\bar{s}}^i$ and $N_{\bar{c}}^i$ anti-quarks, the fugacity can be expressed as
\begin{equation} 
\label{eq:fugacity}
\Upsilon_i = (\lambda_u\gamma_u)^{N_u^i}(\lambda_d\gamma_d)^{N_d^i}(\lambda_s\gamma_s)^{N_s^i}
 (\lambda_{\bar{u}}\gamma_{\bar{u}})^{N_{\bar{u}}^i}(\lambda_{\bar{d}}\gamma_{\bar{d}})^{N_{\bar{d}}^i}(\lambda_{\bar{s}}\gamma_{\bar{s}})^{N_{\bar{s}}^i}\;,
\end{equation} 
where $\gamma_i$ is the phase space occupancy of flavor $i$ and $\lambda_i$ is the fugacity factor of flavor $i$, $i=u,d,s$, extension to charm can be made easily. For quarks and anti-quarks of the same flavor
\begin{equation} 
\gamma_f = \gamma_{\bar{f}}\qquad\text{ and }\qquad \lambda_f = \lambda_{\bar{f}}^{-1},
\end{equation}
which reduces the number of variables necessary to evaluate the fugacity by half. To be specific, for $\Lambda(uds)$ and its antiparticle we have:
\begin{equation} 
\Upsilon_{\Lambda(uds)}= \gamma_u\gamma_d\gamma_s\lambda_u\lambda_d\lambda_s;\qquad
 \Upsilon_{\bar{\Lambda}(\bar{u}\bar{d}\bar{s})}= \gamma_u\gamma_d\gamma_s\lambda_u^{-1}\lambda_d^{-1}\lambda_s^{-1}\;.
\end{equation} 
We now see how a few fugacities along with the source temperature $T$ and the volume $V$ characterize in full particle yields at the time of chemical freeze-out. 

\subsubsection{Threshold and energy dependence of strangeness enhancement}

NA49 and NA61/SHINE experiments at the CERN-SPS continued and continue the exploration of energy and volume thresholds of the onset of deconfinement. The results seen in \rf{fig:onset} show the excitation function for the K$^+/\pi^+$ ratio~\cite{NA61Shine}, which in baryon-rich QGP represents in the numerator the strangeness yield, and in the denominator the entropy content of QGP. 

\begin{figure}[tb]\sidecaption
\centering
\includegraphics[width=0.50\textwidth]{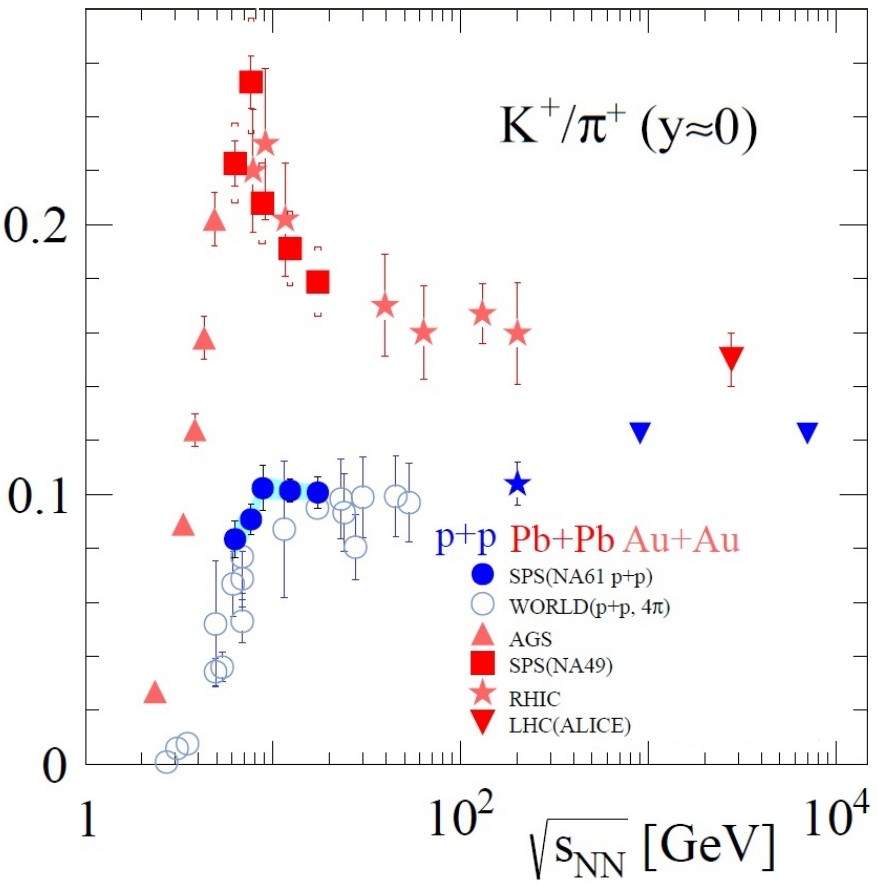}
\caption{The so-called horn structure in energy dependence of the K$^+/\pi^+$ ratio excitation function indicating threshold in strangeness to entropy yield in central Pb+Pb (Au+Au) collisions, from CERN-SPS NA61/Shine collaboration report~\cite{NA61Shine}
}
\label{fig:onset}
\end{figure}

The \lq horn\rq\ in this result could be the searched-for energy threshold to quark deconfinement. This is so since at the onset of QGP formation at the energy above the horn the speed of production of entropy by color bond melting exceeds the speed of strangeness production. One of the unsolved riddles is the mechanism responsible for the rapid production of strangeness seen below the horn peak where strangeness beats entropy. An idea I have been following is the possibility of strange-down quark mixing driven by chiral symmetry restoration. Such an effect could be driven by the ultra strong EM fields present in these reactions.

The nearly obvious question seen the eperimental results is if we can interpret these using SHM. Earlier, I bragged to Marek Ga\'zdzicki that within the chemical nonequilibrium approach I should have no problem to characterize his \lq horn\rq. This was therefore the first project achieved with SHARE~\cite{Letessier:2005qe}. With Jean Letessier I demonstrated that Marek\rq s experimental results were indeed consistent with the SHM chemical non-equilibrium model assuming explosive hadronization of QGP. 

However, the publication of our work was difficult. We submitted to PRC and encountered a wall of resistence. Ultimately, the PRC editors rejected our manuscript. At first we moved on to other projects instead of wasting time convincing referees of another journal, who easily could be the same personalities, as the community of experts is small. However, this arXiv\rq ed work was noted and well cited. Two years later we chose to update our analysis to more recent Marek\rq s experimental data and submit to EPJA. In the year 2007 the horn we were describing was not  a fragment of a unicorn anymore. Our paper reviewed well and was soon published~\cite{Letessier:2005qe}. 

\subsubsection{Addressing small volume effects}\label{sec:canon}

In the statistical hadronization model (SHM), a phase-space based hadron yield evaluation is performed. Such a model cannot be used to describe production of particles that are weakly coupled to the (QGP) fireball source. Thus photons and dileptons have to be obtained using microscopic collision models. SHM can describe production of strongly interacting particles in composite ground state, \ie\ \lq stable\rq\ mesons and baryons, and their resonances with excitation energy measured at comparable scale with the prevailing temperature, \ie\ below GeV-scale. 

The central postulate of the statistical model is that particle yields depend only on the available phase space. However, there are several ways to accomomodate this:
\begin{itemize}
\item 
\underline{Fermi Micro-canonical phase space:}\\ 
has sharp energy and a sharp number of particles. Introduced in study of cosmic ray individual events, this may not be an appropriate approach since laboratory experiments report event-averaged particle abundances.
\item 
\underline{Canonical phase space:} \\
employs an average over an ensemble of systems with temperature $T$ tunable average fireball energy $E$. In this approach at least one of all (quasi) conserved quantum numbers is exactly fixed: these exact numbers include color charge~\cite{Turko:1981nr,Elze:1983du,Elze:1984un}, isospin~\cite{Muller:1982gd}, baryon number\cite{Derreth:1985kk} and, conserved on strong interaction time scale, the number of strange $s$ quark pairs~\cite{Rafelski:1980gk,Rafelski:2001bu}, and similarly charm $c$ or bottom $b$ quark pairs -- a fixed electric charge $Q$ follows if some of the above are conserved exactly, but also can be introduced independently.
\item 
\underline{Grand-canonical ensemble phase space:}\\ 
allows an event ensemble average with regard to energy $E$ and all discrete (quasi) conserved particle numbers. SHARE uses a grand-canonical ensemble allowing the implementation of an average particle abundance constraints. The one exception is the option to conserve strangness exactly: This creates a constraint fixing the value of strangeness chemical potential for any given baryochemical potential and temperature $T$. 
\end{itemize}

Entropy must be conserved or increasing during the hadronization process of QGP. This means that there are, to first approximation, as many hadronic particles produced as there are quarks, antiquarks and gluons present in the fireball of QGP. The production of nonrelativistic (heavy) particles may reduce somewhat the required number of emerging particles. This means that canonical phase space constraints have a weaker impact after hadronization than before. Moreover, hadronic particles without conserved quantum numbers are produced without constraints at all.

One of the cornerstones of the argument that multiply strange particles are direct products of hadronization of QGP arises from the study of the ratio $\bar \Xi(\bar s\bar s q),\ \Xi(ssq)$ compared to $\phi(s\bar s)$, which is practically independent of the energy and centrality of the collision. This is what we expect in the combinant production of these multiply strange hadrons of different mass and strangeness contents, and different quantum nature (Bose/Fermi). Some variance is expected due to need for a light quark in $\bar \Xi, \Xi$ compared to $\phi$, and possibility of interference from other production mechanisms. 

For systems that have a finite baryon density in the fireball, one chooses as a variable~\cite{Petran:2009dc} $\Xi\to \sqrt{\bar \Xi \Xi}$, to neutralize the particle antiparticle asymmetry as shown on the left in \rf{fig:Xiphi}, where we see results compiled in Ref.~\cite{Petran:2009dc} for data available more than decade ago at SPS and RHIC. Note that the bottom (red) line across all lines presents the $\sqrt{\bar \Xi \Xi}/\phi$ ratio. The other results show that the level of variability of other related particle ratios so that one can appreciate that this is a sensitive observable.

On the right in \rf{fig:Xiphi} we see the recent results~\cite{Tripathy:2018ehz,Tripathy:2019flj} obtained by the ALICE collaboration at the CERN LHC collider. Here one can ignore the variance between $\bar \Xi, \Xi$. There are three collision systems that are combined in a single presentation that uses as its variable the mean central pseudo-rapidity $\eta$ charged particle $N_\mathrm{ch}$ density $\langle dN_\mathrm{ch}/d\eta\rangle_{|\eta|<0.5}$. We see the sum of $\bar \Xi+ \Xi$ particle yield divided by $\phi$; hence the horizontal black line is placed at the corresponding value to that seen on left (bottom straight line at 0.281) for SPS and RHIC and earlier LHC results.

\begin{figure}[tb]\sidecaption
\centerline{%
\includegraphics[width=0.52\textwidth]{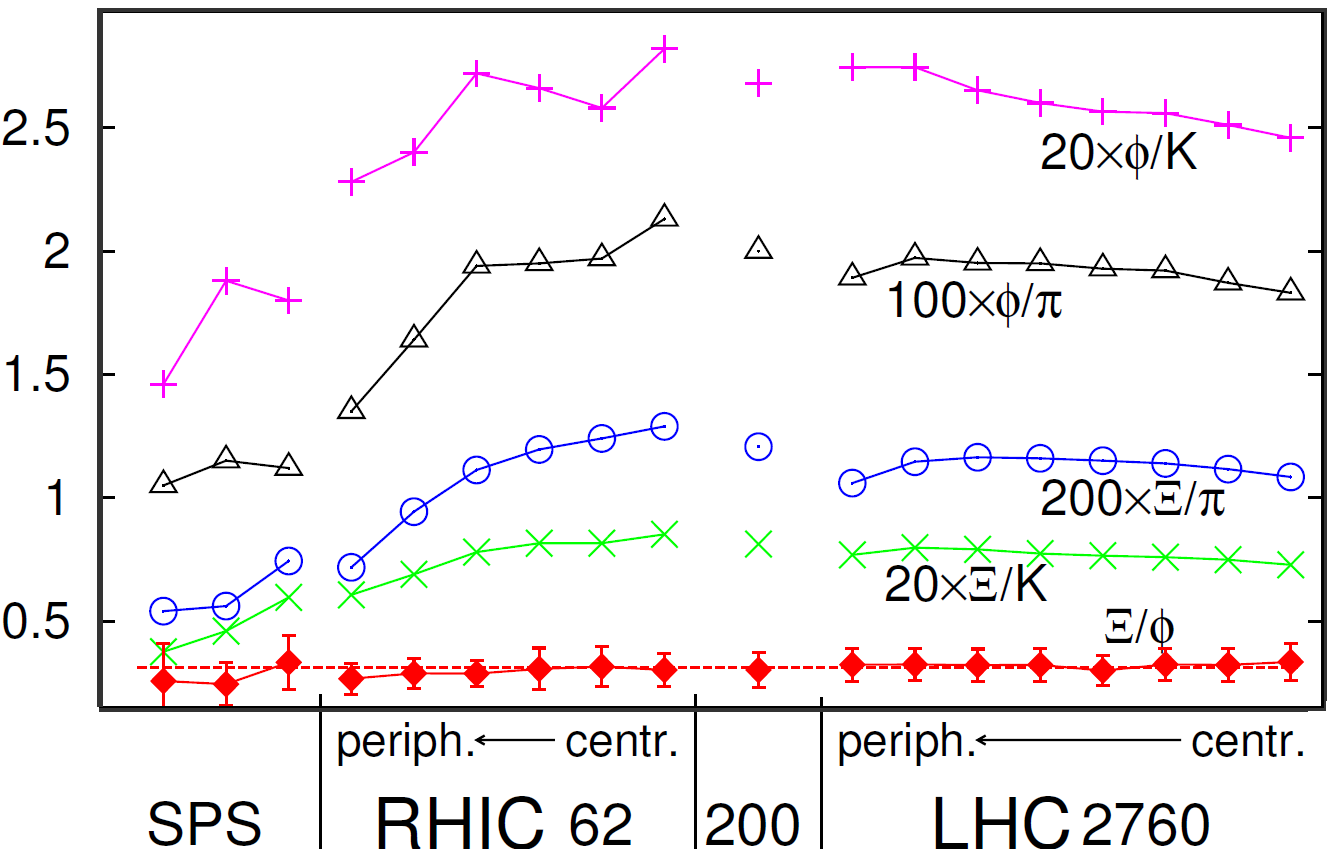}
\includegraphics[width=0.44\textwidth]{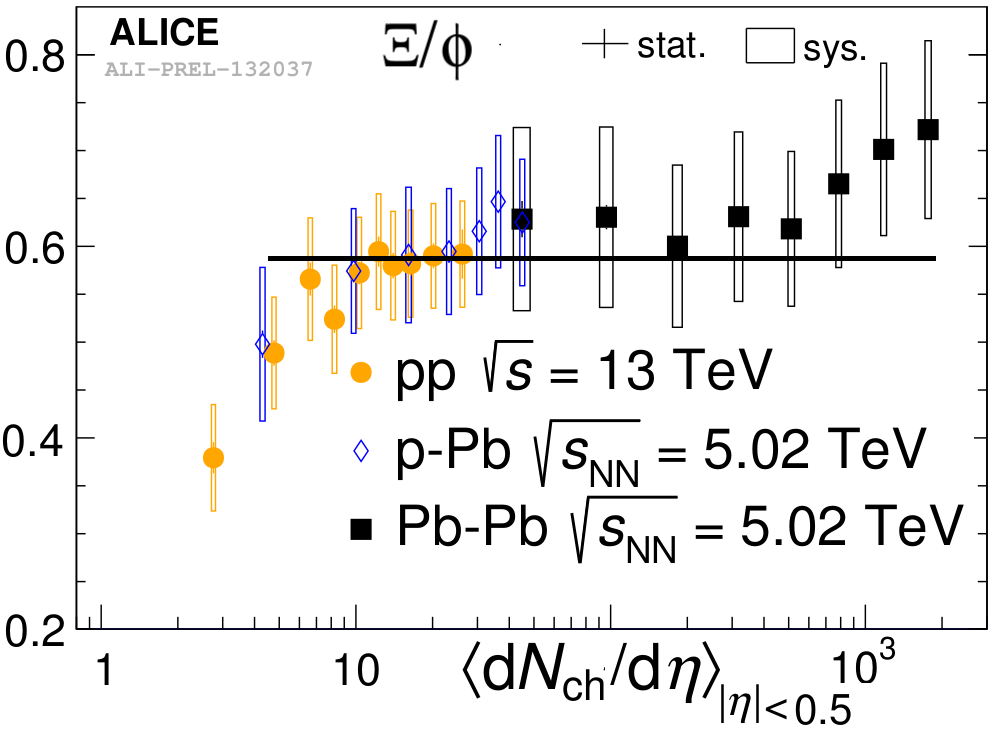}
}
\caption{%
On left: CERN-SPS and RHIC and early \lq low energy\rq\ LHC results for $\sqrt{\overline{ \Xi} \Xi}/\phi$, adapted from Ref.~\cite{Petran:2009dc}, compared in this frame to other more variable ratios: $\Xi$/K, $\Xi/\pi$, $\phi/\pi$. The straight line for $\sqrt{\bar \Xi \Xi}/\phi$, is by definition at half of the LHC-Alice value we see marked on AHS, see text. On right: LHC-Alice results for $(\bar\Xi+\Xi)/\phi$ obtained in three different collision systems at highest available energy as function of charged hadron multiplicity produced, adapted from~\cite{Tripathy:2018ehz,Tripathy:2019flj}
}
\label{fig:Xiphi}
\end{figure}

The particle $\phi(s\bar s)$ has no open strangeness and canonical phase space constraints do not apply to it~\cite{Rafelski:1980gk,Rafelski:2001bu}. However, they apply strongly to the double strange $\Xi$. In the presence of canonical volume effects, the relative yield results seen in \rf{fig:Xiphi} are therefore impossible to attain.

In \rf{fig:Xiphi} we see on the right that for the most peripheral $p$-$p$ Alice collisions with a small ($\simeq 2$--$4$) charged particle multiplicity, the ratio decreases. We note that a QGP fireball picture may not apply here. Considering as source primary parton collisions this is a natural behavior, with $\phi$ requiring one pair of $s$-$\bar s$ quarks, while $\Xi$ requiring that the production of two pairs of strange quarks is suppressed. The rise of the $\Xi/\phi$ ratio seen in most central Pb--Pb collisions, a 1.2sd effect, needs to be confirmed. However, we also note that a high $p_\bot$ enhancement of the $\Xi$ was predicted decades ago, see Sec.\ref{sec:highPT} and Ref.\,\cite{Rafelski:1987un}, and this rise for very large $\langle dN_\mathrm{ch}/d\eta\rangle_{|\eta|<0.5}$ could be due to enhancement at relatively large $p_\bot$.

{To conclude:} The fact that $\phi$ tracks closely the yield of $\bar \Xi, \Xi$ for vastly different collision energies, for different collision systems, in a large domain of collision centrality, demonstrates that these particles emerge directly from a fireball by combinant processes~\cite{Koch:1986ud,Rafelski:1987un}, demonstrating that canonical statistical phase space does not conspire by some mechanism to influence the yields of these particles, see Sec.\ref{sec:SHMwork}.

\subsubsection{Strangness enhancement at collider energies}\label{sec:sS}

Strangeness observable remains experimentally popular at collider (RHIC, LHC) energies, since strange hadrons are produced abundantly and can be measured over a large kinematic domain: All quark flavors can be produced in initial parton collisions. Strangeness differs from the heavier quarks by the relatively low mass threshold. This means that it continues to be produced in ensuing in medium thermal parton processes, dominated in QGP by the gluon fusion process.

This coupling to the gluon degree of freedom implies that the strangeness QGP abundance rises rapidly early on in time line of fireball evolution, but it can also fall, tracking the cooling of the gluon degrees of freedom. Ultimately, when parton temperature and density is sufficiently low, the strangness rich QGP fireball breaks apart. High yield of strangeneess in QGP couples with the self-analyzing decay patterns of strange hadronss. Therefore, a large body of experimental results is available today. 

The total particle yield at colliders, as observed in a small selected centrality interval tracks closely the entropy produced in this rapidity interval. In order to characterize the source of strange particles our bulk QGP fireball target variable is therefore the specific per entropy strangeness-flavor content $s/S$ which we want to track as a function of collision energy and centrality. 

The relative $s/S$ yield measures the number of active degrees of freedom and the degree of relaxation when strangeness production freezes-out. Perturbative expression in chemical equilibrium reads 
\begin{equation}
{s \over S}=\frac{{g_s\over 2\pi^2} T^3 (m_{ s}/T)^2K_2(m_{ s}/T)}
 {(g2\pi^2/ 45) T^3 +(g_s n_{\rm f}/6)\mu_q^2T}\simeq \frac{1}{35}\simeq 0.0286\;.
\end{equation}
When looking closer at this ratio one sees that much of ${\cal O}(\alpha_s)$ QCD interaction effect cancels out. However, for completeness we note that one could argue that $s/S|_{{\cal O}(\alpha_s)}\to 1/31=0.0323$. A stronger effect can occur in presence of QGP nonequilibrium, in this case
\begin{equation}
{s \over S}= { {0.03 \gamma_s^{\rm QGP}} \over 
 {0.4 \gamma_{\rm G} + 
 0.1 \gamma_s^{\rm QGP}\!\!+
 0.5\gamma_q^{\rm QGP}\!\! + 
 0.05 \gamma_q^{\rm QGP} (\ln \lambda_q)^2}}\to 0.03\gamma_s^{\rm QGP}\;.
\end{equation}
Finally, introducing the quantum statistics and doing numerical evaluation produces for $m_s=90$\;MeV the result seen on left in \rf{sSFig} where we also for comparison show this ratio computed in hadron gas. We see that equilibrated QGP is 50\% above equilibrated hadron gas. Actual strangeness production enhancement is larger considering that hadron gas governed reactions are further away from chemical equilibrium. 

\begin{figure}[tb]
\centering
\includegraphics[width=0.51\columnwidth]{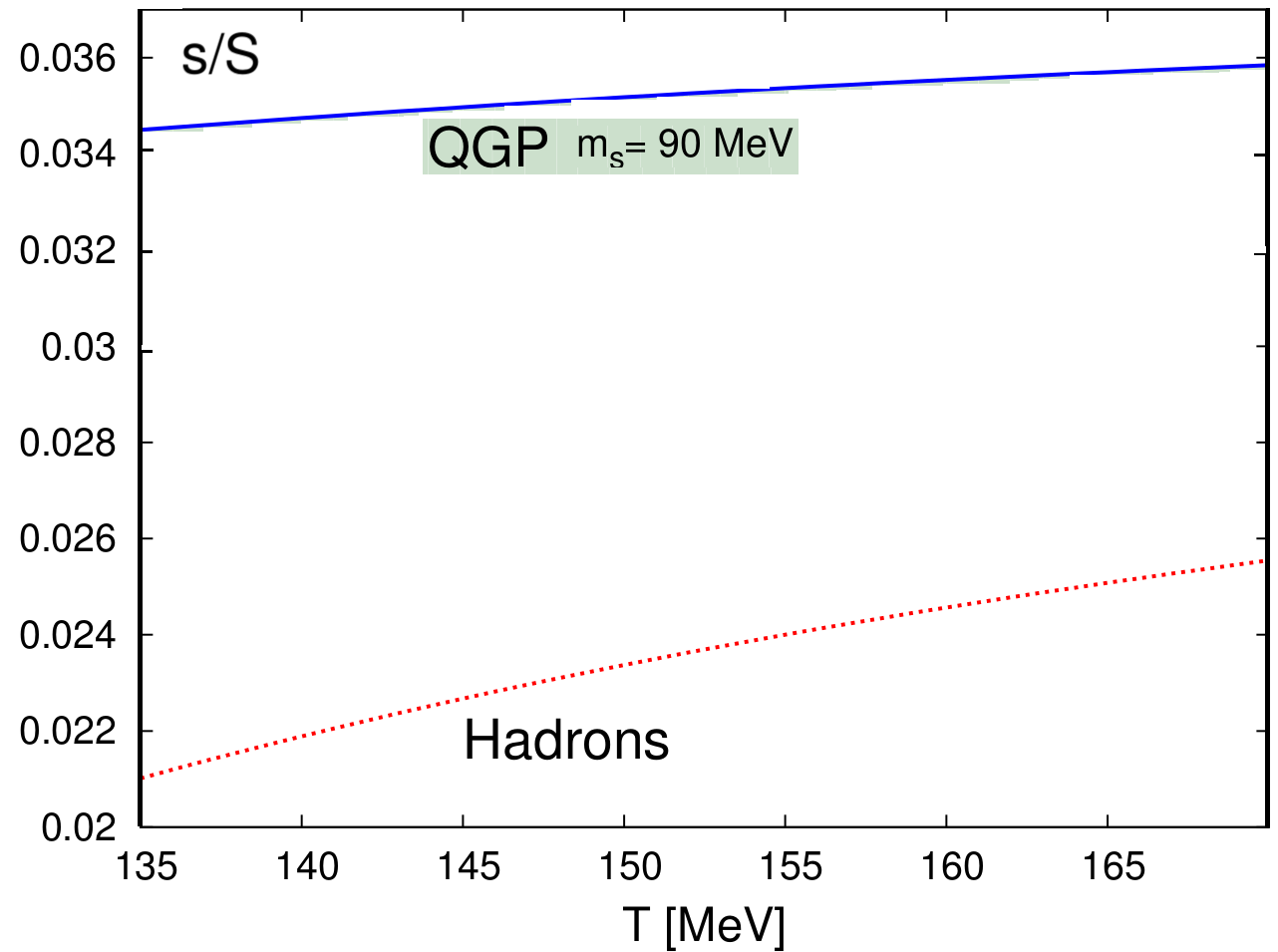}
\includegraphics[width=0.47\columnwidth]{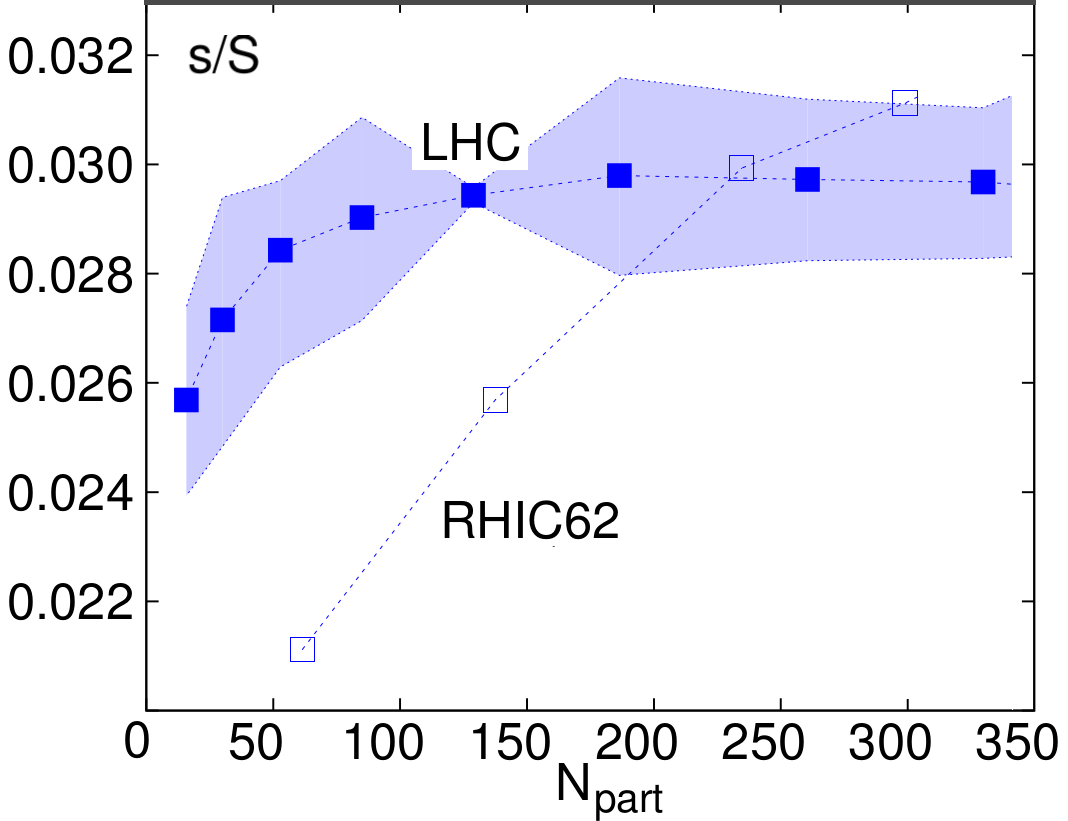}
\caption{Strangeness per entropy $s/S$: On left: as a function of temperature in QGP with $m_s=90$\;MeV, and (red) in the hadron resonance gas as defined by SHARE implemented mass spectrum. On right: Outcome of the fit to ALICE $\sqrt{s_\mathrm{NN}}=2.76$\;TeV results as a function of centrality, expressed by the number of participants. Comparison with RHIC-62 GeV analysis (dotted line) based on STAR data which may contains ROOT-Thermus distortions}
\label{sSFig}
\end{figure}

We show on the right in \rf{sSFig} the centrality dependence for the ALICE and STAR 62 GeV $s/S$ results~\cite{Petran:2013lja}. The ALICE results show a quick rise to saturation in $s/S$ near the perturbative QGP estimated value shown on left in \rf{sSFig}. This can be understood as a piece of evidence that at time of fireball hadronization we study a fireball in which quarks and gluons (but not hadrons) are chemically equilibrated.

We see more fireball properties in \rf{univRL}. Despite a difference in collision energy by a factor 40 we see little, if any, difference, which clearly shows that the same type of QGP fireball is formed at this SPS, RHIC and LHC energies, see Ref.\cite{Petran:2013lja,Rafelski:2009jr,Rafelski:2014cqa,Petran:2013qla}. For the physical properties of the fireball at freeze-out we find the energy density $\varepsilon=0.45\pm 0.05$ GeV/fm$^3$, the pressure of $P=82\pm 2$ MeV/fm$^3$ and the entropy density of $\sigma=3.8\pm 0.3$ fm$^{-3}$ varying little as a function of reaction energy $\sqrt{s_\mathrm{NN}}$, collision centrality $N_\mathrm{part}$.

\begin{figure}[!t]\sidecaption
\includegraphics[width=0.35\textwidth]{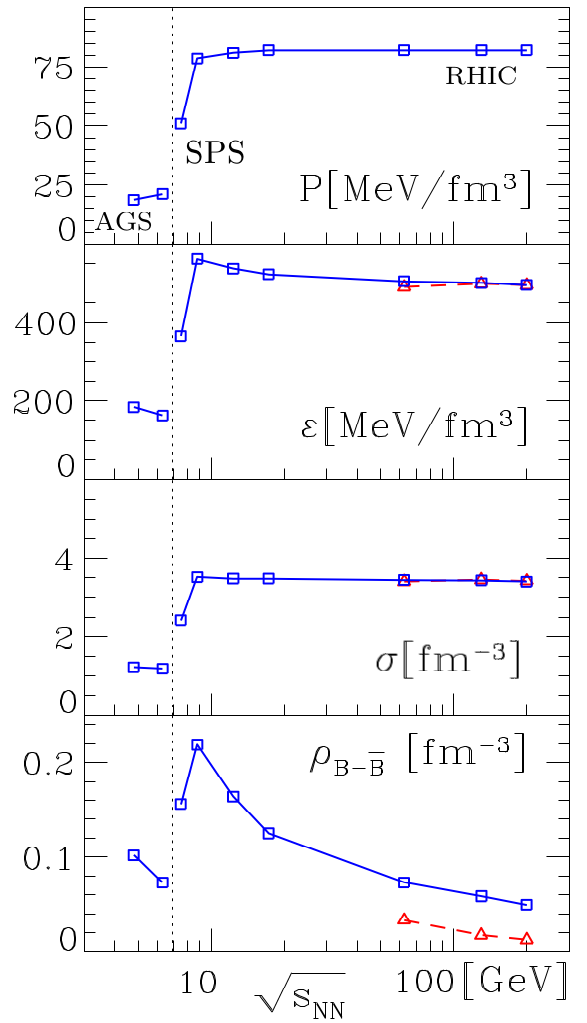}
\includegraphics[width=0.65\textwidth]{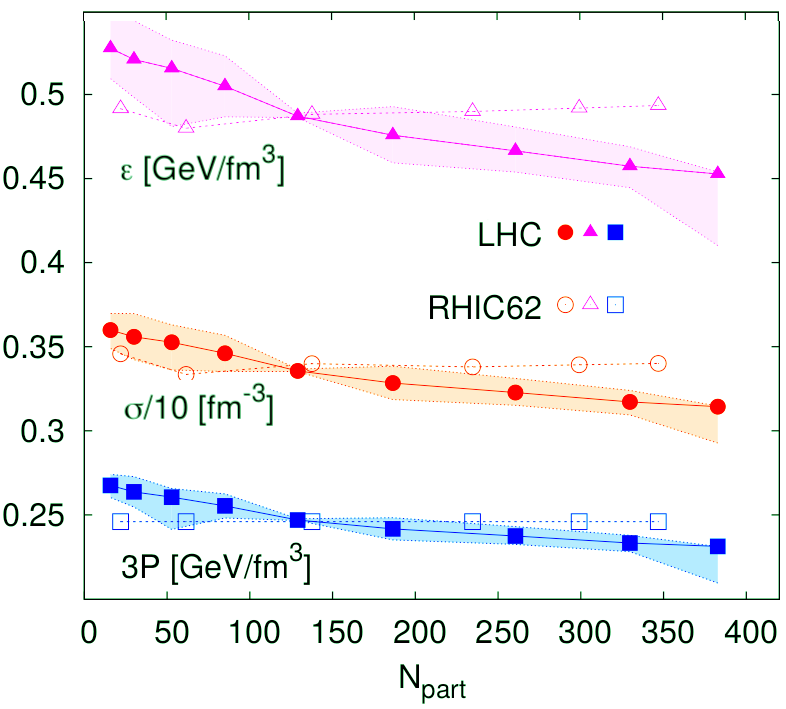}
\caption{The SHARE fit of QGP fireball properties: on left as function of energy for SPS, RHIC (left most point: AGS); on right: as function of the number of participants with comparison of LHC 2.6 TeV results with RHIC 62 GeV, adapted from Refs.\cite{Petran:2013lja,Rafelski:2009jr}}
\label{univRL}
\end{figure}

All of these results are consistent with hadronic particle production occurring from a dense source in which the deconfined strange quarks are already created before hadrons are formed. These (anti-)strange quarks are free to move around or diffuse through the QGP and are readily available to form hadrons. In final state strange hadrons compete in abundancce with nonstrange hadrons. Interpretation of the relation between strange antibaryon production and $s/S$ helps us understand the onset of deconfinement and the appearance of critical point.

\subsubsection{Systematics of ALICE-LHC strangeness results}\label{sec:AliceSys}

In \rf{fig:hypppp} and \rf{fig:hypXO} we see the yields of individual particles, per charged pion, reported by the ALICE-LHC collaboration for all collision systems explored so far. We note in both \rf{fig:hypppp} (on logarithmic scale), and in \rf{fig:hypXO} (on linear scale) the smooth behavior of all shown particle yield results as a function of global $dN_\mathrm{ch}/d\eta$ charged particle multiplicity. The here considered charge multiplicity is measured in the (pseudo)rapidity interval $\eta\in\{-0.5,+0.5\}$. This kinematic domain comprises only a part of the surface surrounding the collision event. Thus the lowest multiplicity bin with $dN_\mathrm{ch}/d\eta\simeq 3\pm 1$ for the most peripheral $p$-$p$ collisions corresponds, allowing for the expected large longitudonal produced particle momentum, to a total charged particle multiplicity that is at least five times larger. Thus we have a sizable, but still a relatively small particle source.

\begin{figure}[tb]\sidecaption
\centerline{%
\includegraphics[width=0.75\textwidth]{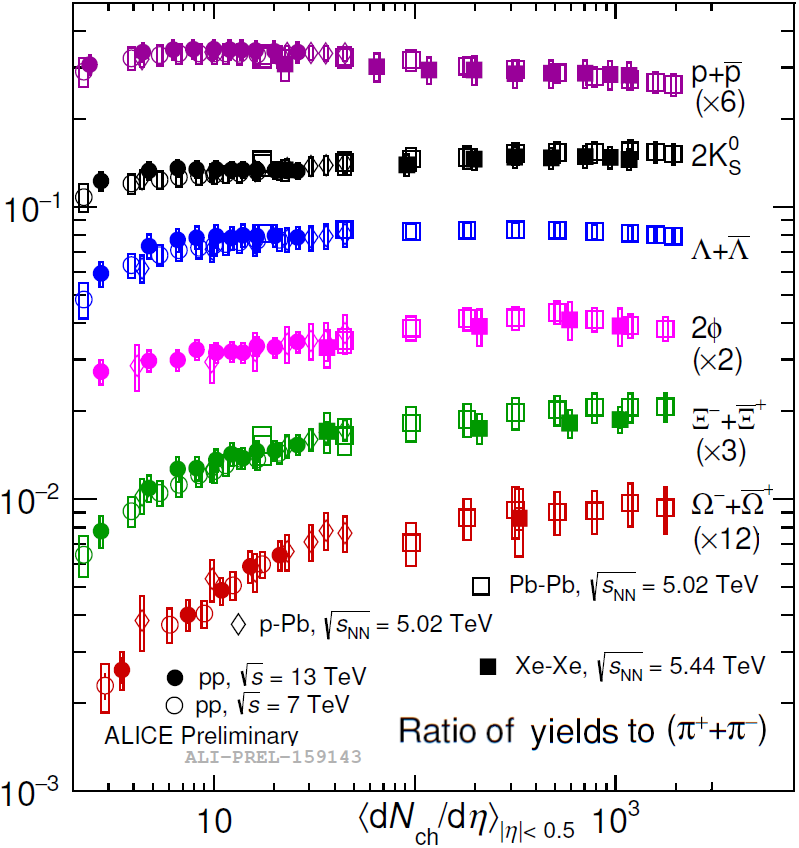} 
}
\caption{%
Universal dependence of $p(uud)$+$\bar p(\bar u\bar u\bar d)$, K$_s(d\bar s+\bar d s$, $\Lambda(uds)+\overline{\Lambda}(\bar u\bar d\bar s)$, $\phi(s\bar s)$, $\Xi^-(dss)+\overline{\Xi}^+(\bar d\bar s\bar s)$, $\Omega(sss)+\overline{\Omega}(\bar s\bar s\bar s)$ multiplicities (scaled by yield of $\pi^+ + \pi^-$) obtained in $p$-$p$, $p$-$A$ and $A$-$A$ by LHC-Alice experiment at LHC collision energies indicated. Adapted from Ref.\cite{Albuquerque:2018kyy}
}
\label{fig:hypppp}
\end{figure}
 
We note three features inherent to these results:
\begin{enumerate}
\item
All results align as a smooth function dependent on the size of the fireball measured by the number of produced charged hadrons $dN_\mathrm{ch}/d\eta$ for the entire LHC energy range. At these ultra high energies the $p$-$p$, $p$-$A$ and $A$-$A$ collisions cannot be clearly distinguished.
\item
All particle yields, shown as a ratio with charged pion yield, do not depend on the energy of the collision in the LHC range. (The 1.5sd disagreement seen between 2.76 and 5.02 TeV results at large multiplicity for $\Xi$ and $\Omega$ is under re-evaluation.) 
\item
For $dN_\mathrm{ch}/d\eta>6$ for all large fireball volumes the ratio 
\begin{equation}\label{eq:RLambda}
R_\Lambda \equiv \frac{\Lambda(uds)+\overline{\Lambda}(\bar u\bar d\bar s)}
{p(uud)+\bar p(\bar u\bar u\bar d)}\ge 1.5
\end{equation}
is greater than unity. 
\end{enumerate}
 
\begin{figure}[tb]\sidecaption
\centerline{%
\includegraphics[width=0.49\textwidth]{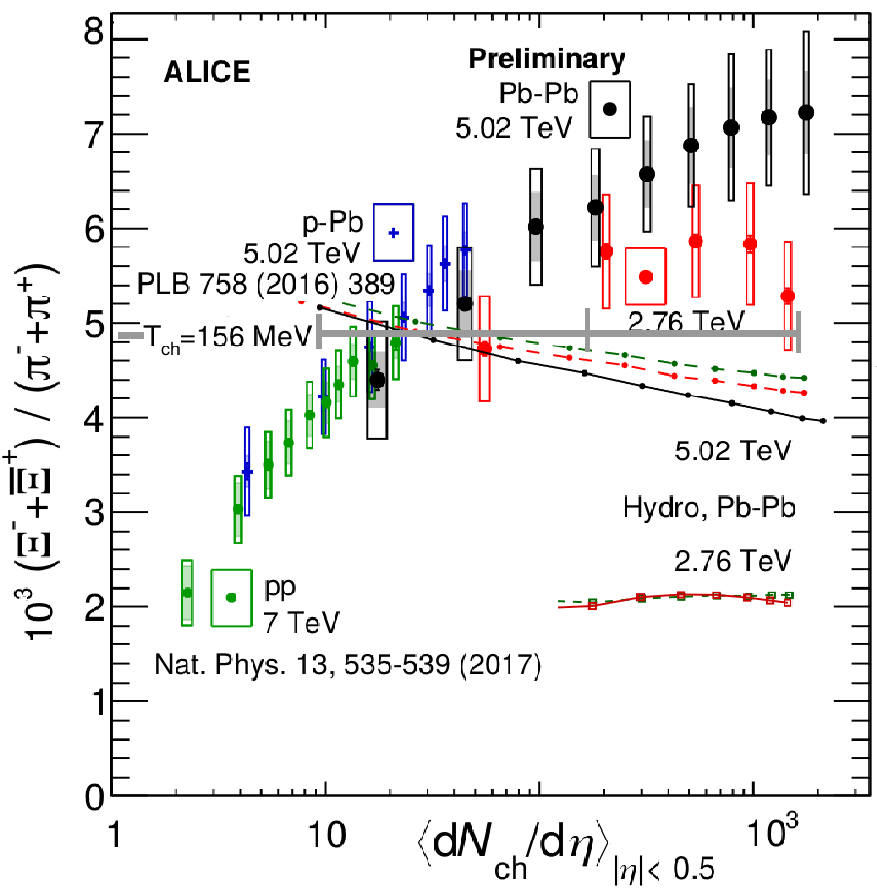} 
\includegraphics[width=0.49\textwidth]{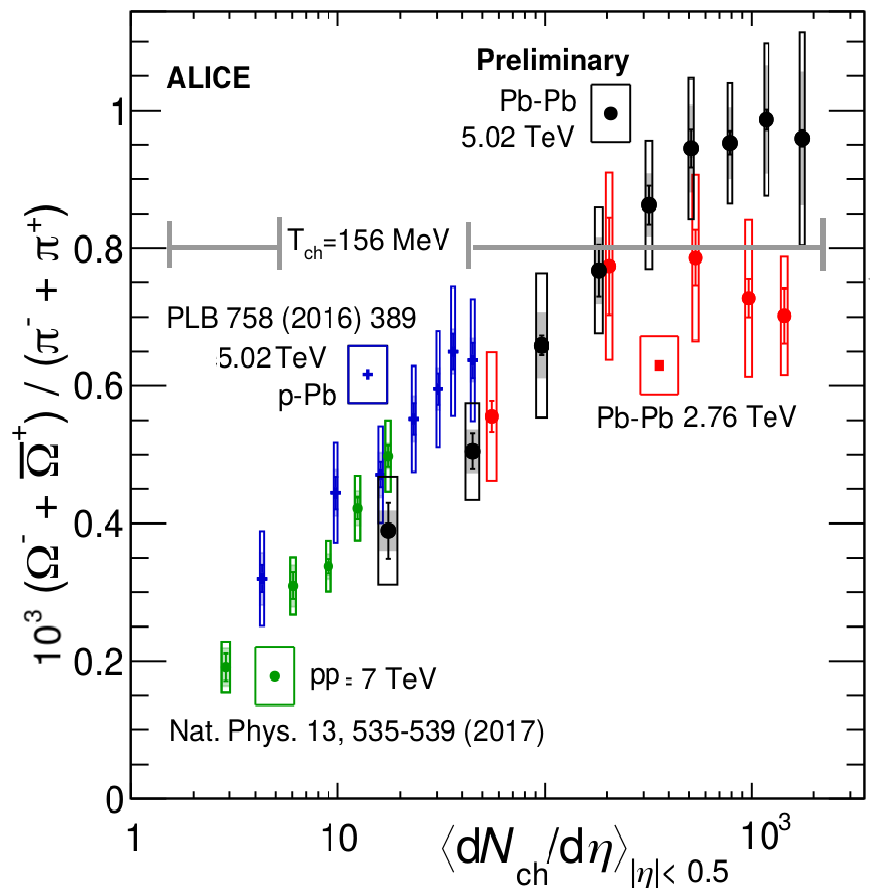} 
}
\caption{%
Results for $\Xi$ (left) and $\Omega$ (right) from \rf{fig:hypppp} are shown on linear scale. The horizontal grey line shows the central result of SHM yields for chemical equilibrium model evaluated at $T=156$\,MeV, for $\Xi$ we see othet models (hydrodynamical computations, dashed including final state re-scattering computed with so called RQMD model). It is generally believed that the high centrality results shown for $2.6$\,TeV (in red) need to be reevaluated. Adapted from Ref.\cite{Albuquerque:PhD} 
}
\label{fig:hypXO}
\end{figure}

The ALICE collaboration associate these results with the formation of the QGP in high multiplicity $p$-$p$ collisions~\cite{ALICE:2017jyt,Albuquerque:2018kyy,Albuquerque:PhD}. In $p$-$p$ and $\alpha$-$\alpha$ collisions studies carried out at the ISR-AFS at 1/100 of the CM energy, a collective effect indicating formation of QGP was not found, see as example the baseline in \rf{RSS}. This indicates that only at sufficiently high energy the small collision system leads to a behavior akin to QGP formation. Similarly, in the scattering of O-Au evaluated at the SPS strangeness enhancement was not observed by the NA35 collaboration, see quote on page \pageref{StockOAu}. Consequently, a boundary of QGP formation must be present both as a function of reaction volume and energy.

Regarding the third result, the ratio \req{RLambda}, this manifestation of strangeness enhancement was presented in our first publications~\cite{Rafelski:1980rk,Rafelski:1980fy}, see bottom paragraph in the quote on page \pageref{FirstPredict}. In the chemical nonequilibrium SHM analysis this result requires that the phase space occupancy in the hadron phase, $\gamma_s/\gamma_q>1.5$ as was have shown in the analysis of the first ALICE 2.6 TeV results\cite{Petran:2013qla,Petran:2013lja}. This result further confirms the universality of the sudden QGP hadronization process~\cite{Rafelski:2000by}, which underpins the principle of universality of hadronization~\cite{Petran:2013qla,Rafelski:2014cqa}. \label{RLam}

The $p$-$p$, $p$-$A$ results agree with the earlier ALICE Pb-Pb results and confirm the need for the full chemical nonequilibrium model with $\gamma_s\ne 1$, and $\gamma_q\ne 1$ when addressing these results within the SHM nonequilibrium approach. In the chemical nonequilibrium SHM analysis these RHIC and LHC hadronization condition results track closely those we reported for the much lower energy at SPS~\cite{Letessier:2005qe} energies. The rise of yields with volume size is well understood as being due to the increased strangeness phase space occupancy factor $\gamma_s$. The relative yield increase seen best in \rf{fig:hypXO} is incompatible with the chemical equilibrium SHM depicted for $T=156$\,MeV, by a grey horizontal line ($\pm5\% $ precision of the model is also indicated).

In the near future a better understanding of QGP formation needs to be developed allowing us to explain why and how small $p$-$p$, ultra-high energy collision systems are capable of forming this new phase of matter. 


\markboth{2. Strangeness  as an Observable of QGP}{Strangeness in QGP: Diaries}
\section{Strangeness in  Quark-Gluon Plasma}
\begin{itemize}
\item
I open the review of the theoretical developments describing the work carried out in the period  1980-1990. For this I will use the first half of the 1992 Il Ciocco Summer School introduction to the topic of strangeness signature of QGP.  
\item
This is followed by the elaboration of strange antibaryon signature, with emphasis on strange antibaryons at \lq high\rq\ momentum dating from 1987. 
\item
A short proposed lecture abstract from 1992 describes the lack of success in presenting this signature of QGP to a wider Nuclear Physics community.  
\item
A report presented  at the 1990  \lq 4th Workshop on  Experiments and Detectors for RHIC\rq\ discusses how these insights influence the experimental landscape under development. 
\item
A few pages from  the 1996 progress report provided to the US funding agency about strangeness production process show refinement of strangness production using QCD properties and show how strangness production and strange antibaryon enhancement depends on collision energy in the SPS energy range. 
\item 
I describe my effort to join the BNL STAR collaboration in 1997-9 where I hoped my theoretical and practical insights gained with the CERN SPS data analysis could be of use, showing additional strangeness production results.
\end{itemize}
Several of these documents are shown in part only, in order to focus on the topic considered and avoid duplications. This is inicated as appropriate, and  in the following section, where methods of experimental result analysis will be introduced, we continue to reprint the contents.

\subsection[Strangeness production
\hfill\textbf{reprinted sections are not numbered}]{Strangeness production}
%
\noindent\textit{The following is part I of the IlCiocco July 12-24, 1992 Summer School~\cite{Gutbrod:1993rp} \lq\lq Particle Production in Highly Excited Matter\rq\rq\ presentation~\cite{Rafelski:1992td} (for part II see page \pageref{Trail1992p2}):}\\[-0.7cm]
%
\begin{mdframed}[linecolor=gray,roundcorner=12pt,backgroundcolor=Dandelion!15,linewidth=1pt,leftmargin=0cm,rightmargin=0cm,topline=true,bottomline=true,skipabove=12pt]\relax%
\centerline{\includegraphics[width=1.0\textwidth]{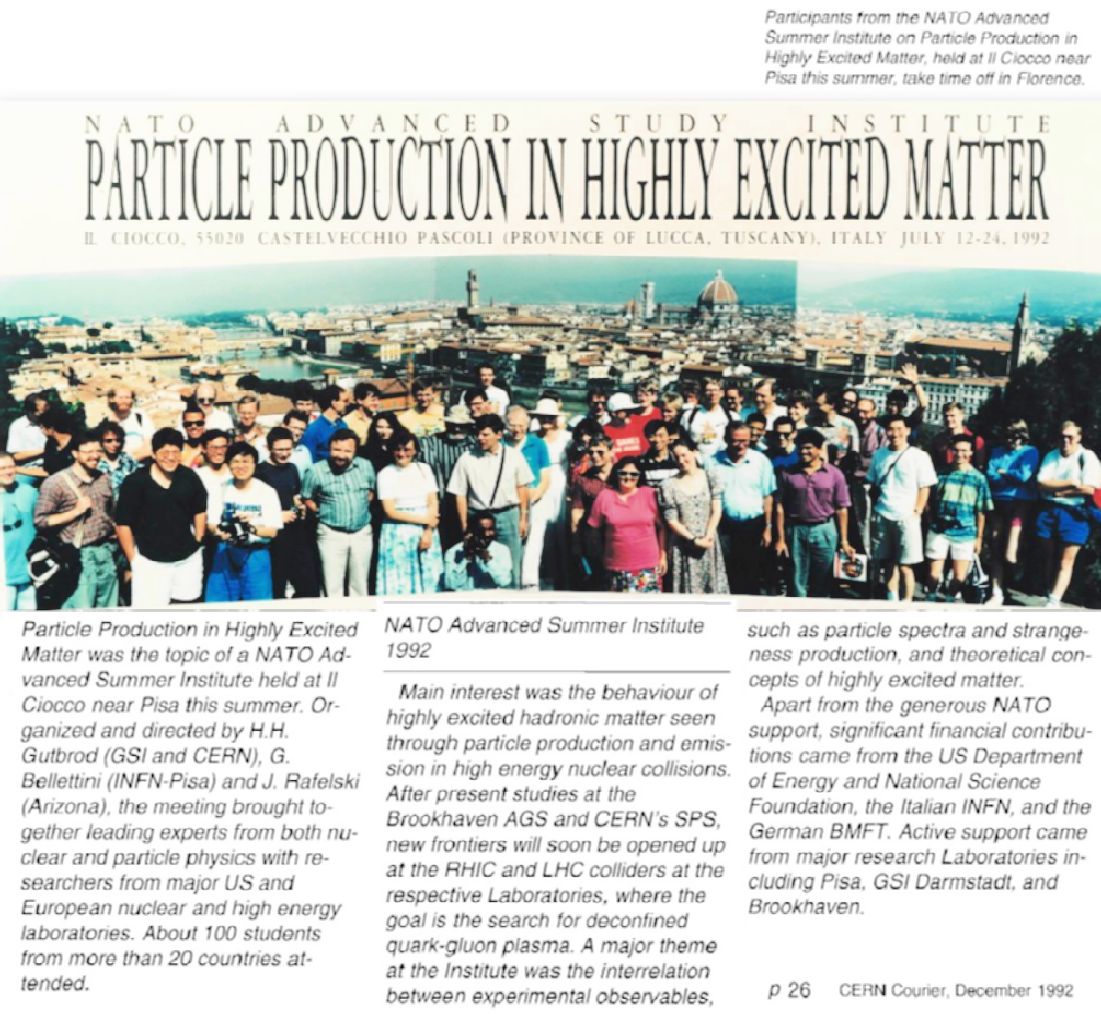}}
\noindent{\small IlCiocco Summer School July 12-24, 1992 Cern Courier Report December 1992: the here shown picture is the color original -- CERN Courier was printed in B/W}\\[0.4cm]
{\Large \bf On the Trail of Quark-Gluon-Plasma:\\
Strange Antibaryons in Nuclear Collisions}
%
\addcontentsline{toc}{subsubsection}{Strangeness as an observable}
\section*{Strangeness as an observable}\label{Trail1992}
There may be no easy laboratory observable of quark-gluon plasma (QGP).
But I hope to convince you that much can be learned about this new form
of nuclear matter studying diligently the third (and still rather light)
strange \lq $s$\rq\ quark flavor. I will in particular address the predicted and observed abundant emission of strange antibaryons in relativistic nuclear collisions which in my opinion constitutes a rather clear indication of new physics$^{1,2,3}$. \ldots
For further theoretical details and numerous references the reader should consult the more comprehensive presentations$^{4,5,6}$ (status 1992).
 
I will address primarily the relative production of strange and
multi-strange antibaryons. As I will mostly address ratios of particle
multiplicities, an important point will escape further attention: there
is a considerable enhancement of the production cross section of these
particles above and beyond the expectations based on hadronic cascading
reactions of the type $p$--$p$. This is in agreement with the naive
expectations based on a scenario involving production of an intermediate
drop of quark-gluon plasma. But why should such an enhancement be
expected in QGP? I will begin by recalling the simple, historic and
somewhat correlated arguments why strange particles in general, and
(multi)strange  (anti)baryons in particular possess a priori a distinct
diagnostic function of the behavior of highly excited nuclear matter:\par
\indent 1: abundance symmetry of $\bar s, \bar u, \bar d$ in statistical
description;\par
\indent 2: strongly differing production rates in different phases of
nuclear matter;\par
\indent 3: high  $s\bar s$--quark pair density in QGP phase;\par
\indent 4: small strange antibaryon background from $p-p$ reactions in
the central region.\par
\noindent I now elaborate on the origin and importance of each of these
points.
 
\subsubsection*{Anti-flavor symmetry} 
Recent CERN-SPS experiments indicate that up to currently available
energies the central rapidity interaction region has a sizable baryon
number and therefore a relatively large baryo-chemical potential
$\mu_{\rm B}$, in CERN experiments with S$\to$W at 200 GeV\,A interaction
we will see that it is about 340 MeV. Statistical models permit
straightforward evaluation of the quark densities in the QGP phase once
$\mu_{\rm B}$ is known. 
 
One easily finds that the heavy quark flavor has a comparable abundance
to the light quarks because of the finite baryon density in the
interaction region. For a central QGP-fireball in chemical equilibrium,
the number of light antiquarks is governed by the factor $e^{-\mu_{\rm
B}/3T}$, while the deconfined strange and antistrange quarks, are not
affected in the QGP by $\mu_{\rm B}$ and are governed in comparison to
$\bar u, \bar d$ quarks only by their non zero mass described by the
factor $e^{m_{\rm s}/T}$ (with $m_{\rm s}\simeq 150-180$ MeV). Both
factors are not very small and also as mentioned, quite similar in
magnitude. Consequently, provided that strangeness production has
saturated the available phase space, the abundance of antiquarks $\bar u, \bar d, \bar s$ will be very similar.
 
\subsubsection*{Production rates} 
Rates for production of $s\bar s$ pairs in the QGP phase were often calculated$^{4,5,7}$ and the relaxation time constant which characterizes the scale of time needed to saturate the phase space is of the order of $10^{-23}$~s results, while in hadronic gas phase it is 10 to 30 times slower$^{4}$ at the same temperature and baryo-chemical potential. This difference is mainly due to the presence of gluons in QGP and to the difference in reaction thresholds. On the other hand the typical time scale for the creation and decay of a central fireball can be estimated as the time to traverse at light velocity, the fireball diameter $2R$~fm i.e. $\simeq 2-4\times 10^{-23}$~s. If the fireball is made of QGP, and is sufficiently large, \textit{e.g.} formed in Pb--Pb collisions, strangeness abundance can reach statistical equilibrium values, in a thermal hadronic gas this is not expected because of the long relaxation time. Should there be some anomalous production mechanisms involving multi-particle scattering, then we have to turn to the next point.
 
\subsubsection*{$s\bar s$--density}
Even in a slow hadronization of an expanding QGP, $s\bar s$ density (now half as high as at its peak due to dilution in expansion) is$^{4}$ about 0.4 strange particle pairs per fm$^3$. In the explosive disintegration \lq\lq Particle Production in Highly Excited Matter\rq\rq\ the discussion of strangness saturation in the QGP fireball. These introductory remarks appeared under title \lq\lq On the Trail of Quark-Gluon-Plasma: Strange Antibaryons in Nuclear Collisions.\rq\rq scenario, the density of strangeness is that of a QGP, thus for a fully saturated phase space up to 0.8 strange particle pairs per fm$^3$. Such a high local strangeness density should favor the formation of multi-strange objects, and particularly multi-strange antibaryons: $\overline{\Omega}=\bar s\bar s\bar s$, $\overline{\Xi}=\bar q\bar s\bar s$. In the hadronic gas phase we lack the strangeness density and therefore we should not expect to form these particles abundantly, as a series of unlikely reactions has to occur in their formation, while their destruction is easily possible in collisions with the hadronic gas constituents
 
\subsubsection*{Direct reaction \lq background\rq} 
It is helpful to recall the magnitude of backgrounds expected for the
production of the multi strange (anti) baryons. The $\overline{\Xi}/
\overline{Y}|_{p_\bot}$ (where $Y=\Lambda,\Sigma$ are the $qqs$ hyperons) ratio seen at ISR at $\sqrt{s}$ = 63 GeV is only 0.06$\pm$0.02 in the central rapidity region$^{8}$. The expected quark-gluon matter result with saturated phase space is up to ten times greater and greatly
enhanced yields have been recently reported by the WA85
collaboration$^9$.
 
The predicted huge strangeness pair density in QGP phase is in my opinion the main point of interest and should be relentlessly pursued by further measurement of the diverse strange particle signatures. It is hard to imagine another scenario in which in particular the strange antibaryons would be abundantly produced: I note that, for example, in a chirally symmetric phase in which the kaon mass could be considerably reduced and hence strangeness could possibly be produced abundantly, there is no  particular reason to expect unusual coagulation of (anti)strangeness into multi-strange antibaryons.

\section*{Kinetics of strangeness production and evolution} 
\addcontentsline{toc}{subsubsection}{Kinetics of strangeness production and evolution} 
\subsection*{Elements of kinetic scattering theory}
Since the time scale in a typical nuclear collision is very short, the
strangeness content of either the QGP phase or the HG phase cannot
without further study be assumed to be saturated (be in `absolute
chemical equilibrium') and it is necessary to determine in a kinetic
approach the rate of strangeness production in alternative scenarios of
fireball evolution. In order to proceed we have to first compute the
typical time required to produce strange quark flavor in the abundance
corresponding to fully saturated phase space. This chemical relaxation
time constant is strongly state dependent, as quite different processes
are contributing in the QGP or HG states of hadronic matter.
 
The strangeness relaxation time constant for the quark-gluon phase is
believed to be in competition with the typical time scale for the
creation and decay of a fireball. Sophisticated calculations which I will
introduce below only in qualitative terms show that we can expect that
strangeness will nearly saturate the available phase space should a
quark-gluon deconfined phase be formed. It is evident that the
coincidental similarity of the computed time constant of strangeness
production with the computed life-span of the QGP has the effect of
making strangeness a quantity very appropriate to the study of the
dynamics of nuclear collisions and QGP. There is furthermore a
considerable impact of the hadronization scenario on strange antibaryon
yield. I like to assume rapid disintegration of the putative QGP phase 
and therefore the strange antibaryon particle yields are representative
of the QGP conditions. An alternative  scenario would be to assume a
relatively slow hadronization which leads to particle yields
characteristic of a (nearly) chemically saturate  hadronic gas$^{10}$.
 
We now briefly consider the theoretical method for the computation of the
rate of the strange quark pair production. Whichever the microscopic
mechanism one adopts for computation of the strange flavor production in
the yet unknown form of high density nuclear matter that has been
generated in these collisions, one can identify the different factors
controlling the yield of the strangeness production processes in a rather
model independent way. Consider two as yet unidentified constituent parts
of centrally interacting nuclei, $A$ and $B$ producing strangeness in
individual collisions. The total number of pairs produced (neglecting
possible strangeness annihilation), leading either to deconfined {\it or}
bound (confined) strange quarks within individual hadrons, is given by
\begin{align*}\tag{1}
N_{\rm s}= V\cdot t\cdot \left(\frac{dN_{\rm s}}{dVdt}\right)\;.
\end{align*}
Here $V$ and $t$ describe the 3+1 dimensional volume in which the
reactions have taken place. The (invariant) rate of production per unit
of time and volume is given by
\begin{align*}\tag{2}
\left(\frac{dN_{\rm s}}{dVdt}\right)\equiv {\cal A}=
\langle\sigma^{\rm s}_{\rm AB}v_{\rm AB} \rangle 
\rho_{\rm A}\rho_{\rm B}\;. \label{Arateis}
\end{align*}
Since  $\rho=N/V$, the specific strangeness yield is:
\begin{align*}\tag{3}
\frac{N_{\rm s}}{n_\pi}=\frac{N_{\rm A}}{n_\pi}\cdot 
\frac{N_{\rm B}}{V} \cdot t\, \cdot 
\langle \sigma^s_{\rm AB} v_{\rm AB}\rangle\;. 
\label{NS}
\end{align*}
The first factor $N_{\rm A}N_{\rm B}/n_\pi$ is rather independent of the
form of {\it proto}matter: the number of components in $A$ or $B$, be
they gluons and quarks or be they pions, will always remain of the same
magnitude as the final multiplicity. This is dictated by the entropy
conservation believed to hold during the evolution of the thermalized
central fireball. Enhancement of strangeness production (as reported by
many CERN and BNL experiments) relative to normal hadronic interactions
(e.g. p+A interactions) is in view of Eq.(\ref{NS}) due to:\\
\indent 1: smaller effective volume $V$ per particle and/or\\
\indent 2: longer interaction time $t$ and/or\\
\indent 3: enhanced microscopic cross section.\\
All these conditions are satisfied in the QGP fireball. Note that the
cross section for formation of particles has the general behavior (note
that $s$ here stands for $(\sqrt{s})^2$, not strangeness):
\begin{align*}\tag{4}
\sigma\simeq k\frac{\alpha^2}{s}\ \sqrt{1-s_{\rm th}/s}\;,
\end{align*}
where the constant $k$ is generally $O(1)$, the threshold $s_{\rm th}$
controls the low energy behavior and the high energy behavior is governed
by the usual $1/s$ form, and $\alpha$ is the strength of the interaction.
These (angle averaged) cross sections for strangeness production have
been studied for many processes involving light quarks, gluons, pions and
so on. They can be parameterized successfully by taking $k\alpha^2\simeq
1$, leading to values of about 0.5~mb for processes at
$\sqrt{s}=2.5T+2.5T\sim1$~GeV. The threshold $s_{\rm th}$ differentiates
to some degree the differing possible processes - in the QGP we expect
$s_{\rm th}=2m_{\rm s}\simeq 350$ MeV, while, in hadronic interactions,
this value is considerably greater on the scale of relevance here: 700
MeV in $\pi+\pi\rightarrow K+K$ reactions. Also, in a confined phase, one
cannot invoke summation over color quantum numbers in the final state,
reducing cross sections still further.
 
\subsection*{Gluons in plasma} 
\addcontentsline{toc}{subsubsection}{Gluons in plasma} 
Including a first-order perturbative effect$^{11}$ the equilibrium gluon
number density in QGP can be estimated using Eq.\,(\ref{fg}) to be:
\begin{equation}\tag{5}
\rho_g \mbox{(fm}^{-3}) = 1.04 \left( {T\over 160\mbox{MeV}} \right)^3
\left( 1- {{15\alpha_{\rm s}}\over{4\pi}} \right)
\end{equation}
giving for a typical temperature of 200 MeV a value of 0.55 fm$^{-3}$ for
$\alpha_{\rm s} = 0.6$ and 0.8 fm$^{-3}$ for $\alpha_{\rm s} = 0.5$. For
a quark-gluon phase volume with radius 4--5 fm, we therefore have
200--300 gluons. Note also that this density rises as the cube of the
temperature. Because gluons can be created and annihilated easily in
interactions with other gluons and light quarks, the gluon density
closely follows the evolution of temperature in the course of the
quark-gluon phase evolution. The equilibrium gluon energy density is
\begin{equation}\tag{6}
\epsilon_g = {8\pi^2\over 15} T^4 \left( 1 - {15\alpha_{\rm s}\over 4\pi}
\right)\;.
\label{fg}
\end{equation}
 
\centerline{\includegraphics[width=0.6\columnwidth]{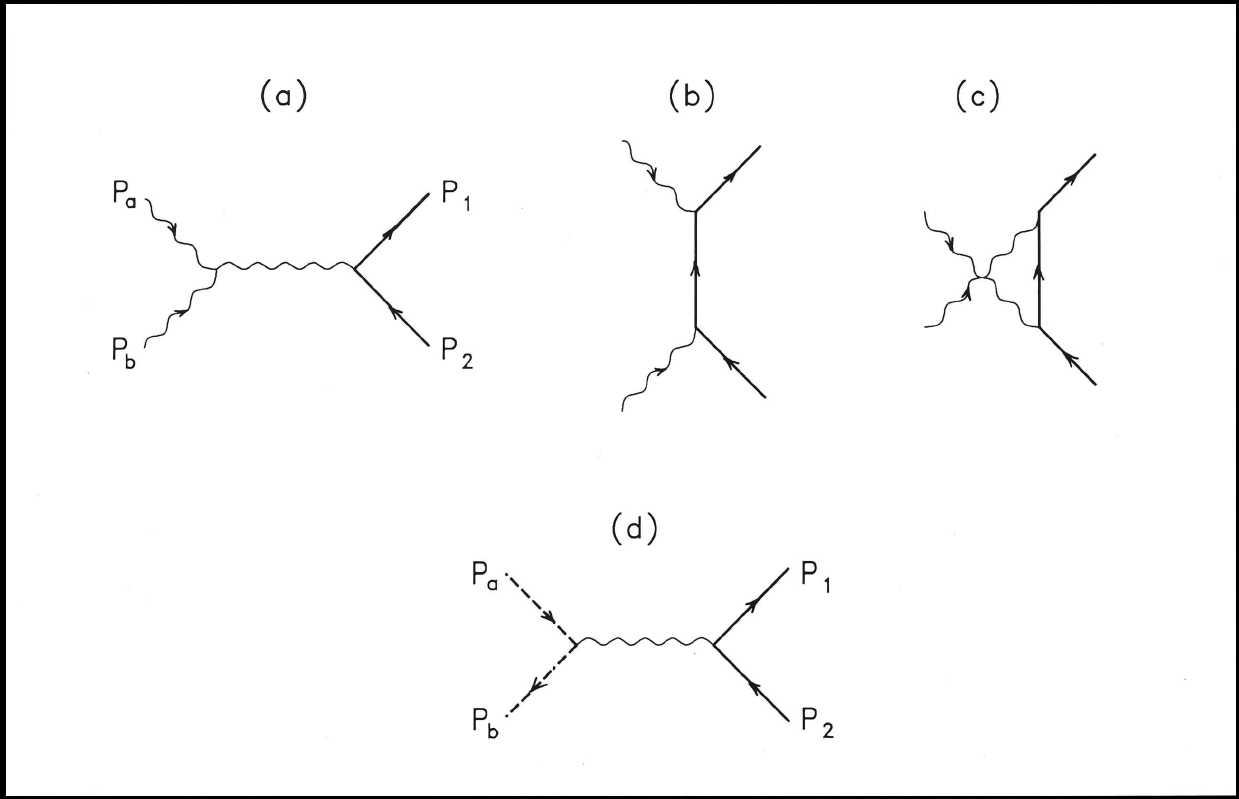} }
\noindent{\small Fig. 1 Lowest-order production of $s\bar s$ by gluons and light quarks
}

Gluons thus play a major role in the dynamics of the quark-gluon
phase-hadronic gas phase transition, also because they carry much of the 
QGP entropy. It is therefore interesting to note that in an indirect way,
strangeness enhancement demonstrates the dynamical presence of glue
degrees of freedom. In QGP strangeness can be formed by processes shown
in Fig.\,1, and higher order corrections of the same basic type.
Calculations show that it is predominantly formed by reactions of gluons,
rather than quarks, despite the fact that the QCD cross sections, shown
in Fig.\,2, are similar for both processes. However the
statistical
factors entering the thermal average will strongly favor the gluon
induced processes: there are simply more glue-glue than quark-antiquark
collisions of suitable quantum number in the plasma. In a scenario of QGP
based fireball practically all of the $s{\bar s}$-pair production results
from collisions of the central gluons, which in a first approximation can
be assumed to be in a thermal distribution. Because of glue dominance of
the production process, the time evolution of strangeness during the
production process is a function of temperature, which solely controls
the glue abundance, but not of the baryo-chemical potential, which
determines the quark densities. Consequently, the detailed baryon number
retained in the plasma (baryon stopping) is of no importance for strange
particle yield; the actual plasma lifetime, volume and temperature (i.e.
gluon density) are the critical parameters  determining the absolute
yield in baryon free and baryon rich environments.
 
\centerline{\includegraphics[width=0.6\columnwidth]{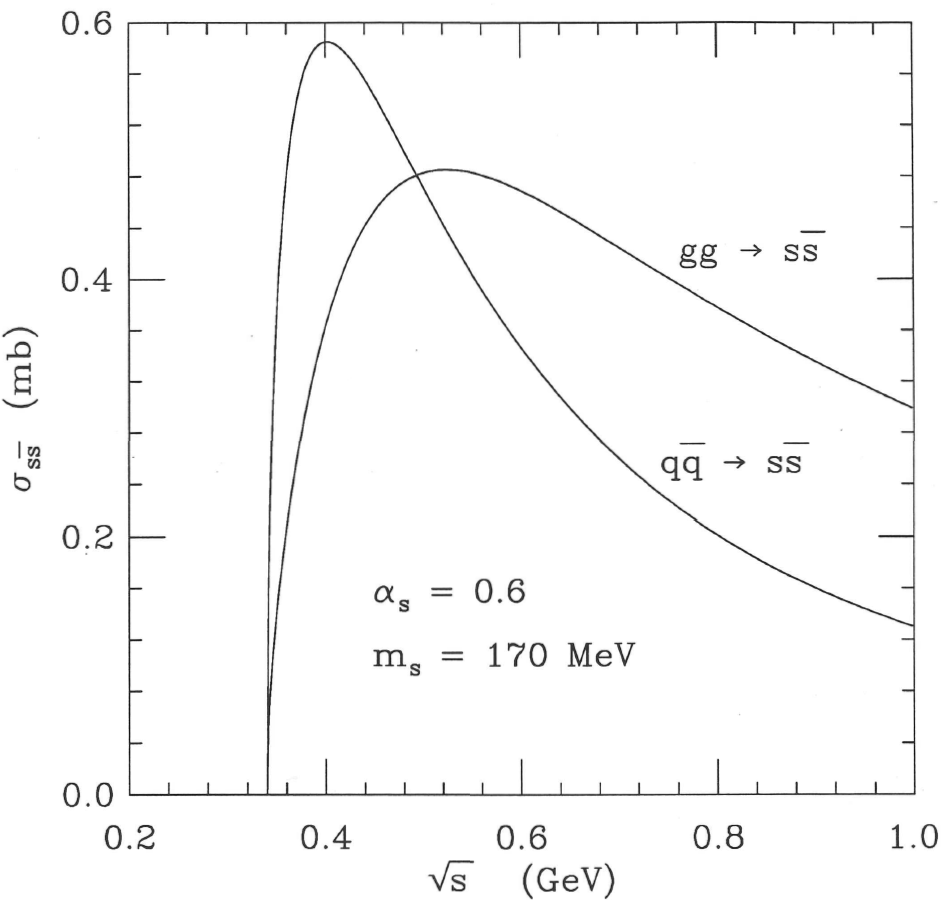}}
\noindent{\small Fig. 2 Strangeness production cross sections in QGP, $\alpha_{\rm s}=0.6$ and $m_{\rm s}=170$ MeV
}

Gluons not only produce strangeness flavor dominantly but they provide
the key distinction between the QGP phase and the HG. The high gluon
abundance and density in the plasma impacts the entire history of the
plasma state, in particular also the process of hadronization at the end
of the quark-gluon phase lifetime, in which appreciable strangeness
production occurs again. Indeed, {\it abundant strangeness should be
viewed  as a signal for presence of gluons} or alternative color charged
objects which are not quarks (strings, ropes etc). 
%
\section*{Approach to absolute chemical equilibrium}
\addcontentsline{toc}{subsubsection}{Approach to absolute chemical equilibrium}
In order to quantify the strangeness production in the dynamical
situation of the rapidly evolving heavy ion collision, it is convenient
to introduce the  quantity:
\begin{equation}\tag{7}
\gamma_{\rm s}\equiv {\int d^3\!p/ 2\pi^3\,
     n_{\rm s}(\vec p,\vec x;t)\over N_0/V}\;,
\end{equation}
which characterizes approach to saturation of the phase-space of strange
particles. $N_0/V$ is the equilibrium particle density. The integration
over the momenta is appropriate if the thermal (kinetic) equilibration 
occurs at a considerably shorter time scale than the (absolute) chemical
equilibration. This observation implies that the factor $\gamma_{\rm s}$
effectively enters the momentum distribution as a multiplicative factor:
\begin{equation}\tag{8}
n_{\rm s}(\vec p,\vec x;t)=\gamma_{\rm s}(t)n_0(\vec p;T(\vec x),\mu_{\rm
s}(\vec x))\;,
\end{equation}
where the $\vec x$ dependence is contained in the statistical parameters.
The absolute chemical equilibrium corresponds to $\gamma_{\rm s}=1$ found
for a fully saturated phase space of QGP or HG. In HG the absolute
density $N_0^{\rm HG}/V$ is smaller by a factor 2-5 (in dependance on the
physical conditions in which the phases are compared) primarily due to
the larger degeneracy following from the liberation of the color degrees
of freedom, and the lower masses of strangeness carrying particles. I
will show below that it is indeed quite easy to measure the value of
$\gamma_{\rm s}$, and thus this discussion is a very important practical
element in understanding the behavior of hadronic phases: $\gamma_{\rm
s}$ can be studied varying a number of parameters of the collision, such
as the volume occupied by the fireball (varying size of the colliding
nuclei and impact parameter), the trigger condition (e.g. the
inelasticity), the energy of colliding nuclei, searching for the
threshold energy of abundant strangeness formation.
 
The theoretical dynamical model to study $\gamma_{\rm s}(t)$ has been
developed to considerable detail. It arises from a standard population
evolution equation. Detailed balance assures that the production dn
annihilation processes are balancing each other as $\gamma_{\rm s}\to 1$:
\begin{align*}\tag{9}
2\tau_{\rm s}\left( {d\gamma_{\rm s}\over dt}+\gamma_{\rm s}
          {dV\over Vdt}\right)=1-\gamma_{\rm s}^2(t)\;,
\label{gamt}
\end{align*}
where 
\begin{align*}\tag{10}
\tau_{\rm s}=0.5n_0/{\cal A}\;,
\label{taus}
\end{align*}
with $\cal A$ as defined by Eq.\,(\ref{Arateis}). The factor $0.5$ is
introduced in this  definition Eq.\,(\ref{taus}) of the strangeness
relaxation time constant to allow for the relation at large times to
assume the standard form: $1-\gamma_{\rm s}\propto e^{-t/\tau_{\rm s}}$.
The second term on the left hand side of Eq.\,(\ref{gamt}) is the
dilution term arising from the possible expansion of the volume occupied
by the system. Ignoring the dilution and assuming that there is no
appreciable change in $\tau_{\rm s}$ with time, a well known solution of
Eq.\,(\ref{gamt}) is:
\begin{equation}\tag{11}
\gamma_{\rm s}(t)=\tanh(t/2\tau_{\rm s})\;.
\end{equation}
 
The first calculation$^{7}$ of $\tau_{\rm s}$ in QGP in which glue
processes were considered without dilution, see Fig.\,IC-3, has shown
that
strangeness can be produced rapidly and abundantly; subsequent
study$^{12}$ has obtained $\gamma_{\rm s}(t\to\infty)$ including the
dilution effect for both QGP and HG phases. It is evident that a
quantitative calculation of the value of $\gamma_{\rm s}$ reaches in the
actual collision requires as input the relaxation time constant
$\tau_{\rm s}$ and the logarithmic derivative of the local volume, i.e.
the dilution of the local density as function of time due to the dynamics
of the collision. Calculations so far performed use perturbative QCD to
obtain $\tau_{\rm s}$ and model the dilution term using dilution $d(\ln
V)/dt=n/t$ with $n=3$ for a spherically expanding fireball and $n=1$ for
longitudinal expansion. The latter case leads to considerably greater
saturation, also because the temperature parameter which enters the
relaxation time decreases as $T\propto t^{-n/3}$ and there is more time
for strangeness production. However, the spherical expansion is probably
a more appropriate model for the situation encountered in S--W or Pb
collisions, and certainly more applicable to the case of forthcoming Pb--Pb experiments.
 
\centerline{\includegraphics[width=0.75\columnwidth]{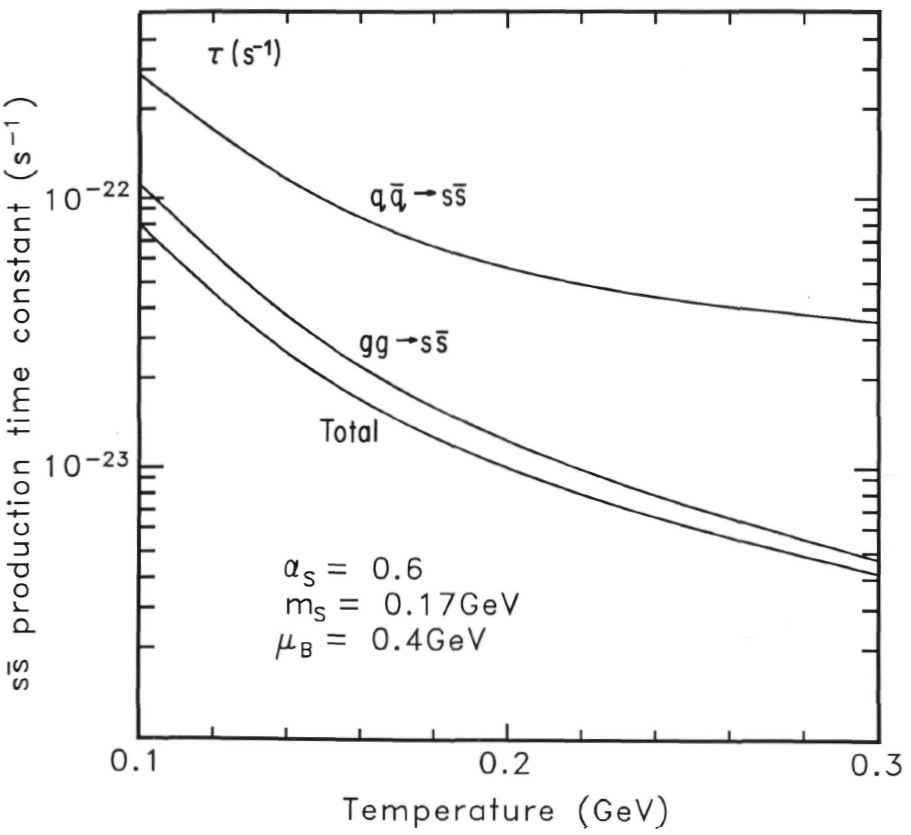}}
\noindent{\small Fig. 3 Relaxation time constants for strangeness production in QGP: total, gluons only ($GG\to s\bar s$) and light quarks only ($q\bar q\to s\bar s$) with $\mu_{\rm B}=400$ MeV,  computed for $\alpha_{\rm s}=0.6$ and $m_{\rm s}=170$ MeV.
}

I will not review in this brief and qualitative talk the detailed results
about the asymptotic value $\gamma_{\rm s}$ assumes in different
scenarios. Suffice here to say that for a $3$ fm radius initial plasma
drop at initial temperature of 250 MeV one estimates $\gamma_{\rm
s}(t\to\infty)\simeq0.5$, with an error as large as 50\% due to the
assumed values of the QCD parameters such as the coupling constant
$\alpha_{\rm s}$ and the strange quark mass $m_{\rm s}$, not to mention
the systematic uncertainty associated with use of perturbative expansion.
Given the current remarkable results on $\gamma_{\rm s}$ it appears
imperative that the models be improved to the level of the experimental
precision which is presently at about 15\%.

\footnotetext{\vspace*{-0.5cm}\begin{itemize} 
\item[1]
J. Rafelski,  Phys. Rep.  C \textbf{88} 331 (1982)
\item[2]
J.~Rafelski and M.~Danos,  Phys. Lett.  B \textbf{192} 432 (1987)
\item[3]
J. Rafelski {\it Phys. Lett.} B262:333 (1991);  Nucl. Phys. A \textbf{544} 279c (1992)
\item[4]
P.~Koch, B.~M\"uller and J.~Rafelski,  Phys. Rep.  C \textbf{142} 167 (1986)
\item[5]
H.C.~Eggers and J.~Rafelski, Int. Journal of Mod. Phys.  A \textbf{6} 1067 
(1991) 
\item[6]
J.~Letessier, A.~Tounsi, U.~Heinz, J.~Sollfrank and J.~Rafelski, {\it
Strangeness Conservation  in Hot Fireballs} Preprint
Paris PAR/LPTHE/92-27, Regensburg TPR-92-28, Arizona 
AZPH-TH/92-23, 1992 (published: Phys. Rev. D \textbf{51} 3408  (1995))
\item[7]
J.~Rafelski and B.~M\"uller,  Phys. Rev. Lett.  \textbf{48} 1066 (1982); and \textbf{56} 2334(E) (1986) 
\item[8]
T.~\AA kesson et al. [ISR-Axial Field Spect. Collab.], Nucl. Phys. B \textbf{246} 1 (1984)
\item[9]
E. Quercigh, this volume; S. Abatzis {\it et~al}.,   Phys. Lett. 
B \textbf{270} 123 (1991)
\item[10]
J. Zimanyi, this volume~\cite{Gutbrod:1993rp}, pp.243-270 
\item[11]
S.A.~Chin,   Phys. Lett.  B \textbf{78} 552 (1978)  
\item[12]
P. Koch, B. M\"uller and J. Rafelski,  Z. Physik  A \textbf{324} 3642 (1986)
\end{itemize}
}
\end{mdframed}

\subsection{Strange antibaryon production}
\subsubsection{High $p_\bot$ recombinant enhancement}\label{sec:highPT}

In the March 1987  lecture~\cite{Rafelski:1987bx} \lq\lq Strange Signals of Quark-Gluon Plasma\rq\rq\ at the Rencontres des Moriond in Les Arcs, France~\cite{TranThanhVan:1987tm}, I extended  the science case for strange antibaryon signatures of quark-gluon plamsa employing the recombinant mechanism~\cite{Koch:1986ud,Rafelski:1987un}. The key result is the enhancement of particle production in  the relatively high $p_\bot$ domain.

These results were confirmed by the RHIC experiments and the data analyis along the line of the recombination model has found general acceptance~\cite{Fries:2003vb,Fries:2003kq} -- the pioneering contributions~\cite{Rafelski:1987un,Rafelski:1987bx}, are not well known. The March 1987 workshop  contents in the pre-web period was seen only by participants -- and few from the RHI community were present at this particle physics meeting. \\ 

\noindent\textit{%
CERN-TH/4716, May 1987 report on high $p_\bot$ strange antibaryons~\cite{Rafelski:1987bx} produced in quark combinant processes:}\\[-0.7cm]
\begin{mdframed}[linecolor=gray,roundcorner=12pt,backgroundcolor=Dandelion!15,linewidth=1pt,leftmargin=0cm,rightmargin=0cm,topline=true,bottomline=true,skipabove=12pt]\relax%
{\Large {\bf Strange Signals of Quark-Gluon Plasma}}\\[0.4cm]
\textbf{Abstract:} It is shown that an overabundance of $\bar\Xi$ is a diagnostic observable of quark-gluon plasma phase of matter. The pertinent physical phenomena are briefly surveyed. New results on $\bar\Xi/\bar\Lambda$ ratios at medium to large transverse momenta are presented. Relevant experiments are discussed.
 
\section*{Introduction}

We address here the question of how the occurrence of otherwise rarely produced multiply strange hadrons can be used to study the formation of the new phase of matter, the quark-gluon plasma\footnotemark[1]\footnotetext{$^1$J. Rafelski, Nucl. Phys. A418 (1984) 215 and references therein} (QGP). At the outset it is important to recognize that the basic subprocess for strange quark production, namely the pair production process is, in principle, the same for both phases of hadronic matter, viz. QGP and Hadronic Gas (HG). But in the latter case of well separated individual hadrons with the nonperturbative (\lq true\rq) QCD vacuum in-between, strangeness production can only take place during the actual constraints in space and time. Furthermore, all initial and final state hadrons are color singlets and the effective number of available degrees of freedom is greatly reduced in comparison to the QGP, in which colored states are permitted.

In the plasma phase there is not only significantly more rapid strangeness production\footnotemark[1]  the higher possible equilibrium strange quark abundance per unit of volume facilitates, in particular, abundant formation of multiply strange antibaryons when the plasma state fragments and recombines to form individual hadrons. In the baryon rich regime of quark-gluon plasma the $\bar s$-quarks are more abundant than the $\bar u$- or $\bar d$-quarks with the consequence that formation of antibaryons with high strangeness content is particularly facilitated during the conversion to the hadronic gas (HG) of the plasma phase\footnotemark[2]\footnotetext{$^2$J. Rafelski and R. Hagedorn, \lq\lq From Hadron Gas to Quark Matter II\rq\rq. CERN-TH 2969 (1980) which 	appeared in \textit{Thermodynamics of Quarks and Hadrons,} (North Holland, 1981) H. Satz, ed.}. One of the highly relevant insights not discussed here is the fact that during the conceivable hadronic reaction time of less than 10$^{22}$ sec, the strangeness produced in HG will not saturate the available (small) phase space. However, any strange particles present will be nonetheless efficiently distributed among various individual hadronic states. These remarks about chemical reequilibration of strange particles apply also to the debris of the QGP after its hadronization. But particles emitted from QGP early on will not be affected by these rescattering phenomena. Strange antibaryons emanating early from the expanding ball of QGP may be far off the relative equilibrium. Upon a brief survey of $s \bar s$-production mechanisms we turn to describe key results on strange particle production with the emphasis being laid on the rare multiply strange baryons.

\section*{Strangeness Production in Quark-Gluon Plasma}
\addcontentsline{toc}{subsubsection}{Strangeness production in quark-gluon plasma}
In QGP the gluonic production rate dominates strangeness production and leads to equilibration times comparable to the expected plasma lifetime\footnotemark[3]\footnotetext{$^3$J. Rafelski and B. M\"uller, Phys. Rev. Lett. \textbf{48} (1982) 1066; and \textbf{55} (1986) 2334(E)}. The averaged cross sections for quark-pair production in lowest order in the QCD coupling constant $\alpha_s$ are used to obtain the rates for strangeness production. In such calculations it is assumed that each perturbative quantum (light quark, gluon) will rescatter many times during the lifetime of the plasma. Hence the required momentum distribution functions are taken to be the statistical Bose, or respectively, Fermi distribution functions, where the temperature $T$ and chemical potential $\mu$ may be functions of $\vec x$, the location of a volume element within the fireball.

\textit{(Repetative Fig.1 showing rates of  strangeness production is omitted.)}\\
The gluon contribution dominates the strange\-ness creation rate, while $q \bar{q} \to  s\bar{s}$ (dashed lines) contributes less than 20 percent to the total rate. The relaxation time $\tau$ is also dominated by the gluonic production mechanism and is falling rapidly with increasing temperature.
\ldots there is virtually no net strangeness annihilation $s \bar s\rightarrow g\bar g$ as the plasma expands, because in the expansion process the temperature and strangeness density both drop rapidly decoupling effectively the strangeness abundance from the statistical equilibrium. This is confirmed in detailed calculations including the dilution term\footnotemark[4]\footnotetext{$^4$\,P. Koch, B. M\"uller and J. Rafelski. Phys. Rep. \textbf{142} 	(1986) 167; P. Koch, Ph.D. Thesis, 1986; P. Koch, 	 contribution to this conference}. It is further found that the strangeness density at hadronization of QCD is in the interval 
$0.15/\mathrm{fm}^3 < \rho_s <  0.3/\mathrm{fm}^3$. This value indicates that clustering of two strange quarks in one hadronic volume $V_h \sim  5 \mathrm{fm}^3$ will be frequent. The high particle density of strange quarks in plasma virtually assures that the numerous $s$ and $\bar s$ quarks will facilitate production of otherwise rare particles such as $\bar \Xi, \bar \Omega$ and particularly important, their antiparticles, instead of being bound in kaons only. Consequently, we will emphasize below our expectations about production of these multistrange hadrons.

\section*{Strange Hadron Formation from Quark-Gluon Plasma}
\addcontentsline{toc}{subsubsection}{Strange hadron formation from quark-gluon plasma}
First we must appreciate that substantial fragmentation of gluons and quarks is required at the transition from QGP to HG. This is easily seen noting that if quarks and antiquarks would recombine into mesons, mainly pions while gluons would vanish in the vacuum, there would be only half as many pions afterwards as there were quarks and antiquarks before which, in turn, are only half of all particles in QGP. Hence the entropy ratio between quark-gluon gas and the hadronic (pion) gas would be 4. In order to conserve entropy during the hadronization process, every gluon and about one third of the quarks must fragment before coalescing into mesons. At finite baryochemical potential the necessity for quark fragmentation is somewhat reduced, since baryon formation accounts for a significant fraction of the total entropy of the hadronic gas. Another aspect of the fragmentation process is that it is producing, with relative strength of about 15\%, further strange quark pairs. This fragmentation value is known from color string breaking considerations.

The flavor composition of all the quarks and antiquarks that finally become constituents of the hadrons produced in the breakup of the plasma is now fully determined. At a given time there are for each flavor, the primary quarks or antiquarks and there are those generated by glue fragmentation. When we combine all these, we obtain the final number of quarks and antiquarks of each flavor, that effectively contribute to hadronization. A combinatoric breakup model to determine the flavor composition of hadrons at the beginning of the evolution of the final hadronic phase can then be used. A further element needed is a model of phase coexistence between HG and QCP. This has been carried through in some detail by P. Koch$^4$. For our purposes it will be sufficient to make a simple estimate. We are particularly interested in the relative abundance of the anticascades ($\bar{\Xi}=\bar{ssq}$) to anithyperon ($\bar Y = \bar{sqq}$). In order to establish this ratio of abundance of $\bar \Xi$ to $\bar Y$, we consider the probabilities $P_i$ of finding the constituent particle in a unit volume $V$. Incorporating gluons which have to fragment when plasma hadronizes into 
$q \bar q$-pairs (85\%) and $s \bar s$-pairs ($f \geq 15$\%), we have
\begin{equation}\tag{3.1}
{{\bar \Xi} \over {\bar Y}}=
{
{(P_{\bar s} + f P_G)^2 (P_{\bar q} +(1-f)P_G)}
\over
{B(P_{\bar s} +f P_G) (P_{\bar q} + (1-f)P_G)^2}
}\;.
\end{equation}
Here B is the branching ratio reflecting the possibility that an $\bar s GG$ system may make not only the desired system $\bar s\bar q\bar q $ + $qq$, but also systems such as $\bar s q + \bar q G$ etc. Clearly  $B<1$. We have from Eq.(3.1)
\begin{equation}\tag{3.2}
B{ {\bar \Xi} \over {\bar Y}}=\displaystyle\frac
{{f \over {1-f}}+ { 1 \over{1-f}} \frac{P_{\bar s}}{P_G}}
{1+{1\over{1-f}}\frac{P_{\bar q}}{P_G}}\;.
\end{equation}
Using the statistical weights for $P_{\bar s}/P_G=3/8$ we obtain 
$\bar \Xi/\bar Y >0.6$(!). 

Including the $\bar q$-density is easy, as 
$P_{\bar q}/P_G=e^{-\mu^s_b/3T} 6/8$, 
 with the last factor again being statistical. In detailed calculations in which various branchings for the different reactions have been allowed\footnotemark[5]\footnotetext{$^5$\,M. Jacob and J. Rafelski, "Longitudinal $\bar\Lambda$Polarization, $\bar\Xi$ Abundance and Quark-Gluon Plasma 	 Formation", CERNTH 4649/87, Phys. Lett. B in 	press (published: \textbf{190} (1987) 173)}, the approximate form Eq.(3.2) given the above parameters is well recovered:
\begin{equation}\tag{3.3}
{{\bar \Xi} \over {\bar Y}} \equiv \frac 3 4
{ 1 \over
{1+0.8e^{-\mu_b /3T}}
}\;.
\end{equation} 

Dashed line in Fig. 2 shows the result Eq.(3.3). It is important to note that a number of these ($\bar \Xi , \bar Y$) particles are expected to be produced in each single nuclear reaction event leading to formation of quark gluon plasma with a volume of several hundred fm$^3$. At this point it is interesting to note that the UA5 collaboration  
(S$  p\bar p  $S) has indeed observed such an anticascade anomaly\footnotemark[6]\footnotetext{$^6$\,G.J. Alver et al (UA5 collaboration) Phys. Lett. B \textbf{151} (1985) 309} in nondiffractive interactions at 
$\sqrt{s}$ = 540 GeV, quoting $\Xi/Y$ ratio of 0.7 at the production point  (note that since baryon density is zero, there is particle-antiparticle symmetry). They further find about 0.1 $\Xi^-$ per event, in which the mean charged particle multiplicity is  $35\pm 4$. We also record that the observed UA5  $\bar \Xi/\bar Y$  ratio is nearly ten times that seen at ISR at $\sqrt{s}=63$\;GeV (central rapidity)\footnotemark[7]\footnotetext{$^7$\,T. Akesson et al (ISR Axial Field Spec. Coll.) 	Nucl. Phys. B \textbf{246} (1984) 1}. It is extremely tempting to conclude that there is quark gluon plasma in $p\bar p$ interactions at  $\sqrt{s}=540$\;GeV, the more so, since the total strangeness yield seems to show anomalous increase as function of 
$\sqrt{s}$ between 500 and 900 GeV.

The discussion presented here confirms the particular suitability of the global abundance of strange antibaryons for diagnosis and study of the quark-gluon plasma state. However, one can also consider the direct effect of the large $s, \bar s$ density in the early plasma before hadronization, and in particular the possibility of strange particle radiation from such a hot plasma state. Two channels of early particle emission may be considered. In the first a fast quark (or diquark) from the plasma impinges on the boundary between the plasma perturbative vacuum and the true vacuum. In the associated color string breaking process, at least one quark-antiquark pair is formed. We will refer to this process as a \lq microjet,\rq\ (not to be confused with the minijet processes). Second, in the prehadronization recombination approach, several constituents of the plasma, clustered into colorless objects, penetrate the surface and hadronize. This latter process is similar to the phase transformation of the plasma to the hadronic gas at the later stages of the plasma life which we have just considered. However, this recombination radiation is, in detail, different in that we may not allow gluon fragmentation. The equilibrium quark abundances in plasma contain the effect of continuous gluon and quark fragmentations and recombinations. Only at phase transition equilibrium is lost and additional microscopic processes such as gluon fragmentation must be explicitly considered.

In Fig. 2 (full lines) we show the predictions based on prehadronization models for the  $\bar \Xi/(\bar Y/2)$ ratio and compare them to global abundance ratios as discussed further above, as functions of the chemical potential (baryon density). The surprise is that in both early emission processes (microjet, recombination) populating medium to high $m_\perp$ portions of the spectra we are led to expect more anticascades than antilambdas. Since  $\bar \Sigma^0 \rightarrow \bar \Lambda +\gamma$ we included $\bar \Sigma^0$ into the \lq lambda\rq\ abundance hence the figure shows only half of the anithyperons abundance  the other half contained in $\bar \Sigma^+$ and $\bar \Sigma^-$ remains invisible in a typical experiment.

\centerline{\includegraphics[width=0.65\columnwidth]{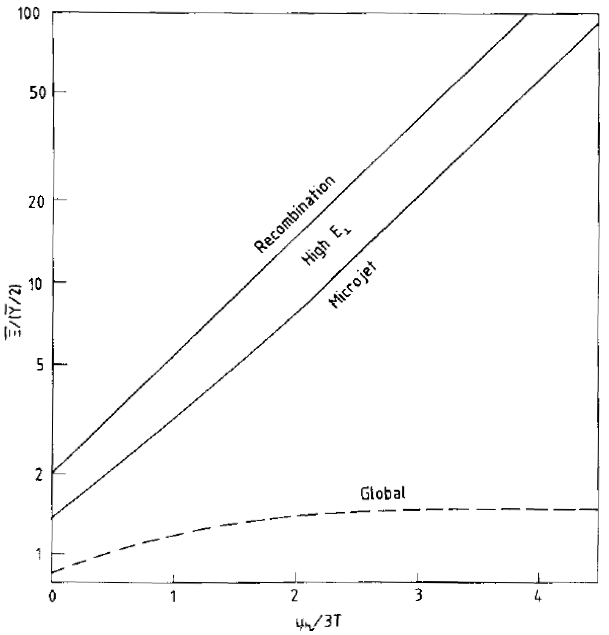}}
\noindent\small{Fig. 2 $\bar \Xi /(\bar Y /2)$ particle ratio as a function of 
$\mu_b/3T$ for the microjet, recombination pictures of particle radiation at medium to high $E_\perp$. Dashed: global ratio}

\section*{Summary}

The results presented here substantiate the expectation that abundances of strange particles, most notably of strange antibaryons, provide a powerful tool to probe the quark-gluon plasma phase of nuclear reactions at very high energy and perhaps even QGP is found in $p \bar p$ annihilation. Strange hadronic particles are expected to emerge from the quark-gluon plasma phase significantly more abundantly than this would be the case in a purely hadronic gas. It is important here to emphasize that strong enhancements of the strange antihyperon $\bar Y$ and anticascade $\bar \Xi$ production is particularly characteristic for the baryon rich plasma phase. In other words: $\bar Y, \bar \Xi$ and $\bar \Omega$ abundance anomalies are characteristic for plasma formation because they will exceed the size of the phase space of individual hadrons.

The source of all these surprising results about strange hadrons from QGP can be traced back to the fact that strange quark-pair production in the plasma phase proceeds at a sufficiently fast rate to permit statistical equilibrium abundance to be established in less than $10\;\mathrm{fm}/c$. This is due to the abundant presence of gluonic excitations, allowing for quark-pair production in gluon-gluon collisions, c.f. Sec. 2. In a way, therefore, abundant strange antibaryon production is indicative of an environment in which gluon collision processes are an essential element of the reaction picture. We have further argued that normally rare strange antibaryon particles with strangeness content provide a very promising experimental signal in the search of the quark-gluon plasma. In particular, abundant strangeness production is indicative of the presence of gluon excitations, a characteristic property of the deconfined QCD phase. Measurement of $\bar \Xi, \bar Y$ particle spectra at medium $p_\perp$ and central rapidity may reveal most notable anomalies.
\end{mdframed}

\subsubsection{Strange antibaryons; has anyone noticed?}
I made many efforts to present the strange antibaryon signature of QGP to the nuclear science community at large. Despite many topical and Summer/Winter school talks, and  contributed parallel sessions of the American Physical Society (APS) Spring, or Divisional Nuclear Physics Fall meetings, neither I, nor that I am aware,  anyone   presented a theoretical lecture  at a major conference on specifically this topic; I did present many times at small workshops, topical events, theoretical schools with 30-200 participants.

Given the importance of the strange antibaryon signature for the understanding of the QGP this is an unusual situation. There were many talks on all other, arguably less relevant, topics; all it takes is a quick look at the proceedings of Quark Matter conference series. To assure fairness of this remark, I note that the organizers of the Venice QM2018 were intending to invite me to present a plenary lecture. This was motivated  by  ALICE results described in Sec.~\ref{sec:AliceSys}. However, the topic was dropped from the program. I was told that two members of the organizing committee objected since I ask too many \lq direct\rq\ questions. A very good friend gave me the advice: \lq\lq You do not want to protest this year.\rq\rq 

The following abstract could be (changing the experimental code number)  submitted today. However it had been submitted to the attention of the 1992 International Nuclear Physics Conference (INPC 92) held 26 July - 1 August 1992 in Wiesbaden, Germany.\\

\noindent\textit{This talk was not selected by INPC 1992 for presentation:}\\[-0.7cm]
\begin{mdframed}[linecolor=gray,roundcorner=12pt,backgroundcolor=Dandelion!15,linewidth=1pt,leftmargin=0cm,rightmargin=0cm,topline=true,bottomline=true,skipabove=12pt]\relax%
\begin{center}
{\bf QGP and Strange Antibaryons} 
\end{center}

Substantial enhancement of production rates of {\it multi}strange {\it anti}baryons in nuclear collisions at central rapidity has been identified as an interesting observable and a potential signature for quark-gluon plasma (QGP) formation in relativistic nuclear collisions\footnote{J. Rafelski, Phys. Lett. {\bf B262} (1991) 333; and \lq\lq Strange and hot matter,\rq\rq\ to appear in Nucl. Phys. {\bf A}, 1992 (and references therein, \textit{published}: Volume \textbf{544}, July 1992, pp. 279-292}. A number of CERN experiments has studied this observable in 200~GeV~A Sulphur collisions with Sulphur and/or heavy nuclei (experiments WA85, NA35, NA36). In the past two years results have been presented which have eluded explanation in terms of models developed for $p$--$A$ scattering processes. These results suggest that the production of $\Lambda, \overline{\Lambda}$ is indeed occurring in a centrally formed fireball reaching temperatures of $T=210\pm10$ MeV\footnote{J. Rafelski, H. Rafelski and M. Danos, \lq\lq Strange Fireballs,\rq\rq\ Preprint AZPH-TH/92-7, \textit{published: Phys. Lett. B \textbf{294}, 5 November 1992, pp 131-138}}. It is possible to consider the properties of a hot hadronic matter fireball source for
$\Lambda, \overline{\Lambda}, \Xi, \overline{\Xi}$ without identifying the nature of the state, viz. if it is deconfined type QGP or normal $\pi,\,N$, etc hot nuclear gas (HG). The difference between these states is than seen in the \lq\lq measured\rq\rq\ values of the parameters:
\begin{itemize}
\item $0<\gamma<1:$ the degree of saturation of the strangeness phase space, with the expectation being $\gamma\simeq 0$ for HG and $\gamma\simeq 1$ for QGP, the enhancement arising from effective glue-based strangeness production processes;
\item $\mu_s:$ the strange quark chemical potential which distinguishes the strange from antistrange hadrons and in general vanishes for the case of QGP, and assumes a wide range of values for the evolving HG fireball;
\item $\mu_b/T:$ determines the density of baryons in the central fireball, viz the degree of their stopping and hence the fireball energy and entropy content.
\end{itemize}
When the data of the experiment WA85 for the abundance ratios of strange baryons and antibaryons are interpreted in such a way~[a] one finds the remarkable set of values: $\gamma=0.7,\ \mu_s=0,\ \mu_b/T=1.5$ with relatively small error bars. These results favor the interpretation of the central fireball in terms of a state which is abundantly producing the strangeness flavor, is symmetric with regard to the formation of both strange and anti-strange quarks, and has considerable baryon content and density; the possibilities that such a result could arise in the context of a QGP or HG fireball are considered\footnote{In preparation, \textit{Published: Jean Letessier, Ahmed Tounsi, and Johann Rafelski \lq\lq Hot hadronic matter and strange anti-baryons,\rq\rq\ Phys. Lett. B \textbf{292}, 15 October 1992, pp. 417-423}}. 
 
It is impossible to come to a definitive conclusion on the basis of a single experimental point. To assert a reaction mechanism we will have to find agreement between the systematic behavior of the measurements and theoretical expectations, while a number of available parameters (such as the mass of the projectile, impact parameter, and the energy per nucleon) are varied. Nevertheless, even today these results imply for a fully strangeness saturated (as would be expected of the larger Pb-Pb originating fireball if formed at similar physical conditions) one aught
to find:
${\overline \Xi}/{\bar \Lambda}={\bar \Lambda}/{\bar p}=1.55\pm0.13 \;,\ 
\Xi/\Lambda=\Lambda/p=0.64\pm0.05 \;. $ 
This result applies to the high transverse momentum sector of the spectrum and relates particle abundances considered at the same {\it transverse energy} in a narrow, central region of rapidity. It displays the anomaly that the more heavy and strange anti-baryon is more abundant. 
\end{mdframed}

\subsection{Soft and strange hadronic observable of QGP at RHIC}
\textit{%
In a lecture~\cite{Rafelski:1990dk}  presented at the July 1990  RHIC-BNL-Workshop~\cite{Fatyga:1990ubn}  I describe strangness flavor related hadronic observables  and evaluate their significance for the observation and identification of QGP:}\\[-0.7cm]
 \begin{mdframed}[linecolor=gray,roundcorner=12pt,backgroundcolor=Dandelion!15,linewidth=1pt,leftmargin=0cm,rightmargin=0cm,topline=true,bottomline=true,skipabove=12pt]\relax%
\centerline{\bf\Large Flavor Flow From Quark-Gluon Plasma}
\addcontentsline{toc}{subsubsection}{Flavor flow from quark-gluon plasma}
\noindent{\bf Abstract:}
I discuss diverse hadronic observable of the reactions between relativistic heavy ions related to the production and flow of flavor, and its significance for the observation and identification of quark-gluon matter. This discussion in particular includes a brief survey of our current understanding of the strange particle signature of quark-gluon plasma.\\[0.5cm]
\noindent\textbf{\large Looking For Quark-Gluon Plasma}\\[0.1cm]
The inherent difficulty of the study of Quark-Gluon matter is its expected fleeting presence when two heavy nuclei collide. Therefore, an important element in theoretical investigation of relativistic heavy ion collisions has been the identification of an observable of this new state of matter. We must from outset realize that an observable can be either \lq characteristic\rq\ and/or\lq descriptive\rq. A characteristic measurement would tell us unequivocally that some time during the nuclear interaction quark-gluon matter has been formed. A descriptive observable will not necessarily be characteristic, but should allow us to study the properties of the quark-gluon matter phase, if we can with certainty assume its formation.  
 
First I note that we can in principle measure as function of rapidity and transverse mass the following simple hadronic observable:
\begin{itemize}
\item the yield of charge;
\item the yield of baryon number;
\item the yield of strange particles and in particular that of:
\begin{itemize}
\item single strange particles ($\bar s q, s\bar q, \bar s qq, s\bar q\bar q$),
\item multi strange baryons ($ssq, \bar s\bar s\bar q, sss, \bar s\bar s\bar s$),
\item $\phi$-meson yield ($\bar s s$),
\item HBT correlations of strange particles,
\item strange exotica.
\end{itemize}
\end{itemize}

In order to present a comprehensive and complete description of the diverse processes occurring, a theoretical interpretation of the data must necessarily account for details of the collision dynamics. This information is at present not available for the energies accessible at RHIC and theoretical models are by necessity dependant on a number of assumptions, in absence of a truly fundamental approach to the collision dynamics. Furthermore, there are additional uncertainties related to carrying through a simulation of the collision dynamics involving a possible phase transformation. Thus it is of essence for the discussion here presented that initial RHIC experiments determine:
\begin{itemize}
\item the \lq\lq stopping power\rq\rq\ of the constituent quarks in the colliding nuclei, as measured by the rapidity distribution of the electrical charge; \item the baryon number stopping power of the nuclear medium, as measured conveniently by rapidity distribution of (strange) baryons; \item the entropy produced in the collision, as measured e.g.~by the particle multiplicity, in particular pion to baryon ratio as function of rapidity; 
\item the characteristic \lq\lq temperature\rq\rq, as measured e.g.~by the slopes of transverse mass spectra;
\end{itemize}
The primary observable we address here is the strange quark flavor and in addition to the above I would like to see a measurement of:
\begin{itemize}
\item the high density, above-equilibrium nature of the over saturated strangeness phase space density, which is noted for by the abundance of multistrange baryons and in particular their anomalous abundance enhancement as compared to singly strange antibaryons, which in turn are enhanced as compared to antiprotons produced; and 
\item the overabundance of strangeness flavor as measured by overabundance of strange particles produced in $A$--$A$ collisions compared to $p$--$p$ and $p$--$A$ reactions;
\item kaon HBT correlations, which should show a smaller source than pionic HBT size of the fireball.
\end{itemize}
The remainder of this lecture is organized as follows: Next, I explain why strangeness flow is viewed as an observable of quark-gluon matter. This is followed by a brief consideration of lessons from the present strangeness data.\\[0.5cm]
\textbf{\large   Why flavor--strangeness?}\\[0.1cm]
\addcontentsline{toc}{subsubsection}{Why flavor--strangeness?}
I proposed about ten years ago$^{1,2}$  strangeness as an
observable of quark-gluon matter. Following on early equilibrium
considerations it became
soon apparent that strangeness production must be treated in a kinetic
approach\footnotemark[3]. Furthermore, in a review prepared for QM\,1982\footnotemark[4] 
\begin{quote}
\ldots measurement of production cross section of strange antibaryons could be already quite helpful in the observation of the phase transition \ldots\\ Measurement of the relative $K^+/K^-$ yield, while indicative for the value of the chemical potential (in hadronic gas phase) may carry less specific information about the plasma. The $K/\pi$ ratio may indeed also contain relevant information - however it will be more difficult to decipher the message ...it appears that otherwise quite rare multistrange hadrons will be enhanced ... hence we should search for the rise of the abundance of particles like $\Xi, {\bar \Xi}, \Omega, {\bar \Omega}, \phi$ and perhaps highly strange pieces of baryonic matter (strangeletts), rather than in the K-channels. It seems that such experiments would uniquely determine the existence of the phase transition to quark gluon plasma \ldots.
\end{quote}
This is in a shell nut my position today, though in the elapsed decade the initial simple ideas have undergone a substantial evolution$^{5,6}$ and have come under intense scrutiny, see Ref.\,[7] and references therein.

I think that those who have been critical of \lq\lq strangeness\rq\rq\ have never taken time to study the detailed ideas related to flavor (strangeness) flow, of which the simplest point of view I have quoted myself above. It seems indeed that we have just gone more than 8 years back, as in the strangeness review at QM\rq\ 90 we can read (see Ref.\, [7]) 
\begin{quote}
Strangeness has been proposed as a signal for quark- gluon plasma formation in RHI collisions. Subsequent to the original proposal several papers appeared which considerably weakened (hic) the early claims (which???) made for strangeness production in heavy ion collisions (references follow from 1985,1986,1988 addressing the question what Kaons can tell us or not.).  I  quote from Ref.[7]  \lq \ldots we conclude that there is no natural large difference in flavor composition between the \ldots QGP and an {\it equilibrium} hadron gas\rq. 
\end{quote} 
The experimentalists working presently in the field investigate the key point which eludes some theorists\footnotemark[8]. The question is not only how much strangeness there is, but {\it what happens to the strange and antistrange quarks}, and how this compares with control data e.g. from $p$--$A$ collisions.

Clearly, the interest to measure strangeness is there  discounting   theoretical controversy, as every experimentalist hopes to see a spectacular phenomenon, a \lq smoking gun\rq\ of the phase transition. Interest in observing strange particles also derives from the second objective of experiments involving relativistic nuclear collisions, the study of equations of state of highly excited nuclear matter. Namely, even without the formation of quark-gluon phase, that is in case that the collision proceeds via the intermediate stage of a fireball consisting only of highly excited hadron gas, the strange particle flow provides essential information about the properties of matter under extreme conditions. However, the relation between observable particle spectra and the equation of state presents many difficulties of detail, and much theoretical modeling will be required; for quark-gluon phase these difficulties are compounded as the observable of a quark-gluon state can be seen only after undergoing a phase transition back into a hadronic form. The phase transition in turn depends on the equations of state. Hence the study of strange particles emanating from collisions at conditions believed not to lead to quark-gluon phase is extremely important as it helps us understand the backgrounds to the quark-gluon phase signatures, at the same time as we learn about {\em confined} nuclear matter.  

A comprehensive survey of the status of the theory of strange particle production and evolution in hadronic collisions before 1985 can be found in Ref.\,[5]. The progress of experiments and theory has been recorded at the Tucson HMIC meeting\footnotemark[5]. An update has been recently prepared by Eggers et al.\footnotemark[9]. \\[0.5cm]
%
\addcontentsline{toc}{subsubsection}{Strange signatures of quark-gluon plasma}
\textbf{\large  Strange Signatures of Quark-Gluon plasma}\\[0.1cm]
Let us consider the situation in some more detail: as is apparent several experimental options for the study of the flavor--strangeness signal of QGP in heavy ion collisions are available. The most obvious measurement is the determination of the multiplicity of various strange hadrons, often represented as ratios to reduce the influence of the experimental bias (trigger). In this class of measurements, however, components originating from all the different production processes are included; for example, strange hadrons may be formed in
\begin{itemize}
\item initial high energy hadronic collisions, 
\item inside the QGP, 
\item during QGP hadronization,
\item in the final expanding hadron gas,
\item rescattering from spectator nuclear matter,
\end{itemize}
or, if the QGP is not formed at all, during the various (equilibrium and non-equilibrium) stages of a hadron gas fireball. This means that the QGP strangeness signal must be evaluated in relation to proton nucleus reactions and detailed conventional wisdom cascade calculations.

Somewhat more specific approach to identify strangeness signal of QGP is to measure strange particle rapidity and transverse energy or momentum spectra. The above mentioned distinct physical processes normally emit particles into different windows of rapidity or transverse energy, making it possible to select particles from a specific process by introducing appropriate cuts in the differential cross section data. Transverse energy spectra are often divided into separate, although overlapping, regions in which a specific physical process dominates\footnotemark[9]. This conjecture is supported by the fact that ratios of different particle species vary strongly with $m_\perp$. At low $m_\perp$, one finds particles formed in the rescattering of the spectator nucleons. At slightly higher $m_\perp$, particles produced in the hadron gas, which decoupled at the freeze-out temperature of the fireball are dominant. Particles emitted with moderately high $m_\perp$ originate from hot and dense form of matter, conceivably the early QGP. A number of mechanisms can be responsible for this sector of the particle abundance. For example in Ref.\,[11] two processes were considered: in the first a quark or diquark from the high-momentum tail of the QGP strikes the phase boundary. It than may create a $q\bar q$ pair e.g. via string-breaking and so a high $m_\perp$ meson or baryon is emitted in such a micro-jet process. Alternatively, a baryon or meson like cluster in the QGP leaves the QGP in unison. In particular it follows from this consideration and the high $\bar s$ density that the differential measurement of multistrange antibaryons should have a good (QGP) signal to (HG) noise ratio. If such multistrange antibaryon yields can be analyzed in terms of their transverse and longitudinal flow, the signature for new phenomena will be clear.  

From this discussion it is clear that the most interesting part of the particle spectrum involves central rapidity, median (e.g. 1-5\,GeV/c) transverse momenta. To sum up the different ways of measuring strangeness, a schematic diagram is shown in Fig.~1.\\

\begin{center}
\setlength{\unitlength}{0.92mm}
\begin{picture}(120,73)(8,2)
\put(38,69){\framebox(60,10){QGP Flavor Signal}}
\put(68,67){\line(0,1){2}}
\put(32,55){\line( 3,1){36}}
\put(104,55){\line(-3,1){36}}
\put(32,53){\line(0,1){2}}
\put(104,53){\line(0,1){2}}
\put(11,43){\shortstack{$s,\bar s$ abundance\\ descriptive QGP signal}}
\put(82,43){\shortstack{$s,\bar s$ density\\ characteristic QGP signal}}
\put(08,41){\framebox(48,12)}
\put(80,41){\framebox(48,12)}
\put(32,39){\line(0,1){2}}
\put(104,39){\line(0,1){2}}
\put(14,33){\line( 3,1){18}}
\put(50,33){\line(-3,1){18}}
\put(86,33){\line( 3,1){18}}
\put(122,33){\line(-3,1){18}}
\put(14,31){\line(0,1){2}}
\put(50,31){\line(0,1){2}}
\put(86,31){\line(0,1){2}}
\put(122,31){\line(0,1){2}}
%
\put(00,20){\framebox(28,11){all $y, m_\perp$}}
\put(36,20){\framebox(28,11)}
\put(42,21){\shortstack{high $m_\perp$\\ central $y$}}
\put(72,20){\framebox(28,11){all $y, m_\perp$}}
\put(108,20){\framebox(28,11)}
\put(114,21){\shortstack{high $m_\perp$\\ central $y$}}
\put(14,16){\line(0,1){4}}
\put(50,16){\line(0,1){4}}
\put(86,16){\line(0,1){4}}
\put(122,16){\line(0,1){4}}
\put(00,00){\framebox(28,16)}
\put(36,00){\framebox(28,16)}
\put(72,00){\framebox(28,16)}
\put(108,00){\framebox(28,16)}
\put(02,1.5){\shortstack{$sqq$ and $\bar s\bar q\bar q$\\global, central\\
$m_\perp$ spectra}}
\put(38,1.5){\shortstack{$ssq$ and $\bar s\bar s\bar q$\\global, central\\
$m_\perp$ spectra}}
\put(75,1.3){\shortstack{multistrange\\correlations\\$K^+K^+$}}
\put(112,02){\shortstack{$\bar s\bar s\bar q$ ratio to\\$\bar s\bar q\bar q$ and
$\bar q\bar q\bar q$}}
\end{picture}

\vspace{5mm}
{\textbf{Fig. 1.} Strange particle quantities for diagnosis of QGP} 
\end{center}
%
\subsubsection*{\bf Arguments for strangeness as a QGP observable}
The correlated factors why strange particles possess a priori a distinct diagnostic function of the behavior of highly excited nuclear matter and are well suited as a signal distinguishing quark-gluon phase from the hadron gas are as follows:
\begin{enumerate}
\item near flavor symmetry for antiquarks $\bar s, \bar u, \bar d$ in all conditions (baryon rich and baryon poor), 
\item strongly differing production rates in different phases and strangeness mass thresholds which are of the same magnitude as temperature;
\item extremely high $s\bar s$--quark pair density in the quark-gluon phase.
\item the predicted strange antibaryon abundance is greater than background $p$--$p$ ISR results.
\end{enumerate}
We now discuss in more detail each of these points. 

{\bf 1. Anti-flavor symmetry:} Recent BNL and CERN experiments indicate that up to currently available energies the fireball usually has a sizable baryon number and therefore a relatively large baryo-chemical potential $\mu_\mathrm{B}$. This means that, for quark-gluon phase in chemical equilibrium, the number of light antiquarks is suppressed. Deconfined strange and antistrange quarks, on the other hand, are not affected by $\mu_\mathrm{B}$ and so are suppressed in quark-gluon phase only by their non zero mass. Consequently, but provided that strangeness production has saturated the available phase space, the abundance of antiquarks $\bar u, \bar d, \bar s$ will be nearly equal. In baryon free region, as possibly established at RHIC, this flavor symmetry of hadronic particles is also in part a result of the fragmentation of the numerous gluons.  

{\bf 2. Production rates and thresholds:} Rates for production of $s\bar s$ pairs in the quark-gluon phase were often calculated, the latest reference being\footnotemark[8]. The strangeness production time constant in the quark-gluon phase is of the order of $10^{-23}$~s, while in hadronic gas phase it is 10 to 30 times slower\footnotemark[12] at the same temperature and baryo-chemical potential. This difference is mainly due to the presence of gluons in QGP and different reaction thresholds. The typical time scale for the creation and decay of a fireball can be estimated as the time to traverse, say, a distance of 15~fm i.e. $\simeq 5\times 10^{-23}$~s, and so strangeness in a thermal hadronic gas will not likely reach equilibrium values, contrary to quark-gluon phase expectations. Thus we expect that any kinetic description of strangeness production involving the usual hadronic particles will give a total strange particle yield significantly below the limits obtained from an equilibrium picture of hadronic gas fireballs. The most accessible reaction (if allowed) is usually the creation of a $\Lambda K$ or $K\bar K$ pair and requires at least 700 MeV. In the quark- gluon phase, on the other hand, the threshold is given by the rest mass of the strange-antistrange quark pair, i.e. only $2 m_s \simeq 350$\,MeV. This difference between the two thresholds though insignificant at the initial high energies, is noticeably impacting the time scale of strangeness production in a \lq\lq thermalized \rq\rq\ glob of hadronic matter. It is anticipated that at RHIC temperatures of $250\pm 50$\,MeV will be reached. Here I note the trivial, though important point that in general strangeness production occurs in the numerous rescattering processes, not in the highly energetic initial parton-parton collisions. From this we expect in particular substantial enhancement of strangeness in Nucleus-Nucleus collisions, as compared to scaled p-Nucleus yield, (this subject to the validity of the hypothesis of formation of a hadronic fireball of any\lq texture\rq\ ). I recall here, however, the discussion of Koch and Rafelski\footnotemark[13] concerning the abundance of strangeness in regular hadronic interactions. It was found so close to the expected equilibrium abundance, that it seems as if quark-gluon plasma like phase were formed, permitting to saturate the available strangeness phase space in most hadronic collisions. However, Wr\\rq\ oblewski\footnotemark[14] determined that regular hadronic interactions are about three times less effective in making strange flavor as compared to light flavors. Since QGP based estimates lead me to expect flavor symmetry in QGP, some strangeness enhancement must be expected in comparison to $p$--$A$ scaled result. 

{\bf 3. $s\bar s$--density:} Even at the time of hadronization, $s\bar s$ density (now half as high as at it peak) is about 0.4 strange particle pairs per fm$^3$. As consequence, most of baryons and antibaryons emerging is strange, and non- strange nucleons are expected to be only 20\% of the total baryon--antibaryon abundance\footnotemark[5]. In the hadronic gas phase, by contrast, all antibaryons are suppressed, particularly those with high (anti)strangeness content\footnotemark[5], leading to the expectation that quark-gluon phase be distinguishable from hadronic gas phase by relatively enhanced numbers of anti-strange hadrons$^{1,4}$. This argument, initially developed for baryon rich quark-gluon matter remains valid without change at RHIC energies at central rapidity region, i.e. in the central fireball. As detailed calculations\footnotemark[5]  have shown. there is an abundance anomaly expected for strange antibaryons arising primarily from the enormous strange pair density in thequark-gluon matter.  

{\bf 4. Expected direct reaction\lq background\rq\ } It is helpful to consider the magnitude of backgrounds expected for the multi strange (anti) baryons. The $\bar \Xi / \bar Y$ ratio seen at ISR at $\sqrt{s}=63$\,GeV is only 0.06$\pm$0.02 in the central rapidity region\footnotemark[15]. The expected quark-gluon matter result at RHIC is predicted to be ten times greater\footnotemark[5], or even up to 50 times greater\footnotemark[11], at relatively high $m_\perp$. The parallel ratio $\bar Y/\bar N$ is 0.27$\pm$0.02 as measured in the same experiment at ISR, my expectation is that $\bar Y/\bar N\vert_{plasma} \sim 2 \pm 0.5$. We thus see that both $\bar \Xi/\bar Y$ and the $\bar Y/\bar N$ ratios a interesting, with the former being characteristic of the new form of matter, as it is more difficult to imagine how an enhancement along the theoretical QGP prediction could be made otherwise.  

{\bf In conclusion:} The enormous strangeness pair density to be expected in RHIC--QGP is in my opinion the main experimental objective of flavor based RHIC experiments. This property of the QGP state is particularly interesting, since the primary production mechanism of strangeness is by gluons present in the deconfined phase. Measurement of strangeness density removes interpretational ambiguities, related to our present ignorance of reaction dynamics, in attempting a comparison of the respective {\em total} strangeness content of quark-gluon phase and hadronic gas phase, as enhancement of quark-gluon phase strangeness may be diluted by the geometry of the ensemble of collisions and can be argued away on the basis of the perpetual ignorance of the lifetime of the hypothetical hadronic gas phase fireball. Thus strange particle abundance per se, though perhaps most interesting\lq barometer\rq\ and\lq thermometer\rq\ of the quark-gluon matter phase, is to be employed to study QGP properties only once the high strangeness density has been established.  

\subsubsection*{\bf Paths to observe multistrange (anti) baryons}
\addcontentsline{toc}{subsubsection}{Paths to observe multistrange (anti) baryons}
Even though at RHIC the \lq\lq common knowledge\rq\rq\ is that the central rapidity region is baryon free, I will not assume here this prejudice and hence refer to the (strange) anti-baryons, which are characteristic for QGP irrespective of the degree of stopping of the baryon number. However, practically every point discussed applies both to baryon flow in baryon free region, and it is of preference if both strange baryons and antibaryons are measured. I will assume that any detector aiming at the measurement of baryon flow will permit the observation and measurement of the charged decay\lq V\rq\ of the neutral $\bar \Lambda$ particles. The decaying $\bar \Lambda$ particles originate in part in the (rapid) electromagnetic decays of the $\bar \Sigma^0$ particles. All anticascades ultimately become $\bar \Lambda$, while only half of all anti-hyperons $\bar Y$ will be in the $\bar \Lambda$-decay chain, of which 64.2\% are giving they typical\lq V\rq\ decay pattern. Assuming full acceptance for the\lq visual\rq\ detector for all V\rq s, the total sample of all seen V-events is 
\begin{equation}\tag{1}
N_{\bar V}=0.642 \bar Y\left({1 \over 2}+{ {\bar \Xi} \over {\bar Y}}\right)\;,
\end{equation} 
and, should the abundance ratio ${\bar \Xi}/{\bar Y} \sim 1/2$, we see that half of the observed V\rq s would be associated with the primordial $\bar \Xi$ abundance.  

The difficulty is that the observable $\Xi^-, {\bar \Xi^+}$ decay over a significantly shorter path (c$\tau = 4.92$ cm)than $\Lambda$ (c$\tau = 7.89$ cm), making necessary a novel detector directly outside the beam pipe. This poses particular instrumental problems, related both to the interface between the two detectors, but more significantly, to the need for extremely high resolution in view of the enormous multiplicity of charged particles, in which the occasional cascade\lq kink\rq\ has to be searched for. Probably this path to the measurement of multi strange (ant) baryons will be ultimately attempted. However, I would like to draw attention to an alternate approach\footnotemark[16]: in order to find out how many $\bar \Lambda$ descend from the cascade decay all that is needed is the measurement of the longitudinal $\bar \Lambda$ polarization.

There is a significant difference in this polarization of the $\bar \Lambda$ descending from the weak $\bar \Xi$ decays. The weak decay polarizes the $\bar \Lambda$-spin longitudinally, the mean value of its helicity being given by the decay asymmetry parameter $\alpha_\Xi$. In the subsequent weak $\bar \Lambda$ decay this polarization is effectively\lq analyzed\rq. The practical approach is to consider the so-called up-down asymmetry of the $\bar \Lambda$ decay with reference to the plane normal to the $\bar \Lambda$-momentum, i.e., to measure how often in the $\bar \Lambda$ rest frame the antiproton appears `above\rq\ as compared to\lq below\rq\;, with respect to a plane normal to the direction of $\bar \Lambda$-momentum.  

The simple criterion which determines the up-down asymmetry is identified boosting the antiproton momentum to the $\bar \Lambda$ rest frame and considering $S$, the vector product between $\bar \Lambda$-momentum and $\bar p$-momentum. I obtain: 
\begin{equation} \tag{2}
S:= \displaystyle\frac{\vec P_{\bar \Lambda} \cdot \vec P_{\bar p}}{P_{\bar \Lambda}^2} -\displaystyle\frac{E_{\bar p}}{E_{\bar \Lambda}}= 
\begin{cases}  \mbox{positive for up}\cr 
\mbox{negative for down}\cr
\end{cases}. 
\end{equation} 
Here we have, as usual, $\vec P_{\bar \Lambda} = \vec P_p + \vec P_\pi$ for the respective particle momenta and similarly for their energies $E = \sqrt{m_i^2 + P_i^2}$. At this point, I note that the longitudinal polarization considered here is of entirely different origin and nature than the transverse polarization of $\bar \Lambda$ associated with hadronic formation processes of these particles. Multiple scattering in the hadronic gas cannot create longitudinally polarized $\bar \Lambda$ out of primordial transverse polarization. However, the longitudinal polarization will be influenced by spin rotation in a magnetic field.  

This up/down asymmetry is given by\footnotemark[16]: 
\begin{equation} \tag{3}
{{N_u-N_d}\over{N_u+N_d}}= {1 \over 2} \alpha_{\bar \Lambda} \wp_{\bar \Lambda}\;, 
\end{equation} 
where $\wp_{\bar \Lambda}$ is the $\bar \Lambda$ polarization and is equal to the $\alpha_\Xi$ decay parameter. This polarization is analyzed by the $\alpha_{\bar \Lambda}$ decay parameter. The different values of the parameters found in the data tables are: $\alpha_\Lambda = - \alpha_{\bar \Lambda} = 0.642 \pm 0.013$; $\alpha_{{\bar \Xi}^0}=-\alpha_{\Xi^0}= 0.413 \pm 0.022$; and $\alpha_{{\bar \Xi}^+}=-\alpha_{\Xi^-}= 0.455 \pm 0.015$. The total up-down asymmetry of all V-events is 
\begin{equation} \tag{4}
{{N_u-N_d} \over {N_u+N_d}}= {{N_{\bar \Xi}} \over {N_{\bar V}}} {1 \over 2} \alpha_\Lambda \alpha_\Xi\;, 
\end{equation} 
where we have included the relative abundance of all polarized $\bar \Lambda$ to the total abundance of V\rq s: $N_{\bar \Xi} / N_{\bar V} = (2 {\bar \Xi}/{\bar Y})/ (1+2 {\bar \Xi}/{\bar Y}).$ With $\bar \Xi/\bar Y$ in the range 1/2( resp. 1/3) we expect a negative up-down asymmetry of 14\% (resp. 11\%). For the\lq normal\rq\ value $\bar \Xi/\bar Y \sim 0.06$ there is the hardly observable asymmetry of only 1.6\%. Hence observation of the longitudinal polarization is QGP specific!  I further note that $\bar \Omega$ weak decays have a negligible influence over the particle abundances and, in particular, their polarizations, since $\bar \Omega$, $\Omega$ are at least five times less abundant\footnotemark[5] than $\bar \Xi$, $\Xi$   and their decay asymmetry parameter ("polarizer" capability) is 5-20 times weaker (depending on the decay channel). The fact that some $\bar Y$, $\bar \Xi$ are descendants of strong decays of $\bar Y$(1385), $\bar \Xi$(1530), etc. is also of no consequence, as abundances of these particles has been considered part of $\bar Y$ resp. $\bar \Xi$ abundance. 

\subsubsection*{\bf Gluons in plasma}
The key role played by gluons in making high strangeness density an important observable is self-evident. Not only do gluons produce strangeness flavor dominantly (see below) but more importantly they provide the key distinction between the quark-gluon phase and the hadron gas. The high gluon abundance and density in the plasma impacts the entire history of the plasma state, in particular also the process of hadronization at the end of the quark-gluon phase lifetime, in which appreciable strangeness production occurs again. Indeed, strangeness can be considered a signal for gluons in the quark-gluon phase. We will briefly summarize here the expectations about the gluonic component in the plasma. We note that since gluons do not carry electrical charge, but only the strong charge, they can be observed (indirectly of course) only by suitable measurement of strongly interacting particles.  

Including a first-order perturbative effect\footnotemark[17] the gluon number density can be estimated from the equilibrium density as 
\begin{equation} \tag{5}
\rho_g [\mbox{fm}^{-3}] = 1.04 \left( {T\over 160\mbox{MeV} } \right)^{3} \left( 1- {15\alpha_s\over 4\pi} \right) 
\end{equation} 
giving for a typical temperature of 200 MeV a value of 0.55 fm$^{-3}$ for $\alpha_s = 0.6$ and 0.8 fm$^{-3}$ for $\alpha_s = 0.5$. For a quark-gluon phase volume with radius 4--5 fm, we therefore have 200--300 gluons. Note also that this density rises as the cube of the temperature. Because gluons can be created and annihilated easily in interactions with other gluons and light quarks, the gluon density closely follows the evolution of temperature in the course of the quark-gluon phase lifetime. The equilibrium gluon energy density is 
\begin{equation}\label{fg} \tag{6}
\varepsilon_g = {8\pi^2\over 15} T^4 \left( 1 - {15\alpha_s\over 4\pi} \right) \;,
\end{equation} 
and the gluon partial pressure is 
\begin{equation}\label{fh} \tag{7}
P_g (\mathrm{ GeV\; fm}^{-3} )  
 = {{1\over 3}} \varepsilon_g 
 =  0.15 \left( {T\over 160 \mathrm{MeV}} \right)^{4} \left( 1 - {15\alpha_s\over 4\pi} \right) 
\end{equation} 
which for $T = 200$\,MeV and $\alpha_s = 0.6$ yields 100MeV,$\,\mbox{fm}^{-3}$  and forms the major component of the quark-gluon phase pressure. (The total quark-gluon phase pressure must, of course, be larger than both the vacuum pressure ${\cal B}^{1/4}$ and the pressure of the hadron gas surrounding it.) 

Gluons also play a major role in the dynamics of the quark-gluon phase-hadronic gas phase transition: they carry much of the quark-gluon phase entropy, contributing an entropy density of about 
\begin{align*}\label{eq:fi} 
\tag{8}\sigma_g \;\mbox{fm}^{-3} 
&=  {32\pi^2\over 45} T^4
   \left( 1 - {15\alpha_s\over 4\pi} \right)
\ms{5}
\\
&=  3.76 \left( {T\over 160 \mbox{MeV}} \right)^{3} 
    \left( 1 -{15\alpha_s\over 4\pi} \right)
\notag
\end{align*}
which for $T = 200$\,MeV and $\alpha_s = 0.6$ is 2 units per $\;\mbox{fm}^{-3}$ (3.6 units per gluon). This large amount of entropy plays a major role in the hadronization phase transition, forcing gluons to fragment into quarks.

\subsection*{Strangeness production in the quark-gluon phase}

Since the time scale in a typical nucleus-nucleus collision is very short, the strangeness content of both quark-gluon phase and hadronic gas phase cannot {\em a priori} be assumed to be in equilibrium: it is necessary to determine explicitly the rate of strangeness production in both phases. The key result was obtained in the work of Rafelski and M\"uller\footnotemark[3]. The plasma initially contains very few, if any, strange quarks as those produced in pre-quark-gluon phase direct hadron-hadron reactions will generally be at higher rapidity than the fireball. Essentially all the $s\bar s$ production is therefore dominated by collisions of the central gluons, which in a first approximation can be assumed to be in a practically thermal distribution; light quark-antiquark collisions, it turns out, play only a minor role. Therefore the time evolution of strangeness density during the production process is only a function of temperature and not of the baryo-chemical potential. I will give here a brief sketch of the theory of strangeness production and show how strangeness density grows with time.


\textit{(Fig 2. presented earlier is omitted here)} \ldots both glue and quark induced processes are of comparable magnitude\footnotemark[9]. However, as we will just see the statistical factors entering the thermal average will strongly favor the gluon induced processes: there are simply more glue-glue than quark-antiquark collisions of suitable quantum number in plasma. In order to identify the energy range contributing to the production of strangeness, it is useful to write the production rate as an integral over the differential rate\footnotemark[9] $dA/ds$ 
\begin{equation}\label{gf}\tag{9}
A_i = \int_{4m^2}^\infty\,ds\,(dA_i/ds) =
\int_{4m^2}^\infty\,ds\,\bar\sigma_i(s)\, P_i(s)\;, \qquad i = g,q\;.
\end{equation}
The weight function $P_g(s)ds$ is the number of (gluon) collisions within the interval ($s,s+\,ds$) per unit time per unit volume, with a similar interpretation for $P_q(s)$. In a thermal system
\begin{equation}\label{gma}\tag{10}
P_g(s) = \int\,{d^3p_a\over (2\pi)^3E_a}\,
{d^3p_b\over (2\pi)^3E_b}\, {s\over 2} \,
\delta[s-(p_a+p_b)^2]\,{1\over 2}g_g^2 f_g(p_a)f_g(p_b)\;.
\end{equation}
In principle, non-equilibrium momentum distribution functions should be used for $f_g$, presumably evolving from the structure functions of the incoming reacting hadrons towards their equilibrium forms. However, because of the high gluon-gluon cross sections, this should happen very quickly\footnotemark[18,19]. In first approximation, one can therefore use the (thermal and chemical) Fermi and Bose equilibrium distributions. In Fig.~3, the product of the weight functions $P_g(s)$ and $P_q(s)$ with the respective cross sections is plotted for $T= 250$\,MeV and $m = 170$\,MeV. In one case, $\alpha_s = 0.6$, in another, the running coupling constant was used with $\Lambda = 200$\,MeV. Note that most $s\bar s$ pairs are made at $\sqrt{s} \simeq 0.5 \mathrm{GeV}$, giving at least some credence to the use of perturbative QCD, and in particular the value $\alpha_s = 0.6$ selected.

\centerline{
\includegraphics[width=0.7\columnwidth]{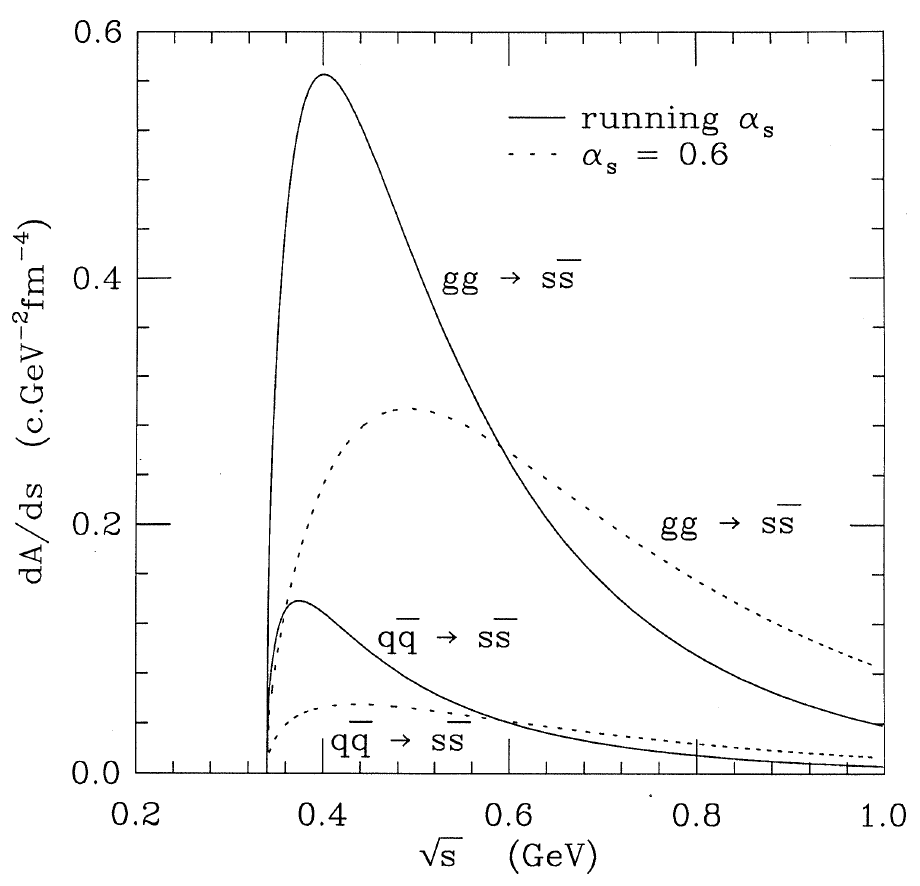}
} 
\noindent{\small Fig. 3 Differential production rate $dA/ds = P(s)\bar\sigma(s)$, with $T$ =250 MeV and $m$ = 170 MeV, for gluons and $q\bar q$ pairs, with $\mu_\mathrm{B}$ = 400 MeV. Solid lines are for running $\alpha_s$ with $\Lambda = 200$\,MeV, dotted lines for $\alpha_s = 0.6$}

In Fig.~4, the time evolution of the density of strange quarks in quark-gluon phase is shown ($\alpha_s = 0.6, m_s = 170$\,MeV). As expected, there is a strong threshold effect at temperatures around 150 MeV. A similar calculation which included an expansion model of the fireball\footnotemark[5] showed that the strong dependence of $s\bar s$ production on the temperature also implies that the strangeness abundance freezes out with a value characteristic of the highest temperatures reached during the collision. No significant strangeness annihilation occurs during the fireball expansion.

\centerline{
\includegraphics[width=0.7\columnwidth]{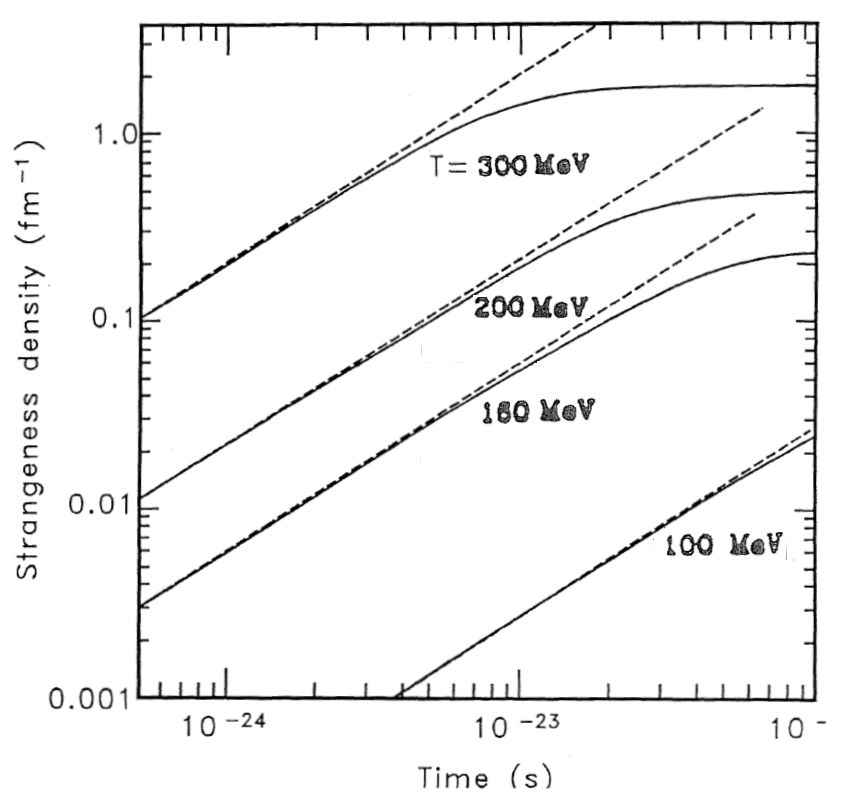}
}
\noindent{\small Fig. 4 Time evolution of the strange quark density in quark-gluon phase for different values of the temperature. Dashed lines: no $s\bar s$ annihilation.}

\section*{Lessons From Present Experimental Results on Strangeness}
\addcontentsline{toc}{subsubsection}{Lessons from early experimental results}
I will focus here on the aspects of current experimental work instructive to the described developments, giving only a schematic interpretation. The experimental method employed to determine the enhancement is to compare the yield of strange particles as a function of the inelasticity of the interaction. In order to demonstrate the kind of analysis we will have to implement for RHIC experiments let me now consider a hypothetical quark-gluon phase fireballs as being at the origin of the latest results in strangeness production. I consider the {\it experimental} situation as it presents itself in May 1990, following on the Quark Matter \rq\ 90 meeting. In all BNL and CERN experiments reported so far strangeness enhancement by a factor $2\pm 0.5$ has indeed been seen, but can not be taken without prejudice to be a signal of quark-gluon plasma.  

In this discussion I will use experimental results to estimate the value of the temperature and chemical potential at which the strange particles are likely to have been born and will try to determine if there is any glaring inconsistency of the present data with such a hypothesis. Alas, as we will see, total strangeness data of from BNL can not point to a particular phase of matter, much as expected. Nevertheless, in order to test the consistency, rather than two parameters $(T,\mu_\mathrm{B})$ we must consider three quantities characterizing the average thermodynamical properties of the fireball, e.g.:
\begin{enumerate}
\item The temperature $T$, as obtained directly from particle transverse energy spectra. Here we must take care to distinguish the projectile and target rapidity regions from the central region which of greatest interest to us here. For most particles with large cross sections such as pions, the observed slopes of transverse mass particle spectra provide us with the temperature $T^{H}$ at freeze-out of the particular particle species in the hadron gas. However, strange particles can exhibit higher temperatures as their interaction length is larger. For the thermal picture to be applicable, a similar temperature should be found in the corresponding rapidity spectrum.
\item As a direct measure of the baryo-chemical potential, we can consider the entropy per baryon $S/B$, which we assume here to be mostly produced during the initial stages of the nuclear collision. During the subsequent, in particular hydrodynamical evolution of the quark-gluon phase or hadronic gas phase, no entropy is produced and the baryon number is also constant. For the case of the perturbative QCD equation of state, constant-$S/B$ implies $T \propto \mu$ with a determined constant\footnotemark[20,21]. The value of entropy per baryon reached in the reaction is obtained under the assumption that a gas of nucleons and pions dominates all central fireball secondaries (counting all mesons as pions, all baryons as nucleons)\footnotemark[22]:
\begin{equation}\label{el}\tag{11}
{S\over B} \simeq {S_N\over B} + {S_\pi \over \langle{n_{\pi}}\rangle}\;{\langle{n_{\pi}}\rangle \over B}\;,
\end{equation}
where the entropy per pion is about 4.05 and the entropy per nucleon outside of the degeneracy region is $S/B = 2.5 + (m_N-\mu_\mathrm{B})/T$.
\item The baryo-chemical potential $\mu_\mathrm{B}^H$ in the (final stage) hadron gas phase can be determined conveniently from the $K^+/K^-$ or $K_s/\Lambda$ ratios, which are sensitive to $\mu_\mathrm{B}^H$\footnotemark[23], because of large strangeness exchange cross sections which rapidly establish the so-called relative chemical equilibrium between different species of strange particles. This is true even if absolute chemical equilibrium is not attained for strangeness in the hadronic gas phase\footnotemark[5]. If the values of $T$ and $\mu_\mathrm{B}^H$ do not disagree to much with the entropy based QGP constraint (see above), this can be taken as a first indication that we are possibly close to the quark-gluon phase. 
\end{enumerate}

\subsection*{BNL--RHI results}
There are two experiments at BNL measuring strange particle spectra, of which the more ambitious TPC-based E810 has just begun to collect data\footnotemark[24], while the magnetic spectrometer experiment E802 has essentially completed its data taking\footnotemark[25]. Both experiments see an appreciable strangeness signal in 14.6 A\,GeV/c Si--Au collisions (the beam rapidity is 3.44), with a central collision trigger. The common result of both experiments is that strange particles have a rather \lq\lq thermal\rq\rq\ shape in the central rapidity region, and that the temperature is in the vicinity of 150 MeV, but with a statistical error which is presently 15 MeV. While E810 expects to measure the abundances of diverse multi strange baryons and antibaryons in the near future, experiment E802 provides already today precise data on ratio of meson abundances\footnotemark[25]. Additional data including in particular the antiproton spectra has recently been presented at the BNL-HIPAGS workshop\footnotemark[26]. It therefore seems justified to assess the results of E802 with a simple fireball model in mind. We will need just the most naive of the pictures for further discussion: the tube model of the nucleus-nucleus collision leads to the formula for the number of participating target nucleons $A_\mathrm{t} = 1.5 A_{\rm projectile}^{2/3}A_{\rm target}^{1/3}$ predicts $A_\mathrm{t} \simeq 80$ for the Si--Au case and hence a total baryon content of the fireball $B \simeq 108$. This corresponds to a theoretical rapidity $y_{lab} \simeq 1.23$ for a fireball made out of $(A_p + A_\mathrm{t})$ nucleons, closely corresponding to the experimentally inferred central rapidity $y = 1.2$. This assumes complete stopping so that the accessible CM energy $\sqrt{s} = 261$\,GeV is mostly transferred into the internal excitation of the fireball,suggesting an energy content of 2.42 Gev per baryon, less energy in excitation of spectator matter. 

From rapidity particle densities we can now derive the central pion to baryon ratio: for $1.1 < y < 1.6$, the proton rapidity density $dN/dy$ is $16.2 \pm 0.3$, implying a baryon rapidity density $38 \pm 0.7$ (given the baryon to proton ratio of 2.35 in the tube model for Si-Au collisions). Both the $\pi^+$ and $\pi^-$ rapidity density is quoted at $16 \pm 1$. Allowing for an equal number of neutral pions the pion central rapidity density is $48 \pm 1.8$. This results in a pion to baryon ratio $1.25 \pm 0.05.$ For $T \simeq 125$\,MeV (observed pion temperature), the pion gas entropy per pion is $\simeq 4.3$, and hence the pion entropy per baryon is 5.4 units of entropy. This implies that we are still in a rather degenerate nuclear gas phase, and hence the entropy contribution of baryon gas is, relatively speaking, small. For $\mu_\mathrm{B} = 840, T = 125$\,MeV, the baryon gas contributes about 3.7 units of entropy, while at $\mu_\mathrm{B} = 500, T= 150$\,MeV we have 5.4 units additional entropy. Hence we are at $S/B \simeq 8 - 10 $, the lower value for the higher range of baryo-chemical potential.  

The observed relative abundance $K^+/\pi^+ = 0.203 \pm 0.019$ is obtained by ignoring the possible distortions of the low energy spectra due to\lq low energy\rq\ phenomena; both $K$ and $\pi$ spectra are extrapolated assuming the Boltzmann form controlled by the fireball properties. Similarly, the K-ratio $K^-/K^+ = 0.19 \pm 0.03 $ is found - with same limitations as described above. The question now is if these results on particle ratios, temperature, and other inferred fireball properties are consistent with the assumption of a particular phase of hadronic matter and the above constraints. Before beginning this discussion we note that the first of the particle ratios is indeed a lower limit, in the sense that whatever the reaction mechanism, we do not expect to saturate the strangeness phase space fully, and hence the preliminary {\it equilibrium} picture we develop should predict a larger value than is actually observed.  

Let us first make the hypothesis that hadronic gas was made\footnotemark[27]. I fix the $K^-/K^+$-ratio at 0.2. There are two options: 
\begin{itemize} 
\item I take T = 125 MeV as the freeze out temperature. I infer following Koch et. al\footnotemark[23] a value of baryo-chemical potential of 520 MeV; the expected $K^+/\pi^+$ ratio is about 0.26, allowing for pions from $\Delta$-decays, and assuming that the strangeness phase space has been saturated. 
\item Taking instead as basis the strange particle temperature T= 150 MeV (under the tacit assumption that pion spectra are distorted by $\Delta$ decays and rescattering on spectator matter) the K - ratio implies a slightly lower baryo-chemical potential of just below 500 MeV and the $K^+/\pi^+$ ratio is slightly higher at 0.334. 
\end{itemize}
Thus with the proviso that \lq only\rq\ 80, resp. 60\% of the strangeness phase space is saturated both temperature hadronic gas scenario seems fully consistent with the data, with the exception that we do not understand how so much strangeness could be made by hadronic gas processes. At this point I note that this discussion disagrees in its detail with Ref.\footnotemark[8], which assumes fully saturated strangeness phase space. Therefore a rather low temperature is found, incompatible with the transverse spectra, or said differently (allowing for flow effects), with the mean energy per particle. Interestingly, the difference between our (and L\'evai\rq s\footnotemark[27]) analysis and Ref.[8] is the predicted d/p ratio which is highly sensitive to the entropy per baryon: assuming to much strangeness, additional pions are needed in order to `dilute\rq\ the strange particle abundance, an effect which I estimate at about 3 units of entropy per baryon. By implication the expected value for Ref.[8] of the $d/p = 0.05$, our discussion suggests 2--3 times larger value.  

For both above considered choices $T,\mu_\mathrm{B}$, the energy per baryon, which in this region of parameters is mainly controlled by the $K^-/K^+$-ratio, turns out to be below 1.9\,GeV. To get a slightly higher value, as it may seem required within the simple fireball model presented above, we should have set the Kaon ratio to a larger value, allowing for an unseen low energy fraction of $K^-$. Taking a value 0.25 at $T = 150$\,MeV would lead to energy per baryon somewhat above 2\,GeV and at the same time a $\pi^+/p$ ratio near 1.3, in agreement with the value reported at central rapidity. 

Next, let us see how the data fare under the assumption of quark-gluon plasma phase. Naturally, the advantage of this assumption is that we have little difficulty swallowing the saturation of strangeness phase space, thanks to the described rapid strangeness production. Furthermore it turns out that in region of $(\mu_\mathrm{B},T) = (850,130)$ MeV there would be a similar amount of entropy in the quark-gluon phase, as in the hadronic gas phase at $(\mu_\mathrm{B},T) = (500,150)$\,MeV. The supposition is that in the phase transition of the isolated glob some reheating from about 130 to 150 MeV takes place, and there is corresponding reduction of the chemical potential. As any pre--transition emission from the plasma would in such environment be covered by the soft component of the hadronic gas phase, we should not expect any visible quark matter effects in kaon spectra. Thus solely from the observation of singly strange particles we can not make a definitive statement about the presence of QGP in nuclear collisions at BNL. However, it is interesting to note that the BNL conditions are near to the baryon-rich quark-gluon phase domain. This conjecture is supported by the recent finding of antiproton multiplicity\footnotemark[25], which in the central rapidity region is less than one part in thousand of the proton multiplicity.  

But presently the only argument one could make in favor of QGP at Brookhaven is that the values of the parameters estimated above imply that even at BNL energies strangeness production in the quark-gluon phase will be rapid and will nearly saturate the available phase space. It is therefore most interesting to look at BNL for strange antibaryons, which without quark-gluon phase formation should hardly be produced at these energies. Given the suppression of antiprotons, which is expected for a baryon-rich fireball consisting of either hadronic gas phase or quark-gluon phase, observation of a {\em greater} strange antibaryon yields would strongly suggest that already at BNL energies this state of matter may be formed. It is to be hoped that the results from experiment E810 will allow us to conclude this issue.  

\subsection*{CERN--RHI results}

At CERN the available energy is much greater and ranges from 60 up to 200 GeV per nucleon. However, the laboratory has not taken full advantage of the available machine resources as yet, by limiting its main experimental runs to the highest available energy. In the asymmetric reactions such as the S--W collisions studied by\footnotemark[28,29,30]  there is the advantage over the S--S collisions studied by\footnotemark[31,32] NA35 of the much greater baryon number stopping. But there are difficulties in interpreting the data, which are associated with overlap of the different kinematic regions (target, central and projectile). In this regard, one has here in principle less of a problem than at BNL since the rapidity window is almost twice as large as at BNL: the projectile rapidity at 200 A\,GeV/c is 6, compared to 3.4 for BNL. The particular advantage of the S--S data is the symmetry of the kinematics, permitting a much better understanding of particle flows. The disadvantage is the likely presence of significant transparency at 200\,GeV per nucleon. However, the central rapidity region is 3 (for symmetric collisions), making a particle in the laboratory very fast. Consequently an experiment similar to E802 is impossible, as the time of flight does not permit particle identification. Thus the small aperture spectrometer experiment at CERN, NA34, is concentrating on the target fragmentation region. In view of the currently available results and this discussion it would seem that it would be of considerable advantage to study the symmetric S--S collisions at lowest available CERN energy, viz. 60 A\,GeV, in expectation of the lead beam run initially at a similar energy.  

Points of importance to our work in the most recent results of NA35 are:
\begin{itemize}
\item The $\Lambda$--$\bar\Lambda$ rapidity distribution, which shows two
pronounced peaks within the projectile and target rapidity regions, an
indication of a severe depletion in the central region. This shows that
much of the $\Lambda$ signal derives from re-scattering in the baryon rich
projectile and target fragmentation region.
\item The ${\bar\Lambda}$ multiplicity is sharply confined to the central
region $y = 3 \pm 0.5$. The rate of ${\bar\Lambda}$ production in S--S
collisions is about 120 times greater than in $p$--$p$ collisions (the error
quoted is large). The per trigger event multiplicity of ${\bar\Lambda}$ is
given to be 1.5! This (120-fold) enhancement has to be confronted with the
36-fold enhancement of the negatively charged tracks (i.e. pions). This
truly surprising result cannot even remotely be explained by cascading in
hadronic gas, as the probability of ${\bar\Lambda}$ formation is
decreasing during the moderation of the beam energy.
\item The general strangeness flavor production is up by a factor 2.5 on
top of the factor 36 for negatives: the $K/\pi$-ratio at mid-rapidity
$y=3$ is 0.15, to be compared to 0.06 for similar energy $p$--$p$ system. 
\end{itemize}

All these results remind us of the quark-gluon phase. Unfortunately, we do not have comparable data on production of $\bar p$ or ${\bar\Xi}$ and thus cannot conclude that the expected systematic signal of quark-gluon phase has been found. The lack of data on the essential $\bar p$ and ${\bar \Xi}$ production is being filled by the large aperture $\Omega\rq\ $-spectrometer WA85 experiment, which has presented the first results from the study of S--W collisions at 200 A GeV. Because of complex Monte Carlo studies required to understand the relative sensitivity of the experiment to ${\bar\Lambda}$ and ${\bar\Xi}$, this ratio is not known as yet, though WA85 has already reported first observation of $\bar \Xi$. The following has now been reported\footnotemark[28,29,30] by WA85 
\begin{itemize}
\item The temperature (inverse slope) of negatives, $\Lambda$ and ${\bar\Lambda}$ is the same and is 227 MeV, i.e.~higher than the temperature seen in S--S collisions. Because of the greater stopping expected, this result can be taken as a confirmation that the highest energy and baryon densities were reached in this experiment. Unfortunately, we cannot determine the baryo- chemical potential for this experiment as yet, nor can we determine the entropy per baryon. To this end we would need data on kaon $(K_s)$, pion (negatives) and also positive particle (protons and positive kaons) spectra in relation to the strange baryons and antibaryons.
\item The yield of both $\Lambda$ and ${\bar\Lambda}$ per negative track in the central rapidity region $2.4 < y < 2.65 $ is enhanced by a factor 1.7 in comparison to the control $p$--W data. Both enhancements are similar and the ratio of $\Lambda$ to ${\bar\Lambda} \simeq 0.2$ does not change. 
\item There seems to be an enhancement in the anticascade to cascade ratio in S--W collisions $(\sim 0.43 \pm 0.07$) as compared to the control $p$--W run $(\sim 0.27 \pm 0.06)$. Clearly, more statistics are needed to reconfirm this result. Also, it is important to know by how much the $\bar\Xi/\bar\Lambda$ ratio is enhanced in S--W reactions with reference to $p$--W reactions.
\end{itemize}

In CERN data, we hence once again see a clear strangeness enhancement, accompanied now by a highly significant enhancement of strange antibaryon yield. We cannot imagine how to interpret this data other than in terms of quark-gluon plasma. However, the data are still fragile at the level of only a few standard deviations, and require some improvement in the statistics. Also, we need a more complete evaluation of all available data in order to be able to give more detailed characterization of the conditions reached in the S--W and S--S collisions.

\section*{Concluding Comments}

Without a substantial interaction between experiment and theory, the most spectacular measurements remain, especially in this subject matter without much concrete insight. The situation is further complicated by numerous superficial if not wrong publications (as exemplified above) relating to the subject, as well as the process of \lq\lq reinventing the wheel\rq\rq, which so often leads not only to the repetition of the old mistakes. The particular reason why flavor flow experiments are very attractive in the beginning of any nuclear collider operation is the fact that the high expected strangeness production allows event by event analysis. Even if event rate should initially be small, strangeness will be clearly visible. The experiments suggested here are based on the following key observations:
\begin{enumerate}
\item At sufficiently high energy densities, heavy ion collisions may lead to formation of a deconfined phase of strongly interacting nuclear matter, the quark-gluon phase, in which flavor symmetry is partially restored and strangeness becomes abundant. The full event characterization is needed to fix the thermodynamic variables of essence for the basic understanding of reaction kinematics needed in understanding (strange) particle flows.
\item Compared to a hadron gas, in quark-gluon phase strangeness is produced faster and strangeness density is greatly higher. Also, strangeness is produced in quark-gluon phase almost totally by glue-glue processes. Uncertainty about the hadronization process makes global strangeness measurements less attractive as a signal of quark-gluon phase than observation of specific (multi)strange particles.
\item Anomalous (large) strange and multistrange antibaryon multiplicities can be viewed as the clearest signal that something unusual is happening in central collisions, particularly when viewed in specific windows of $(p_\perp,y)$. 
\item Multistrange antibaryons can provide crucial information as they are predominantly formed in phase space regions characterized by a very high strangeness density. 
\item As the theory of strangeness production and hadronization relies on key parameters of QCD, these will become accessible to measurement in heavy-ion collision induced reactions, through the measurement of diverse flavor and particle flows and detailed comparison of experiment with theory.
\end{enumerate}
{\it Acknowledgement} I would like to thank G. Odyniec, S. Lindenbaum and L. Madanski for their interest in this work and pertinent comments about important experimental aspects. I thank T. Ludlam for his kind hospitality at BNL.
\footnotetext{\vspace*{-0.5cm}\begin{enumerate}
\item J.~Rafelski and R.~Hagedorn, in:  \textit{Statistical Mechanics of
Quarks and Hadrons,} ed. H.~Satz, ed. North Holland, Amsterdam (1981)

\item J.~Rafelski, in:  \textit{Workshop on Future Relativistic Heavy Ion
Experiments,} eds. R. Bock and  R. Stock, GSI 81-6, Darmstadt 1981

\item J.~Rafelski and B.~M\" uller, Phys.~Revs.~Lett. {\bf 48}  1066 (1982); ibid. {\bf 56}, 2334(E) (1986) 

\item J.~Rafelski, ``Strangeness in Quark-Gluon Plasma'',
S. ~African~J.~Phys. {\bf 6} 37, (1983)  

\item P.~Koch, B.~M\" uller and J.~Rafelski, Phys.~Rep. {\bf 142} 167, (1986)
 
\item  P.~Carruthers and J.~Rafelski,  eds. \textit{Hadronic Matter in Collision 1988,} proceedings of a meeting held in Tucson, Arizona, October 6-12, 1988; (World Scientific, 1989)

\item  J.~Cleymans, ``Strangeness Production in Relativistic Ion
Collisions - Theoretical Review'', UCT-TP 142/90, to appear in proceedings
of QM'90 meeting, Menton, France, May 1990.

\item J.~Cleymans, H.~Satz, E.~Suhonen and D.W.~von Oertzen,
Phys.\ Lett.\ B {\bf 242} 111 , (1990)

\item  H.C.~Eggers and J.~Rafelski, ``Strangeness and Quark-Gluon
Plasma: Aspects of Theory and Experiments'', Preprint AZPH-TH/90-28,
submitted to J. Mod. Phys.  A (published: \textbf{6}  1067 (1991))

\item  E. Shuryak, in this volume

\item  J.~Rafelski and M.~Danos, Phys.\ Lett.\ B {\bf 192} 432  (1987)

\item  P.~Koch and J.~Rafelski, Nucl.\ Phys.\ A {\bf 444} 678  (1985) 

\item  P.~Koch and J.~Rafelski, S. ~African~J.~Phys. {\bf 9} 8  (1986)

\item  A.~Wr\'oblewski, Act.\ Phys.\ Pol. B {\bf 16} 379  (1985)

\item  T.~Akesson et al. [ISR-Axial Field Spectrometer Collaboration],
Nucl.\ Phys.\ B {\bf 246} 1, (1984)

\item  M.~Jacob and J.~Rafelski Phys.\ Lett.\ B {\bf 190} 173  (1987)

\item S.A.~Chin, Phys.\ Lett.\ B {\bf 78} 552  (1978)

\item S.~Mr\'owczy\'nski and J.~Rafelski, Phys.\ Rev.\ C {\bf 40} 1077  (1989)

\item  B. M\"uller, in this volume

\item J.~Rafelski and A.L. Schnabel, in: Conference on Intersections
between Particle and Nuclear Physics, Rockport 1988, AIP\# 176 

\item J.~Rafelski and A.L. Schnabel, Phys.\ Lett.\ B {\bf 207} 6, (1988)

\item N.K.~Glendenning and J.~Rafelski, Phys.\ Rev.\ C {\bf 31} 823, (1985)

\item P.~Koch, J.~Rafelski and W.~Greiner, Phys.\ Lett.\ B {\bf 123} 151  (1983)

\item  S.E. Eiseman et al. [E810 collaboration], ``Neutral $V$
production with 14.6 $x$ A GeV/c Silicon Beams'', BNL-44716, submitted to
Phys. Lett. B

\item  T.~Abbott et al. [E802 collaboration], Phys.~Revs.~Lett. {\bf 54} 847  (1990)
 
\item  J.B.~Costales [E802 Collaboration], HIPAGS Workshop,
Brookhaven, March 1990

\item  P.~L\'evai, B.~Luk\'acs and J.~Zim\'anyi, J.~Phys.~G {\bf 16} , 1019  (1990)

\item  N.J.~Narjoux et al. [WA85 Collaboration], Lecture at Quark
Matter '90, Menton, France, May 7--11, 1990 

\item D.~Evans et al. [WA85 Collaboration], Lecture at Quark Matter '90, Menton, France, May 7--11, 1990 

\item E.~Quercigh, in: ``Hadronic Matter in Collision 1988'', World Scientific, 1989, eds. P.~Carruthers and J.~Rafelski
 
\item R.~Stock et al. [NA35 Collaboration], Lecture at Quark Matter
'90, Menton, France, May 7--11, 1990

\item  H.~Str\"obele et al. [NA35 Collaboration], Lecture at Quark Matter '90, Menton, France, May 7--11, 1990

\end{enumerate}
}
 
\end{mdframed}

\subsection{Strangeness production with running QCD parameters}\label{QCDrunning}

Many things were happening in 1995, advancing the strangness signature of QGP:
\begin{enumerate}
\item
In January 1995 a first Strangeness in Quark Matter meeting took place in Tucson, and the proceedings were published rapidly~\cite{Rafelski:1995zq}. This meeting grew into a conference series which gathers today several hundred participants, see listing at \url{https://sqm2019.ba.infn.it/index.php/previous-editions/}. As remarked maybe QGP discovery should have been announced at this venue, see page \pageref{SQM95an}.
\item
The Hagedorn conference of Summer 1994  was readied for publication~\cite{Letessier:1995ic}.
\item 
I completed a review addressing the theoretical developments advancing QGP formation in heavy ion collisionsj and strangness~\cite{Rafelski:1996hf}, including many SHM fit results.
\item
Refinement of the production rates of strangness was achieved considering the QCD-running of the strange quark mass and coupling constant, and allowed more detailed comparison with experiment~\cite{Letessier:1996ad}.  
\end{enumerate}

\noindent\textit{The March 1996 research progress report to the  US-DoE--Office of Science includes:}\\[-0.7cm]
\begin{mdframed}[linecolor=gray,roundcorner=12pt,backgroundcolor=Dandelion!15,linewidth=1pt,leftmargin=0cm,rightmargin=0cm,topline=true,bottomline=true,skipabove=12pt]\relax%
%
{\large \textbf{From:} March 1996 progress report prepared for:}\\ 
\indent\indent\indent  The U.S.-D. of E.--Office of Science\\
\textbf{Abstract:} \ldots we were primarily  engaged in an effort to substantiate our suggestion concerning the  formation of deconfined and nearly statistically equilibrated QGP phase in 160-200 A GeV  ($\sqrt{s_{\rm NN}}\simeq 8.6 + 8.6$ GeV) interactions. We continued the exploration of the energy dependence of the observables, refining the understanding of the nonequilibrium parameters and hadronization models, developing strangeness production descriptions free of ad hoc parameters and assumptions, and comprising more adequate  description of the fireball development. 
%
\section*{Strangeness in Dense Hadronic Matter} 
Strange particle production is  recognized as one of the interesting hadronic observables$^1$ of dense, strongly interacting matter and much of the current theoretical and experimental effort in study of relativistic nuclear collisions is devoted to  this topic$^2$. Our great interest in the subject arises from the realization that the experimentally observed anomalous production of (strange) antibaryons$^{3,4,5.6}$ cannot be interpreted without introduction of some new physical phenomena. 

Quantum-Chromodynamics (QCD) is accepted as the theoretical foundation of strong interactions and we can expect that large regions of strongly interacting highly excited matter would obey the laws of perturbative thermal QCD, as is seen in lattice gauge calculations$^7$. There is little doubt about the existence of the deconfined phase (QGP) at high temperature, say  $T=1$ GeV in which physical processes are governed by perturbatively interacting quarks and gluons. The practical issue is, how extreme are the conditions required to form this new phase of matter? There are also quite intricate issues related to the short lifespan of the relativistic nuclear collision, which put in question the use of statistical physics methods, and suggest that elements of relativistic (quantum) transport theory must also be incorporated into the description of the physical phenomena occurring. For a recent survey of the physics of quark-gluon plasma we refer to review of Harris and M\"uller$^{8}$  and the many references cited there.
 
While near to the phase boundary of QGP with the confined hadronic gas (HG) phase quite complex phenomena may occur, involving in principle still other forms of hadronic matter, in a first approximation our effort  concentrates  on finding a suitable  extrapolation of the properties of perturbative QGP phase to this domain in order to be able to understand hadronic particle spectra  and abundances emerging in relativistic heavy ion collisions. From such a full description than emerges a diagnostic element of our work, as we seek to  correlate the properties of the source with  anomalies of particle abundances. We have proposed long ago$^{9}$ that strange antibaryons are the best tools in such a study of the QGP phase, and we are finding today$^{10,11,12}$ that the experimental results we mentioned above can be  interpreted using the properties of QGP computed for $T=250$ MeV, but that it is very difficult and outright impossible$^{13}$   to do so within completely conventional  pictures  of  nuclear reactions considering experimental data collected at 160--200 A GeV.

Aside of the study of strange antibaryon abundances, which in their production yield comprise knowledge of both the initial state and the freeze-out conditions, we also study the total final state strangeness yield, which is mostly dependent on the conditions prevailing in the dense hadronic matter in its most extreme initial moments. 

\subsection*{Production of Strangeness}
\addcontentsline{toc}{subsubsection}{Production of strangeness}
The production of strangeness flavor in deconfined QGP arises in its dominant fraction in gluon fusion processes$^{14,15}$ $gg\to s\bar s$,  and to a lesser extend in light quark fusion$^{16}$  $ q\bar q\to s\bar s$. While the first order free space (perturbative vacuum) strangeness production processes at fixed values of $\alpha_{\rm s}=0.6$ and $m_{\rm s}= 160$--200 MeV, have been considered for some time, non-perturbative effects were more recently explored. The thermal production rates in medium, incorporating temperature dependent non-perturbative particle masses$^{17}$  have lead to the total strangeness production rate which was found little changed compared to the free space rate. This finding was challenged$^{18}$, but a more recent revaluation of this work$^{19}$ confirmed that the rates obtained with perturbative glue-fusion processes when compared with thermal perturbative results are describing adequately the strangeness production rates in QGP. For further discussion of the current situation regarding thermal rates we refer  to reference$^{20}$.

Uncertainty in the value of  of the strong interaction coupling $\alpha_{\rm s}^2$ introduces considerable systematic error into the computed thermal rates. Recent experimental and theoretical studies of $\alpha_{\rm s}$ have allowed us to eliminate this as an ad-hoc parameter from our description of strange quark production$^{21,22}$. Using nonperturbative techniques of the QCD renormalization group we were able to obtain the strangeness production cross sections and thermal production rates in QGP using $\alpha_{\rm s}(M_Z)$ as input. Specifically, running QCD renormalization group is employed$^{21,22}$ to resum even-$\alpha_{\rm s}$  Feynman-diagrams involving two particles in initial and final states.  We used these results to extend our study of the two generic strangeness observables as function of the impact parameter (baryon content) and collision energy:
\begin{itemize}
\item Specific (with respect to baryon number $B$) strangeness yield  $\langle \bar s\rangle/B$\\ {\it Once produced strangeness escapes, bound in diverse hadrons, from the evolving fireball and hence  the total abundance observed is characteristic for the initial extreme conditions;}
\item Phase space occupancy $\gamma_{\rm s}$\\ {\it Strangeness freeze-out conditions at particle hadronization time $t_{\rm f}$, given  the initially produced abundance, determine the final state observable phase space occupancy of strangeness $\gamma_{\rm s}(t_{\rm f})$.} 
\end{itemize}
\centerline{
\includegraphics[width=0.6\columnwidth]{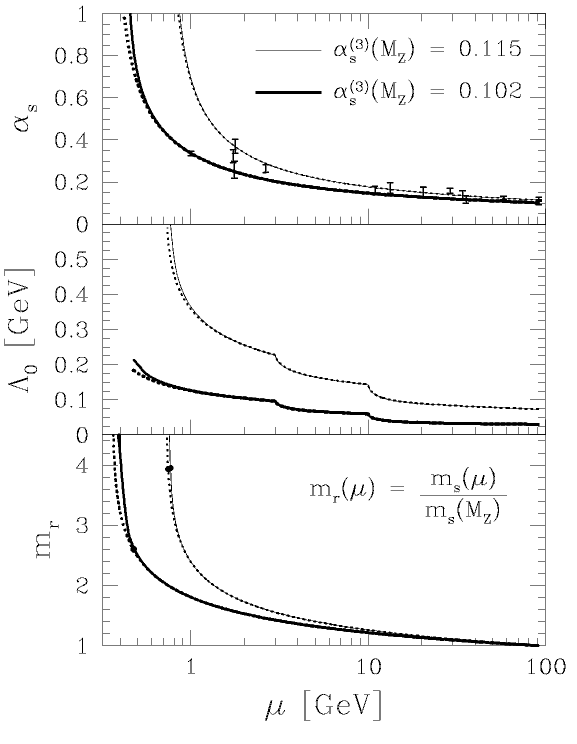} 
} 
\noindent Fig.\;1: $\alpha_{\rm s}(\mu)$, the $\Lambda$-parameter $\Lambda_0$ and $m_{\rm r}(\mu)=m(\mu)/m(M_Z)$ as function of energy scale $\mu$. Thick lines correspond to initial value $\alpha_{\rm s}(M_Z)=0.102$, thin lines are for the initial value $\alpha_{\rm s}(M_Z)=0.115$. Dotted lines are results obtained using the perturbative expansion for the renormalization group functions, full lines are obtained using  Pad\'e approximant of the $\beta$ function. Experimental results for $\alpha_{\rm s}$ selected from recent experimental work and Ref.\,[23]. In bottom portion the dots indicate the pair production thresholds for $m_{\rm s}(M_Z)=$~90~MeV (from Ref.[22]).\\

The QCD renormalization group equations for the running coupling constant $\alpha_{\rm s}$ and quark mass are:
\begin{equation}
\tag{1}
\mu \frac{\partial\alpha_{\rm s}}{\partial\mu}
=\beta(\alpha_{\rm s}(\mu))=-\alpha_{\rm s}^2\left[\ b_0
   +b_1\alpha_{\rm s} +b_2\alpha_{\rm s}^2 +\ldots\ \right] 
\,,
\end{equation}
\begin{equation}
\tag{2}
\mu {\frac{\partial m}{\partial\mu}} =-m\,
\gamma_{\rm m}(\alpha_{\rm s}(\mu))=-m\alpha_{\rm s}\left[\ c_0
+c_1\alpha_{\rm s} + \ldots\ \right]
\,,
\end{equation} 
and the coefficients $b_i,\,c_i$ are given in Ref.[23]. We introduce a Pade approximant of Eq.\,(1) and integrate these equations, using the precise determination of $\alpha_{\rm s}$ at the scale $M_Z=91.2$ GeV. The solutions in Fig.\,1 are obtained for $\alpha_{\rm s}(M_Z)=0.102$\, (thick lines) and $\alpha_{\rm s}(M_Z)=0.115$\,, (thin lines). In the top section of Fig.\,1 we show the variation of $\alpha_{\rm s}$, which is in here relevant 1GeV energy range not well characterized by the first order result often used. This is shown in the middle section of Fig.\,1 where the value $\Lambda_0$ based on a first order result, defined by the implicit equation:
\begin{equation} 
\tag{3}
\alpha_{\rm s}(\mu)\equiv\frac{2b_0^{-1}
(n_{\rm f})}{\ln(\mu/\Lambda_0(\mu))^2}\,,
\end{equation} 
is shown. We see that $\Lambda_0(1\mbox{\,GeV})=240\pm100$ MeV, assuming that the solid lines provide a valid upper and lower limits on $\alpha_{\rm s}$. However, the variation of $\Lambda_0(\mu)$ is significant for $\mu<3$~GeV, questioning the use of first order expressions. 
 
Because  Eq.\,(2) is linear in $m$, we consider the universal multiplicative quark mass scale factor
\begin{equation}  
\tag{4}
\hfill m_{\rm r}=m(\mu)/m(\mu_0)\,.
\end{equation}
Since we refer to  $\alpha_{\rm s}$ at the scale of $\mu_0= M_Z$  we use this reference point also for quark masses. As seen in the bottom portion of Fig.\,1, the change in the quark mass factor is highly relevant, since it is driven by the rapidly changing $\alpha_{\rm s}$ near to $\mu\simeq 1$~GeV. For each of the two different functional dependences $\alpha_{\rm s}(\mu)$ we obtain a different function $m_{\rm r}$. Note that the difference between Pade approximant result (solid lines) and perturbative expansion (dotted lines) in Fig.\,1 amounts to a slight `horizontal' shift of $\alpha_{\rm s}$ and $m_{\rm r}$ as function of~$\mu$.
 
\centerline{
\includegraphics[width=0.8\columnwidth]{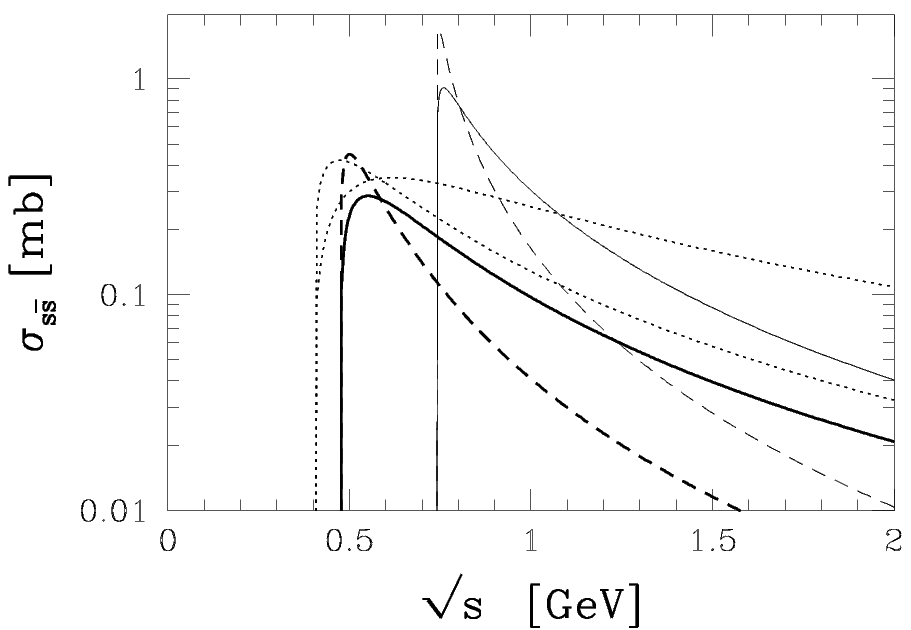} 
} 
\noindent Fig.\;2: QCD strangeness production cross sections  obtained for running $\alpha_{\rm s}(\protect\sqrt{\rm s})$ and $m_{\rm s}(\protect\sqrt{\rm s})$.  Thick lines correspond to initial value $\alpha_{\rm s}(M_Z)=0.102$, thin lines are for the initial value $\alpha_{\rm s}(M_Z)=0.115$. Dotted: results for fixed  $\alpha_{\rm s}=0.6$ and $m_{\rm s}=200$ MeV. solid lines $gg\to  s\bar s$; dashed lines  $q\bar q\to s\bar s$ (adapted from Ref.\,[21])\\ 

In Fig.\,2, the strangeness production cross sections are shown with $m_{\rm s}(M_Z)=90$~MeV. For the two choices of the running coupling constant considered in Fig.\,1 we depict the cross sections for the processes $gg\to s\bar s$ (solid lines, upper dotted line) and $q\bar q\to s\bar s$ (dashed lines, lower dotted line). Dotted are cross sections computed with fixed $\alpha_{\rm s}=0.6$ and $m_{\rm s}=200$ MeV cross sections shown here for comparison. We note that the glue based flavor production dominates at high $\sqrt{s}$, while near threshold the cross sections due to light quark heavy flavor production dominate. We note the different thresholds for the two values of $\alpha_{\rm s}(\mu)$ used. It is apparent that the cross sections are `squeezed' away from small $\sqrt{s}$ as we increase the value of $\alpha_{\rm s}(M_Z)$, but that the energy integrated cross sections ($\simeq$ rates) are relatively little changed.

\centerline{
\includegraphics[width=0.6\columnwidth]{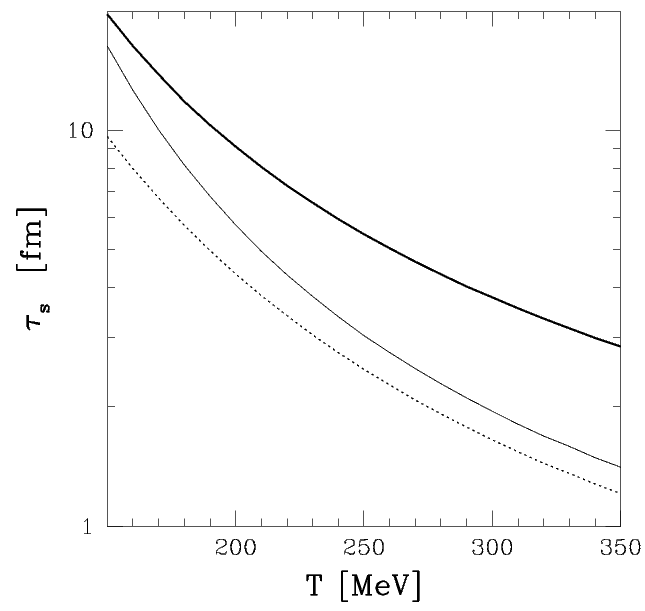}
} 
\noindent Fig.\;3: QGP strangeness relaxation time obtained using running $\alpha_{\rm s}$ cross sections shown in Fig.\,2.  Thick lines correspond to initial value $\alpha_{\rm s}(M_Z)=0.102$, thin lines are for the initial value $\alpha_{\rm s}(M_Z)=0.115$. Dotted: results for fixed  $\alpha_{\rm s}=0.6$ and $m_{\rm s}=200$ MeV (adapted from Ref.\,[22])\\

Using the QCD cross sections we have established, we compute the invariant strangeness production rate $A_{\rm s}$:
\begin{equation}
\notag 
A_{\rm s}= A_{gg}+A_{q\bar q}=\int_{4m_{\rm s}^2}^{\infty}ds
2s\delta (s-(p_{\rm A}+p_{\rm B})^2)
\int{d^3p_{\rm A}\over(2\pi)^32E_{\rm A}}\int{d^3p_{\rm
B}\over(2\pi)^32E_{\rm B}} 
\end{equation}
\begin{equation}
\tag{5}
\times\left[{1\over 2} g_g^2f_g(p_{\rm A})f_g(p_{\rm B})
{\sigma_{gg}}(s) + n_{\rm f}g_q^2 f_q(p_{\rm A})
f_{\bar q}(p_{\rm B}){\sigma_{q\bar q}}(s)\right]\,,
\end{equation}
and we obtain the relaxation time constants $\tau_{\rm s}$ shown in
Fig.\,3:
\begin{equation}
\tag{6}
\tau_{\rm s}\equiv
{1\over 2}{\rho_{\rm s}^\infty\over{(A_{gg}+A_{qq}+\ldots)}}\,,
\end{equation}
where the dots indicate that other mechanisms may contribute to strangeness production, reducing the relaxation time, obtained here considering the processes of gluon and quark fusion. Solid lines correspond to the two cases $\alpha_{\rm s}(M_Z)$ considered here, dotted line shows for comparison  the result of earlier studies with fixed   $\alpha_{\rm s}=0.6$ and $m_{\rm s}=200$ MeV.   

In Fig.\,4 both strangeness observables of interest here, are shown as function of the laboratory energy of the beam for central collisions.  To the right we see the observable most related to the initial conditions$^{12}$  
of the fireball: the ratio of the total strangeness produced $\langle s \rangle$ to the number of baryon participants $B$ in the fireball. By taking the ratio, we eliminate the explicit dependence of the QGP fireball volume. The result is a sensitive measure of the initial conditions. The experimental result we note is $\langle s \rangle/B=0.86\pm0.14$, reported in Ref.\,[24] for the collisions of S--Ag at 200 A GeV. This is in remarkable agreement with the results we obtained. Since the yield of strangeness per baryon is primarily determined  by the initial thermal properties of the QGP fireball and the early fireball evolution, we must presume that we have an appropriate description not only of strangeness formation rate, but also of the initial conditions (temperature) and the early evolution of the fireball \ldots 

\centerline{
\includegraphics[width=0.75\columnwidth]{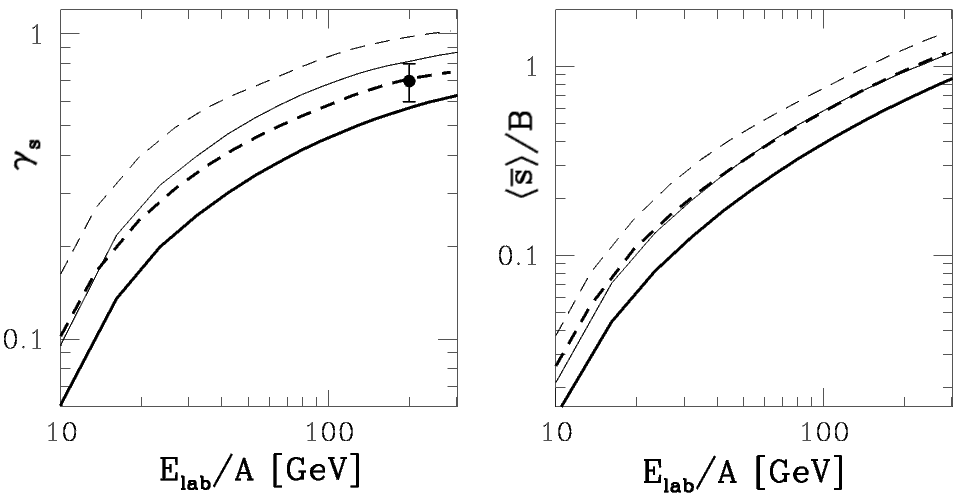}
} 
\noindent Fig. 4: $\gamma_{\rm s}(t_{\rm f})$ and $\langle \bar s\rangle/B$ as function of beam energy for central  S--W/Pb collisions (solid lines) and  Pb--Pb collisions (dashed lines) assuming $m_{\rm s}(M_Z)=90$ MeV, three dimensional expansion of the fireball with $v=c/\protect\sqrt(3)$, and stopping 50\% (S--W/Pb), 100\% (Pb--Pb). For $\gamma_{\rm s}$ we  take freeze-out at $T_{\rm f}=140$ MeV --- the vertical bar corresponds to the value of  $\gamma_{\rm s}$ found in S--W data analysis$^{11}$ (adpted from Ref.\,[1])\\


In a wide energy range we find that the specific strangeness yield rises linearly with the (kinetic) fireball energy content, reaching $\langle \bar s\rangle/B=0.8\pm0.15$ for S--W/Pb collisions at 200A GeV. It would be quite surprising to us, if other reaction models without QGP would find this linear behavior with similar coefficients, which we can determine in our case as function of the properties of QGP and its dynamical evolution. We therefore believe that this result is an interesting characteristic feature of our QGP thermal fireball model.

The phase space occupancy $\gamma_{\rm s}(t_{\rm f})$, shown to left in Fig.\,4 is influenced by the initial condition, the fireball evolution and the freeze-out conditions. Since the initially produced strangeness abundance does not reannihilate, strangeness can even overpopulate the final available phase space at plasma disintegration, so for large and long lived fireball scenarios, strange antibaryon abundances in Pb--Pb collisions could show $\gamma_{\rm s}>1$, and thus lead to spectacular enhancement of some particle ratios such as $\overline{\Xi}/\overline{\Lambda}\propto \gamma_{\rm s}$. However, results shown here, suggest that we just reach $\gamma_{\rm s}=1$ in Pb--Pb collisions up to 300 A GeV.

\subsection*{Fireball Dynamics and Initial Conditions}
\addcontentsline{toc}{subsubsection}{Fireball dynamics and initial conditions}
A key input in the discussion has been the initial conditions and evolution$^{12}$ of  the fireball which in our approach are fully described in a dynamical reaction picture, and nothing is left to arbitrary assumption. However, the picture of the reaction is based on the conventional wisdom and comprises barely proven assumptions, such as that the fireball expansion is adiabatic, the freeze-out occurs at 140 MeV, in addition to a courageous extrapolation of the QGP equations of state to the temperature range of importance here. The initial conditions are determined from the requirement that the dynamic pressure imparted on the fireball in the collision should be balanced by the {\it internal parton pressure}, which we assume to be given in terms of the initial temperature. The energy per baryon content determines the baryon density in the fireball and thus also the chemical potentials, once the degree of chemical equilibration is  known. The energy per baryon is derived from the stopping fractions that can be extracted form global features of the experimental results. \ldots

\centerline{
\includegraphics[width=0.6\columnwidth]{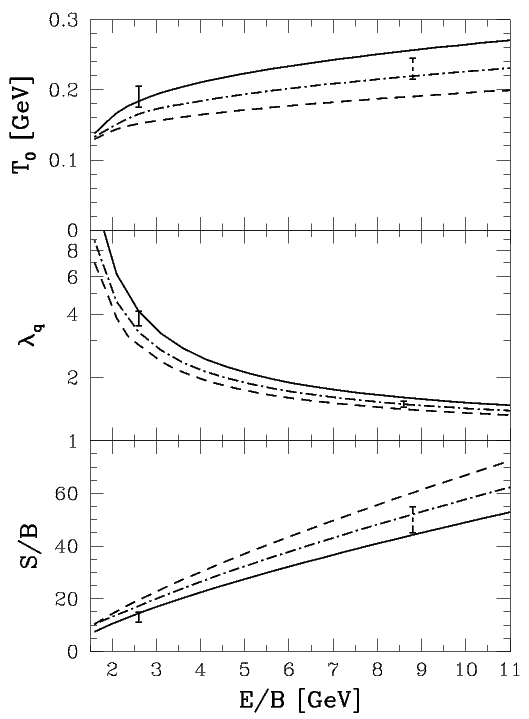} 
}
\noindent Fig.\;5:  Temperature $T_0$, light quark fugacity $\lambda_{\rm q}$ and entropy per baryon $S/B$ at the time of full chemical equilibration  as function of the QGP-fireball energy content $E/B$. Results for momentum stopping $\eta=1$ (solid line), 0.5 (dot-dashed line) and 0.25 (dashed line) are shown. Experimental `data' points are derived from our interpretation of experimental data$^{11,25}$ (adapted from Ref.\,[12])\\

In  Fig.\,5 we show, as function of the specific energy content $E/B$,  the expected behavior of temperature $T_0$,  the light quark fugacity $\lambda_{\rm q}$ and the entropy per baryon $S/B$ at the time of full chemical equilibration in the QGP fireball.  The range of the possible values as function of $\eta$ is indicated by showing  results, for $\eta=1$ (solid line), 0.5 (dot-dashed line) and 0.25 (dashed line). The experimental bars on the right hand side  of the  Fig.\,5 show for high (8.8 GeV) energy the result of analysis$^{11}$ of the WA85 data$^{3}$. The experimental bars  on the left hand side of the  Fig.\,5 (2.6 GeV) are taken from our analysis of the BNL-AGS data$^{25}$, but note that in this case we had found $\lambda_{\rm s}=1.7$ and not $\lambda_{\rm s}=1$ as would be needed for the QGP interpretation at this low energy. 

Among the key features in the  Fig.\,5, we note that, in qualitative terms, the drop in temperature with decreasing energy and stopping is intuitively as expected. At low (BNL-AGS) energies there is relatively rapid variation in $\lambda_{\rm q}$ which drives much of the variation in $T_0$. The value of $\lambda_{\rm q}$ is relatively insensitive to the stopping power.  This implies that even when different trigger conditions lead to different stopping fractions $\eta_i$, the resulting value of $\lambda_{\rm q}$ which is determining  the strange particle (baryon/antibaryon) ratios, is rather independent of different trigger conditions. 
 
Since we have now obtained as function of collision energy both the particle fugacities and strangeness phase space occupancy, we can combine these to derive as function of energy the  antibaryon ratios $\overline{\Lambda}/\overline{p}$ and $\overline{\Xi^-}/\overline{\Lambda}$.  The result, shown in Fig.\,6, (we show here the ratio of integrated yields with $m_\bot\ge 1.7$ GeV.) suggests that for a QGP fireball these ratios essentially stay constant, as the energy is lowered, since the increase associated with an increase in baryochemical potential is just compensated by the decrease in $\gamma_{\rm s}$ arising from lower gluon  collision frequency. We know that these ratios are much smaller, essentially zero,  in transport models involving conventional hadronic matter. Consequently, a way to explore the possibility that deconfinement has been present at high energies is  to seek a  substantial decrease in in this ratio, and thus a change in the reaction mechanism, as the collision energy is lowered. We thus believe that in order to ascertain the possibility that indeed the QGP phase is formed at energies available today (up to 9 GeV per nucleon in the CM frame) a more systematic exploration as function of collision energy of two above discussed strange particle observables $\gamma_{\rm s}(t_{\rm f})$ and $\langle \bar s \rangle /B$, and in particular the strange antibaryon ratios, is needed. \ldots

\centerline{
\includegraphics[width=0.56\columnwidth]{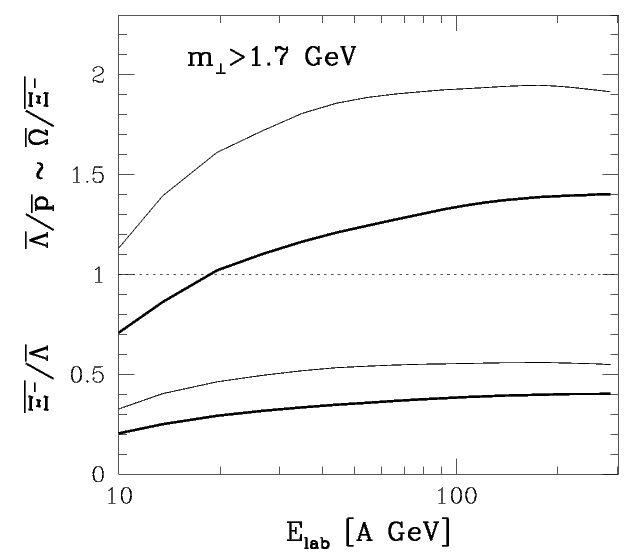}
} 
\noindent Fig.\;6: Fixed $m_\bot$ $\overline{\Lambda}/\overline{p}$ and $\overline{\Xi^-}/\overline{\Lambda}$ as function of $E/B$ in the Pb--Pb fireball, taking into account variation of $\gamma_{\rm s}$ shown in Fig.\,4. Thick lines correspond to initial value $\alpha_{\rm s}(M_Z)=0.102$, thin lines are for the initial value $\alpha_{\rm s}(M_Z)=0.115$ (from Ref.1)\\ 

\footnotetext{\vspace*{-0.5cm}\small
\begin{enumerate}
\item
J. Rafelski, J. Letessier, and A. Tounsi, {\it Strange Particles from 
Dense Hadronic Matter}, survey prepared for {\it Acta Phys. Pol. B} June (1996) (published Acta Phys.\ Polon.\ B {\bf 27}, 1037 (1996))

\item
{\it Strangeness in Hadronic Matter: S'95} Proceedings of Tucson
workshop, January 1995, American Institute of Physics Proceedings Series
Vol.\,340, (New York 1995), J. Rafelski, editor 
 
\item
WA85 collaboration presentations in  Ref.\,[2];
see also:\\
D. Evans {et~al.} (WA85 collab.), {Nucl. Phys.} A
{\bf 566}, 225c (1994);\\
S. Abatzis {et~al.} (WA85 collab.), {Phys. Lett.} B
{\bf 259}, 508 (1991);\\
S. Abatzis {et~al.} (WA85 collab.), {Phys. Lett.} B {\bf
270}, 123 (1991) 
  
\item
F. Antinori, in  Ref.\,[2];\\
S. Abatzis {et al.} (WA85 collab.), {Phys. Lett.} B
{\bf 316}, 615 (1993);  {Phys. Lett.} B {\bf 347}, 158 (1995) 
 
\item
Th. Alber et al. (NA35 collab.), Z. Phys. C{\bf 64}, 195 (1994), 
and references therein;\\
M. Ga\'zdzicki et al. (NA35 collab.), Nucl. Phys. A{\bf 590},
 197c (1995);\\
P. Foka and the NA35 collaboration and
M. Ga\'zdzicki and the NA35 collaboration, in Ref.\,[2] 
 
\item
J. G\"unther for the NA35
collaboration, to appear in proceedings of QM'95, Monterey, January
1995 (edited by A. Poskanzer et al.); and private communication.
 
\item
C. DeTar, {\it QGP in Numerical Simulations of Lattice
QCD}, in {\it Quark-Gluon Plasma 2}, p.1,  R.C. Hwa, Editor, World
Scientific, (Singapore 1995);\\
T. Blum, L. K\"arkk\"ainen, D. Toussaint, S. Gottlieb, 
Phys. Rev. D {\bf 51}, 5153 (1995)

\item
J. W. Harris and B. M\"uller, {\it The search for the Quark-Gluon Plasma},
to appear in {\it Ann. Rev. Nuc. Science}, preprint: hep-ph/9602235 (1996) 

\item
J. Rafelski, {\it Phys. Rep.} {\bf 88} (1982) 331;\\
J. Rafelski and R. Hagedorn, in: {\it Statistical
Mechanics of Quarks and Hadrons}, edited by H. Satz, North Holland, 
Amsterdam 1981, pp. 253--272;\\ 
J. Rafelski, in: {\it Workshop on Future Relativistic
Heavy Ion Experiments}, edited by  R. Bock and R. Stock, GSI-Orange 
Report 81-6, Darmstadt 1981, pp. 282--324 

\item
J. Rafelski, {\it Phys. Lett. B} {\bf 262}, 333 (1991); \\
J. Rafelski,{\it Nucl. Phys. A} {\bf 544}, 279c, (1992).
 
\item
J. Letessier, A. Tounsi, U. Heinz, J. Sollfrank and J. Rafelski, 
{Phys.\ Rev.} D {\bf 51}, 3408 (1995);\\
J. Letessier, J. Rafelski and  A. Tounsi,  {Phys. Lett.} B {\bf 321}, 394
(1994);\\
J. Sollfrank, M. Ga\'zdzicki, U. Heinz and J. Rafelski,  {Z. Physik} C
{\bf 61}, (1994) 
 
\item
J. Letessier, J. Rafelski and A. Tounsi, 
{\it Energy dependence of strange particle yields from
a quark-gluon plasma fireball} Preprint AZPH-TH/95-13
and PAR/LPTHE/95-24, submitted to {\it Phys. Rev. C};\\
J. Letessier, J. Rafelski,  and A. Tounsi,
{\it Quark-gluon plasma formation and strange antibaryons}, 
Preprint AZPH-TH/95-14R and PAR/LPTHE/95-36R,
submitted to {\it Phys. Lett. B} ;\\
J. Letessier, J. Rafelski,  and A. Tounsi, A., 
{\it Phys. Lett. B} {\bf 333}, 484, (1994) 

\item
L. Csernai, private communication;\\
N.S. Amelin, L.V. Bravina, L. P. Csernai, V.D. Toneev, K.K.
Gudima, S.Yu. Sivoklokov, {Phys. Rev.} C {\bf
47} (1993) 2299 
 

\item
J. Rafelski and B. M\" uller, Phys. Rev. Lett.  {\bf 
48}, 1066, (1982); {\bf 56}, 2334E, (1986) 

\item
T. Matsui,B. Svetitsky and L.D. McLerran, 
{\it Phys. Rev. D} {\bf 34}, 783, (1986) 

\item
T. Bir\'o and J. Zim\'anyi, {Phys. Lett. B} {\bf 113}, 6,
(1982); {\it Nucl. Phys. A} {\bf 395}, 525, (1983)

\item
T.S. Bir\'o, P. L\'evai and B. M\"uller, 
{\it Phys. Rev. D} {\bf 42}, 3078, (1990) 

\item
T. Altherr and D. Seibert, {\it Phys. Lett. B}
{\bf 313}, 149, (1993), and {\it Phys. Rev. C} {\bf 49}, 1684, (1994) 
 
\item
N. Bili\'c, J. Cleymans, I. Dadi\'c and D. Hislop,
{\it Phys. Rev. C}  {\bf 52}, 401, (1995).  

\item
J. Sollfrank and U. Heinz {\it The Role of
Strangeness in Ultrarelativistic Nuclear Collisions}, Helsinki
preprint HU-TFT-95-27, in {\it Quark-Gluon Plasma 2},
R.C. Hwa (Eds.), World Scientific, Singapore (1996).

\item
J. Letessier,  J. Rafelski,  and A. Tounsi,
{\it Impact of QCD and QGP properties on strangeness production},
{\it submitted to  Phys. Lett. B}, Preprint 
AZPH-TH/96-08, PAR/LPTHE/96-10,  (1996),
 
\item
J. Rafelski, J. Letessier,  and A. Tounsi,
{\it $\alpha_{\rm s}(M_Z)$ and Strangeness Production}
to appear in proceedings of Snowbird Nuclear 
Dynamics Workshop, February 1996, eds. W. Bauer 
and G. Westphal,  Plenum Press, New York (1996)

 
\item
I. Hinchliffe, {Quantum Chromodynamics}, 
update (URL:http://pdg.lbl.gov/) August 1995, in  
L. Montanet {et al.}, 
{\it Phys. Rev. D} {\bf 50}, 1173, (1994);\\
T. Muta, ``Foundations of Quantum Chromodynamics'', 
World Scientific, Singapore (1987).

\item
M. Ga\'zdzicki and D. R\"ohrich, 
{\it Strangeness in Nuclear Collisions}
preprint IKF-HENPG/8-95, {\it Z. Physik C} (1996).

\item
J. Rafelski and M. Danos {Phys. Rev.} C {\bf 50}, 1684 (1994);\\  
J. Letessier, J. Rafelski and A. Tounsi, {Phys. Lett.} B {\bf 328}, 499
(1994).
\end{enumerate}
}
\end{mdframed}
\vskip 0.5cm

\subsection{A picture with  STAR at RHIC}
One of the outgrowths of the Hagedorn 1995 meeting was further strengthening of my good relationship with Hans Gutbrod: he became a SUBATECH  lab director and I became engaged in a collaboration  with the RHI research group at SUBATECH in Nantes, France. This in turn  resulted in  my search for a more active role in the research program of the STAR collaboration, taking with me the SUBATECH group. In  1997 I prepared  an individual theoretical proposal how I could contribute to the STAR collaboration work. I show in the following a few pages only from this lengthy document.

At the February 1999  STAR collaboration meeting  held at BNL, I was  not admitted to the membership. The  STAR collaboration saw my effort that paralleled SUBATECH entry as a maneuver, probably an effort to gain early insight into the forthcoming STAR experimental data -- which were delayed  by technical RHIC problems to mid-2000. I was told to consider creating an experimental group at Arizona that would have to apply to join STAR, a long-term diplomatic delay, if feasible.  

This decision  was  of disadvantage for STAR, since they did not have my data analysis support. On the other hand I could  present, beginning with 2004/5, my own analyzis of their available data~\cite{Rafelski:2004dp,Letessier:2005kc}. These results were demonstrating the universality of RHIC and SPS bulk fireball properties. This insight was complemented after LHC-ALICE date became available by demonstration that the more precise lower energy results from STAR at RHIC and the very high energy results from ALICE at LHC also originate in the same bulk fireball properties~\cite{Petran:2013qla,Rafelski:2014cqa}. 

There is another possible consequence of this fateful decision. I  was wired into the CERN-SPS context. I could have helped to coordinate between the CERN work on the QGP announcement that followed exactly the February  events at BNL. On the other hand I was now perhaps mistrusted at CERN and, at the same time I could not help the flow of information between the labs. As consequence, CERN coasted to the new phase of matter announcement without much of communication with STAR.  

I believe the STAR decision  was the event that severed any chance of possible joint QGP announcement between RHIC and SPS groups. It should be remembered that the STAR  collaboration  made the decision to go alone, arguably in the belief they were holding in February 1999 the key to the QGP discovery.\\

\noindent\textit{The February, 1999 RHIC-STAR collaboration photo is followed by a few paragraphs from the 1998 proposal I presented, when attempting to join STAR, this short segment from a very long document introduces SHM analysis method and updates kinetic theory of strangeness production, more details of SHM presenteed in this document  are found on  page~\pageref{SHM-STAR} below:}
%

\begin{mdframed}[linecolor=gray,roundcorner=12pt,backgroundcolor=Dandelion!15,linewidth=1pt,leftmargin=0cm,rightmargin=0cm,topline=true,bottomline=true,skipabove=12pt]\relax%
\centerline{
\includegraphics[width=1.0\columnwidth]{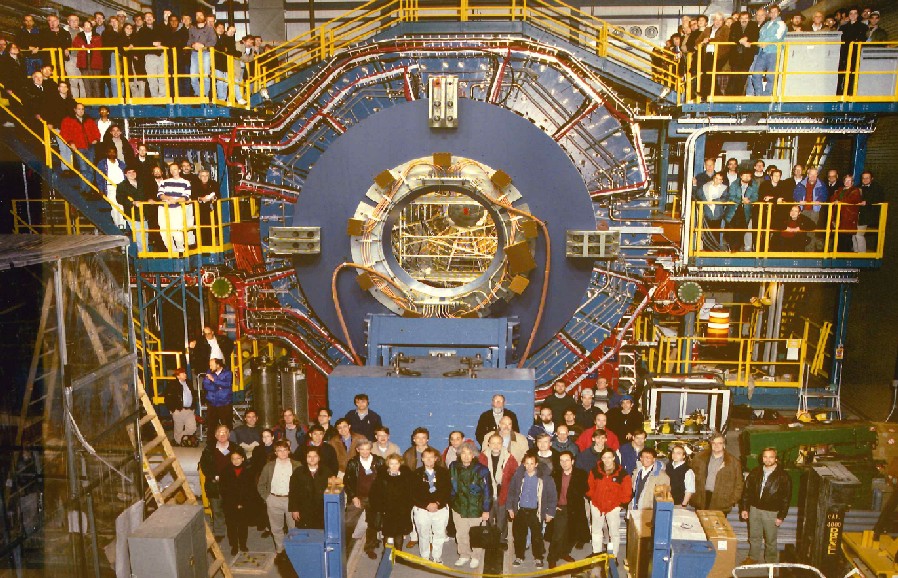} 
} 
\noindent February 1999 STAR collaboration picture. Front raw center: John Harris, spoksperson; the author is third person on his right also fourth person in picture from left

\section*{Hadronic Probes of QGP}
\addcontentsline{toc}{subsubsection}{Hadronic probes of QGP}
\subsection*{Strangeness as observable of QGP}
\ldots  We would like to refine the capability of the STAR detector such that certain longer lived strange particles can be detected with greater efficiency and precision, creating a more effective diagnostic tool of the dense state created in nuclear collisions at 100 A GeV center of mass energy.

Strangeness has been predicted already 15 years ago to be abundantly produced should the deconfined QGP phase be formed. Further study has confirmed that it is not reannihilated in rapid decomposition of the dense matter state and that the pattern of strange particle production is specific for the state of matter formed, its evolution and hadronization process.  Because there are many different strange particles, we have a very rich field of observables with which it is possible to explore diverse properties of the particle source. \ldots  

Today, it has been seen  in SPS experiments up to 200 A GeV that overall particles containing  strangeness are indeed produced more abundantly in relativistic nuclear collisions. Through the diligent work of the NA35/NA49 collaboration, which has developed a complex `$4\pi$'-detectors strangeness production excess of about factor two over  expected yields based on simple scaling of $N$--$N$ reactions has been today established$^{1}$. 

Equally significantly, the work of WA85/94/97 collaborations$^{2,3,4}$  as well as that of NA35/49$^{5}$ for the ratio $\overline\Lambda/\bar p$, shows that the abundance of strange antibaryons is rather unusual in that production pattern of these particles appears to arise in manner expected from the evaporation from the deconfined QGP phase. For example, the remarkable observation that the yield of strange antibaryons $\overline{\Lambda}+\overline{\Sigma}$ exceeds the yield of antipartons$^{5}$ is quite unexpected for a reaction picture involving confined particles. 

On the other hand it is a rather natural consequence for particles evaporated from deconfined QGP phase in which strange and non-strange quarks are have reached chemical equilibrium. Unfortunately, this spectacular result of the NA35 collaboration, is marred by lack of precision, one can still argue that the surprise is comprised in a few standard deviations, and thus could still go away. It is  our hope that the fully developed STAR tracking will allow to see such anomalous  effects at the level of precision which  will allow to perform detailed comparison with theoretical models, establishing the production mechanisms and thus the formation of deconfined phase.

\subsection*{Strangeness signatures of deconfinement}
\addcontentsline{toc}{subsubsection}{Strangeness signatures of deconfinement}
As implied in above qualitative discussion, there are two generic strangeness  observables which allow further diagnosis of the physical state produced in relativistic heavy ion interactions:
\begin{itemize}
\item 
{\bf absolute yield} of strangeness: Once produced in hot and dense hadronic matter, {\it e.g.},  the QGP phase, strangeness/charm is not reannihilated in the evolution of the deconfined state towards freeze-out, because in the expansion and/or cooling process the rate of  production/annihilation rapidly diminishes and becomes negligible.  Therefore the massive flavor yield is characteristic of the initial, most extreme conditions, including the approach to chemical equilibrium of gluons in the deconfined phase.
\item 
{\bf phase space occupancy$^{6}$  $\gamma_{i}$}: $\gamma_{i}$ describes how close the flavor  yield per unit of volume ($i={\rm s,c}$) comes to the chemical equilibrium expected; $\gamma_i$  impacts strongly the distribution of flavor among final state  hadronic particles.
\end{itemize}

This rise of $\gamma_s\to 1$, (in QGP phase) which in the collision occurs rapidly as function of time, and leads to a large freeze-out value seen in experiment, as observed in nuclear collisions at 160--200 A GeV, is believed to be at the origin of the significantly enhanced abundance of multistrange particles.

\ldots\ldots

\ldots the invariant rate the strangeness relaxation time
$\tau_{\rm s}$ shown in Fig.\,1, as function of
temperature is obtained$^{9,10}$:
\begin{equation}\tag{4}\label{tauss}
\tau_{\rm s}\equiv
{1\over 2}{\rho_{\rm s}^\infty(\tilde m_{\rm
s})\over{(A_{gg}+A_{qq}+\ldots)}}\,. 
\end{equation}
Note that any so far unaccounted strangeness production processes would add to the production rate incoherently, since they can be distinguished by the presence of incoming/outgoing gluons. Thus the current calculation offers an upper limit on the actual relaxation time, which may still be smaller. In any case,  the result shown in Fig.\,1   suffices to confirm that strangeness will at the end of QGP evolution at RHIC be very near to chemical equilibrium, assuming that the lifespan of QGP is at least given by the size of the colliding system.

We  see in Fig.\,1 also the impact of a 20\% uncertainty in $m_{\rm s}(M_{{Z}})$, indicated by the hatched areas. The calculations made$^{11,12}$ at fixed values $\alpha_{\rm s}=0.5$ and $m_{\rm s}=200$~MeV (dotted line in Fig.\,1) are well within the band of values related to the uncertainty in the strange quark mass. \\
 
\centerline{
\includegraphics[width=0.7\columnwidth]{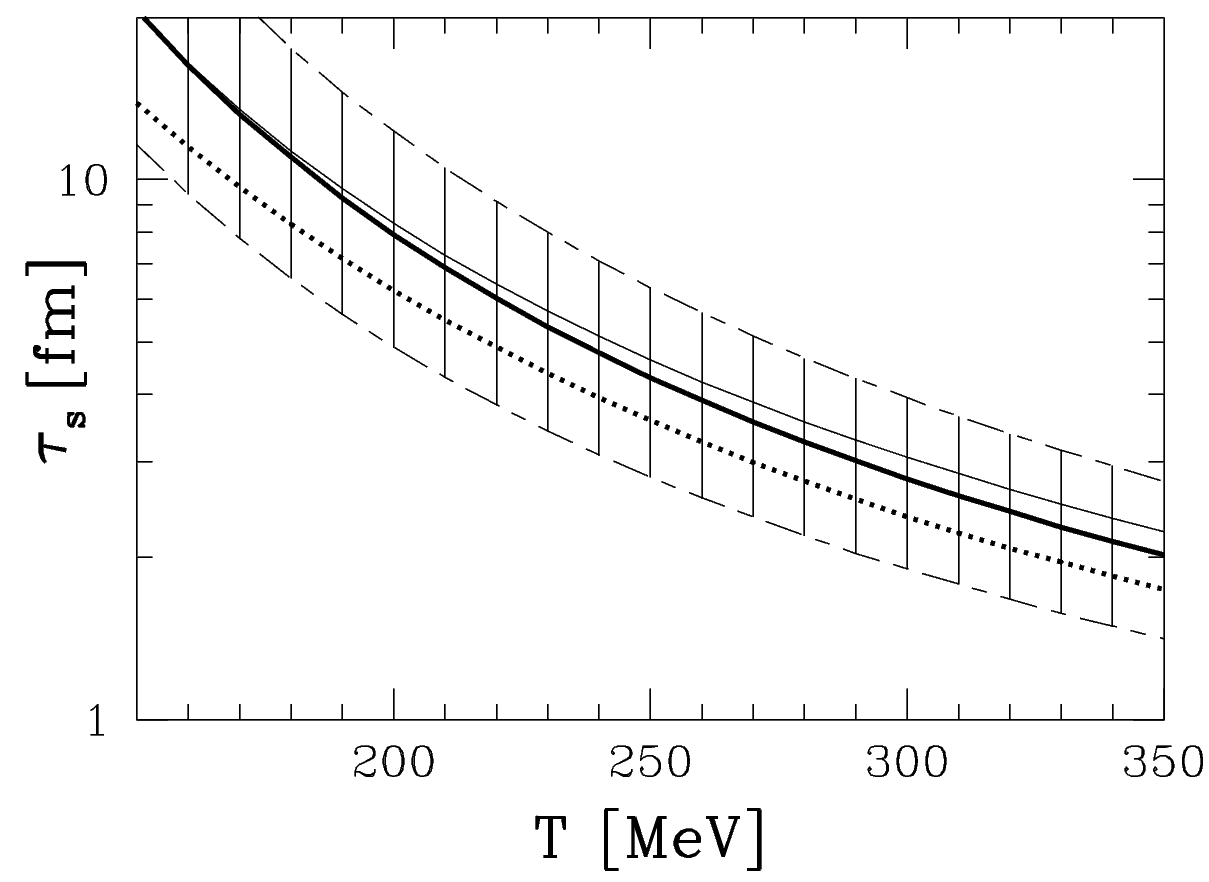} 
}
\noindent{\small Fig. 1: QGP strangeness relaxation time, for $\alpha_{\rm s}(M_{Z})=0.118$, (thick line) and = 0.115 (thin line); $m_{\rm s}(M_{{Z}})=90$~MeV. Hatched areas: effect of variation of strange quark mass by 20\%. Dotted: comparison results for fixed  $\alpha_{\rm s}=0.5$ and $m_{\rm s}=200$ MeV (adapted from Refs.[9,10])\\
}

Allowing for dilution of the phase space density 
in expansion, the dynamical equation describing 
the change in $\gamma_{s}(t)$ is:
\begin{equation}\tag{5}\label{dgdtf}
\hspace*{-0.2cm}{{d\gamma_{\rm s}}\over{dt}}\!=\!
\left(\!\gamma_{\rm s}{{\dot T m_{\rm s}}\over T^2}
     {d\over{dx}}\ln x^2K_2(x)\!+\!
{1\over 2\tau_{\rm s}}\left[1-\gamma_{\rm s}^2\right]\!\right).
\end{equation}
Here K$_2$ is a Bessel function and $x=m_{\rm s}/T$. With the relaxation constant $\tau_{\rm s}(T(t))$, these equations can be integrated numerically, leading to the values of the two observables, $\gamma_{\rm s}$ and $N_{\rm s}$ that control the yields of strange particles$^{13}$ evaporated from the expanding and hadronizing QGP blob, \ldots
 
{\it The proposed analysis methods using particle ratios   appears in the following section of these diaries.}

\subsection*{Concluding remarks}
The relative total abundance of strangeness is most related to the initial condition, the \lq hotter\rq\ the initial state is, the greater the production rate, and thus the final state relative yield, to be measured with respect to  baryon number or global particle multiplicity (entropy). The phase space occupancy of strangeness $\gamma_{s}$ depends aside of the initial production rate, on the final state dilution characterized by dynamics of the expansion and the freeze-out temperature. We believe that we will be able to use observed  features of strange mesons,  baryon and antibaryon production to see the formation of the deconfined state and to study some QCD properties and parameters. Experience with the analysis at SPS energies confirms that validity of this method to the study of the deconfined phase, though the precision of the SPS results so far has not sufficed to convince everyone that indeed a QGP phase has been formed. However, the strange particle production results obtained at 160--200 A GeV are found to be well consistent with the QGP formation hypothesis. 

\footnotetext{\vspace*{-0.5cm}
\begin{enumerate}

\item
G. Odyniec, invited lecture at the QM'97-Tsukuba, to be published in
proceedings {\it Nucl. Phys. } {\bf A}, (1998).

\item
D. Evans, (WA85 Collaboration), p.\,79 in \cite{S96}. 
M. Venables, (WA94 Collaboration), p.\,91 in \cite{S96}.

\item
S. Abatzis et al. (WA94 Collaboration)
Phys. Lett. B {\bf 354}, 178 (1995); D. Evans {et~al.} (WA85 Collaboration),
{Nucl. Phys.} A {\bf 566}, 225c (1994).
  
\item
S. Abatzis {et al.} (WA85 Collaboration), 
{Phys. Lett.} B {\bf 347}, 158 (1995); 
{Phys. Lett.} B {\bf 316}, 615 (1993).  
 
\item
I. Kr\'alik, (WA97 Collaboration), 
{\it Hyperon and Antihyperon Production in 
Pb--Pb Collisions at 158A GeV/c}, in \cite{SQM97};\\
QM'97presentation, to be published in
proceedings {\it Nucl. Phys. } {\bf A}, (1998).

\item
T. Alber et al, (NA35 Collaboration)
{\it Phys. Lett} {\bf B 366}, 56 (1996).

\item
J. Rafelski, {\it Phys. Lett.} {\bf B 262}, 333 (1991);
{\it Nucl. Phys.} {\bf A544}, 279c (1992).

\item
J. Letessier, J. Rafelski, and A. Tounsi,
 {\it Phys. Lett.} {\bf B389}, 586 (1996).
 
\item
M. Schmelling, 
{\it Status of the Strong Coupling Constant}, p91, in
proceedings of: 28th International Conference on High
Energy  Physics, Warsaw, July 1996, Z. Ajduk and 
A.K. Wr\'oblewski , Eds.,  (World Scientific, Singapore 1997)

\item
J. Rafelski, J. Letessier and A. Tounsi,
{\it Thermal Flavor Production and Signatures of Deconfinement},
p.971 in proceedings of: 28th International Conference on High
Energy  Physics, Warsaw, July 1996, Z. Ajduk and 
A.K. Wr\'oblewski , Eds.,  (World Scientific, Singapore 1997)

\item
J. Rafelski, J. Letessier and A. Tounsi,
{\it Acta Phys. Pol.} {\bf B},  (1998) (in press); hep-ph/9710340.

\item
J. Rafelski and B. M\"uller, {\it Phys. Rev. Lett}
{\bf 48}, 1066 (1982); {\bf 56}, 2334E (1986).

\item
P.~Koch, B.~M\"uller, and J.~Rafelski, {\it Phys. Rep.} {\bf 142},
167 (1986); Z. Phys. {\bf A324}, 453 (1986).

\item
J. Rafelski, J. Letessier and A. Tounsi,
{\it Acta Phys. Pol.} {\bf B27}, 1035 (1996).

\end{enumerate}
}
\end{mdframed}

\markboth{3. Soft Hadron Data Analysis and Interpretation}{Strangeness in QGP: Diaries}
\section{Soft Hadron Data Analysis and Interpretation}
\subsection{Statistical hadronization model (SHM)}
We return to the \lq\lq On the Trail of Quark-Gluon-Plasma: Strange Antibaryons in Nuclear Collisions\rq\rq discussion~\cite{Rafelski:1992td} presented at the IlCiocco July 12-24, 1992 Summer School. In this part, the just invented  statistical hadronization model~\cite{Rafelski:1991rh} (SHM) was introduced  in more detail for the first time. Note that SHM, as the name of the model, only appeared several years later.\\

\noindent\textit{Part II of Ref.\cite{Rafelski:1992td} \lq\lq On the Trail of Quark-Gluon-Plasma: Strange Antibaryons in Nuclear Collisions.\rq\rq, for Part I see page~\pageref{Trail1992}:}\\[-0.7cm]
\begin{mdframed}[linecolor=gray,roundcorner=12pt,backgroundcolor=Dandelion!15,linewidth=1pt,leftmargin=0cm,rightmargin=0cm,topline=true,bottomline=true,skipabove=12pt]\relax%
\centerline{\includegraphics[width=1.0\textwidth]{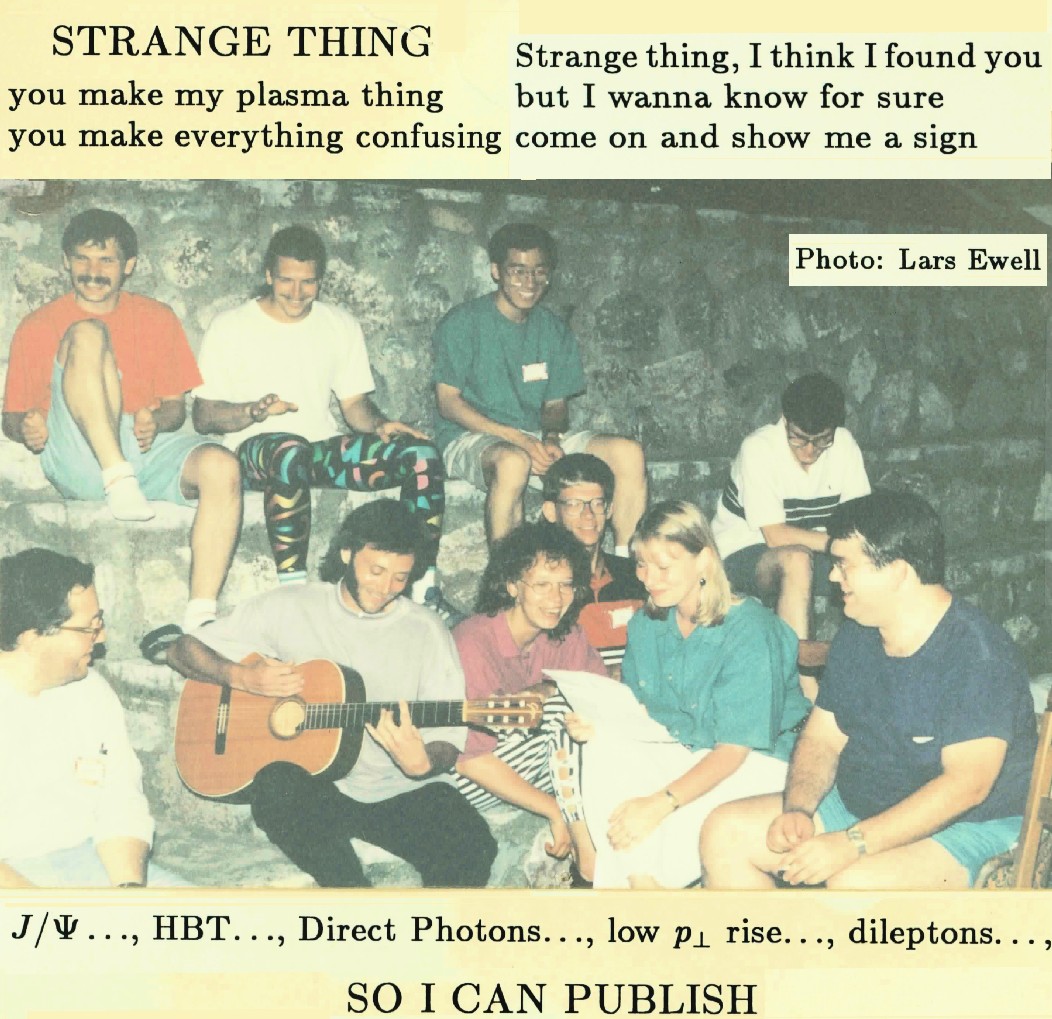}}
\noindent{\small Singing students at the Il Ciocco  Summer School July 12-24, 1992   \lq celebrating\rq\ strangeness and QGP}
%
%
\section*{The central fireball}
\subsection*{Particle spectra}
\label{Trail1992p2}
\addcontentsline{toc}{subsubsection}{Particle spectra}
The favorite scenario as suggested above looks like this: very rapid thermalization of the fireball energy in a central high energy nuclear collisions in which numerous radiation quanta, gluons are formed, followed by glue based formation of strange quark pairs. Next step is the formation of final state hadrons, either in the process of general QGP decomposition or in radiative emission from QGP. It is in this step that particles are formed that are ultimately observed in the experiment. I will now describe how we can use the observations to obtain information about the proto-phase of the reaction. In this I will develop a method which is equally suitable for the case that no QGP has been formed and that the reaction has proceeded by the way of usual hadronic interactions (HG). However, my approach rests on the presence in the reaction of a locally equilibrated fireball of dense nuclear matter, and the question arises if there is evidence today for such a reaction mechanism of strange particle production in relativistic nuclear collisions.
 
How can we argue that the strangeness enhancement originates in a central fireball or another similar high density state? First note that the relative probability to find a composite particle per unit of phase space volume $d^3\vec x \, d^3\vec p/(2\pi)^3$ becomes 
\begin{equation} 
\tag{12} \frac{d^6N}{d^3\vec x \, d^3\vec p /(2\pi)^3}= \prod_{\rm i}
~g_{\rm i}\ \lambda_{\rm i} \ \gamma_{\rm i} \ 
{\rm e}^{-E_{\rm i}/T_{\rm f}} \;.
\end{equation}
The overall normalization of the yield is not easily accomplished, but the relative yields should be well described by Eq.\,(12). We will return to discuss in detail the important pre-exponential factors in next section. We will solely concentrate on the exponential (Boltzmann) term: for a composite particle at energy $E=\sum_i E_i$, Eq.(12) becomes simply a phase space factor times the Boltzmann exponential $e^{-E/T}$ factor.
 
In order to arrive at measured rapidity spectra in the fireball model, an integration of the Boltzmann spectrum, Eq.~(12) over $m_\bot$ is required. With $p_\parallel = m_\bot \sinh (y-y_{\rm f})$, $E=m_\bot \cosh (y-y_{\rm f})$ we have:
\begin{equation}
\tag{13} \frac{dN}{dy} ~=~ C
\int_{m_\bot^{\rm min}}^{\infty}m_\bot\;dm_\bot\;\; m_\bot
\cosh(y-y_{\rm f})~{\rm e}^{-m_\bot\cosh(y-y_{\rm f})/T_{\rm f}}
 \;,
\end{equation}
with the normalization constant $C$ being dependent on the volume and
other intrinsic properties of the fireball.
These spectra are shown in Fig.\,4. 

\centerline{\includegraphics[width=0.75\columnwidth]{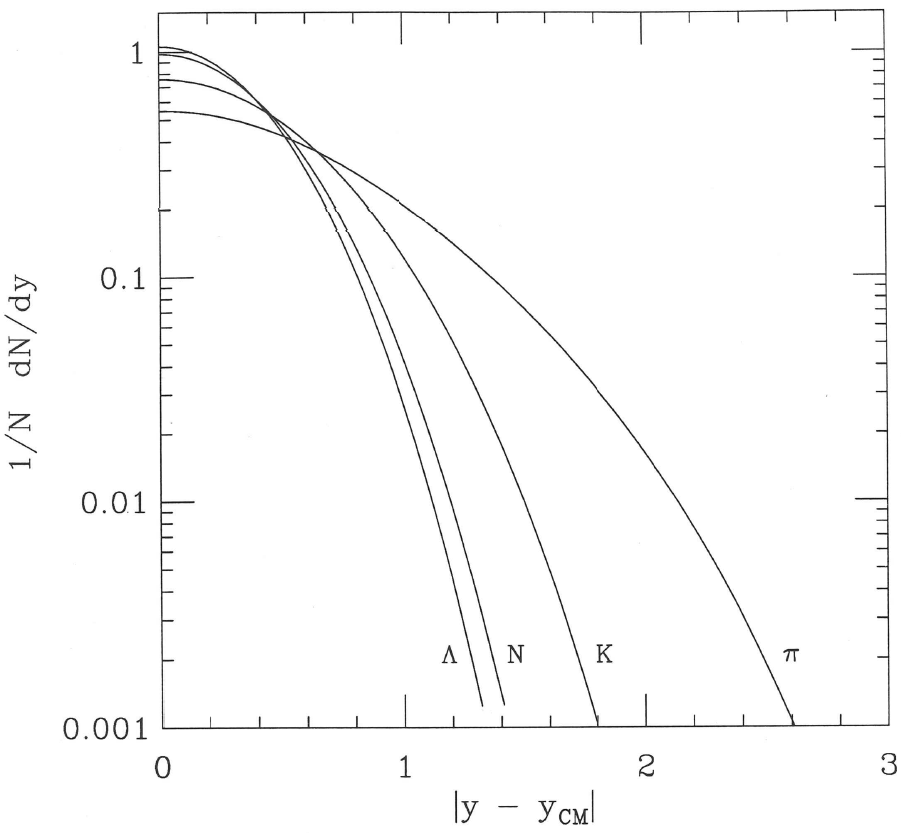}}
\noindent{\small Fig.\;4 The rapidity spectra of particles according to Eq.\,(13) 
}

For the rapidity of the
central fireball $y_{\rm f}$, using simple relativistic kinematics I
obtain:
\begin{equation}
\tag{14} y_{\rm f} ~=~ {\textstyle 1\over2} ~y_{\rm P}~-~{\textstyle
1\over2} ~\ln(A_t/A_{\rm P})
\end{equation}
where $y_{\rm P}=5.99$ is the sulphur projectile $(A_{\rm P}=32)$
rapidity at the CERN sulphur beam with energy 200~GeV\,A, and $A_t$ is
the number of participating target nucleons. In a simple geometric model
one finds that the number of participants from the target nucleus is
$A_t=1.32~A_{\rm T}^{1/3} ~A_{\rm P}^{2/3}$ for the asymmetric case such
as the S-Pb collision. This would imply that the central fireball is
shifted from the symmetric position at $y_{\rm f}=3$ to $y_{\rm f}=2.54$. 
Clearly, the assumption of complete stopping can not be made with
certainty, nor can we assume that all matter in the path of the
projectile participates fully in the inelastic reaction forming the
fireball; such effects tend to make the collision system more symmetric,
resulting in a fireball rapidity being closer to the symmetric value
$y_{\rm f}^{\rm s}=3$. 
 
We can easily obtain explicitly the shape of Eq.\,(13) for
$m_\bot=m_0$ or any lower cut off suitable to the experimental conditions
of an experiment - the limit $m_0<<T$ can be done analytically. I wish to
note here the analysis$^{13}$ of the NA36 experiment$^{14}$ which
strongly supports the hypothesis of the central fireball: the rise of the
spectra seen$^{14}$ near to the expected central rapidity region, as well
as indications that the width of the peak in rapidity is in general
agreement with the form Eq.\,(13), which for the heavy $\Lambda$
is of the magnitude 1. Similarly, the $m_\bot$ spectral distribution is
in substantial agreement with the thermal model if a temperature of the
magnitude 205 Mev is used. This last observation is true not only for the
NA36 spectra, it results from a study$^{13}$ of (strange) particle
transverse mass spectra reported for S -- A collisions, with A$\sim 200$
and which suggests $T=215\pm10$ MeV.
 
I wish to stress that relatively narrow strange particle rapidity
distributions are not in contradiction to the relatively wide pion
rapidity spectra: while pions can be produced in many processes and are
easily rescattered on heavier particles, shifting their rapidity, the
strange particles, in particular strange anti-baryons are predominantly
produced in the central region only and are destroyed in rescattering.
Another important issue is the longitudinal flow stemming either from
remembrance of the entrance momentum, or from the expansion of the
compressed matter which than contributes in hadronization and strongly
widens the already relatively wide pion distribution. I thus conclude
that the observation of the central rapidity production of strange
particles and in particular of $\overline{\Lambda}$ by the experiment
NA36 in S -- Pb collisions at 200 GeV\,A collisions strongly supports a
fireball as the source of these particles. Similar behavior was already
reported for the S -- S collisions at 200 GeV\,A by the NA35 
collaborations, but it is hard to ignore the relatively strong 
longitudinal flow visible in these relatively small size collisions. It
will be interesting to see the result of the Pb -- Pb collisions.
 
\section*{Counting (strange) particles in hot matter}
\addcontentsline{toc}{subsubsection}{Counting (strange) particles in hot matter}
As in an experiment strange particles consisting of a few constituents,
we have to understand how the abundance of composite particles is
governed by the thermal parameters of the fireball. I will now show that
statistical counting rules allow to describe the relative abundances of
strange baryons and antibaryons at fixed $m_\bot$.
 
\subsection*{Chemical potentials} 
The statistical variables of the fireball system are aside of the
temperature $T$ the chemical potentials $\mu_i$ of the different
conserved quark flavors $i=u,d,s$. Chemical potentials, as we shall see
in detail, are introduced to allow to set a prescribed abundance of
particles of the kind `$i$'. Akin to temperature which characterizes the
mean energy per particle, the chemical potential is generally related to
the energy expanded in the change of the number of particles. For
example, the cost in energy to replace the particle of kind $i$ by the
particle $j$ is $\mu_j-\mu_i$. The chemical potentials are related to the
so called fugacities $\lambda_i$ in the usual way:
\begin{equation}
\tag{15} \lambda_i=e^{\mu_i/T}
\end{equation} 
Thus the factors in Eq.\,(12) which control the formation of
composite particles in dense matter are: the Boltzmann exponential,
statistical multiplicity factors $g_i$, referring to the degeneracy of
the $i(=u,d,s)$ component, and characterizing also the likelihood of
finding among randomly assembled quarks, the suitable spin-isospin of the
particle; chemical fugacities see Eq.\,(15) which define the
relative abundance of quarks and anti-quarks with $(\lambda_{\bar q} =
\lambda_q^{-1})$. The factors $\gamma_i$ allow for the absence of
chemical equilibrium $(0 \le \gamma_i \le 1)$ for each quark flavor. The
difference between $\gamma$ and $\lambda$ is that $\gamma$ is the same
for {\it both} quarks and antiquarks of the same flavor:
$\gamma_q=\gamma_{\bar q}$. I will assume that, for light flavors, the
$\gamma$-factor is effectively unity, and will consider only the
possibility that strange quarks are not in absolute chemical equilibrium:
$0\le\gamma_{\rm s}\le1$, as is suggested by the dynamical models of
strange flavor production, discussed above in Eq.\,(9).
 
The fugacity of a composite particle is the product of the fugacities of
the components, that is of the fugacities for the `valance' quarks, since
the contributions of the see quark pairs cancel and eventual glue content
has no chemical attribute. Thus we will use as the chemical variables the
strange quark chemical potential $\mu_{\rm s}$ and the light quark
chemical potential $\mu_q$ with:
\begin{align*}
\tag{16}\mu_q&=(\mu_d+\mu_u)/2 \;,\quad\mu_{\rm B}=3\mu_q \;,\\
\tag{17}\delta\mu&=\mu_d-\mu_u \;;
\end{align*}
$\mu_{\rm B}$ is the baryo-chemical potential and $\delta\mu$ describes
the (small) asymmetry in the number of up and down quarks due to the
neutron excess in heavy ion collisions. The magnitude of $\delta\mu$
depends on the $u,d$ asymmetry, that is neutron - proton asymmetry in the
fireball. It has been obtained for both the QGP and HG models of the
fireball. In QGP a simple analytical relation between the ratio of $u$
and $d$ content arises from the perturbative expressions for the quark
density:
\begin{equation}
\tag{18} R_{\rm f}^{\rm QGP}\equiv {{\langle d \rangle- \langle{\bar d} \rangle}
\over{ \langle u\rangle - \langle{\bar u} \rangle}}
={{2-{Z_{\rm f}/A_{\rm f}}}\over {1+{Z_{\rm f}/ A_{\rm f}}}}
={{\mu_d/T\Big(1+\Big({\mu_d\over{\pi T}}\Big)^2\Big)}\over
{\mu_u/T\Big(1+\Big({\mu_u\over{\pi T}}\Big)^2\Big)}}
\simeq{\mu_d\over\mu_u}, 
\end{equation}
where the last equality arises because $(\mu_q/\pi T)^2<<1$. In the tube
model, in which all nucleons in the target in the path of the isospin
symmetric projectile participate in the fireball, $R_{\rm f}$ is 1.08 for
the Sulphur-Tungsten collision and 1.15 for Pb-Pb collisions. From
Eq.\,(18) arises the simple relationship:
\begin{equation}
\tag{19} {\delta\mu\over T} \approx {\mu_{\rm q}\over T}(R_{\rm f}-1)\ .
\end{equation}
I have not included above the superscript \lq QGP\rq\ since detailed
calculations show$^6$ that in a standard model of HG this results also
approximately holds in the domain of $T,\mu_{\rm B}$ associated with the
source of strange antibaryons. In view of the expected smallness of the
effect of the neutron - proton asymmetry I will mostly ignore $\delta\mu$
here, or employ the theoretical model, Eq.\,(19) to fix its
value.
 
Returning to the discussion of the strange quark chemical potential I
first note that despite the fact that $\mu_{\rm s}$ was introduced into
the nomenclature in the manner described here, a certain confusion is
possible with some recent work using instead the Kaon chemical potential
to characterize strangeness. Since the quark content of Kaons is $q\bar
s$ the chemical potential of Kaons denoted $\mu_{\rm S}=\mu_{\rm
q}-\mu_{\rm s}$. There is considerable advantage in the use of the
strange quark chemical potential, as one can directly compare the
properties of the QGP phase with the HG phase. For example since the
production of strange pairs is not influenced by presence of $u,d,\bar
u,\bar d$ quarks in the QGP phase, independent of the baryon number
content we always have (as long as $\langle s \rangle =\langle \bar s
\rangle$) $\mu_{\rm s}^{\rm QGP}=0$. On the other hand the HG, when
constrained to zero strangeness, implies in general a non vanishing value
of $\mu_{\rm s}$. In consequence, if we demand $\mu_{\rm s}^{\rm HG}=0$,
this establishes a constrain in the $\mu_{\rm q}$ -- $T$ plane, which
turns out to be a simple line akin in its form to the expected phase
transition boundary between HG and QGP. However, the values are very
different$^{15}$: at temperatures of the order 150 MeV the baryochemical
potential is $\mu_{\rm B}\sim900$ MeV, and $\mu_{\rm B}=0$ arises at
temperature $T\sim230$ MeV. The letter value is somewhat dependent on the
number of strange hadronic resonances included, which may still be
undiscovered, or their statistical factors (spin etc) which are either
unknown, or assumed with some degree of confidence in the structure
models of hadrons. A thorough discussion of the values of $\mu_{\rm s}$
is contained in Ref.\,[6] and we refrain here from entering into a more
detailed discussion which requires a rather thorough study of the
properties of HG -- the presented details will fully suffice to
understand the points addressed presently.
 
\subsection*{Measuring chemical potentials} 
\addcontentsline{toc}{subsubsection}{Measuring chemical potentials} 
All baryons considered have spin $1/2$, but they include spin $3/2$
resonances which become spin $1/2$ states through hadronic decays. This
is implicitly contained in the counting of the particles by taking the
product of the quark spin degeneracies; since in all ratios to be
considered this factor is the same, I shall ignore it, even though a
slight change results$^6$. As the method of measurement distinguishes the
flavor content, I keep explicit the product of $\lambda$-factors;
$\gamma_{\rm s}$ will enter when one compares particles with different
number of strange quarks and antiquarks.
 
When considering hyperons two different charge zero states of different
isospin must be counted: the experimental abundances of $\Lambda$ and
$\overline{\Lambda}$ (I=0) implicitly include, respectively, the
abundance of $\Sigma^0$ and ${\overline{\Sigma^0}}$ (I=1, I$_3$=0),
arising from the decay $\Sigma^0\rightarrow\Lambda^0+\gamma(74$ MeV),
and similarly for ${\overline{\Sigma^0}}$. Thus the true abundances must
be corrected by nearly a factor 2 (exactly 2 when the small difference in
mass $m_\Sigma - m_\Lambda=77$ MeV is neglected). 
 
The method of approach is very simple: I compare {\it spectra} of
particles within overlapping regions of $m_\bot$ and find that in
suitable ratios most statistical and spectral factors cancel allowing to
determine the conditions prevailing in the source. For example the
ratios:
\begin{align*}
\tag{20} R_\Xi&={\overline{\Xi^-}\over {\Xi^-}} =
{{\lambda_d^{-1}\lambda_{\rm s}^{-2}}\over 
{\lambda_d\lambda_{\rm s}^2}}\;,\\
\tag{21} R_\Lambda&={{{{\bar \Lambda}}\over{\Lambda}}} =
{{\lambda_d^{-1}\lambda_u^{-1}\lambda_{\rm s}^{-1}}\over 
{\lambda_d\lambda_u\lambda_{\rm s}}} .
\end{align*}
determine the quark fugacities. Indeed, the cascade and lambda ratios can
easily be related to each other, in a way which shows explicitly the
respective chemical asymmetry factors and strangeness fugacity
dependance. Eqs.(20,21) imply, in view of the definition
Eq.\,(15):
\begin{align*}
\tag{22} R_\Lambda&=R_\Xi^2 \cdot e^{2\delta\mu/T}e^{6\mu_{\rm s}/T}\;,\\
\tag{23} R_\Xi&=R_\Lambda^2 \cdot e^{-\delta\mu/T}e^{6\mu_q/T}\;.
\end{align*} 
Eqs.(22,23) are generally valid, irrespective of the state
of the system (HG or QGP). They fix the value of the chemical potentials,
subject to the tacit assumption that the particles considered are emitted
from the central fireball.
 
\subsection*{Experimental particle ratios} 
\addcontentsline{toc}{subsubsection}{Experimental particle ratios} 
In order to determine the values of the chemical potentials as enter
Eqs.\,(22,23) we recall that the $\overline{\Xi^-}/ \Xi^-$
ratio has been reported as$^9$:
\begin{equation}
\tag{24} R_\Xi:={\overline{\Xi^-}}/ \Xi^-|_{m_\bot} 
 = 0.39\pm 0.07\ \quad \mbox{ for }
 y\in(2.3,3.0) \mbox{ and } m_\bot>1.72\ \mbox{GeV}.
\end{equation}
Note that, in p--W reactions in the same $(p_\bot,y)$ region, a smaller
value for the $R_\Xi$ ratio, namely $0.27\pm 0.06$, is found. The ${\bar
\Lambda}/\Lambda $ ratio is:
\begin{equation}
\tag{25} R_\Lambda:={{\overline \Lambda}/\Lambda}|_{m_\bot} 
 = 0.13\pm 0.03\ \quad \mbox{ for } 
 y\in(2.3,3.0) \mbox{ and } m_\bot>1.72\ \mbox{GeV}.
\end{equation} 
In Eq.\,(25), corrections were applied to eliminate hyperons
originating from cascade decays, but not those originating from decays of
$\Omega \to \Lambda + \overline{K}$ or $\overline{\Omega} \to
\overline{\Lambda} + K$ which are of little signification for the high 
$m_\bot$ considered here. The ratio $R_\Lambda$ for S--W collisions is
slightly smaller than for p--W collisions in the same kinematic range.
 
From these two results, together with
Eqs.\,(22,23,19) and the value of $u$ -- $d$
asymmetry I obtain the following values of the chemical potentials for
S--W central collisions at 200 GeV A:
\begin{align*}
\tag{26} {\mu_{\rm q}\over T} & = {\ln R_\Xi/R_\Lambda^2 \over 5.92}
    = 0.53 \pm 0.1\;;\\ 
\tag{27} {\delta\mu\over T} & = {\mu_{\rm q}\over T}(R_{\rm f}-1)
    = 0.042 \pm 0.008\;;\\
\tag{28}{\mu_{\rm s}\over T} & = {\ln R_\Lambda/R_\Xi^2\ -0.084\over 6}
    = -0.018 \pm 0.05\;.
 \end{align*}
The last result translates into the value $\lambda_{\rm s} = 0.98 \pm
0.05$ for the strange quark fugacity. It turns out that many physical
properties of the fireball (such as e.g. entropy per baryon in the QGP
phase) depend only on the dimensionless values given above, and hence do
not depend on the determination of temperature. Under the assumption that
the transverse mass slope of the produced particles is entirely due to
the thermal motion leads to the temperature $T=210\pm10$ MeV, and
therefore:
\begin{equation} 
\tag{29} \mu_{\rm B} = 340 \pm 20 \mbox{ MeV}\ ,\quad 
 \delta\mu = 9 \pm 2 \mbox{ MeV}\ ,\quad 
 \mu_{\rm s} = -3.8 \pm 10 \mbox{ MeV}\ .
\end{equation} 
To considerable surprise we see that the strange particle ratios imply
that the strange chemical potential is very small and perfectly
compatible with zero. Another way to note this surprising result is to
observe the square of $R_\Xi$ is nearly equal to $R_\Lambda$. Is this
behavior characteristic for collisions involving the large nuclei at
these high energies, or is this a chancy coincidence? This is in this
field a big question which will have considerable impact on how we
understand the physical processes involving strangeness production in
relativistic heavy ion collisions.
 
\subsection*{Phase space saturation} 
\addcontentsline{toc}{subsubsection}{Phase space saturation} 
We turn our attention now to the determination of $\gamma_{\rm s}$. A
complete cancellation of the fugacity factors occurs when I consider the
product of the abundances of baryons and anti-baryons. Furthermore I can
take advantage of the cancellation of the Boltzmann factors by comparing
this product for two different particle kinds, e.g. consider:
\begin{equation}
\tag{30} \Gamma_{\rm s} \equiv \left. {\Xi^-\over\Lambda}
 \cdot {\overline{\Xi^-}\over \overline{\Lambda} 
 }\right\vert_{m_\bot>m_\bot^{\rm cut}} \, .
\end{equation} 
If the phase space of strangeness, like that of the light flavors, were
fully saturated, the fireball model would imply $\Gamma_{\rm s}=1$.
However, any deviation from absolute chemical equilibrium as expressed by
the factor $\gamma_{\rm s}$ will change the value of $\Gamma_{\rm s}$. 
\begin{equation}
\tag{31} \Gamma_{\rm s}=\gamma_{\rm s}^2\, .
\end{equation}
In principle, the measurement of $\gamma_{\rm s}$ can be done with other
particle ratios, in the absence of resonance feed-down we have
\begin{equation}
 \tag{32} \gamma_{\rm s}^2 = \left. 
 {\Lambda\over p} \cdot {\overline{\Lambda}\over \overline p} 
   \right\vert_{m_\bot>m_\bot^{\rm cut}}
   = \left. 
 {\Xi^-\over\Lambda} \cdot {\overline{\Xi^-} \over
\overline{\Lambda}}
   \right\vert_{m_\perp>m_\perp^{\rm cut}} 
   = \left. 
 {\Omega^-\over 2\Xi^-} \cdot {\overline{\Omega^-} \over 
      2\overline{\Xi^-}}
   \right\vert_{m_\perp>m_\perp^{\rm cut}} \, ,
\end{equation}
where in the last relation the factors 2 in the denominator correct 
for the spin-3/2 nature of the $\Omega$.
 
In the kinematic domain of Eqs.\,(24,25) the
experimental results reported by the WA85 collaboration are:
\begin{equation}
\tag{33} \frac{\overline{\Xi^-}}{\overline{\Lambda}+\overline{\Sigma^0}}
 = 0.6 \pm 0.2\, , \quad
 \frac{\Xi^-}{\Lambda+\Sigma^0} = 0.20 \pm 0.04\, .
\end{equation}
If the mass difference between $\Lambda$ and $\Sigma^0$ is neglected,
this implies in the framework of the thermal model that an equal number
of $\Lambda$'s and $\Sigma^0$'s are produced, such that
\begin{equation}
\tag{34} \frac{\overline{\Xi^-}}{\overline \Lambda}
 = 1.2 \pm 0.4\, , \hspace{1.15cm} 
 \frac{\Xi^-}{\Lambda} = 0.40 \pm 0.08\, .
\end{equation}
The fact that the more massive and stranger anti-cascade exceeds at fixed
$m_\bot$ the abundance of the anti-lambda is most striking. These
results are inexplicable in terms of cascade models for the heavy-ion
collision$^{16}$. The relative yield of $\overline{\Xi^-}$ appears to be
5 times greater than seen in the $p$--$p$ ISR experiment$^{8}$ and all
other values reported in the literature$^9$.
 
Combining the experimental result Eq.\,(34) with 
Eqs.\,(30,31), we find the value
\begin{equation}
\tag{35} \gamma_{\rm s}=0.7 \pm 0.1\, .
\end{equation} 
A more detailed discussion$^6$ including the resonance decays leaves
this result intact, and only if significant flow component is assumed
such that the fireball temperature drops to zero, there is an increase in
$\gamma_{\rm s}$ to a value near 0.9. However such a flow model is
somewhat inconsistent with the current understanding of the
hadronization, as the observed value $\mu_{\rm s}=0$ and the requirement
for strangeness conservation.

\section*{Entropy of the fireball} 
\addcontentsline{toc}{subsubsection}{Entropy of the fireball} 
The properties of the HG and QGP fireballs are considerably different in
particular with regard to the entropy content. Both states are easily
distinguishable in the regime of values $\mu_{\rm B},T$ of interest here.
the specific entropy per baryon in the hadronic gas is ${\cal S}^{\rm
HG}/{\cal B} =21.5\pm1.5$ for $T=215$ MeV and $\mu_{\rm B}=340$ MeV. 
This is less than half of the QGP based expectations ${\cal S}^{\rm
QGP}/{\cal B} =50\pm5$ and which are as shown somewhat dependent on the
value of the QCD parameters. Clearly the difference is considerable in
terms of experimental sensitivity, as it implies different final state
multiplicity. Note that charged particle multiplicity {\it above 600} in
the central region has been seen$^{17}$ in heavy ion collisions
corresponding possibly to a total particle multiplicity of about 1,000,
as required in the QGP scenario for the central fireball we described
above.
 
In order to study the relationship between the specific entropy and
particle multiplicity it is best to consider the quantity:
\begin{equation}
\tag{36} D_{\rm Q}= {{N^+-N^-}\over{N^++N^-}} \ ,
\end{equation}
since on the experimental side it is straightforward to measure it, while
on theoretically it is closely related to the yield of baryon number per
pion. Indeed, if only pion number $N_\pi$ and nucleon number $N$ is
considered:
\begin{equation}
\tag{37} D_{\rm Q}^{\pi,N} = 0.75 {N\over N_\pi} {1\over 1+0.75 N/N_\pi}
\end{equation}
where $N=2p$ is the total number of nucleons in the source, twice as
large as the proton number $p$, and $N_\pi=3N_{\pi^-}$ is the total
number of pions, which includes the three different charge components.

I note that in the central region of 200 GeV A S-Ag interactions$^{17}$
with the \lq central\rq\ trigger being the requirement for the total charged
multiplicity $> 300$ all up to date scanned (15) events yield $D_{\rm
Q}(\eta=2.5\pm0.5)=0.088\pm0.007$. It is rather simple to find in a
theoretical model that the specific entropy per baryon ${\cal S/B}\propto
D_{\rm Q}^{-1}$ for here interesting conditions, with the proportionality
constant being rather $T,\mu_{\rm B}$ independent. This result is rather
model independent as long as there is only limited production of entropy
in the hadronization. ${\cal S/B}\cdot D_{\rm Q}$ is essentially the
nearly constant entropy content per pion, with strange and heavy
resonance effects largely balancing out. The importance of this
observation is that for the observed value of $D_{\rm Q}=0.88\pm0.007$ we
find an entropy content${\cal S/B}=50\pm5$ as would be expected from the
QGP fireball. 

On the other hand in the conventional model of HG the relationship
between $D_{\rm Q}$ and $\mu_{\rm B}$ is found to be$^{6}$:
\begin{equation}
\tag{38}\hfil D_{\rm Q}= {\mu_{\rm B}\over\mbox{1.3 GeV}}\ \mbox{for }
\mu_{\rm B}<0.6 \mbox{ GeV}\ .
\end{equation}
where the temperature for each $\mu_{\rm B}$ is selected to assure
strangeness conservation. We see that the experimentally compatible value
$\mu_{\rm B}=340$ MeV implies a multiplicity ratio $D_{\rm Q}=0.26$,
which is incompatible with the data$^{17}$ of the experiment EMU05.
 
The source of the strange antibaryons is not a simple hadronic gas. The
source has entropy per baryon enhancement by the factor two expected from
the QGP equations of state.
 
\section*{Final Remarks}
I have shown that in studying the formation of rare strange particles,
one can obtain very precise and detailed information about the highly
excited nuclear matter formed in relativistic heavy ion collisions. Full
event characterization with considerable precision is needed to fix the
parameters of the system essential to a basic understanding of the state
of matter formed. Measurement of excitation functions for quantities such
as $\gamma_{\rm s}$ and possibly $\mu_{\rm s}$ would lead to a definitive
understanding of the high density source of these particles. Noteworthy
is the fact that the entropy content of the central interaction region
seems to favor a high entropy phase with properties as expected of a QGP
source of the observed antibaryons.
 
The observed enhancement of (relative) production rates of {\it
multi-}strange {\it anti-}baryons $\overline \Xi$ in nuclear collisions,
in particular at central rapidity and at highest transverse masses,
cannot be obtained so far in microscopic reaction models. After some
considerable effort to the contrary$^6$ I still cannot imagine how to
interpret these data other than in terms of a explosively evaporating
drop of quark-gluon plasma, in particular considering the substantial
hadronic multiplicity seen. Thus my tentative conclusion first put
forward a year ago$^3$ still holds: the source of the high $m_\bot$
centrally produced anti-cascades is the primordial and/or explosive QGP
state of matter with $T\simeq 215\pm10$ MeV and $\mu_{\rm B}\simeq
340\pm20$ MeV. 
\footnotetext{\vspace*{-0.5cm}\begin{itemize} 
\item[1]
J. Rafelski, Phys. Rep. C \textbf{88} 331 (1982) (not cited in this text fragment)
\item[2]
J.~Rafelski and M.~Danos, Phys. Lett. B \textbf{192} 432 (1987) (not cited in this text fragment)
\item[3]
J. Rafelski {\it Phys. Lett.} B262:333 (1991); Nucl. Phys. A \textbf{544} 279c (1992)
\item[4]
P.~Koch, B.~M\"uller and J.~Rafelski, Phys. Rep. C \textbf{142} 167 (1986) (not cited in this text fragment)
\item[5]
H.C.~Eggers and J.~Rafelski, Int. Journal of Mod. Phys. A \textbf{6} 1067 
(1991) (not cited in this text fragment)
\item[6]
J.~Letessier, A.~Tounsi, U.~Heinz, J.~Sollfrank and J.~Rafelski, {\it
Strangeness Conservation in Hot Fireballs} Preprint
Paris PAR/LPTHE/92-27, Regensburg TPR-92-28, Arizona 
AZPH-TH/92-23, 1992 (published: Phys. Rev. D \textbf{51} 3408 (1995))
\item[7]
J.~Rafelski and B.~M\"uller, Phys. Rev. Lett. \textbf{48} 1066 (1982); and \textbf{56} 2334(E) (1986) (not cited in this text fragment)
\item[8]
T.~\AA kesson et al. [ISR-Axial Field Spect. Collab.], Nucl. Phys. B \textbf{246} 1 (1984)
\item[9]
E. Quercigh, this volume; S. Abatzis {\it et~al}., Phys. Lett. 
B \textbf{270} 123 (1991)
\item[10]
J. Zimanyi, this volume~\cite{Gutbrod:1993rp}, pp.243-270 (not cited in this text fragment)
\item[11]
S.A.~Chin, Phys. Lett. B \textbf{78} 552 (1978) (not cited in this text fragment)
\item[12]
P. Koch, B. M\"uller and J. Rafelski, Z. Physik A \textbf{324} 3642 (1986) (not cited in this text fragment)
\item[13]
J. Rafelski, H. Rafelski and M. Danos, Phys. Lett. B: in press
(1992) (published: \textbf{294} 131 (1992))
\item[14]
R. Zybert and E. Judd, this volume, p.545-553; E. Andersen {\it et~al}. Phys. Lett B in press (1992) (published: \textbf{316} 603 (1993) 
\item[15]
J. Letessier, A. Tounsi and J. Rafelski, Phys. Lett. B: in press
(1992) (published: \textbf{292} 417 (1992))
\item[16]
L. Csernai, N.S. Amelin, E.F. Staubo and D. Strottman, Bergen University
Report 1991-14, table 4 and private communication
\item[17]
Y. Takahashi et al, CERN-EMU 05 collaboration, private communication (added for this republication: the results here quoted appear in\href{https://cds.cern.ch/record/295506/files/SC00000469.pdf}{CERN/SPSLC 93-18, Fig. 28, p.39} \lq\lq A Research Proposal submitted to the CERN SPSC For the Lead-Beam Experiments,Isospin Correlations in High Energy Pb+Pb Interactions,\rq\rq\ Submitted by the EMUO5 Collaboration, Y. Takahashi et al.)
\end{itemize}
}
\end{mdframed}
\vskip 0.5cm

\subsubsection{Proposal to STAR collaboration (continued)}
\noindent\textit{In the following we return to the final part of the report I prepared in 1998 to represent my potential input into the RHIC-STAR collaboration work. This segement addresses the particle ratio method for determining the SHM model parameters:} \\[-0.7cm]
\begin{mdframed}[linecolor=gray,roundcorner=12pt,backgroundcolor=Dandelion!15,linewidth=1pt,leftmargin=0cm,rightmargin=0cm,topline=true,bottomline=true,skipabove=12pt]\relax%
\label{SHM-STAR}
\textbf{Relative particle yields}\\
\addcontentsline{toc}{subsubsection}{Relative particle yields and SHM parameters}
Precise measurement of the multistrange particle
production is being used to determine the chemical properties of
the source$^{1,2,3}$, and we expect to be able to
also perform a similar analysis at RHIC. 

The relative number of particles of same type 
emitted at a given instance by a locally equilibrated, 
thermal hot source is obtained by
noting that the probability to find all the $j$-components
contained within the $i$-th emitted particle is
\begin{equation}\tag{1}\label{abund}
N_i\propto \gamma_{\rm s}^k\prod_{j\in i}\lambda_je^{-E_j/T}\,,
\end{equation}
and we note that the total energy and fugacity of the particle is:
\begin{equation}\tag{2}
E_i=\sum_{j\in i}E_j,\qquad \lambda_i=\prod_{j\in i}\lambda_j\,.
\end{equation}
The strangeness occupancy $\gamma_{\rm s}$ enters
Eq.\,(\ref{abund}) with power $k$, which equals the number of
strange and antistrange quarks in the hadron $i$. 
With $E_i=\sqrt{m_i^2+p^2}=\sqrt{m_i^2+p_\bot^2}\cosh y $ 
the transverse momentum range 
as constrained in the experiment (here $p_\bot>0.6 $ GeV)
and taking central rapidity region $y\simeq 0$, is integrated 
over to obtain the relative strengths of particles produced. All
hadronic resonances are allowed to disintegrate 
in order to obtain the final relative multiplicity of `stable'
particles required to form the observed particle ratios. 

As we can see, the relative abundance of particles emerging from
the thermal fireball is controlled the chemical (particle
abundance) 
parameters, the particle fugacities which allow to
conserve flavor quantum numbers. The fugacity of each hadronic 
particle species is the product of the valence quark fugacities, 
thus, for example, the hyperons have the fugacity 
$\lambda_{\rm Y}=\lambda_{\rm u}\lambda_{\rm d}\lambda_{\rm s}$.
Fugacities are related to the chemical potentials $\mu_i$ by:
\begin{equation}\tag{3}
\lambda_{i} =e^{\mu_{i}/T}\,,\quad 
\lambda_{\bar{\imath}}=\lambda_i^{-1}\qquad 
i={u,\,d,\,s}\, . \label{lam}
 \end{equation}
Therefore, the chemical potentials for particles and 
antiparticles are opposite to each other, provided that there is
complete chemical equilibrium, and if not, that the deviation from
the full phase space occupancy is accounted for by introducing a 
 non-equilibrium chemical parameter $\gamma$.
 
In many applications it is sufficient to combine the
light quarks into one fugacity 
\begin{equation}\tag{4}
\lambda_{\rm q}^2\equiv\lambda_{\rm d}\lambda_{\rm u}\,,\quad 
\mu_{\rm q}=({\mu_{\rm u}+\mu_{\rm d}})/2\,. \end{equation}

Since a wealth of experimental data can be described with just a
few model parameters, this leaves within this approach a
considerable predictive power and a strong check of the internal
consistency. In the directly hadronizing 
off-equilibrium QGP-fireball there are 5
particle multiplicity parameters characterizing all particle
yields. aside of the usual temperature $T$ and
$\lambda_q,\lambda_s$ (we expect $\lambda_{\rm
s}=1$ because of strangeness conservation in the QGP phase) it is
advisable to introduce two
particle abundance non-equilibrium parameters: the strangeness 
occupancy $\gamma_{s}$ and the ratio $R^{\rm s}_{\rm
C}$, of meson to baryon phase space abundance$^{2}$. The
last of these parameters is related to the mechanism 
governing the final state hadronization process. It does not appear
in any if we only consider baryon yields.

The ratios of strange antibaryons to strange baryons {\it
of same particle type\/}: 
\begin{equation}\tag{5}R_\Lambda=\overline{\Lambda}/\Lambda\,,\quad 
R_\Xi=\overline{\Xi}/\Xi\quad \mbox{and}\quad
R_\Omega=\overline{\Omega}/\Omega\,,\end{equation}
are in our approach simple functions of the quark 
fugacities. For example one has specifically
 \begin{align}\tag{6}
 R_\Xi &= {{\overline{\Xi^-}}\over {\Xi^-}} =
 {{\lambda_{\rm d}^{-1} \lambda_{\rm s}^{-2}} \over
 {\lambda_{\rm d} \lambda_{\rm s}^2}} \, ,
\qquad \label{ratioL}\\
\tag{7} R_\Lambda &= {\overline{\Lambda}\over \Lambda} =
{{\lambda_{\rm d}^{-1} \lambda_{\rm u}^{-1}
    \lambda_{\rm s}^{-1}} \over
 {\lambda_{\rm d} \lambda_{\rm u} \lambda_{\rm s}}} \, . 
\label{ratio}
 \end{align}

Only the ratios between antibaryons with
different strange quark content are dependent on
the strangeness saturation factor $\gamma_s$. 
At fixed $m_\bot$ and up to cascading corrections 
a complete cancelation of the fugacity and Boltzmann 
factors occurs when we form the product of the abundances 
of baryons and antibaryons, comparing this product 
for two different particle kinds$^{1}$, { e.g.}:
 \begin{equation}\tag{8}
 \left. {\Xi^-\over\Lambda}
 \cdot {\overline{\Xi^-}\over \overline{\Lambda}
 }\right\vert_{m_\perp>m_\perp^{\rm cut}} =\gamma_{\rm s}^2 \,, 
 \label{gam1}
 \end{equation}
where we neglected resonance feed-down contribution in first 
approximation, which are of course considered in numerical studies$^{2}$. Similarly we have 
\begin{equation}\tag{9}
 \gamma_{\rm s}^2 = \left.
 {\Lambda\over p} \cdot {\overline{\Lambda}\over \overline p} 
 \right\vert_{m_\perp>m_\perp^{\rm cut}}
   = \left.
 {\Omega^-\over 2\Xi^-} \cdot {\overline{\Omega^-} \over 
     2\overline{\Xi^-}}  
  \right\vert_{m_\perp>m_\perp^{\rm cut}} \, ,
\label{gam3}
 \end{equation}
where in the last relation the factors 2 in the denominator correct
for the spin-3/2 nature of the $\Omega$.

The evaluation of the resonance 
decay effect is actually not simple, since resonances at
different momenta and rapidities contribute to a given daughter 
particle $m_\bot$. The measurements sum 
the $m_\bot$ distributions with $m_\perp \geq m_\perp^{\rm cut}$
and it is convenient to consider this integrated abundance for
particle `i' at a given (central) rapidity $y$:
 \begin{equation}\tag{10}
 \left. {dN_i \over dy} \right\vert_{m_\perp \geq
     m_\perp^{\rm cut}}
 = \int_{m^{\rm cut}_\perp}^\infty dm_\perp^2
 \left\{ {dN_i^{0}(T) \over dy\, dm_\perp^2} +
 \sum_R b_{R\to i} {dN_i^R(T) \over dy\, dm_\perp^2} \right\}\, ,
\label{reso}
 \end{equation}
showing the direct `0' contribution and the daughter
contribution from decays into the observed channel $i$) 
of resonances $R\to i$\,, with branching ratio $b_{R\to i}$\,, see
Ref.[2]. Extracting the degeneracy factors
and fugacities of the decaying resonances, we write shortly 
\begin{equation}\tag{11}
 N^R_i \equiv \gamma_R \lambda_R \tilde N^R_i\,,
 \label{ntilde}
 \end{equation}
and imply that particles of same quantum numbers are comprised in
each $N^R_i$. Here $\gamma_R $ is the complete non-equilibrium
factor of hadron (family) $R$. Between particles and anti-
particles 
we have the relation
 \begin{equation}\tag{12}
N_{\bar i}^{\bar R} = \gamma_R\, \lambda_R^{-1}\, {\tilde N}_i^R 
  = \lambda_R^{-2}\, N_i^R \, .
 \label{ntildea}
 \end{equation}
Thus the above considered particle ratios now become:
 \begin{align}\tag{13}
 R_\Xi &= 
\left. {\overline{\Xi^-} \over \Xi^-}
\right\vert_{m_\perp\geq m_\perp^{\rm cut}}
  &= {\gamma_{\rm s}^2 \lambda_{\rm q}^{-1} 
\lambda_{\rm s}^{-2} 
 {\tilde N}_\Xi^{\Xi^*} +
  \gamma_{\rm s}^3 \lambda_{\rm s}^{-3}
  {\tilde N}_\Xi^{\Omega^*} \over
  \gamma_{\rm s}^2 \lambda_{\rm q} \lambda_{\rm s}^2 
 {\tilde N}_\Xi^{\Xi^*} +
  \gamma_{\rm s}^3 \lambda_{\rm s}^3
  {\tilde N}_\Xi^{\Omega^*} }\ ,
 \label{rxi}\\
\tag{14} R_\Lambda &= 
\left. \hphantom{^-}
  {\overline{\Lambda} \over \Lambda}
  \right\vert_{m_\perp\geq m_\perp^{\rm cut}}
&=
{  \lambda_{\rm q}^{-3}{\tilde N}_\Lambda^{N^*}+ 
\gamma_{\rm s} \lambda_{\rm q}^{-2} \lambda_{\rm s}^{-1}  
 {\tilde N}_\Lambda^{Y^*} +
 \gamma_{\rm s}^2 \lambda_{\rm q}^{-1} \lambda_{\rm s}^{-2} 
 {\tilde N}_\Lambda^{\Xi^*}
\over
  \lambda_{\rm q}^3 {\tilde N}_\Lambda^{N^*} +
  \gamma_{\rm s} \lambda_{\rm q}^2 \lambda_{\rm s} 
 {\tilde N}_\Lambda^{Y^*} +
  \gamma_{\rm s}^2 \lambda_{\rm q} \lambda_{\rm s}^2 
 {\tilde N}_\Lambda^{\Xi^*}
}\, , \label{rlam}\\
\tag{15} R_{\rm s} &= 
\left. {\Xi^- \over \Lambda}
  \right\vert_{m_\perp\geq m_\perp^{\rm cut}}
&=
{  \gamma_{\rm s}^2 \lambda_{\rm q} \lambda_{\rm s}^2 
   {\tilde N}_\Xi^{\Xi^*} +
  \gamma_{\rm s}^3 \lambda_{\rm s}^3
  {\tilde N}_\Xi^{\Omega^*}
\over
  \lambda_{\rm q}^3{\tilde N}_\Lambda^{N^*}+
  \gamma_{\rm s} \lambda_{\rm q}^2 \lambda_{\rm s} 
 {\tilde N}_\Lambda^{Y^*} +
  \gamma_{\rm s}^2 \lambda_{\rm q} \lambda_{\rm s}^2 
 {\tilde N}_\Lambda^{\Xi^*}
}\, .\label{rs}
 \end{align}
$\tilde N_\Lambda^{Y^*}$ contains also (in fact as its most
important contribution) the electromagnetic decay $\Sigma^0 \to
\Lambda + \gamma$.

This approach allows to compute the relative
strengths of strange (anti)baryons both in case of 
surface emission and equilibrium disintegration of a particle gas
since the phase space occupancies are in both cases properly
accounted for by Eq.\,(\ref{abund}). The transverse flow phenomena
enter in a similar fashion into particles of comparable mass and
are not expected to
influence particle ratios. Therefore detailed information about the
chemical and thermal composition of the particle source is derived,
provided that precise input particle abundances are used in the 
analysis. Presence of longitudinal flow in the dense matter from
which observed particles are emitted has no impact on the relative
particle ratios considered here, but it will need to be considered
for full evaluation of the dynamics of hadronic particle
production.

\footnotetext{\vspace*{-0.5cm}
\begin{enumerate}
\item
J. Rafelski, {\it Phys. Lett.} {\bf B 262}, 333 (1991);
{\it Nucl. Phys.} {\bf A544}, 279c (1992).
 
\item
J. Letessier, A. Tounsi, U. Heinz, J. Sollfrank and J. Rafelski,
{\it Phys.\ Rev.} {\bf D51}, 3408 (1995);\\
J. Letessier, J. Rafelski and A. Tounsi, {\it Phys. Lett.} {\bf
B321}, 394 (1994); {\bf B323}, 393 (1994); {\bf B333}, 484 (1994);
{\bf B390}, 363 (1997); {\bf B 410}, (1997) 315. 
 
\item
J. Rafelski, J. Letessier and A. Tounsi,
{\it Acta Phys. Pol.} {\bf B27}, 1035 (1996).

\end{enumerate}
}
\end{mdframed}
\vskip 0.5cm

\subsection{Fireball of QGP in Pb-Pb collisions at CERN-SPS}\label{PRL2000}

\subsubsection{Fit to data and bulk fireball properties}

In the Winter 1998/9 I completed in collaboration with Jean Letessier an analysis of CERN-SPS Pb--Pb 158$A$\;GeV particle production data. On 8 March 1999 the PRL editorial office acknowledged the submission of our manuscript LC7284. The manuscript in v1, v2 and final v3 format is today available on arXiv~\cite{Rafelski:1999xv} as manuscript nucl-th/9903018. To best of my knowledge there was/is nothing wrong with this unpublished work in every version. 

Below, after this work is presented in its v3 format, I will show the pertinent correspondence with PRL. The reader should remember that had our paper been published  in the Summer 1999, this would have been a strong support of the CERN QGP announcement. Thus rejection of the publication of our work was of essence for those at CERN who were in opposition to QGP CERN announcement, see page \pageref{Heinz2000}. We recall that one of the coauthors of the  QGP discovery at CERN document was just that person, and that this document contains quite flawed theoretical phrases, compare Sec.\ref{Berk2000}. At the time in 1999/2000 this individual was also the divisional associate editor of Physical Review Letters (PRL): he very likely he was consulted by editors with regard to choices of referees, and he was the judge who terminated the publication process.\\
 
\noindent \textit{The manuscript LC7284 was received 8 March 1999 by editors of PRL. The process terminated on 13 January 2000, 4 weeks before CERN announces QGP discovery. The following shows the unpublished PRL manuscript~\cite{Rafelski:1999xv} {\bf LC7284} in final version v3, the other two versions can be found on arXiv:}\\[-0.7cm]
\begin{mdframed}[linecolor=gray,roundcorner=12pt,backgroundcolor=Dandelion!15,linewidth=1pt,leftmargin=0cm,rightmargin=0cm,topline=true,bottomline=true,skipabove=12pt]\relax%
\begin{center}
{\large {\bf On hadron production in\\[0.2cm] Pb-Pb collisions at 158$A$\;GeV\\[0.2cm]}}
\end{center}

{\bf Abstract:} \textit{A Fermi statistical model analysis of hadron abundances and spectra obtained in several relativistic heavy ion collision experiments is utilized to characterize a particle source. Properties consistent with a disintegrating, hadron evaporating, deconfined quark-gluon plasma phase fireball are obtained, with a baryochemical potential $\mu_{B}=200$--210\,MeV, and a temperature $T_f\simeq 140$--150\,MeV, significantly below previous expectations.}

Discovery and study of quark-gluon plasma (QGP), a state consisting of mobile, color charged quarks and gluons, is the objective of the relativistic heavy ion research program$^1$ underway at Brookhaven National Laboratory, New York and at CERN, Geneva. Thermalization of the constituents of the deconfined phase created in high energy large nuclei collisions is a well working hypothesis, as we shall see. The multi-particle production processes in 158$A$ GeV Pb--Pb collisions carried out at CERN-SPS will be analyzed in this paper, using the principles of the statistical Fermi model$^2$: strongly interacting particles are produced with a probability commensurate with the size of accessible phase space. Since the last comprehensive review of such analysis has appeared$^3$, the Pb-beam experimental results became available, and model improvements have occurred: we implement refinements in the phase space weights that allow a full characterization of the chemical non-equilibria with respect to strange and light quark flavor abundances$^{4,5}$. Consideration of the light quark chemical non-equilibrium is necessary in order to arrive at a consistent interpretation of the experimental results of both the wide acceptance NA49-experiment$^{6,7,8,9,10}$ and central rapidity (multi)strange (anti)baryon WA97-experiment$^{11,12,13}$.

We further consider here the influence of collective matter flow on $m_\bot$-particle spectra and particle multiplicities obtained in a limited phase space domain. The different flow schemes have been described before$^{14}$. We adopt a radial expansion model and consider the causally disconnected domains of the dense matter fireball to be synchronized by the instant of collision. We subsume that the particle (chemical) freeze-out occurs at the surface of the fireball, simultaneously in the CM frame, but not necessarily within a short instant of CM-time. Properties of the dense fireball as determined in this approach offer clear evidence that a QGP disintegrates at $T_f\simeq$\,144\,MeV, corresponding to energy density$^{15}$ $\varepsilon=\cal O$(0.5) GeV/fm$^3$. Our initial chemical non-equilibrium results without flow have been suggestive that this is the case$^{16}$, showing a reduction of the chemical freeze-out temperature from $T_f=180$\,MeV$^{17}$; an earlier analysis could not exclude yet higher hadron formation temperature of 270\,MeV$^{18}$. 

The here developed model offers a natural understanding of the systematic behavior of the $m_\bot$-slopes which differs from other interpretations. The near equality of (inverse) slopes of nearly all strange baryons and antibaryons arises here by means of the sudden hadronization at the surface of an exploding QGP fireball. In the hadron based microscopic simulations this behavior of $m_\bot$-slopes can also arise allowing for particle-dependent freeze-out times$^{19}$.

In the analysis of hadron spectra we employ methods developed in analysis of the lighter 200$A$ GeV S--Au/W/Pb system$^{5}$, where the description of the phase space accessible to a hadronic particle in terms of the parameters we employ is given. Even though we use six parameters to characterize the hadron phase space at chemical freeze-out, there are only two truly unknown properties: the chemical freeze-out temperature $T_{f}$ and light quark fugacity $\lambda_q$\, (or equivalently, the baryochemical potential $\mu_\mathrm{B}=3\,T_{f}\ln \lambda_q$) -- we recall that the parameters $\gamma_i,\,i=q,s$ controls overall abundance of quark pairs, while $\lambda_i$ controls the difference between quarks and anti-quarks of given flavor. The four other parameters are not arbitrary, and we could have used their tacit and/or computed values:\\ 1) the strange quark fugacity $\lambda_s$ is usually fixed by the requirement that strangeness balances$^{4}$ $\langle s-{\bar s}\rangle=0$. The Coulomb distortion of the strange quark phase space plays an important role in the understanding of this constraint for Pb--Pb collisions$^{16}$, see Eq.\,(\ref{lamQ});\\ 2) strange quark phase space occupancy $\gamma_s$ can be computed within the established kinetic theory framework for strangeness production$^{20,21}$;\\ 3) the tacitly assumed equilibrium phase space occupancy of light quarks $\gamma_q=1$\,; and \\ 4) assumed collective expansion to proceed at the relativistic sound velocity$^{21}$, $v_c=1/\sqrt{3}$.\\ However, the rich particle data basis allows us to find from experiment the actual values of these four parameters, allowing to confront the theoretical results and/or hypothesis with experiment. 

The value of $\lambda_s$ we obtain from the strangeness conservation condition $\langle s-{\bar s}\rangle=0$\ in QGP is, to a very good approximation$^{16}$:
\begin{equation}\label{lamQ}
\tag{1}\tilde\lambda_s\equiv \lambda_s \lambda_{\rm Q}^{1/3}=1\,,\qquad
\lambda_{\rm Q}\equiv
\frac{\int_{R_{\rm f}} d^3r e^{\frac V{T}} } {\int_{R_{\rm f}} d^3r}\,.
\end{equation}
 $\lambda_{\rm Q}<1$ expresses the Coulomb deformation of strange quark phase space. This effect is relevant in central Pb--Pb interactions, but not in S--Au/W/Pb reactions. $\lambda_{\rm Q}$ is not a fugacity that can be adjusted to satisfy a chemical condition, since consideration of $\lambda_i,\ i=u,d,s$ exhausts all available chemical balance conditions for the abundances of hadronic particles. The subscript ${R_{f}}$ in Eq.\,(\ref{lamQ}) reminds us that the classically allowed region within the dense matter fireball is included in the integration over the level density. Choosing $R_{\rm f}=8$\,fm, $T=140$\,MeV, $m_s=200$\,MeV (value of $\gamma_s$ is practically irrelevant), for $Z_{\rm f}=150$ the value is $\lambda_s=1.10$\,.

In order to interpret particle abundances measured in a restricted phase space domain, we study abundance ratios involving what we call compatible hadrons: these are particles likely to be impacted in a similar fashion by the not well understood collective flow dynamics in the fireball. The available particle yields are listed in table~1, top section from the experiment WA97 for $p_\bot>0.7$ GeV within a narrow $\Delta y=0.5$ central rapidity window. Further below are shown results from the large acceptance experiment NA49, extrapolated to full $4\pi$ phase space coverage. There are 15 experimental results. The total error $\chi^2_{\rm T}\equiv\sum_j({R_{\rm th}^j-R_{\rm exp}^j})^2/ ({{\Delta R _{\rm exp}^j}})^2$ for the four theoretical columns is shown at the bottom of table~1 along with the number of data points `$N$', parameters `$p$' used and (algebraic) redundancies `$r$' connecting the experimental results. For $r\ne 0$ it is more appropriate to quote the total $\chi^2_{\rm T}$, with a initial qualitative statistical relevance condition $\chi^2_{\rm T}/(N-p)<1$. The four theoretical columns refer to results with collective velocity $v_c$ (subscript $v$) or without ($v_c=0$). We consider data including \lq All\rq\ data points, and also analyze data excluding from analysis four $\Omega,\,\overline\Omega$ particle ratios, see columns marked \lq No-$\Omega$\rq. Only in letter case we obtain a highly relevant data description. Thus to describe the $\Omega,\,\overline{\Omega}$ yields we need an additional particle production mechanism beyond the statistical Fermi model. We noted the special role of these particles, despite bad statistics, already in the analysis of the S-induced reactions$^5$.\\

\noindent{\small Table 1: WA97 (top) and NA49 (bottom) Pb--Pb 158$A$ GeV particle ratios and our theoretical results, see text for explanation.}
\begin{center}
\begin{tabular}{|lcl|ll|ll|}
\hline
 Ratios & $\!\!\!\!$Ref. & Exp.Data & All & All$|_v$& No-$\Omega$ & No-$\Omega|_v$ \\
\hline
${\Xi}/{\Lambda}$ & [12] &0.099 $\pm$ 0.008 & 0.107 & 0.110 & 0.095 & 0.102 \\
${\overline{\Xi}}/{\bar\Lambda}$ & [12] &0.203 $\pm$ 0.024 & 0.216 & 0.195 & 0.206 & 0.210 \\
${\bar\Lambda}/{\Lambda}$ & [12] &0.124 $\pm$ 0.013 & 0.121 & 0.128 & 0.120 & 0.123 \\
${\overline{\Xi}}/{\Xi}$ & [12] &0.255 $\pm$ 0.025 & 0.246 & 0.225 & 0.260 & 0.252 \\
${\Omega}/{\Xi}$ & [12] &0.192 $\pm$ 0.024 & 0.192 & 0.190 &0.078$^*$&0.077$^*$\\
${\overline{\Omega}}/{\overline{\Xi}}$ & [11] &0.27 $\pm$ 0.06 & 0.40 & 0.40 &0.17$^*$ &0.18$^*$ \\
${\overline{\Omega}}/{\Omega}$ & [12] &0.38 $\pm$ 0.10 & 0.51 & 0.47 &0.57$^*$ &0.60$^*$ \\
$(\Omega+\overline{\Omega})\over(\Xi+\bar{\Xi})$& [11] &0.20 $\pm$ 0.03
  & 0.23 & 0.23 &0.10$^*$ &0.10$^*$ \\
\hline
$(\Xi+\bar{\Xi})\over(\Lambda+\bar{\Lambda})$& [6] &0.13 $\pm$ 0.03
  & 0.109 & 0.111 & 0.107 & 0.114 \\
${K^0_{\rm s}}/\phi$ & [7] & 11.9 $\pm$ 1.5\ \ & 16.1 & 15.1 & 9.89 & 12.9 \\
${K^+}/{K^-}$ & [8] & 1.80$\pm$ 0.10 & 1.62 & 1.56 & 1.76 & 1.87 \\
$p/{\bar p}$ & [6] &18.1 $\pm$4.\ \ \ \ & 16.7 & 15.3 & 17.3 & 17.4 \\
${\bar\Lambda}/{\bar p}$ &[24] & 3. $\pm$ 1. & 0.65 & 1.29 & 2.68 & 2.02 \\
${K^0_{\rm s}}$/B & [23] & 0.183 $\pm$ 0.027 & 0.242 & 0.281 & 0.194 & 0.201 \\
${h^-}$/B & [23] & 1.83 $\pm $ 0.2\ \ & 1.27 & 1.55 & 1.80 & 1.83 \\
\hline
 & & $\chi^2_{\rm T}$ & 19 & 18 & 2.1 & 1.8\\
 & & $ N;p;r$ &15;5;4 & 16;6;4 & 11;5;2 & 12;6;2\\
\hline
\end{tabular}
\end{center}

Considering results obtained with and without flow reveals that the presence of the parameter $v_c$ already when dealing only with particle abundances improves our ability to describe the data. However, $m_\bot$ spectra offer another independent measure of the collective flow $v_c$: for a given pair of values $T_{f}$ and $v_{\rm c}$ we evaluate the resulting $m_\bot$ particle spectrum and analyze it using the spectral shape and kinematic cuts employed by the experimental groups. To find the best values we consider just one `mean' strange baryon experimental value ${\bar T}_{\bot}^{\rm Pb}=260\pm10$, since within the error the high $m_\bot$ strange (anti)baryon inverse slopes are overlapping. Thus when considering $v_c$ along with ${\bar T}_{\bot}$ we have one parameter and one data point more. Once we find best values of $T_{\rm f}$ and $v_{\rm c}$, we study again the inverse slopes of individual particle spectra. We obtain an acceptable agreement with the experimental $T_{\bot}^j$ as shown in left section of table~2\,. For comparison, we have also considered in the same framework the S-induced reactions, and the right section of table~2 show a good agreement with the WA85 experimental data$^{25}$. We used here as the `mean' experimental slope data point ${\bar T}_{\bot}^{\rm S}=235\pm10$. We can see that within a significantly smaller error bar, we obtained an accurate description of the $m_\bot^{\rm S}$-slope data. This analysis implies that the kinetic freeze-out, where elastic particle-particle collisions cease, cannot be occurring at a condition very different from the chemical freeze-out. However, one pion HBT analysis at $p_\bot<0.5$ GeV suggests kinetic pion freeze-out at about$^{26}$ $T_k\simeq120$ MeV. A possible explanation of why here considered $p_\bot>0.7$ GeV particles are not subject to a greater spectral deformation after chemical freeze-out, is that they escape before the bulk of softer hadronic particles is formed.\\

\noindent{\small Table 2: Experimental and theoretical $m_\bot$ spectra inverse slopes $T_{\rm th}$. Left Pb--Pb results from experiment$^{10}$ NA49 for kaons and from experiment$^{13}$ WA97 for baryons; right S--W results from$^{25}$ WA85}
\begin{center}
\begin{tabular}{|l|cc|cc|}
\hline
 & $T_{\bot}^{\rm Pb}$\,[MeV]&$T_{\rm th}^{\rm Pb}$\,[MeV]&$T_{\bot}^{\rm S}$\ [MeV]&
  $T_{\rm th}^{\rm S}$\ [MeV]\\
\hline
$T^{{\rm K}^0}$ & 223 $\pm$ 13& 241& 219 $\pm$ \phantom{1}5 & 215\\
$T^\Lambda$ & 291 $\pm$ 18& 280& 233 $\pm$ \phantom{1}3 & 236\\
$T^{\overline\Lambda}$ & 280 $\pm$ 20& 280& 232 $\pm$ \phantom{1}7 & 236\\
$T^\Xi$ & 289 $\pm$ 12& 298& 244 $\pm$ 12& 246\\
$T^{\overline\Xi}$ & 269 $\pm$ 22& 298& 238 $\pm$ 16& 246\\
\hline
\end{tabular}
\end{center}

The six statistical parameters describing the particle abundances are shown in the top section of table~3, for both Pb- and S-induced reactions$^{5}$. The errors shown are one standard deviation errors arising from the propagation of the experimental measurement error, but apply only when the theoretical model describes the data well, as is the case here, see the header of each column --- note that for the S-induced reactions (see Ref.[5]) the number of redundancies is large since same data comprising different kinematic cuts is included in the analysis. We note the interesting result that within error the freeze-out temperature $T_{\rm f}$ seen in table~3, is the same for both the S- and Pb-induced reactions. The collective velocity rises from $v_c^{\rm S}=0.5c$ to $v_c^{\rm Pb}\simeq c/\sqrt{3}=0.58$. We then show the light quark fugacity $\lambda_{q}$, and note $\mu_\mathrm{B}^{\rm Pb}=203\pm5 >\mu_\mathrm{B}^{\rm S}=178\pm5$\,MeV. As in S-induced reactions where $\lambda_{s}=1$, now in Pb-induced reactions, a value $\lambda_{s}^{\rm Pb}\simeq 1.1$ characteristic for a source of freely movable strange quarks with balancing strangeness, {\it i.e.,} $\tilde\lambda_{s}=1$ is obtained, see Eq.\,(\ref{lamQ}).\\

\noindent{\small Table 3: Top section: statistical parameters, and their $\chi^2_{\rm T}$, which best describe the experimental results for Pb--Pb data, and in last column for S--Au/W/Pb data presented in Ref.[5]. Bottom section: specific energy, entropy, anti-strangeness, net strangeness of the full hadron phase space characterized by these statistical parameters. In the middle column we fix $\lambda_s$ by requirement of strangeness conservation and choose $\gamma_q=\gamma_q^c$, the pion condensation point.}
\vspace{-0.2cm}\begin{center}
\begin{tabular}{|l|cc|c|}
\hline
 & Pb--No-$\Omega|_v$& Pb--No-$\Omega|_v^*$ & S--No-$\Omega|_v$ \\
$\chi^2_{\rm T};\ N;p;r$&1.8;\ 12;\,6;\,2 & 4.2;\ 12;\,4;\,2 & 6.2;\ 16;\,6;\,6 \\
\hline
$T_{f}$ [MeV] & 144 $\pm$ 2 & 145 $\pm$ 2 & 144 $\pm$ 2 \\
$v_c$ & 0.58 $\pm$ 0.04 & 0.54 $\pm$ 0.025 & 0.49 $\pm$ 0.02\\
$\lambda_{q}$ & 1.60 $\pm$ 0.02 & 1.605 $\pm$ 0.025 & 1.51 $\pm$ 0.02 \\
$\lambda_{s}$ & 1.10 $\pm$ 0.02 & 1.11$^*$ & 1.00 $\pm$ 0.02 \\
$\gamma_{q}$ & 1.7 $\pm$ 0.5 & $\gamma_q^c=e^{m_\pi/2T_f}$ & 1.41 $\pm$ 0.08 \\
$\gamma_{s}/\gamma_{q}$& 0.86 $\pm$ 0.05 & 0.78 $\pm$ 0.05 & 0.69 $\pm$ 0.03 \\
\hline
$E_{f}/B$ & 7.0 $\pm$ 0.5 & 7.4 $\pm$ 0.5 & 8.2 $\pm$ 0.5 \\
$S_{f}/B$ & 38 $\pm$ 3 & 40 $\pm$ 3 & 44 $\pm$ 3 \\
${s}_{f}/B$ & 0.78 $\pm$ 0.04 & 0.70 $\pm$ 0.05 & 0.73 $\pm$ 0.05 \\
$({\bar s}_f-s_f)/B$ & 0.01 $\pm$ 0.01 & 0$^*$ & 0.17 $\pm$ 0.02\\
\hline
\end{tabular}
\end{center}

$\gamma_q>1$ seen in table~3 implies that there is phase space over-abundance of light quarks, to which, {\it e.g.,} gluon fragmentation at QGP breakup {\it prior} to hadron formation contributes. $\gamma_q$ assumes in our data analysis a value near to where pions could begin to condense$^{27}$, $\gamma_q=\gamma_q^c\equiv e^{m_\pi/2T_f}$\,. We found studying the ratio $h^-/B$ separately from other experimental results that the value of $\gamma_q\simeq\gamma_q^c$ is fixed consistently and independently both, by the negative hadron ($h^-$), and the strange hadron yields. The unphysical range $\gamma_q>\gamma_q^c$ can arise, since up to this point we use only a first quantum (Bose/Fermi) correction. However, when Bose distribution for pions is implemented, which requires the constraint $\gamma_q\le\gamma_q^c$, we obtain practically the same results, as shown in second column of table~3. Here we allowed only 4 free parameters, {\it i.e.} we set $\gamma_q=\gamma_q^c$\,, and the strangeness conservation constraint fixes $\lambda_s$\,. We then show in table~3 the ratio $\gamma_s/\gamma_q$, which corresponds (approximately) to the parameter $\gamma_s$ when $\gamma_q=1$ had been assumed. We note that $\gamma_s^{\rm Pb}>1$. This strangeness over-saturation effect could arise from the effect of gluon fragmentation combined with early chemical equilibration in QGP, $\gamma_s(t<t_f)\simeq 1$. The ensuing rapid expansion preserves this high strangeness yield, and thus we find the result $\gamma_s>1$\,, as is shown in figure 33 in Ref.[21].

We show in the bottom section of table~3 the energy and entropy content per baryon, and specific anti-strangeness content, along with specific strangeness asymmetry of the hadronic particles emitted. The energy per baryon seen in the emitted hadrons is nearly equal to the available specific energy of the collision (8.6 GeV for Pb--Pb, 8.8--9 GeV for S--Au/W/Pb). This implies that the fraction of energy deposited in the central fireball must be nearly the same as the fraction of baryon number. The small reduction of the specific entropy in Pb--Pb compared to the lighter S--Au/W/Pb system maybe driven by the greater baryon stopping in the larger system, also seen in the smaller energy per baryon content. Both collision systems freeze out at energy per unit of entropy $E/S=0.185$ GeV. There is a loose relation of this universality in the chemical freeze-out condition with the suggestion made recently that particle freeze-out occurs at a fixed energy per baryon for all physical systems$^{28}$, since the entropy content is related to particle multiplicity. The overall high specific entropy content we find agrees well with the entropy content evaluation made earlier$^{29}$ for the S--W case.

Inspecting figure 38 in Ref.[21] we see that the specific yield of strangeness we expect from the kinetic theory in QGP is at the level of 0.75 per baryon, in agreement with the results of present analysis shown in table~3. This high strangeness yield leads to the enhancement of multi-strange (anti)baryons, which are viewed as important hadronic signals of QGP phenomena$^{30}$, and a series of recent experimental analysis has carefully demonstrated comparing p--A with A--A results that there is quite significant enhancement$^{13,31}$, as has also been noted before by the experiment$^{32}$, NA35.

The strangeness imbalance seen in the asymmetrical S--Au/W/Pb system (bottom of table~3) could be a real effect arising from hadron phase space properties. However, this result also reminds us that though the statistical errors are very small, there could be a considerable systematic error due to presence of other contributing particle production mechanisms. Indeed, we do not offer here a consistent understanding of the $\Omega,\,\overline\Omega$ yields which are higher than we can describe. We have explored additional microscopic mechanisms. Since the missing $\Omega,\,\overline\Omega$ yields are proportional (13\%) to the $\Xi,\overline\Xi$ yield, we have tested the hypothesis of string fragmentation contribution in the {\it final state}, which introduces just the needed `shadow' of the $\Xi,\overline\Xi$ in the $\Omega,\overline\Omega$ abundances. While this works for $\Omega,\,\overline\Omega$, we find that this mechanism is not compatible with the other particle abundances. We have also explored the possibility that unknown $\Omega^*,\,\overline{\Omega^*}$ resonances contribute to the $\Omega,\,\overline\Omega$ yield, but this hypothesis is ruled out since the missing yield is clearly baryon--antibaryon asymmetric. Thus though we reached here a very good understanding of other hadronic particle yields and spectra, the rarely produced but greatly enhanced $\Omega,\,\overline\Omega$ must arise in a more complex hadronization pattern. 
We have presented a comprehensive analysis of hadron abundances and $m_\bot$-spectra observed in Pb--Pb 158$A$ GeV interactions within the statistical Fermi model with chemical non-equilibrium of strange and non-strange hadronic particles. The key results we obtained are: $\tilde \lambda_s=1$ for S and Pb collisions\,; $\gamma_s^{\rm Pb}>1, \ \gamma_q>1$\,; $S/B\simeq 40$\,; $ s/B\simeq 0.75$\,; all in a remarkable agreement with the properties of a deconfined QGP source hadronizing without chemical re-equilibration, and expanding not faster than the sound velocity of quark matter. The universality of the physical properties at chemical freeze-out for S- and Pb-induced reactions points to a common nature of the primordial source of hadronic particles in both systems. The difference in spectra between the two systems arises in our analysis from the difference in the collective surface explosion velocity, which for larger system is higher, having more time to develop. Among other interesting results which also verify the consistency of our approach we recall: good balancing of strangeness $\langle \bar s-s\rangle=0$ in the Pb--Pb case; increase of the baryochemical potential as the collision system grows; energy per baryon near to the value expected if energy and baryon number deposition in the fireball are similar. We note that given the magnitude of $\gamma_q$ and low chemical freeze-out temperature, most (75\%) of all final state pions are directly produced, and not resonance decay products. Our results differ significantly from an earlier analysis regarding the temperature at which hadron formation occurs. Reduction to $T_f=140$--$145$\,MeV becomes possible since we allow for the chemical non-equilibrium and collective flow, and only with these improvements in analysis our description acquires convincing statistical significance, which e.g. a hadronic gas scenario does not offer$^{33}$. Because we consider flow effects, we can address the central rapidity data of the WA97 experiment at the required level of precision, showing the consistency in these results with the NA49 data discussed earlier$^{17}$. 

In our opinion, the only consistent interpretation of the experimental results analyzed here is that hadronic particles seen at 158$A$ GeV nuclear collisions at CERN-SPS are formed directly in hadronization of an exploding deconfined phase of hadronic matter, and that these particles do not undergo a chemical re-equilibration after they have been produced.

\footnotetext{\vspace*{-0.5cm}
\begin{enumerate}

\item
J. Harris and B. M\"uller, {\it Ann. Rev. Nucl. Part. Sci.}
{\bf 46}, pp71-107, (1996); and references therein.

\item
E. Fermi, {\it Progr. Theor. Phys.} {\bf 5} 570 (1950);
{\it Phys. Rev.} {\bf 81}, 115 (1950);
{\it Phys. Rev.} {\bf 92}, 452 (1953).

\item
J. Sollfrank, {\it J. Phys. } G {\bf 23}, 1903 (1997).

\item
J. Rafelski, {\it Phys. Lett. }B {\bf 262}, 333 (1991);
{\it Nucl. Phys.} A {\bf 544}, 279c (1992).

\item
J. Letessier and J. Rafelski, {\it Phys. Rev.} C {\bf 59}, 947 (1999).

\item
G.J.\,Odyniec, {\it Nucl. Phys.} A {\bf 638}, 135, (1998).

\item
F.\,P\"uhlhofer, NA49, 
{\it Nucl. Phys.} A {\bf 638}, 431,(1998).

\item
C.\,Bormann, NA49, 
{\it J. Phys.} G {\bf 23}, 1817 (1997).

\item
H. Appelsh\"auser {\it et al.}, NA49,
{\it Phys. Lett.} B {\bf 444}, 523, (1998).

\item
S. Margetis, NA49,
{\it J. Physics} G {\bf 25}, 189 (1999).

\item
A.K.\,Holme, WA97, {\it J. Phys.} G {\bf 23}, 1851 (1997).

\item
I.\,Kr\'alik, WA97, 
{\it Nucl. Phys.} A {\bf 638},115, (1998).

\item
E. Andersen {\it et al.}, WA97,
{\it Phys. Lett.} B {\bf 433}, 209, (1998);
{\bf 449}, 401 (1999).

\item
K. S. Lee, U. Heinz and E. Schnedermann, 
{\it Z. Phys.} C {\bf 48}, 525 (1990). 

\item
F. Karsch, and M. L\"utgemeier, 
{\it Nucl. Phys.} B {\bf 550}, 449 (1999).

\item
J. Letessier and J. Rafelski, {\it Acta Phys. Pol.};
{\bf B30}, 153 (1999);
{\it J. Phys.} Part. Nuc. {\bf G25}, 295, (1999)\,.

\item
 F. Becattini, M. Gazdzicki and J. Sollfrank,
{\it Eur. Phys. J.} C {\bf 5}, 143-15, (1998). 

\item
J. Letessier, J. Rafelski, and A. Tounsi, 
{\it Phys. Lett.} B {\bf 410}, 315 (1997);
{\it Acta Phys. Pol.} B {\bf 28}, 2841 (1997).

\item
H. van\,Hecke, H. Sorge and N. Xu,
{\it Phys. Rev. Lett.} {\bf 81}, 5764 (1998).

\item
{J. Rafelski and B. M\"uller}, {\it Phys. Rev. Lett}
{\bf 48}, 1066 (1982); {\bf 56}, 2334E (1986);
{P.~Koch, B.~M\"uller and J.~Rafelski},
{\it Phys. Rep.} {\bf 142}, 167 (1986).

\item
{J. Rafelski, J. Letessier and A. Tounsi},
{\it Acta Phys. Pol.} B {\bf 27}, 1035 (1996), and references therein.

\item
G.J.\,Odyniec, NA49, 
{\it J. Phys.} G {\bf 23}, 1827 (1997).

\item
P.G.\,Jones, NA49, 
{\it Nucl. Phys.} A {\bf 610}, 188c (1996).

\item
D. R\"ohrig, NA49,
``Recent results from NA49 experiment on Pb--Pb collisions at 158 A GeV'',
see Fig. 4, in proc. of EPS-HEP Conference, Jerusalem, Aug. 19-26, 1997.

\item
D. Evans, WA85,
{\it Heavy Ion Physics} {\bf 4}, 79 (1996).

\item
D. Ferenc, U. Heinz, B. Tomasik, U.A. Wiedemann, and J.G. Cramer,
{\it Phys. Lett.} B {\bf 457}, 347 (1999).

\item
U. Heinz, private communication.

\item
J. Cleymans and K. Redlich, {\it Phys. Rev. Lett.} {\bf 81}, 5284 (1998);
and references therein.

\item
J. Letessier, A. Tounsi, U. Heinz, J. Sollfrank and J. Rafelski
{\it Phys. Rev. Lett.} {\bf 70}, 3530 (1993); 
{\it Phys.\ Rev.} D {\bf 51}, 3408 (1995).

\item
J. Rafelski, pp 282--324, in 
{\it Future Relativistic Heavy Ion Experiments}, 
R. Bock and R. Stock, Eds., GSI Report 1981-6; 
in {\it New Flavor and Hadron Spectroscopy},
 J. Tran Thanh Van, Ed. p 619, Editions Frontiers (Paris 1981);
and in {\it Nucl. Physics} A {\bf 374}, 489c (1982).

\item
F. Antinori {\it et al.}, WA85,
{\it Phys. Lett.} B {\bf 447}, 178 (1999).

\item
Th. Alber {\it et al.}, NA35,
{\it Z. Phys.} C {\bf 64}, 195 (1994).

\item
P. Braun-Munzinger, I. Heppe, and J. Stachel, {\it Chemical
Equilibration in Pb+Pb collisions at the SPS}, [nucl-th/9903010],
submitted to {\it Phys. Lett. B}, March 1999. 

\end{enumerate}
}
\end{mdframed}
\vskip 0.5cm 

\subsubsection{Echos of forthcoming new state of matter CERN announcement}
The publication effort of the above manuscript terminated  on 13 January 2000, just 4 weeks before CERN announced its new phase of matter discovery. In my letter below the reader sees the context of this announcement mentioned which included some scientific arguments surrounding the CERN preparations for release of the QGP announcement: I refer to disputes between NYC Columbia University based Dr. Miklos Gyulassy and Maurice Jacob regarding the CERN announcement scheduled for early February.\\

\noindent \textit{My following letter to PRL editor in chief Jack Sandweiss of January 13, 2000 with whom I had a personal and freindly relation reads:}\\[-0.7cm]
%
\begin{mdframed}[linecolor=gray,roundcorner=12pt,backgroundcolor=Dandelion!15,linewidth=1pt,leftmargin=0cm,rightmargin=0cm,topline=true,bottomline=true,skipabove=12pt]\relax%
January 13, 2000\\
Dear Jack,\\

\ldots If in your judgment this work is not PRL suitable, so be it, with all the ensuing consequences -- these have just begun. The endless delay of our work has muffled a scientific discussion (for others were standing by and watching what happens) and what you thus see today are draft (CERN) press releases fought off by Guylassy. It would have been nicer to have PRL papers arguing the matter of QGP at SPS energies. Somewhere things went bad.\\

\small{CLARIFICATION ABOUT CONTENTS:}\\
Permit me to notice that it is impossible to add to the paper within the prescribed length the items that the referees are asking for (predictions for example).

I wish also to set straight one impression: The key results and methods were at the time of submission to PRL completely original and who claims otherwise may be in need of an alibi.

Best wishes JAN
\end{mdframed}

\subsubsection{Conversation with referees about QGP fireball at CERN SPS}
Returning to the timeline: The manuscript LC7284 was sent out to referees on 24 March 1999. In my publication file I have only  two relevant reports, with second iterations, as well as follow-up correspondance.\\

\noindent \textit{On 21 May 1999 I received the following letter from PRL with one referee report; a second referee report follows by fax on 26 May 1999:}\\[-0.7cm]
\begin{mdframed}[linecolor=gray,roundcorner=12pt,backgroundcolor=GreenYellow!15,linewidth=1pt,leftmargin=0cm,rightmargin=0cm,topline=true,bottomline=true,skipabove=12pt]
\relax%
%


{\bf Referee A:}\\
This work discusses very interesting measurements from both the NA49 and the WA97 collaborations at the CERN-SPS. However, this variant of the fireball model is based on methods already well developed (see ref. 16, refs. 3,4,5 and ref. 14) with only incremental refinements as noted in the first paragraph. In addition, the point of this paper proported by the authors does not bring any new physical insight with regard to these measurements (see their paper, ref. 16, for example). Furthermore,  their claim of a QGP at the SPS is at best controversial.

Interestingly, this article completely skirts the intriguing observations made by WA97 of the small inverse $m_t$ slopes and large abundances of the multi-strange baryons. The $m_t$ dependent assumptions of this fireball model for the baryons which result in larger inverse $m_t$ slopes with more strangeness are inconsistent with the measured smallar inverse $m_t$ slopes with more strangeness. This unique dependence has been discussed in light of the early freezeout of the multi-strange baryons by H. Van Hecke, H. Sorge and N. Xu, Phys.Rev.Lett.81:5764-5767,1998, (a reference that should be included in this work).

Furthermore, understanding particle production with the Omegas is critical since the enhanced production of multi-strange baryons is predicted as one of the signals of the QGP (see Ref. 21). The poor description by this fireball model of particle ratios when including the Omegas are in contradiction of the authors\rq\ initial claim of a QGP scenario. In fact, as emphasized in the conclusion, only by relaxing chemical equilibration and introducing dynamical flow parameters can the \lq fit\rq\ be improved. However, the theoretical origin of such non-equilibrium features requires transport dynamical approaches which are totally ignored here.

In light of the above discrepancies and the lack of new methods or insights, I can not recommend this work for publiciation in Phys. Rev. Lett.
\end{mdframed}
\vskip 0.5cm

\noindent {\it This prompted on 14/15 June, 1999 the response along with arXiv-ing of v2 of the manuscript \url{https://arxiv.org/pdf/nucl-th/9903018v2 dated 25 June 1999}:}\\[-0.7cm]
\begin{mdframed}[linecolor=gray,roundcorner=12pt,backgroundcolor=Dandelion!15,linewidth=1pt,leftmargin=0cm,rightmargin=0cm,topline=true,bottomline=true,skipabove=12pt]
 \relax%
Dear \ldots (PRL)\\


\noindent a) \small{GENERAL REMARKS OF REFEREE B:}\\
In response to the general concern if the manuscript can be understood by the wider audience we did work hard to word the paper better making minute but frequent changes in the English. We believe that this short paper now meets the stated criterion.

To stress the theoretical nature of our study, pursuant to next remark of referee B in second paragraph of the review we have in particular\\
 i) replaced in the title \lq Hadrons from..\rq\ by \lq On hadron production in...\rq\ and \\
 ii) In the abstract, we added in the first phrase \lq ...obtained in several relativistic heavy ion experiments is utilized..\rq\\

\noindent b) \small{GENERAL REMARKS OF REFEREE A:}\\
We agree with the referee that the methods we have been developing since 1991 [4] are today well established and are widely used. Indeed this is the strength of our work. The refinement we here introduce are not trivial, even if these can be mentioned in one simple phrase: we have in full incorporated the radial flow, and its impact on the $m_t$ spectra, see old second paragraph. 

We have now refined that (part of our) discussion and as suggested by referee A, we include in the short third paragraph the new reference to the work of vanHecke, Sorge, Xu (new reference 19) on transverse slopes. The paragraph reads: \ldots The here developed model offers a natural understanding of the systematic behavior of the $m_\bot$-slopes which differs from other interpretations. The near equality of (inverse) slopes of nearly all strange baryons and antibaryons arises here by means of the sudden hadronization at the surface of an exploding QGP fireball. In the hadron based microscopic simulations this behavior of $m_\bot$-slopes can also arise allowing for particle-dependent freeze-out times [19]. \ldots






\noindent h) In response to referee A we separate and extend the discussion of the Omega particles into a separate paragraph (4th from the end):\\
ldots we do not offer here a consistent understanding of the $\Omega,\,\overline\Omega$ yields which are higher than we can describe. We have explored additional microscopic mechanisms. Since the missing $\Omega,\,\overline\Omega$ yields are proportional (13\%) to the $\Xi,\overline\Xi$ yield, we have tested the hypothesis of string fragmentation contribution in the {\it final state}, which introduces just the needed `shadow' of the $\Xi,\overline\Xi$ in the $\Omega,\overline\Omega$ abundances. While this works for $\Omega,\,\overline\Omega$, we find that this mechanism is not compatible with the other particle abundances. We have also explored the possibility that unknown $\Omega^*,\,\overline{\Omega^*}$ resonances contribute to the $\Omega,\,\overline\Omega$ yield, but this hypothesis is ruled out since the missing yield is clearly baryon--antibaryon asymmetric. Thus though we reached here a very good understanding of other hadronic particle yields and spectra, the rarely produced but greatly enhanced $\Omega,\,\overline\Omega$ must arise in a more complex hadronization pattern. \ldots \\


Overall there are no major changes in the contents of our paper, though some improvement in presentation has been reached, due to thorough review and constructive comments of the referees.

We hope that the attached manuscript will be accepted for publication in PRL.
\end{mdframed}
\vskip 0.5cm
 
\noindent \textit{This was not to be. Jumping forward to 27 August 1999, PRL writes:}\\[-0.7cm]
\begin{mdframed}[linecolor=gray,roundcorner=12pt,backgroundcolor=GreenYellow!15,linewidth=1pt,leftmargin=0cm,rightmargin=0cm,topline=true,bottomline=true,skipabove=12pt]\relax%
\ldots I had sent your manuscript do a Divisional Associate Editor of Physical Review Letters for advice on your appeal. However, we recently received an updated report from the second referee B. In view of the attached additional remarks I would like to give you an opportunity to consider the updated report before I continue the appeals process.

Please let us know how you wish to process. We are holding your manuscript in this office awaiting your response.\\

Second report of Referee B: Updated report (meaning that there was a 2nd report we have not seen? JR)

Re: Manuscript LC7284

The ammendments made by J.Rafelski and J.Letessier in response to my comments on their manuscript \lq\lq On Hadron Production in Pb-Pb Collisions at 158 A GeV\rq\rq\ are satisfactory.

Not withstanding the opinion of the other referee, I believe the modifications to the fireball model introduced by Rafelski and Latessier are significant (e.g. the treatment of radial flow). Also I should point out that the measurements of Omega yields are relatively more uncertain than all the other ratios used in the R-L model analysis. Therefore, the two sets of calculations, with and without the omega ratio, are very informative.

However (and, unfortunately), I did overlook it when writing my original report) this model has a technical flaw in its treatment of the pion Bose statistics, resulting in the oversaturation of light quarks and the divergence (pion lasers) at $p_t=0$ ($\mu_\pi=156$ MeV was obtaind, inspite of Bose condensate at $\mu_\pi=$pion mass). This point clearly needs clarification before the paper can be recommended for publication.
\end{mdframed}
\vskip 0.5cm

Here a word of explanation regarding the last paragraph is needed: I had in good faith discussed this work as I believe by phone with Ulrich Heinz of later QGP rejection fame; in this discussion the question of the relevance of the Bose statistic\label{HeinzBose} for pions came up. This was  not addressed in v1 and v2 of the manuscript.  We  know by means of precise HBT analysis that the pion emission lasts about 2\;fm/c. This means that even if the total pion abundance is large, there is no need for Bose statistics as this characteristic time for pion emission suffices to cumulate overabundance observed in experiment by means of sequential emission. 

However, in 1999 one could argue that the creation of the pion yield could be truly sudden. I was asked to investigate this for the purpose of completeness of our work and that we did for v3 of the manuscript. Consistency of the sudden hadronization model demands  that we show that Bose statistics describing the pion yield will work, so that perfectly sudden hadronization can be demonstrated -- even if it has little if any true physics meaning.

As above correspondance quote shows, somehow referee B learned about my related conversation.  This allowed   the PRL divisional associate editor (same person), to find a reason to reject our work. Without this reason the situation would be that we answered objections of referee A, implemented comments and received a nod from referee B; he would need to accept the paper that claims QGP was found at SPS. Clearly, by doing this the associate editor would have contradicted his objections to the CERN announcement of QGP. This is what he in essence told me and I still hear his words in my mind 20 years later.\\

\noindent \textit{I made a further effort with PRL without considering what was inevitable: That hostile  associate editor   would be the final judge and not the referee B. I respond to PRL on Sept 8, 1999:}\\[-0.7cm]
\begin{mdframed}[linecolor=gray,roundcorner=12pt,backgroundcolor=Dandelion!15,linewidth=1pt,leftmargin=0cm,rightmargin=0cm,topline=true,bottomline=true,skipabove=12pt]\relax%
In response to the request for a clarification made by referee B, we have recomputed our results using the Bose statistics for pions. We are sure that the astute referee will be convinced by the stability of our results subject to this technical refinement.

In order to properly explain the calculational contents, we have expanded as follows a phrase in the old manuscript to a longer comment now located at the end of page 2 of the current PRL style printout. The full paragraph reads:

\lq\lq $\gamma_q>1$ seen in table~3 implies that there is phase space over-abundance of light quarks, to which, {\it e.g.,} gluon fragmentation at QGP breakup {\it prior} to hadron formation contributes. $\gamma_q$ assumes in our data analysis a value near to where pions could begin to condense [27], $\gamma_q=\gamma_q^c\equiv e^{m_\pi/2T_f}$\,. We found studying the ratio $h^-/B$ separately from other experimental results that the value of $\gamma_q\simeq\gamma_q^c$ is fixed consistently and independently both, by the negative hadron ($h^-$), and the strange hadron yields. The unphysical range $\gamma_q>\gamma_q^c$ can arise, since up to this point we use only a first quantum (Bose/Fermi) correction. However, when Bose distribution for pions is implemented, which requires the constraint $\gamma_q\le\gamma_q^c$, we obtain practically the same results, as shown in second column of table~3. Here we allowed only 4 free parameters, {\it i.e.} we set $\gamma_q=\gamma_q^c$\,, and the strangeness conservation constraint fixes $\lambda_s$\,. We then show in table~3 the ratio $\gamma_s/\gamma_q$, which corresponds (approximately) to the parameter $\gamma_s$ when $\gamma_q=1$ had been assumed. We note that $\gamma_s^{\rm Pb}>1$. This strangeness over-saturation effect could arise from the effect of gluon fragmentation combined with early chemical equilibration in QGP, $\gamma_s(t<t_f)\simeq 1$. The ensuing rapid expansion preserves this high strangeness yield, and thus we find the result $\gamma_s>1$\,, as is shown in figure 33 in Ref. [21].\rq\rq\ \ldots 
 
To make space for this explanation of the procedure we conclude our paper now with the abreviated conclusions which contain just one remark as follows:

\lq\lq In our opinion, the only consistent interpretation of the experimental results analyzed here is that hadronic particles seen at 158$A$ GeV nuclear collisions at CERN-SPS are formed directly in hadronization of an exploding deconfined phase of hadronic matter, and that these particles do not undergo a chemical re-equilibration after they have been produced. \rq\rq\

We have also added in penultimate paragraph the phrase: \lq\lq We note that given the magnitude of $\gamma_q$ and low chemical freeze-out temperature, most (75\%) of all final state pions are directly produced, and not resonance decay products.\rq\rq\\
b) updated references 15, 26, shortened references 14 and 29 (now 30) 
c) smoothed English\\
 i) in paragraph below equation 1,\\
 ii) in the paragraph above the new paragraph and \\
iii) the second last paragraph.\\

We sincerely hope that you will consider our manuscript 
now suitable to be published in PRL.
\end{mdframed}
\vskip 0.5cm

The editors of PRL consulted  the  divisional associate editor, who did not recuse himself despite his well-known public position against QGP at CERN. Through PRL channels, he rejected our work on September 28, 1999. His letter was full of inaccuracies as he mixed up two manuscript files that he was both rejecting for different reasons, thus a further delay applied before the manuscript was rejected as noted previously.  

\noindent \textit{The scientific argument presented  by the PRL  divisional associate editor on September 28, 1999 were:}\\[-0.7cm]
\begin{mdframed}[linecolor=gray,roundcorner=12pt,backgroundcolor=GreenYellow!15,linewidth=1pt,leftmargin=0cm,rightmargin=0cm,topline=true,bottomline=true,skipabove=12pt]\relax%
RE: LC7284\\
\ldots  I am thus sure that I understand very well what the authors have done and achieved in their paper.

My own assessment of the paper is as follows: By restricting their attention to the abundance ratios between only a fraction of the measured hadrons, leaving out the Omega and Anti-Omega baryons, and by allowing the pions to develop a Bose condensate in order to absorb the measured large pion abundance at the suggested low freeze-out temperature, the authors leave more questions open than they solve. Since the methods used in the paper are not new, its importance must be judged by its results and the implied physical picture. The claim for fame of the paper is based on the very low chemical freeze-out temperature found by the authors (contradicting all other published values) and the high statistical significance of their fit. But the resulting physical picture is not convincing: of 15(-4) measured particle abundance ratios, 11(-2) can be fit very nicely with 5 parameters, but only if 4(-2) others are excluded from the fit, by postulating (but not successfully identifying) a different creation mechanism. (In brackets I counted the redundant ratios.) The authors emphasize that the slopes of the kaon, Lambda and Xi transverse mass spectra are fit well if strong transverse flow is allowed for, but their model is unable to explain the steeper Omega spectra, and the expected strong effects of the Bose condensation of pions (as implied by their fit and mentioned in the latest version of the MS) on the pion spectrum are not discussed. Finally, even if they don't stress this in words, the authors make the dramatic prediction of a pion condensate in heavy ion collisions,  without discussing the many other observable effects which should result from such a phenomenon, nor the fact that it contradicts the findings in Ref. [26]. 
  
\end{mdframed}
\vskip 0.5cm

Regarding the argument seen above: 
\begin{enumerate}
\item We did not predict pion condensates, see page \pageref{HeinzBose}. We demonstrated answering referee after-thaught demand that Bose statistics of emitted pion density does not alter our results, earlier obtained without. 
\item
Seen from twenty years historical perspective: 
Everything the associate editor said and used was in essence a personal opinion,  a mix-up with another opinion he was preparing in parallel for another PRL paper.
\item
What we presented  was a result of a model that withstood the test of time as this long article has demonstrated.  
\end{enumerate}
To conclude: associate editor of PRL has now admitted in the 2019 interview, see page~\pageref{Heinz2019}, to have been mistaken in his rejection of the possibility of QGP at CERN-SPS. I extend these remarks to include this evaluation of our work.

\markboth{4. Epilogue: Using QGP}{Strangneess in QGP: Diaries}
\section{Epilogue: Using QGP}
One natural use of the QGP understanding is the connection between the QGP phase in the early Universe, and the matter surrounding us, see page \pageref{UniverseR1}. In a first step one must understand the particle content in the Universe once quark-gluon  radiation turns into hadron matter. I submitted a manuscript~\cite{Fromerth:2002wb} for publication in 2002. The initial reviews were  outright helpful. 

The problem  was, they conflicted with editorial procedures. As we struggled to include the requested additional material at the cost of deleting something else, in order to meet the constraints of the four page limit, the publication process  derailed. I show the referee requests from the first round of review of~\cite{Fromerth:2002wb}, and reprint the manuscript below in response format v3. The final arXiv\rq ed v4  presents yet  a different remix of results. In the end I through up my arms. Our valuable set of insights remained unpublished -- a decade later some of the material was embedded into a longer conference report~\cite{Fromerth:2012fe}. \\

\subsection{Quark-hadron Universe}
\subsubsection{Good or bad advice?}
\noindent\textit{First round proposals of referees of the  Physical Review Letters manuscript LY8289 \lq Hadronization of the Quark Universe\rq\  by \textbf{Michael J. Fromerth and Johann Rafelski} received  on December 23,  2002.}\\[-0.7cm]
\begin{mdframed}[linecolor=gray,roundcorner=12pt,backgroundcolor=GreenYellow!15,linewidth=1pt,leftmargin=0cm,rightmargin=0cm,topline=true,bottomline=true,skipabove=12pt]\relax%
\textbf{Referee A}\\
\indent \ldots need to revise their manuscript substantially to make clear their goals and accomplishments and why it is that their readers should be interested and care about their results. \ldots  Why should I (the reader) care about the d-quark chemical potential \ldots Is it the chemical potential, or the RATIO of it to the temperature that is the physically intereting quantity? \ldots  Much of the early universe cosmology described by the authors may not really be needed in their text. \ldots The authors seem confused by the differences between annihilation in a symmetric and in an asymmetric universe. \ldots PRODUCTS of cross sections and velocity become constant. This is true for exothermic processes, such as s-wave annihilation. However, none of these numbers really play a crucial role in what - I think - the authors are interested in. \ldots The use of \ldots  an unrefereed book, to justify details or assumptions seems inappropriate to me. \ldots There are schemes for the creation of baryon number in which the baryon number and the lepton number can be dissimilar. \ldots  I am confused by the issue of charge distillation.\\

\noindent\textbf{Referee B:}\\
\indent The authors trace back chemical potentials in the early Universe. They find that during the quark-hadron phase transition the two phases have differing and non-zero charge densities on the average. I think that the phenomenon of charge distillation is interesting, though quantitatively the effect is not so large. In my opinion the manuscript in general meets the criteria of Physical Review Letters. I recommend that it would be published after \ldots  There is some entropy generated. But presumably only a tiny amount, beacause of the weakness of the transition. I think that the effect for the present results should be in practice negligible. \ldots  the authors require that B-L = 0. I agreee that this makes a lot of sense for practical purposes. However, it is not unconditionally dictated by baryogenesis. \ldots It would be interesting to understand how the these results would incorporate into the dynamics of the (first-order) phase transition \ldots
\end{mdframed}
\vskip 0.5cm 

\subsubsection{Hadronization of the quark universe}
\noindent\textit{This is v3 of the arXiv\rq ed manuscript  astro-ph/0211346 of December 31, 2002, which contains changes made to accomodate the first round of  referee comments.}\\[-0.7cm]
%
%
\begin{mdframed}[linecolor=gray,roundcorner=12pt,backgroundcolor=Dandelion!15,linewidth=1pt,leftmargin=0cm,rightmargin=0cm,topline=true,bottomline=true,skipabove=12pt]\relax%
\textbf{\Large Hadronization of the Quark Universe}\\[0.2cm]
\textbf{Abstract}
{\small Recent advances in the understanding of equations of state of strongly interacting matter allow exploration of conditions in which matter (protons, neutrons) formed. Using the recently solidified knowledge about photon to baryon ratio, and neutrino oscillations, we are able to trace out the evolution of particle chemical potentials, beginning in quark-gluon phase (QGP) when the Universe was 1\,$\mu$s old, through hadronization and matter-antimatter annihilation toward onset of nucleosynthesis.  In  the mixed hadron-quark phase a significant hadron sector  electric  charge distillation is found given non-zero chemical potentials.}\\

In the standard big-bang model, the large primordial baryon abundance formed at hadronization of the deconfined quark-gluon plasma (QGP) disappears due to mutual annihilation, exposing a slight net baryon number observed today~[1]. The annihilation period began after the phase transformation from the QGP to a hot hadronic gas (HG), roughly $20$--$30\,\mu$s after the big bang when the Universe was at a temperature of $\sim 160$~MeV. In the ensueing evolution the energy fraction in baryons and antibaryons in the  Universe dropped from $\sim 10\%$ when Universe was about 40\,$\mu$s old to less than $10^{-8}$ when it was one second old. 

Tracing the evolution  of particle chemical potentials with temperature in the hadronic domain allows us to connect this picture to ongoing laboratory relativistic heavy ion collision experimental work, and to verify our understanding of the hadronic matter behavior in the early Universe. We will show how the study of the Universe chemical composition opens up the possibility of amplification of a much smaller and preexisting matter-antimatter asymmetry in a  matter-antimatter distillation separation process. 

The observational evidence about the antimatter non-abundance in the Universe is supported by the highly homogeneous cosmic microwave background derived from the period of photon decoupling~[2]. This has been used  to argue that the  matter-antimatter domains on a scale smaller than the observable Universe are unlikely~[3]; others see need for further experimental study to confirm this result~[4]. The current small value of the baryon-to-photon ratio is the result of  annihilation of the large matter-antimatter abundance. Considering several observables, the range $\eta \equiv n_{\rm B}/n_\gamma = 5.5 \pm 1.5 \times 10^{-10}$ is established~[5]. The  matter-antimatter asymmetry is expressed by non-zero values of the chemical potentials. Our objective is to  quantify the values of  chemical potentials  required to generate the observed value of $\eta$, and to use this to quantify the electrical charge distillation occurring during hadronization. 

To compute the thermodynamic properties of the QGP and HG phases, we study the partition functions $\ln{Z_{\rm QGP}}$ and $\ln{Z_{\rm HG}}$ as described in Ref.~[6,7,8]: we use a phenomenological description of QGP equations of state developed in~[7], which agrees well with properties of quantum chromodynamics (QCD) at finite temperature obtained in lattice QCD  for the limit of vanishing particle density~[9], and at finite baryon density~[10]. This approach  involves quantum gases of quarks and gluons with perturbative QCD corrections applied, and a confining vacuum energy-pressure component ${\mathcal{B}}= 0.19$~GeV~fm$^{-3}$. In the HG partition function, we sum partial gas contributions including all hadrons  from Ref.~[11] having mass less than 2~GeV, and apply finite volume corrections~[8].

Our use of partition functions assumes that local thermodynamic equilibrium (LTE) exists. Considering the particle  spectra and yields measured at  the Relativistic Heavy Ion Collider at Brookhaven National Laboratory (RHIC-BNL), it is observed that a thermalization timescale on the order of $\tau_{\rm th}\lesssim 10^{-23}$~s is present in the QGP at hadronization~[6]. The microscopic mechanisms for such rapid thermalization are at present under intense study. We expect this result to be valid qualitatively in the primordial  QGP phase of matter. This then assures us of LTE being present in the evolving Universe. The local chemical equilibrium (LCE) is also approached at RHIC, indicating that this condition also prevails in the early Universe. 

To  apply these experimental results we recall that the size scale $R$  of the radiative Universe  evolve as $R \propto t^{1/2}$. Furthermore, if the expansion is adiabatic and energy conserving, then: $R \propto T^{-1}$. The thermalization timescale is roughly: $\tau_{\rm th}\approx   {1}/{n\, \sigma\, v}$, with $n$ the particle number density, $\sigma$ the cross section for (energy-exchanging) interactions, and $v$ the mean velocity. For a roughly constant value of $\sigma$, we can expect an increase in $\tau_{\rm th}\ $ as we cross from the relativistic QGP ($v \simeq c = 1$) to the HG  phase having strong non-relativistic components ($v \propto T^{1/2}$). Allowing for the change in relative velocity and  a reduction in density, we expect $\tau_{\rm th}\lesssim 10^{-22}$~s for the HG at $T = 160$~MeV. Since the thermalization time scales as $T^{-7/2}$  in the cooling HG phase, considering $R(T)$ and $v(T)$, its value increases to $\tau_{\rm th}\lesssim 10^{-14}$~s at $T=1$~MeV. At this point, the Universe is already one second old, so our assumption of LTE (and also of LCE) has a large margin of error and is in our opinion fully justified throughout the period of interest.

Chemical equilibration timescales are longer, due to  significantly smaller cross sections, than thermalization timescales. When chemical equilibrium cannot be  maintained in an expanding Universe, we find particle yield freeze-out. Near the phase transformation from HG to QGP, chemical equilibrium for hadrons made of $u,d,s$ quarks is established. Hadron abundance evolution in the early Universe and deviations from the local equilibrium at lower temperature have not yet been studied in great detail. In a baryon symmetric Universe there is a freeze-out of nucleon and antinucleon densities. On the other hand, in a locally asymmetric Universe  baryon annihilation reactions essentially cease at a temperature near 35 MeV, and baryon density at lower temperature is  determined by Universe expansion. However, the antinucleons keep annihilating  until there are none left. 

In order to further the  understanding of all particle and in particular hadron abundances, we obtain the values of chemical potentials describing particle abundances beginning near to QGP hadronization through the nucleosynthesis period. In a system of non-interacting particles, the chemical potential $\mu_i$ of each species $i$ is independent of the chemical potentials of other species, resulting in a large number of free parameters. The many chemical particle interactions occurring in the QGP and HG phases, however, greatly reduce this number.

First, in thermal equilibrium, photons assume the Planck distribution, implying a zero photon chemical potential; i.e., $\mu_\gamma = 0$. Next, for any reaction $\nu_i A_i = 0$, where $\nu_i$ are the reaction equation coefficients of the chemical species $A_i$, chemical equilibrium occurs when $\nu_i \mu_i = 0$, which follows from a minimization of the Gibbs free energy. Because reactions such as $f + \bar{f} \rightleftharpoons 2 \gamma$ are allowed, where $f$ and $\bar{f}$ are a fermion -- antifermion pair, we immediately see that $\mu_f = -\mu_{\bar{f}}$ whenever chemical and thermal equilibrium have been attained.

Furthermore, when the system is chemically equilibrated with respect to weak interactions, we can write down the following relationships~[12]:
\begin{equation}\tag{1}\label{dmul}
\mu_e - \mu_{\nu_e}=\mu_\mu - \mu_{\nu_{\mu}}=\mu_{\tau} - \mu_{\nu_{\tau}}\ 
\equiv\ \Delta \mu_l,
\end{equation}
along with $\mu_u=\mu_d - \Delta \mu_l$, and $\mu_s=\mu_d$, with the chemical equilibrium of hadrons being equal to the sum of the chemical potentials of their constituent quarks; {\it e.g.}, $\, \Sigma^0\,$({\it uds}) has chemical potential $\mu_{\Sigma^0}=\mu_u + \mu_d + \mu_s=3\, \mu_d - \Delta \mu_l$, and the baryochemical potential is:
\begin{equation}\tag{2}
\mu_b=\frac32(\mu_d +\mu_u)=3 \mu_d -\frac32 \Delta \mu_l.
\end{equation}
  
We will use $\mu_d$ in both QGP and HG phases to characterize hadron abundances and note that $\mu_b\simeq 3\mu_d$ in the HG phase. Finally, if the experimentally-favored ``large mixing angle'' solution~[13] is correct, the neutrino oscillations $\nu_e \rightleftharpoons \nu_\mu \rightleftharpoons \nu_\tau$ imply that~[14]: $\mu_{\nu_e} = \mu_{\nu_{\mu}} =\mu_{\nu_{\tau}} \equiv \mu_\nu$, which reduces the number of independent chemical potentials to three, where we assume that in dense matter oscillations occur rapidly. We choose these potentials  to be $\mu_d$, $\mu_e$, and $\mu_\nu$. To determine these in a homogeneous (i.e., single phase) Universe, we seek to satisfy the following three criteria:
\begin{itemize}
\item[i.] {\it Charge neutrality} ($Q = 0$) is required to eliminate  Coulomb energy.  This implies that:
\begin{equation}\tag{3}\label{Q0}
n_Q\equiv \sum_i\, Q_i\, n_i (\mu_i, T)=0, 
\end{equation}
where $Q_i$ and $n_i$ are the charge and number density of species $i$, and the summation is carried out over all species present in the phase. \item[ii.] {\it Net lepton number equals net baryon number} ($L = B$) is required in baryogenesis.  This implies that:
\begin{equation}\tag{4}\label{LB0}
n_L - n_B\equiv \sum_i\, (L_i - B_i)\, n_i (\mu_i, T)=0 ,
\end{equation}
where $L_i$ and $B_i$ are the lepton and baryon numbers of species $i$. This expression can be generalized  for schemes in which $L\ne B$. A modified lepton density would require an increase in lepton chemical potential which is not essential to the understanding of the hadron behavior, given the results we obtain.  \item[iii.] {\it Constant entropy-per-baryon} ($S/B$).  This is the statement that the Universe evolves adiabatically, and is equivalent to:
\begin{equation}\tag{5}\label{SperB}
\frac{s}{n_B}\equiv\frac{\sum_i\, s_i(\mu_i, T)}{\sum_i\, B_i\, n_i(\mu_i, T)}
  ={\rm constant,}
\end{equation}
where $s_i$ is the entropy density of species $i$.
\end{itemize}

The value of $S/B$ can be estimated from the value of $\eta$. In the low temperature era ($T \ll 1$~MeV), the entropy is dominated by photons and nearly massless neutrinos. It is straightforward to compute the entropy densities of these species from the partition function, and then convert $\eta$ to $S/B$ using the known photon number density. In doing so, we obtain a value of $S/B = 4.5^{+1.4}_{-1.1} \times 10^{10}$.

With $S/B = s/n_B$ fixed, Eqs.~(\ref{Q0})--(\ref{SperB}) constitute a system of three coupled, non-linear equations of three unknowns ($\mu_d$, $\mu_e$, and $\mu_\nu$) for a given temperature. These equations were solved numerically using the Levenberg-Marquardt method~[15] to obtain Fig.~1, which shows the values of $\mu_d$, $\mu_e$, and $\mu_\nu$.

The bottom axis of Fig.~1 shows the age of the Universe, while the top axis shows the corresponding temperature. The error bars arise from ``experimental'' uncertainty in the value of $\eta$. Note that the value of the chemical potentials required to generate the current matter-antimatter asymmetry are significantly smaller than 1~eV (horizontal line in Fig.~1) at the time of hadronization.

As the temperature decreases, the value of $\mu_d$ asymptotically approaches weighted one-third of the nucleon mass ($(2m_n-m_p)/3\simeq 313.6$~MeV). This follows because the baryon partition functions are in the classical Boltzmann regime at these temperatures, and the residual baryon number is dominated by the proton and neutron degrees of freedom.

\centerline{\includegraphics[width=0.8\columnwidth]{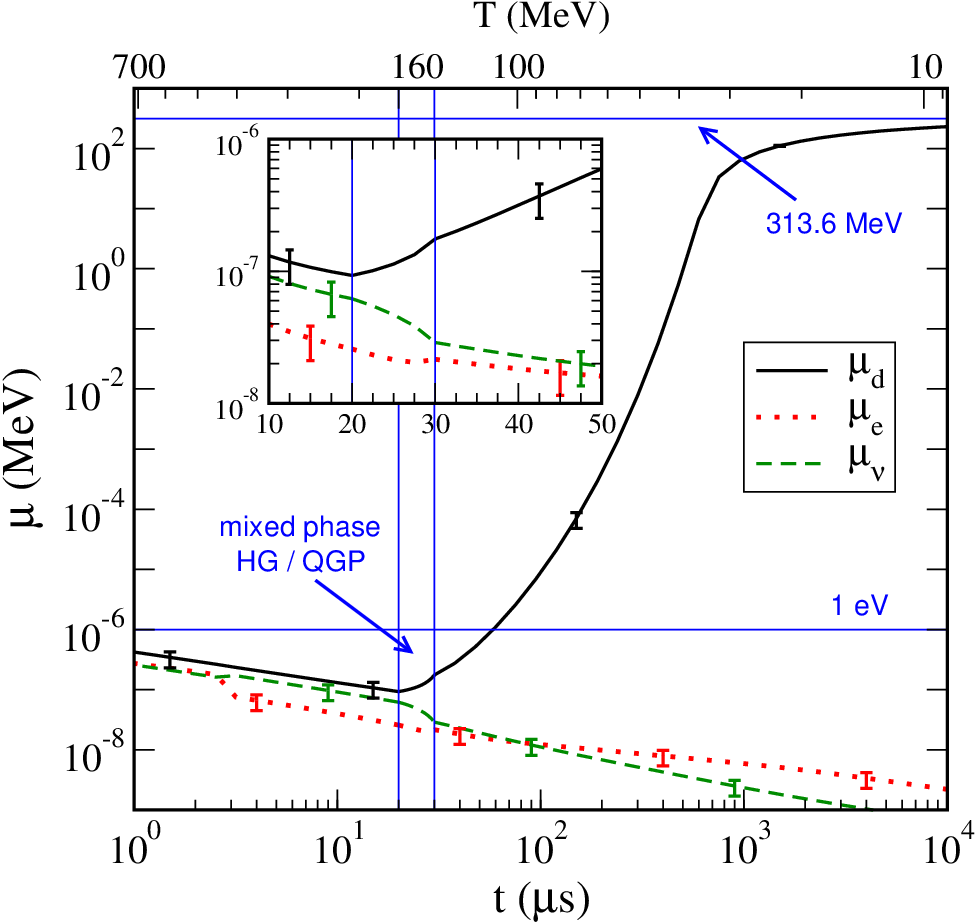}}
\noindent {\small Fig. 1: Chemical potentials $\mu_d$, $\mu_e$, and $\mu_\nu$ around the time of the QGP-HG phase transformation.  The error bars arise from the uncertainty in $\eta\,$.  Insert --- expanded view around the phase transformation. Horizontal and vertical lines inserted to guide the eye.} 

Figure~2 shows the hadronic energy content in the Universe as a function of temperature. The fraction of energy in baryons and antibaryons is roughly 10\% at the QGP-HG phase transformation, but rapidly vanishes, becoming significant again only when the Universe has cooled and enters its atomic era. In today's matter-dominated Universe, the large nucleon rest mass overwhelms completely the background radiation.

During the QGP to HG phase transformation, when both phases co-exist, the macroscopic conditions i.~--~iii.~above must be satisfied for the system as a whole, but may be violated locally. This means that Eqs.~(\ref{Q0})--(\ref{SperB}) are no longer valid within either the QGP or HG phases individually, and that the correct expressions must contain combinations of the two phases.

We therefore parameterize the partition function during the phase transformation as $\ln{Z_{\rm tot}}=f_{\rm HG} \ln{Z_{\rm HG}}+(1 - f_{\rm HG}) \ln{Z_{\rm QGP}}$, where the factor $f_{\rm HG}$ represents the fraction of total phase space occupied by the HG phase. The correct expression analogous to Eq.~(\ref{Q0}) is:
\begin{align}
Q  = & n_Q^{\rm QGP}\, V_{\rm QGP}+n_Q^{\rm HG}\, V_{\rm HG} 
\notag \\ 
\tag{6}
   = & V_{\rm tot} \left[ (1-f_{\rm HG})\, n_Q^{\rm QGP}+f_{\rm HG}\, n_Q^{\rm HG} \right]=0,
\end{align}
where the total volume $V_{\rm tot}$ is irrelevant to the solution. Analogous expressions can be derived for Eqs.~(\ref{LB0}) and (\ref{SperB}). These expressions were used to obtain Fig.~3, which shows the fraction of the total baryon number in the QGP and HG phases as a function of $f_{\rm HG}$.
 
\centerline{\includegraphics[width=0.65\columnwidth]{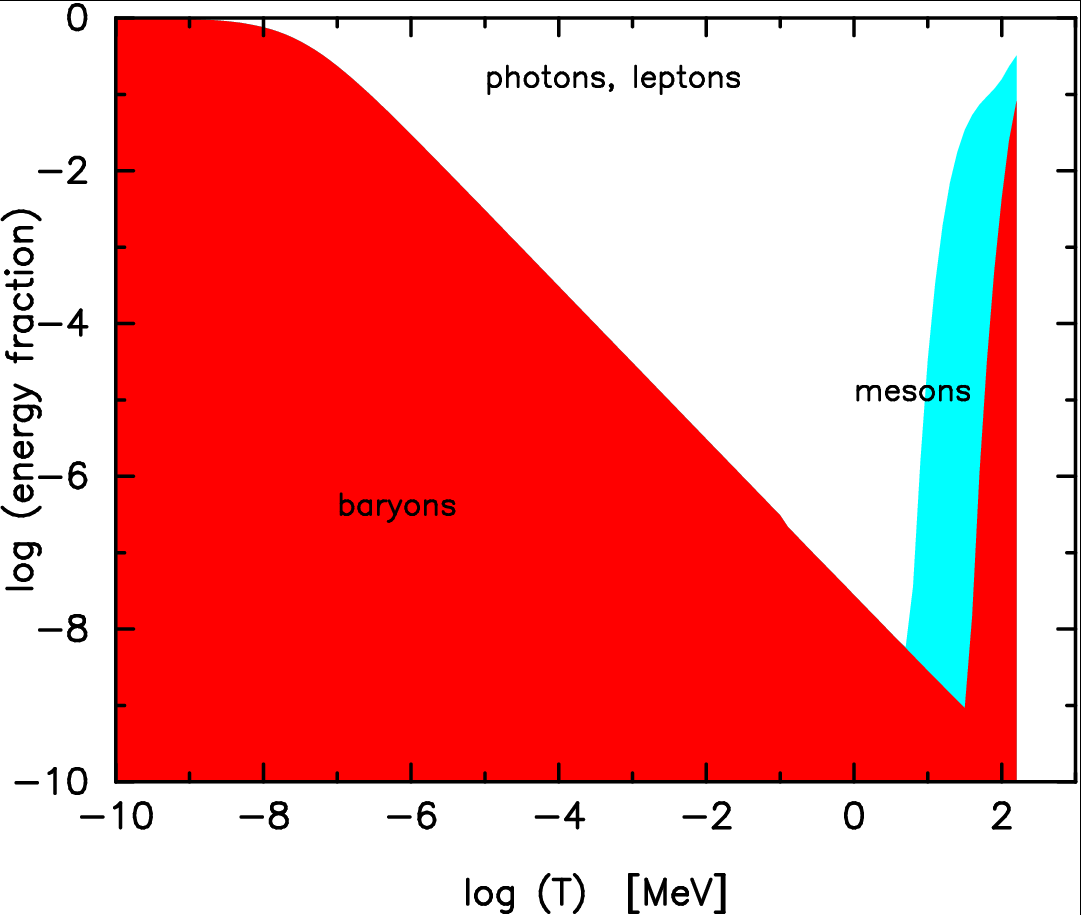}}
\noindent {\small Fig. 2:  The hadronic energy content of the luminous matter in the Universe as a function of temperature assuming a constant entropy-per-baryon number of $4.5 \times 10^{10}$.} 

In Fig.~1, it was assumed that the value of $f_{\rm HG}$ evolved linearly in time and that the duration of the phase transformation was $\tau_h=10\,\mu$s. In reality, these quantities are sensitive to the properties of the equations of state and the dynamics of the phase transformation. Our value of $\tau_h$ is an estimate discussed in~[6], and is consistent with the phase dynamics considered in~[16].

Figure~4 shows the net charge per baryon in each phase as a function of $f_{\rm HG}$. At its onset the small region of HG phase takes on an initial positive charge density, which can be attributed to the proton-neutron bias toward positive charge. As a result, the QGP domain takes on a (initially tiny) negative charge density. The charge asymmetry cannot be avoided, since in general it is impossible to obtain at given values $T,\mu_i$ the vanishing of both $n_Q^{\rm QGP}$ and $n_Q^{\rm HG}$. Such distilled dynamical asymmetry in particle yields was previously explored for strangeness separation and associated strangelet formation~[17,18].

\centerline{\includegraphics[width=0.65\columnwidth]{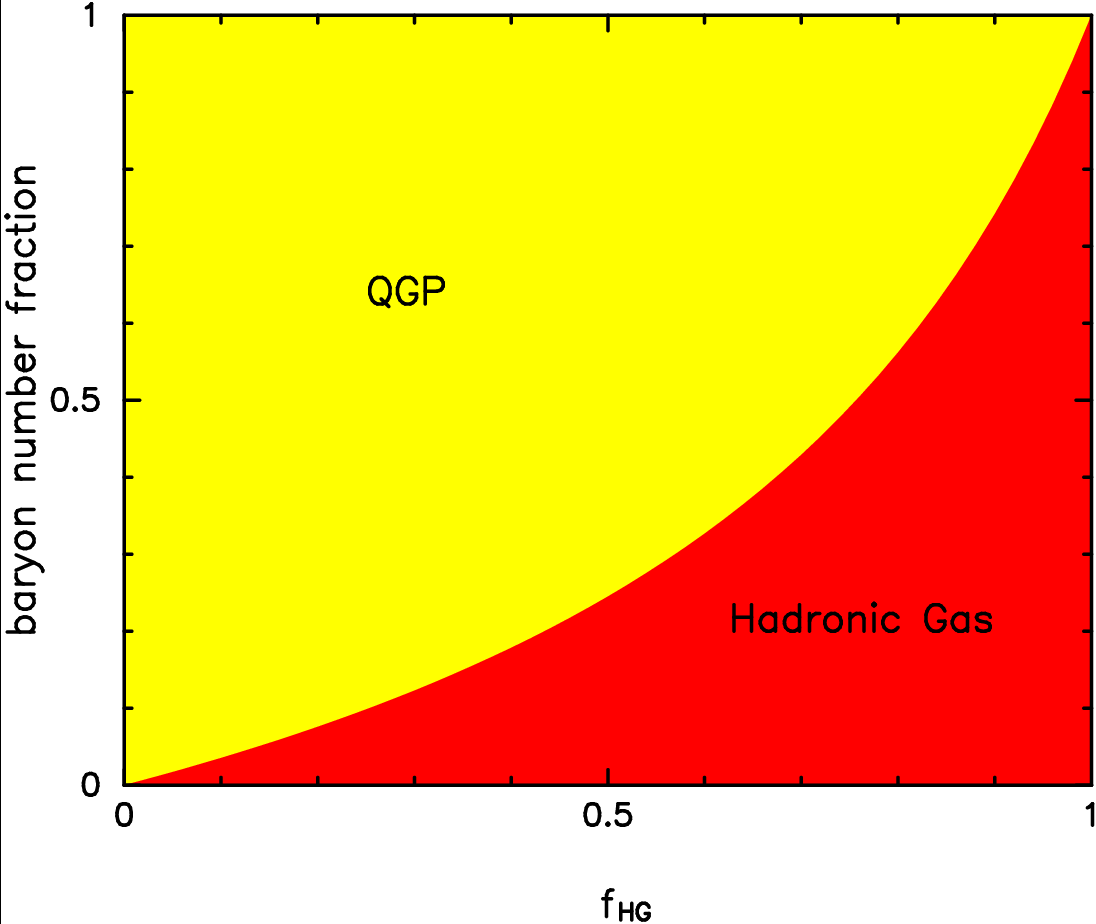}}
\noindent {\small Fig. 4: The fraction of baryons in the HG and QGP during phase transformation. The parameter $f_{\rm HG}$ is the fraction of total phase space occupied by the hadronic gas phase.} 

Since the sign of the effect seen in Fig.~4 is the same across the entire hadronization region, the total charge of the remaining QGP domains is ever-increasingly negative and one would expect development of electromagnetic potential, which effectively alters the values of chemical potentials for charged species. It is evident  that the process of charge distillation will have a feed-back effect on the QGP-HG transformation, and that flows of particles will occur that will alter the uniformly small net baryon density~[19]. This can affect (during the phase transformation) any local initial baryon-antibaryon asymmetry, and may also serve as a mechanism for generating magnetic fields in the primordial Universe~[20]. Evaluation of this baryon asymmetry enhancement effect in greater precision  requires  methods of advanced transport theory beyond the scope of this work. We note that a  separation of baryons and antibaryons into domains could maintain a homogeneous zero charge density Universe, a phenomenon which could, {\it e.g.}, play a significant role in amplifying a pre-existent, much smaller net baryon yield.   

\centerline{\includegraphics[width=0.65\columnwidth]{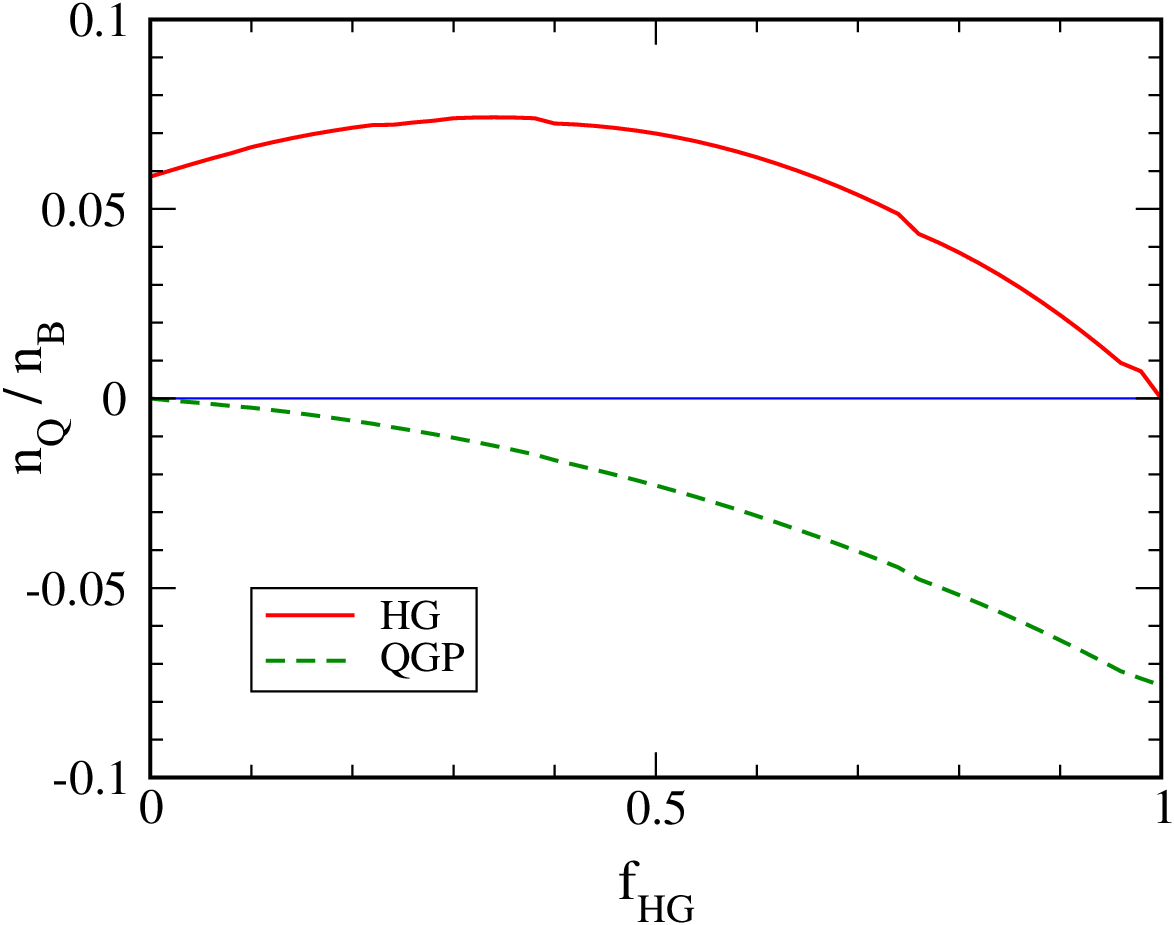}}
\noindent {\small Fig. 3: Net charge (including leptons) per net baryon number in the HG and QGP during phase transformation.  Horizontal line at zero inserted to guide the eye.}

In summary, we have determined the chemical potentials required to generate the matter-antimatter asymmetry in the Early Universe. The baryochemical potential reaches its minimal value at the begin of  hadronization of the quark Universe, where $\mu_b=0.33^{+0.11}_{-0.08}$\,eV. Our quantitative results derive from the known entropy content per baryon in the Universe. Other than a  small and most probably negligible increase of entropy in the phase transition of QGP to HG, during nucleosynthesis, and electron-ion recombination, the entropy to baryon ratio is a constant of motion, allowing us to trace out the chemical potentials in the early Universe. There remains some theoretical uncertainty in the behavior of the equations of state at the QGP/HG phase boundary which, along with the time dependence of the mixing fraction $f_HG$ and the dynamics of phase transition or transformation (e.g., expanding HG bubbles, shrinking QGP droplets, and the here uncovered distillation process), will need to be addressed in future work.
 
\footnotetext{\vspace*{-0.5cm}
\begin{enumerate}

\item 
 E. W. Kolb and M. S. Turner, \textit{The Early Universe}
(Addison-Wesley, 1990).

\item 
 D. J. Fixsen, E. S. Cheng, J. M. Gales, J. C. Mather,
R. A. Shafer, and E. L. Wright,  Astrophys. J. \textbf{473}, 576
(1996).
\item 
A. G. Cohen, A. de Rujula, and S. L. Glashow, Astro-
phys. J. \textbf{495}, 539 (1998).

\item 
 W. H. Kinney, E. W. Kolb, and M. S. Turner, Phys. Rev. Lett. \textbf{79}, 2620 (1997).

\item 
G. Steigman, Fortsch. Phys. \textbf{50}, 562 (2002).

\item 
J. Letessier and J. Rafelski, \textit{Hadrons and Quark-Gluon
Plasma} (Cambridge, 2002). 


\item  
S. Hamieh, J. Letessier, and J. Rafelski, Phys. Rev. C \textbf{62}, 064901 (2000).

\item 
R. Hagedorn and J. Rafelski, Phys. Lett. B \textbf{97}, 136 (1980).


\item 
F. Karsch, E. Laermann, and A. Peikert, Phys. Lett. B \textbf{478}, 447 (2000);\\
Nucl. Phys. B \textbf{605}, 579 (2001).

\item 
Z. Fodor, S.D. Katz, and K.K. Szabó, hep-lat/0208078,
Phys. Lett. B, in press (2003).

\item 
C. Caso et al., Eur. Phys. J. C \textbf{3}, 1 (1998).

\item 
N. K. Glendenning, \textit{Compact Stars: Nuclear Physics,
Particle Physics, and General Relativity} (Springer, 2000).
 
\item 
 Q. R. Ahmad et al., Phys. Rev. Lett. \textbf{89}, 11302 (2002).
 
\item 
G. G. Raffelt, hep-ph/0208024 (2002); and: \\
A.D. Dolgov, S.H. Hansen, S. Pastor, S.T. Petcov, G.G. Raffelt, and
D.V. Semikoz, Nucl. Phys. B632, 363 (2002).

\item 
W. H. Press, S. A. Teukolsky, W. T. Vetterling, and B. P.
Flannery, \textit{Numerical Recipes in C. The Art of Scientific
Computing} (Cambridge, 1992).

\item 
J. Ignatius, and D.J. Schwarz, Phys. Rev. Lett. \textbf{86}, 2216
(2001).

\item 
C. Greiner, P. Koch, and H. Stocker, Phys. Rev. Lett.
\textbf{58}, 1825 (1987).

\item 
C. Greiner and H. Stocker, Phys. Rev. D \textbf{44}, 3517 (1991).

\item 
E. Witten, Phys. Rev. D \textbf{30}, 272 (1984);

\item 
 B. Cheng and A. V. Olinto, Phys. Rev. D 50, 2421 (1994); and:\\
 G. Sigl, A. V. Olinto, and K. Jedamzik, Phys. Rev. D \textbf{55}, 4582 (1997).

\end{enumerate}
}
\end{mdframed}
\vskip 0.5cm 

\noindent{\bf Acknowledgements:} Some of the here reprinted arXiv\rq ed work was  supported in part by the U.S. Department of Energy, Office of Science, Office of Nuclear Physics grant DE-FG03-95ER40937.  
I thank Victoria Grossack for her support and editorial help.

\markboth{References}{Strangneess in QGP: Diaries}
\addcontentsline{toc}{section}{References}

\end{document}